\documentclass[12pt]{report}
\usepackage{graphicx,amsmath,amssymb}
\usepackage{indentfirst}
\usepackage[usenames]{color}
\usepackage[colorlinks=true, urlcolor=navyblue, linkcolor=navyblue, citecolor=navyblue]{hyperref}
\usepackage{epstopdf}
\usepackage{enumerate}
\usepackage{appendix}
\usepackage{lineno}
\usepackage{setspace}

\definecolor{navyblue}{rgb}{0,0.08,0.45}

\begin{document}


{
\begin{flushright}
{\small JLAB-PHY-16-2199 \\ \vspace{3pt}
}
\end{flushright}
\begin{flushright}
{\small SLAC--PUB--16448 \\ \vspace{3pt}
}
\end{flushright}

\vspace{40pt}

\centerline{\huge \bf The QCD Running Coupling}

\vspace{80pt}

\centerline{Alexandre~Deur}

\vspace{5pt}

\centerline {\it Thomas Jefferson National Accelerator Facility, Newport News, VA 23606, USA}

\vspace{15pt}

\centerline{Stanley J. Brodsky}

\vspace{5pt}

\centerline {\it SLAC National Accelerator Laboratory, Stanford University, Stanford, CA 94309, USA}

\vspace{15pt}

\centerline{Guy F. de T\'eramond}

\vspace{5pt}

{\centerline {\it Universidad de Costa Rica, San Jos\'e, Costa Rica}

\vspace{20pt}

{\small \centerline{\today}}

\vspace{60pt}

{\small
\centerline{\href{mailto:deurpam@jlab.org}{\tt deurpam@jlab.org},}

\vspace{1pt}

{\small
\centerline{\href{mailto:sjbth@slac.stanford.edu}{\tt sjbth@slac.stanford.edu}, \, \href{mailto:gdt@asterix.crnet.cr}{\tt gdt@asterix.crnet.cr},}
}}

 \vspace{40pt}

 \centerline{ \it (Invited review article to be published in Progress in Particle and Nuclear Physics)}

\newpage

{\centerline{ \bf \large Abstract}}

\vspace{10pt}
We review the present theoretical and empirical knowledge for $\alpha_{s}$, the fundamental  coupling underlying the  interactions of quarks and gluons in Quantum Chromodynamics (QCD).  The dependence of  $\alpha_s(Q^2)$ on momentum transfer $Q$ encodes the underlying dynamics of hadron physics --from color confinement 
in the infrared domain to asymptotic freedom at short
distances.  We review constraints 
on $\alpha_s(Q^2)$  at high $Q^2$, as predicted by 
perturbative QCD, and its analytic behavior  at small $Q^2$, based on models of nonperturbative dynamics.
In the introductory  part of this review, we explain  
the phenomenological  meaning of the coupling, the reason for its running, and
the challenges facing a complete understanding of its analytic behavior in the infrared domain. 
In the second, more technical, part of the review, we discuss the behavior of $\alpha_s(Q^2)$ 
in the high momentum transfer domain of QCD. We review how $\alpha_s$ is defined,
including its renormalization scheme dependence, the definition
of its renormalization scale,  the utility of  effective charges, as
well as  ``Commensurate Scale Relations'' which connect the various
definitions of the QCD coupling without   renormalization-scale ambiguity. 
We also report recent significant measurements and advanced theoretical analyses which have led to precise QCD predictions
at high energy. As an example of an important optimization procedure, we discuss the
``Principle of Maximum Conformality'',  which enhances
QCD's predictive power by removing the dependence
of the predictions for physical observables on the choice of theoretical conventions such as the renormalization scheme.
In the last part of the review, we discuss the challenge of understanding the analytic behavior $\alpha_s(Q^2)$ in the low momentum transfer 
domain. We  survey various theoretical models for the nonperturbative strongly coupled regime, such as the light-front 
holographic approach to QCD.
This new framework predicts the form of the  quark-confinement potential underlying  hadron spectroscopy and dynamics, and it gives a 
remarkable connection between the perturbative QCD scale $\Lambda$  and
hadron masses. One can also identify a specific scale $Q_0$ which demarcates the
division between perturbative and nonperturbative QCD.
We also  review other important methods for computing the QCD coupling, including 
lattice QCD, the Schwinger--Dyson equations
and the Gribov--Zwanziger analysis.
After describing these approaches and enumerating  their conflicting predictions, we
discuss the origin of these discrepancies
and how to remedy them. Our aim is not only to review the advances
in this difficult area, but also to suggest what could
be an optimal definition of $\alpha_s(Q^2)$  in order to
bring better unity to the subject.

\tableofcontents

\chapter{Preamble}

We review the status of the coupling $\alpha_s(Q^2)$, the function which
sets the strength of the interactions involving quarks and gluons in quantum 
chromodynamics (QCD), the fundamental gauge theory of
the strong interactions, as a function of the momentum transfer $Q$. It
is necessary to understand the behavior and magnitude of the QCD coupling
over the complete $Q^2$ range in order to describe hadronic interactions at 
both long and short distances.  At high $Q^2$ (short distances) precise
knowledge of $\alpha_s(Q^2)$ is needed to match the growing accuracy
of hadron scattering experiments as well as to test high-energy models 
unifying strong and electroweak forces. For example, uncertainties
in the value of $\alpha_s({Q^2})$ contribute  to the total theoretical
uncertainty in the physics probed at the Large Hadron Collider, such
as Higgs production via gluon fusion. It is also necessary to know
the behavior of $\alpha_{s}$ at low $Q^2$ (long distances), such as the scale of the
proton mass, in order to understand hadronic structure, quark confinement and
hadronization processes. For example, processes involving the production of 
heavy quarks near threshold require the knowledge of the QCD coupling at low   momentum scales. 
Even reactions at high energies may involve the integration of the behavior of 
the strong coupling over a large domain of momentum scales.   
Understanding the behavior of $\alpha_{s}$  at  low  $Q^2$ also 
allows   us to reach a long-sought goal of hadron physics: to establish an explicit 
relation between the long distance domain 
characterized by quark confinement, and the short-distance 
regime where perturbative calculations are feasible.
For such endeavor, a major challenge is to relate the parameter   $\Lambda$, 
which controls the predictions of perturbative QCD (pQCD) at short distances,  to the mass scale of hadrons. 
We will   show later in this review how new theoretical insights into the behavior of QCD at large distances
lead to an analytical relation between hadronic masses and  
 $\Lambda$. This problem also 
involves one of the fundamental questions of QCD:   How the mass scale for hadrons, such as the 
proton and $\rho$ meson, can emerge   in the limit of zero-quark masses since
no explicit mass scale appears in the QCD Lagrangian.

The origin and phenomenology of the behavior of $\alpha_s({Q^2})$ at small
distances, where asymptotic freedom appears,  is well understood and 
explained in many textbooks on Quantum Field Theory and
Particle Physics. Numerous reviews exist;  see   {\it e.g.}    Refs.
\cite{the:PDG 2014, the:Prosperi alpha_s review, the:Altarelli review}, some of which cover 
the long-distance behavior as well,   {\it e.g.}   Ref. \cite{the:Prosperi alpha_s review}.
However, standard explanations often create an apparent conundrum, as will be addressed  in this
review.  Other questions remain even in this well understood regime: a significant issue 
is how to identify the scale $Q$ which controls a given hadronic process, especially if it 
depends on many physical scales. A  fundamental requirement 
--called ``renormalization group invariance''-- is that physical observables cannot depend 
on the choice of the renormalization  scale. We will briefly describe a new method, 
``the Principle of Maximum Conformality''  (PMC),  which can be used to  set the 
 \emph{renormalization scale} unambiguously order-by-order in pQCD, while satisfying renormalization group invariance.    
The PMC method reduces, in the Abelian limit $N_C\to 0$, to the standard 
Gell-Mann--Low method which is used to obtain precise predictions for QED.
Although $\alpha_{s}$ is well understood at small distances, 
it is much less so  at long distances where it is related to color confinement. 
As Grunberg has emphasized \cite{the:Grunberg}, the QCD coupling can 
be defined from any physical observable which is perturbatively calculable. 
This is in analogy with QED, where the conventional Gell-Mann--Low coupling 
$\alpha(Q^2)$ is defined from the potential underlying the scattering 
amplitude of two heavy charged particles.  Couplings defined following Grunberg's 
prescription are called  ``effective charges''. They can be related to each other analytically via ``Commensurate Scale Relations''
and can be used at low $Q^2$. We will also review other 
possible definitions. An important example of  a well measured effective charge, 
is the coupling $\alpha_{g_1}(Q^2) $, which is defined from the Bjorken sum rule 
for polarized deep inelastic lepton--proton scattering.

We will start this review with a phenomenological
description of the behavior of $\alpha_s({Q^2})$.  We will then continue with a theoretical discussion of its behavior
at small distances based on  the renormalization group equations and various approaches
which can improve the standard formalism.    We conclude this first section
with the experimental and numerical measurements of $\alpha_{s}$ at short distances.
We then discuss $\alpha_{s}$ at long distances,
and survey several nonperturbative approaches which have been used to predict
its behavior. We  conclude
this section with a comparison and discussion of the various approaches. 

The level of this review is aimed for the main part at non-specialists and
advanced graduate students. In that spirit, a lexicon  specific to
the study of the QCD running coupling and related topics is given in the
Appendix. Words from this list are italicized. 

We have endeavored to be as extensive as possible in this review,  but since the study of $\alpha_{s}$ 
has been a very broad and active subject of research, the thousands
of articles relevant to $\alpha_{s}$ quickly quench any
ambition of an exhaustive review. Some topics closely related to $\alpha_{s}$,
such as studies of the gluon propagator, renormalon phenomena, and  \emph{higher-twist}
effects will only be mentioned;  each of these topic  deserves a review on its
own. Likewise, $\alpha_{s}$ at finite temperature or in the  time-like
domain will not be discussed. Our choice of topics is subjective, and we apologize in advance to the specialists
whose work has not been covered; there is no implication regarding
their importance and validity.  As usual, natural units $\hbar=c=1$
are used.


\chapter{Phenomenological overview: QCD and the behavior of $\alpha_{s}$
\label{sec:Phenomenological-introduction:}}

In this chapter, we will present an introduction to the behavior of $\alpha_{s}(Q^2)$ from the soft to  the hard domain.
A reader already familiar with this topic can skip this 
chapter;  its purpose is to initiate the curious reader into the subject. 

In Abelian quantum electrodynamics (QED), the running coupling 
$$\alpha = {\alpha(0)\over 1- \Pi(Q^2)}$$ 
modifies the bare coupling ${\overline \alpha }= {e^2\over 4\pi}$ appearing in the  QED Lagrangian by incorporating the renormalization of the photon propagator with virtuality $q^2=-Q^2$.
It sums all  vacuum polarization insertions $\Pi(Q^2)$ starting with the familiar Serber--Uehling contribution from lepton pairs.  
The logarithmic ultraviolet (UV) divergence arising from 
vacuum polarization loop integration is conventionally regulated by the Pauli--Villars (massive-photon subtraction) method. 
The UV divergence is eliminated when one normalizes the coupling by experiment at a specific momentum $Q_0$.
The UV divergences from  the vertex renormalization factor $Z_1$ and  the fermion propagator renormalization factor $Z_2$ cancel by the QED Ward--Takahashi identity. Consequently, only $Z_3$ from the virtual photon propagator requires renormalization.

In a general renormalizable quantum field theory (QFT), the coupling constant 
controlling the  strength of the interactions described by the Lagrangian acquires a
scale dependence after regularization of UV divergent integrals  
and the renormalization procedure.  Typically, one uses dimensional regularization 
to define the divergent integrals  since the procedure is gauge invariant.
The UV  cut-off dependence of the coupling is then eliminated 
by allowing the couplings and masses which appear in the Lagrangian
to acquire a scale dependence and by normalizing them to a known (measured)
value at a given scale. This ``renormalization''  procedure defines the running  couplings
(and running masses); in effect, they are  \emph{effective couplings. } 

The origin of the ultraviolet divergences is often interpreted as a 
manifestation that a QFT is a low energy effective theory of more fundamental
yet unknown theory. The UV cut-offs shield the  very short distance domain where the QFT perhaps ceases to be valid. 
After normalizing the coupling to a measured value, the \emph{effective coupling}  is not
sensitive to the  ultraviolet (UV) cut-off and unknown phenomena arising beyond this scale.   Thus
the scale dependence of the coupling can be well understood formally
and phenomenologically.

We illustrate this  behavior for the coupling that arises in the static case of heavy sources and which provides a simple physical 
picture. Historically, and in the case of linear theories with massless
force carriers, a force coupling constant is a universal coefficient
that links the force to the {}``charges'' of two bodies ({\it e.g.}, the
electric charge for electricity or the mass for gravity) divided by
the distance dependence $1/r^{2}$. The $1/r^{2}$ dependence was
classically interpreted as the weakening of the force flux as it spreads
uniformly through space. In QFT this is interpreted as the manifestation
in coordinate space of the propagator of the force mediator; the propagator
is proportional to $1/q^{2}$ in momentum space in the 
first Born approximation ({\it i.e.}, one boson exchange, with $q$ the 4-momentum of
the exchanged boson). For weak enough forces, the first Born approximation
dominates higher order contributions and the $1/q^2$ propagator in momentum yields 
the familiar $1/r^2$ factor in coordinate space. However, higher orders do contribute
and deviations from the $1/r^{2}$ law thus occur. This extra $r$-dependence
is folded in the coupling which then acquires a scale dependence.

In QED, the 
contributions to the $Q^2$-dependence of the coupling 
only come from the vacuum polarization graphs.  In QCD,   the vacuum polarization (Fig. \ref{Flo:QCD anti-screening}),
the quark self-energy (Fig. \ref{QCD_vertex}(a),   the vertex corrections, 
and  the gluon loop corrections to the elementary three-gluon and four-gluon couplings,
(Fig. \ref{QCD_vertex}(b) all contribute. (The separation
of the amplitude in different graphs depends on convention. Hence,
the statement of which graphs contribute has some arbitrariness.
For example, the vertex corrections, Fig. \ref{QCD_vertex}(b),
do not contribute in the Landau gauge, $\partial_\mu A^{\mu} =0$.) Tadpole graphs (Fig. \ref{QCD_vertex}(c)
do not contribute. Virtual emissions, where the gluon is reabsorbed
by the rest of the hadron 
do not contribute to the coupling renormalization if they are UV finite.
Since higher order quantum effects are
responsible for the deviation from the $1/r^{2}$ law, we expect the
scale dependence to be significant only at   a microscopic scale. Superficially, this discussion 
based on a perturbative framework seems
inapplicable to QCD, since it is a confining theory irrelevant to macroscopic
scales.  However, the non-Abelian nature
of QCD induces a weakly-coupled QCD regime at small distances in which
the first  Born approximation is relevant and where
the above discussion applies.

\begin{figure}[ht]
\centering
\includegraphics[width=12.0cm]{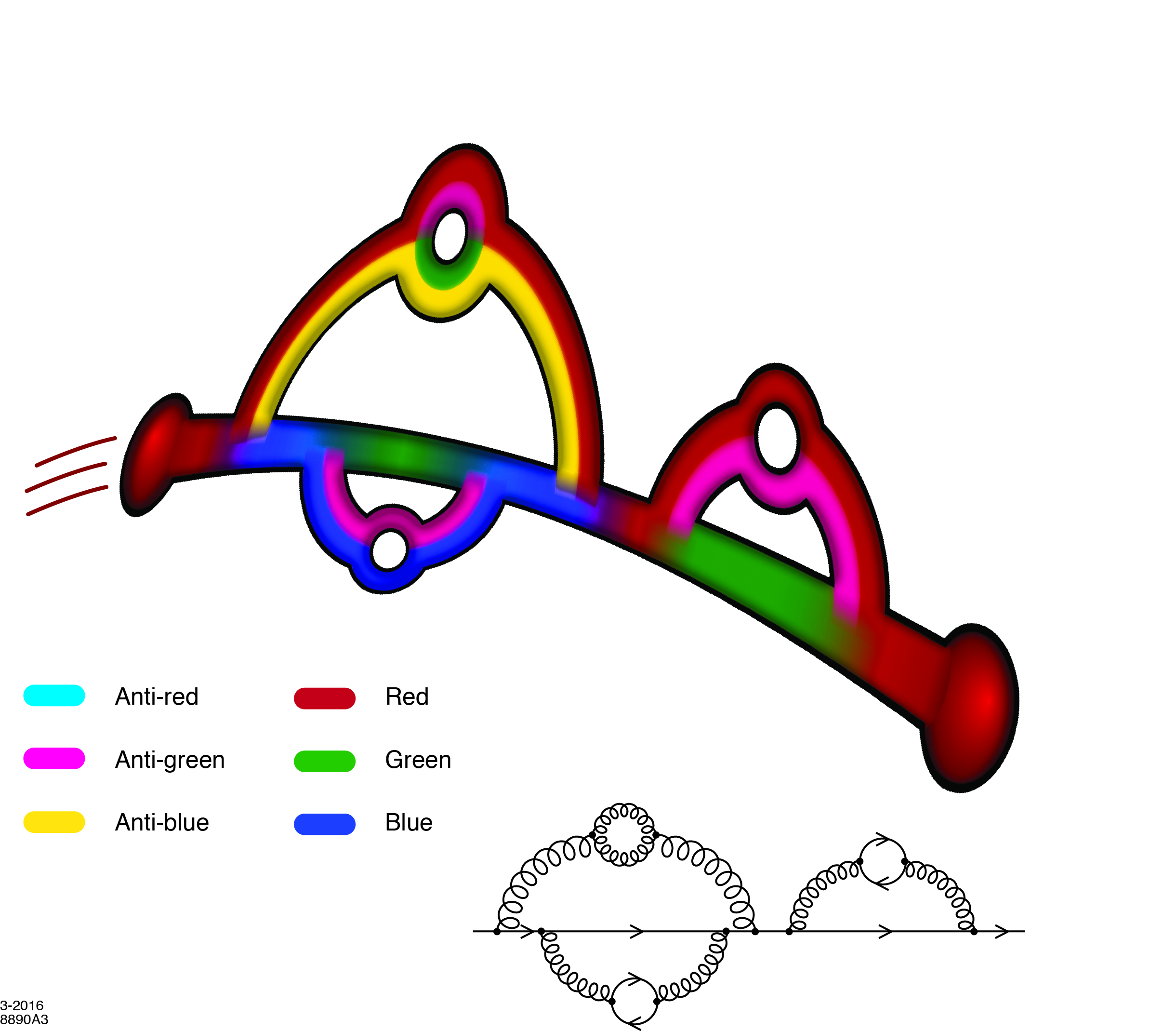}
\caption{\label{Flo:QCD anti-screening} \small  
{\it Color spread and the physics of asymptotic freedom and confinement}.  In contrast with QED the force carriers of  QCD, the gluons, carry color charges which leads to spread of color.  The dilution of the initial color charge  is responsible for the weakening of the QCD coupling $\alpha_s$ at small distances; {\it i. e.}, when the quark experiences a large momentum transfer.   Since there is no intrinsic length scale, values of $\alpha_s$ at different momentum transfer scales are  related by a logarithmic function: $\alpha_s$ decreases continuously with increasing momentum transfer (shorter distances) leading to asymptotic freedom.  Conversely $\alpha_s$ increases logarithmically at larger distances. However, this growth with large distance cannot continue indefinitely since the proton has a finite size: color is confined. Confinement implies that  long wavelengths of quarks and gluons are cut off at a typical hadronic size. Consequently, the effects of quantum loops responsible for the logarithmic dependence of $\alpha_s$ disappear  and $\alpha_s$ should  freeze to a constant value at hadronic scales~\cite{Brodsky:2007hb, the:Brodsky & Shrock}. The evolution of a  quark's color is illustrated in the figure. Here, an initially red quark converts to either a green or blue quark when its red color charge is carried away by the emission of a gluon.  
The anti-red color is symbolized as cyan, anti-green  as magenta and anti-blue as yellow. In the accompanying diagram, the straight line represents a color-triplet quark (or an anti-quark), and a curved line represents a color-octet gluon carrying both color and anti-color.}
\end{figure}

\begin{figure}[ht]
\centering
\includegraphics[width=8.0cm]{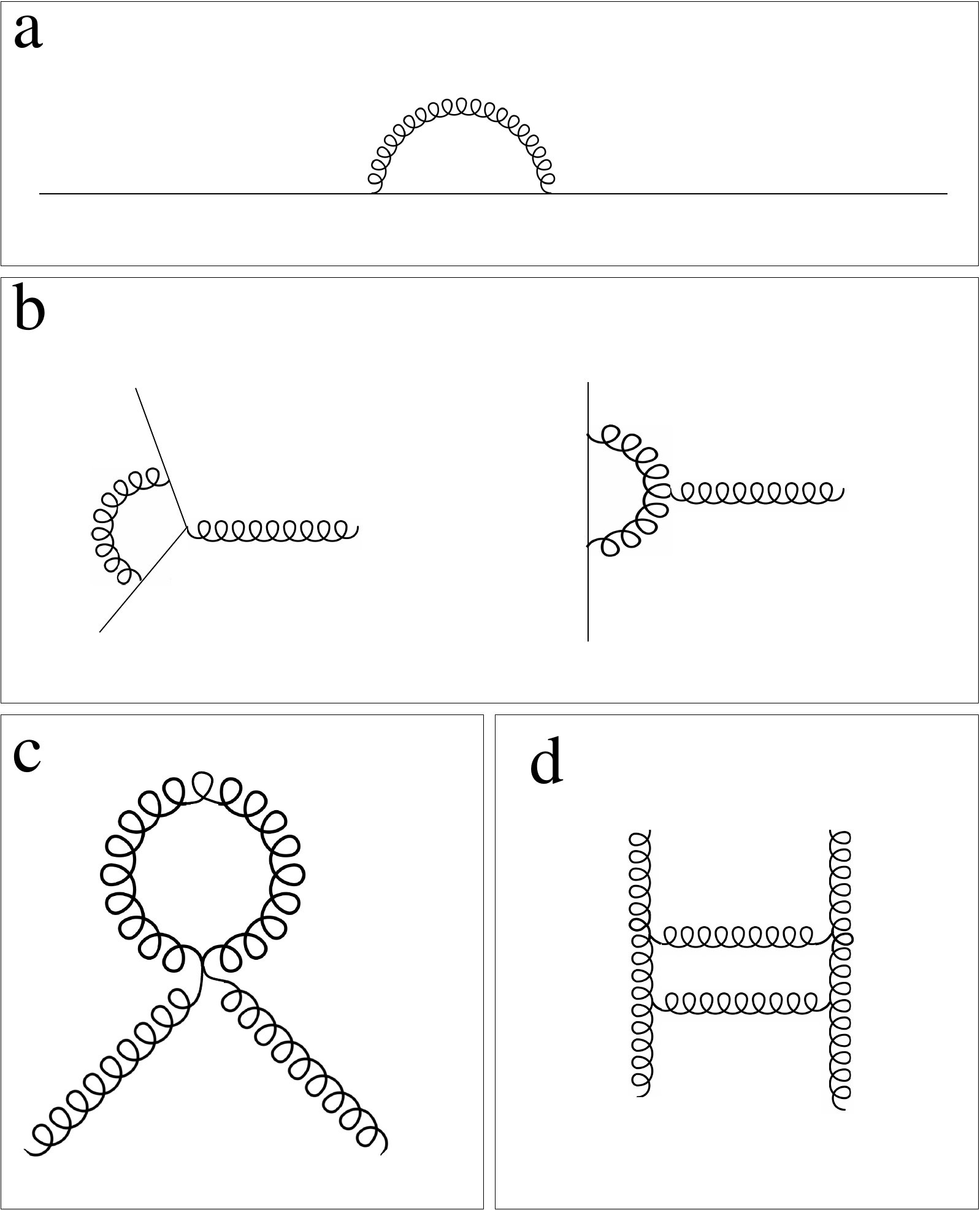}
\caption{\label{QCD_vertex}\small  
 Panel (a): virtual gluon correction for the quark propagator.
Panel (b) vertex corrections.
Panel (c): Tadpole graph. Panel (d): H (or ladder) graph.}
\end{figure}

Having linked the deviation of the force from the $1/r^{2}$ law to
vacuum polarization and other processes, we can go further and explain
the direction toward which a force deviates from $1/r^{2}$. In QED
pairs of particle-antiparticle are created around
a test charge $q_c$. The particles of charge opposite to $q_c$  will tend to
be closer to the test charge, see Fig. \ref{Flo:QED screening}(a).
This is analogous to electric charge screening in a dielectric
medium, see Fig. \ref{Flo:QED screening}(b). In both cases, by Gauss'
law, the total charge inside a sphere of radius $r$ is smaller. The
larger $r$, the more the total charge will tend to the test charge value $q_c$. From this,
one sees that the magnitude of the running coupling in QED will decrease
at larger distances, tending to its macroscopic value $\alpha(0) \simeq 1/137$,
 and get stronger at small distances.
For QCD the situation is different because gluons carry color charges.
The gluon linking the test color charge to a particle-antiparticle
pair will have carried away the initial color of the test charge,
see Fig. \ref{Flo:QCD anti-screening}. Thus, the gluons tend to
spatially dilute the initial test charge: the initial red quark in
Fig. \ref{Flo:QCD anti-screening}(a) spends most of its time as a green
or blue quark, thus invisible to any high resolution gluons carrying
an anti-red color that would normally interact with it. As a result,
the opposite effect to QED happens (anti-screening). Since this effect
dominates over the QED-like screening effect from quark--antiquark
loops, the QCD running coupling tends to decrease with small distances.
This gives physical insight into  the phenomenon of ``asymptotic freedom'' 
({}``ultraviolet'' UV regime) at high momentum transfer and
to a strong coupling regime in the low momentum transfer regime ({}``infrared''
IR regime). 

\begin{figure}[ht]
\centering
\includegraphics[width=9.0cm]{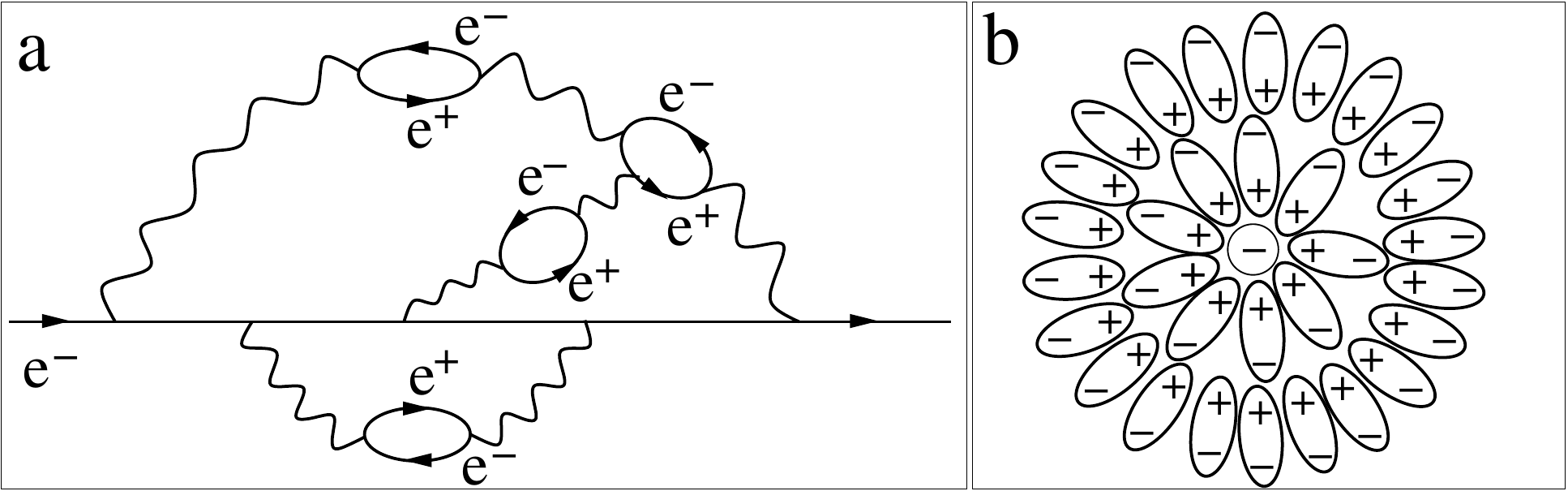}
\caption{\label{Flo:QED screening} \small Panel (a): screening of the QED charge.  Panel (b): analogy
to the charge screening occurring in a dielectric material.}
\end{figure}

Formally, the running of a coupling  originates from
the renormalization procedure. However, predictions for observables must be independent of the choice of renormalization
scheme (RS). This invariance under  the choice of procedures
forms a symmetry group, which allows techniques from group theory
to be used to establish the behavior of $\alpha_{s}$.  However, $\alpha_{s}$ itself
 needs not to be an observable, apart from particular cases such as the
measure of the static QCD potential.  Usually $\alpha_{s}$ is an expansion
parameter in the perturbative series describing an observable $R$:
Typically 

\begin{equation}
R=\sum_{n}r_{n}\left(\frac{\alpha_{s}}{\pi}\right)^{n}.
\end{equation}

Although $R$
is RS-independent, the series coefficients, and thus $\alpha_{s}$,
will depend on the scheme choice. However, because of the asymptotic freedom, $R\rightarrow r_{0}$
when the distance scale at which the observable is measured tends
to zero. Thus the first coefficient of the series must be RS-independent.
Formally at very high scales, the RS-dependence becomes minimal. Conversely,
one anticipates that the dependence becomes increasingly important
as larger distances are probed. 

As we shall discuss in  Section \ref{ss:PMC}, one can use an analytic procedure, the Principle of Maximum Conformality, to 
make predictions for physical observables which are independent of the choice of RS.

The fact that  $\alpha_{s}$ is not in general an observable
and that it becomes increasingly dependent on the RS as the distance
scale increases are important facts that must be kept in mind, especially
since these effects in the prototypical quantum field theory --QED--
are small or irrelevant. In general, the QCD \emph{effective coupling} is
considered to be only an intermediate quantity without precise physical
meaning; if the coupling varies, {\it e.g. } by a factor of three depending
on the RS choice, how can it be associated to the physical strength
of the strong force? The advantages of viewing $\alpha_{s}$ as an
intermediate non-physical quantity are that unintuitive features such
as RS-dependence or divergences (\emph{Landau pole}) can be ignored,
as long as observables are studied at small enough distances. However,
this causes a conundrum: if $\alpha_{s}$ is not an observable and
depends on arbitrary choices, how relevant is the phenomenological
description given previously? This will be explained in Section \ref{sub:Effective charges and CSR}
where we will see how $\alpha_{s}$ can be restored to the status
of an observable. Advantages of such approach are that $\alpha_{s}$
re-acquires a physical and intuitive meaning, and that $\alpha_{s}$
can be defined at long distances. 

So far, the discussion applies to 
the short-distance, weak-coupling,
regime of QCD, where pQCD is applicable. What happens
at large distances? 
Perturbative QCD suggests that at the confinement scale $\Lambda$,
$\alpha_{s}$ becomes infinite, the so-called Landau  pole. 
In the past, this behavior was viewed 
responsible for inducing the so-called {}``IR slavery''
which keeps  the quarks confined. In other words, quark confinement has
been seen as a direct consequence of the \emph{Landau pole}. However, the modern
view is that, while UV freedom thrives, IR slavery should be abolished:
the pQCD prediction of the \emph{Landau pole} is irrelevant to the physical
behavior of $\alpha_{s}$ since pQCD breaks down when $\alpha_{s}/\pi$
becomes close to unity. It may be useful to remember here that the coupling
may not need to become infinite for a theory to be confining. For example
the coupling is finite in QCD(1+1) which is a confining theory.

The subject of the IR-behavior of $\alpha_{s}$
is interesting and difficult. It will be discussed in the second part
of this review. Instead of displaying a Landau pole, we expect (with the insight of numerous studies)
that the strong increase of $\alpha_{s}$ should stop  at some infrared scale,  since its running
stems from the colored particle/antiparticle loop creations, but the
wavelengths of those particles cannot exceed the physical hadron size
\cite{the:Brodsky & Shrock}. Hence, loops become suppressed in the
 infrared (IR) limit and $\alpha_{s}$ should become scale independent. This IR
behavior is called the ``\emph{freezing}  of $\alpha_{s}$''. The following
simple example illustrates the naive misconception that the divergence
of the perturbative coupling at the \emph{Landau pole} is responsible for
quark confinement: the static $\mbox{Q--}\overline{\mbox{Q}}$ potential displays a linear
dependence  up to a distance of 1.3 fm in the physical case,
or larger distances for \emph{pure field} theory calculations without light quarks,
see Section \ref{sub:Potential-approach}. However, as will be
discussed in Section \ref{sec:alpha_s in the IR}, the same theoretical calculations
indicate that $\alpha_{s}$ becomes independent of the distance above
0.4 fm. It thus cannot be directly responsible for the continuing
linear rise of the potential which keeps quarks and gluons confined. Indeed, 
rather than vacuum polarization loops increasing the coupling characterizing
a one-gluon exchange, multi-gluon exchange between quarks, such as the 
H-graphs (see Fig. \ref{QCD_vertex}(d) which are infrared  divergent~\cite{the:Appelquist, the:gluon H graph divergence}, 
seems to be relevant to confinement.

To summarize this phenomenological description, $\alpha_{s}$ behaves as follows: 
the coupling exhibits weak (inverse  logarithmic) scale dependence at distances
much smaller than  $\Lambda^{-1}$, typically smaller than $10^{-16}$~m. 
This is due to color charge spreading (asymptotic freedom properties).
This weak behavior grows into a strong scale dependence at distances larger than 
a  tenth of  a Fermi. Finally,  scale independence is restored
at larger distances due to the confinement of  quarks and gluons. The fact
that there  is a region where the strong coupling $\alpha_{s}$ has minimal scale dependence (\emph{conformal} window) allows the 
 use of methods based on the gauge-gravity duality \cite{the:Maldacena duality} to hadronic physics.
 Its application to  nonperturbative QCD calculations will
be discussed in Sections \ref{sub:Holographic-QCD large Q} and \ref{sub:Holographic-QCD Lowq}.


\chapter{The strong coupling $\alpha_{s}$ in the perturbative domain}

In this chapter, 
we  give a more formal
discussion of the strong coupling $\alpha_{s}$ in the perturbative domain.
Although the content of this chapter
is well explained in textbooks and reviews, we will give a brief review 
in order to define notations  as well as to set the basis for a proper understanding of
the behavior of $\alpha_{s}$ in the IR domain, which we postpone until 
chapter  \ref{sec:alpha_s in the IR}.

We first recall the equations ruling  the behavior of $\alpha_{s}$ in the perturbative domain. Then we discuss various techniques
which can be used to optimize the perturbative series in order to improve
pQCD's predictive power. Finally, we summarize the current experiment/Lattice
QCD status on $\alpha_{s}$. The momentum transfer scale at which $\alpha_{s}$
is  determined is $Q^{2}=-q^{2}$, with $q$ the four-momentum flow
of the process. We have $Q^{2}>0$ for space-like processes.
One can work in a given RS, such as  the  one following from dimensional 
regularization~\cite{Bollini:1972ui}: The minimal-subtraction (MS)  renormalization scheme \cite{the:msbar scheme} or  the 
$\overline {MS}$ RS~\cite{the: bardeen 1978}. In the  $\overline {MS}$ RS one 
subtracts ${2\over\epsilon} + \log 4\pi - \gamma_E$ (where $\gamma_E$ is the 
Euler constant)  after regularizing UV divergent integrals in $4- \epsilon$ dimensions. In 
the MS and $\overline {MS}$ renormalization schemes 
$\alpha_{s}$ is gauge invariant. It is  convenient to choose to  
work in the Landau gauge, $\partial_{\mu}A^{\mu}=0$. Thus, the gauge 
parameter $\xi$, see {\it e.g.}, Eq. (\ref{eq:gluon loop}), is fixed at $\xi=0$ in 
this section.

\section{Purpose of the running coupling\label{Purpose of the running coupling}}

We shall assume that quark masses can be neglected at the typical  scale
$Q^{2}$, the value
characterizing a given reaction. The QCD Lagrangian density then contains
no mass or energy scale and is classically \emph{conformally} invariant:
\begin{equation}
\mathcal{L}=i\overline{\psi}\gamma^{\mu}D_{\mu}\psi-\frac{1}{4}G_{\mu\nu}^{a}G_{a}^{\mu\nu},\label{eq:QCD Lagrangian}
\end{equation}
where $\psi$ is the quark field, $G_{\mu\nu}^{a}$ the gluon field
strength and $a$ the color index. The gluon field strength is given
by $G_{\mu\nu}^{a}\equiv\partial_{\mu}A_{\nu}^{a}-\partial_{\nu}A_{\mu}^{a}+gf^{abc}A_{\mu}^{b}A_{\nu}^{c}$,
with $A_{\mu}^{a}$ the gluon field, $f^{abc}$ the SU(3) structure
constant and $g$  the dimensionless coupling constant.
The bare coupling 
is $\overline{\alpha_{s}}=g^{2}/4\pi$. Although there are
no dimensionful parameters in Eq. (\ref{eq:QCD Lagrangian}), a mass
scale $\mu$ is acquired during the renormalization procedure. The
emergence of $\mu$ from a Lagrangian without explicit scale is called
\emph{dimensional transmutation} \cite{the: dim. trans.}. The value of $\mu$ is arbitrary
and is the momentum  at which the UV divergences are subtracted. Hence
$\mu$ is called the \emph{subtraction point} or \emph{renormalization
scale. } A dimensionless observable $R(Q^{2},x_f )$, where $x_f $ represents
any dimensionless kinematic variables, must be independent of the
arbitrary value of $\mu$. The purpose of making $\alpha_{s}$ scale-dependent
is to transfer to $\alpha_{s}$ all terms involving $\mu$ in the
perturbative series of $R(Q^{2},x_f )$.

In the perturbative domain, observables are expressed in perturbative 
expansions of $\overline{\alpha_{s}}$: 
\begin{equation}
R\left(Q^{2},\mu^{2},\overline{\alpha_{s}},x_f\right)=\sum_{n=0}^{n_{max}\sim\pi/\alpha_{s}}r_{n}\left(Q^{2},\mu^{2},x_f \right)  
 \left(\frac{\overline{\alpha_s}}{\pi}\right)^n.
\label{eq:pert. exp of an observable with coupling CONSTANT}
\end{equation}
Since $R$ is dimensionless and since there is no mass scale in the
QCD Lagrangian, the $Q^{2}$-dependence of $\alpha_{s}$ can only
be a function of the $Q^{2}/\mu^{2}$ ratio. Except for $r_{0}$ and
$r_{1}$ which are independent of $Q^{2}/\mu^{2}$, the coefficients
$r_{n}$ are polynomials of $\mbox{ln}\left(Q^{2}/\mu^{2}\right)$ with highest
power $n-1$. The independence of $R$ with  respect to $\mu$ is given by the
Callan--Symanzik relation for QCD \cite{the: beta-function}: 
\begin{eqnarray}
\label{eq:Callan-Symanzik}
\frac{d}{d~\mbox{ln}\left(\mu^{2}\right)}R\left(Q^{2}/\mu^{2},\alpha_{s},x_f \right)=\mu^{2}\frac{d}{d\mu^{2}}R\left(Q^{2}/\mu^{2},\alpha_{s},x_f \right)=  \\
\left(\mu^{2}\frac{\partial}{\partial\mu^{2}}+\mu^{2}\frac{\partial\alpha_{s}}{\partial\mu^{2}}\frac{\partial}{\partial\alpha_{s}}\right)R\left(Q^{2}/\mu^{2},\alpha_{s},x_f \right)=0. \nonumber
\end{eqnarray}
 The \emph{$\beta$-function} is defined from Eq. (\ref{eq:Callan-Symanzik})
as $\mu^{2}\frac{\partial\alpha_{s}}{\partial\mu^{2}}=\beta(\alpha_{s})$.
Setting $t=\mbox{ln}\left(Q^{2}/\mu^{2}\right)$, Eq. (\ref{eq:Callan-Symanzik})
becomes 
\begin{equation}
\left(-\frac{\partial}{\partial t}+\beta\frac{\partial}{\partial\alpha_{s}}\right)R\left(e^{t},\alpha_{s},x_f \right)=0.\label{eq:Callan-Symanzik2}\end{equation}
We remark that if we had not yet fixed the gauge fixing parameter to $\xi=0$, Eqs. (\ref{eq:Callan-Symanzik}) and (\ref{eq:Callan-Symanzik2}) would also have a term stemming from the variation of  $\xi$.  Expressing the observable as a perturbative series now in $\alpha_{s}(\mu^{2})$
defined by $\alpha_{s}(\mu^{2})\equiv\alpha_{s}$ and $t=\int_{\alpha_{s}}^{\alpha_{s}(Q^{2})}\frac{1}{\beta(y)}dy$,
one has: 
\begin{equation}
R\left(e^{t},\alpha_{s},x_f\right)=\sum_{n=0}^{n_{max}\sim\pi/\alpha_{s}}r_{n}\left(1,\alpha_{s},x_f \right)
\left( \frac{\alpha_s\left(\mu^2\right)}{\pi}\right)^n.
\label{eq:pert exp of an observable with eff. alpha_s}
\end{equation}
One can check that it is a solution of Eq. (\ref{eq:Callan-Symanzik2}).
Setting $e^{t}=1$, {\it i.e.}, setting the physical scale $Q$ equal to the 
\emph{renormalization scale} $\mu$, $Q^{2}=\mu^{2}$, makes the coefficients
$r_{n}$ independent of $Q^{2}/\mu^{2}$; the $\mu$-dependence has
been folded into $\alpha_{s}(\mu^{2})$. Choosing $Q^{2}=\mu^{2}$
thus yields the simplest form for the perturbative expansions of given
observables. There are however disadvantages to this choice, such  as an unintuitive
interpretation of the coupling and convergence issues with the series
(\emph{renormalon problem}). Other choices, while leading to more
complex perturbative expressions, can be more physically motivated.
This will be discussed in Sections \ref{sub:Effective charges and CSR}
and \ref{sub:Other optimization procedures}. Another choice would
be similar to a different choice of RS, see Section \ref{sub:Effective charges and CSR}
and Fig. \ref{Flo:strong-coupling in diff schemes}. One generally 
keeps $Q^{2}\sim\mu^{2}$ so that the  coefficients $r_{n}$ remain small
enough for the perturbative expansion of $R\left(t,\alpha_{s},x_f \right)$
to be valid.

\section{The evolution of $\alpha_{s}$ in perturbative QCD \label{sub:pQCD evolution-equation}}

The scale dependence of the strong coupling  is controlled by the
\emph{$\beta-$}function which can be expressed in the UV as a perturbative
series:
\begin{eqnarray}
Q^{2}\frac{\partial}{\partial Q^{2}} \frac{\alpha_{s}}{4\pi} & =\beta\left(\alpha_{s}\right)=-\left(\frac{\alpha_{s}}{4\pi}\right)^{2}\sum_{n=0}\left(\frac{\alpha_{s}}{4\pi}\right)^{n}\beta_{n}.\label{eq:alpha_s beta series}
\end{eqnarray}
The values of the first terms of the $\beta$-series are~\cite{the:beta_0 calculation}:
\begin{equation}
\beta_{0}=11-\frac{2}{3}n_{f},
\end{equation} 
and~\cite{the:Caswell beta_1 calculation -1, the:beta_1 calculation}
\begin{equation}
\beta_{1}=102-\frac{38}{3}n_{f},
\label{eq:beta_1}
\end{equation} 
with $n_{f}$ the number of quark
flavors active at the scale $Q^{2}$. These terms are RS-independent
(as long as quark masses are neglected, see Section \ref{sub:Quark-thresholds})
because of the UV renormalizability of QCD  --only the first two loops
are dominated by small distance processes: the 1 and 2-loops integrals
diverge in the $Q^{2}\rightarrow\infty$ limit, while higher loops
are finite in this limit. The higher order $\beta$-terms are known
but are RS-dependent. Since they also encompass  nonperturbative
 contributions not included in the perturbative series  Eq. (\ref{eq:alpha_s beta series}),
there is  an inherent ambiguity for these terms. In schemes based on   the dimensional 
regularization method   the higher order terms in the minimal-subtraction scheme 
$\overline{MS}$  (the most common RS used ~\cite{the:msbar scheme, the: bardeen 1978}) are~\cite{the:beta_2 calculation MSbar}
\begin{equation}
\beta_{2}=\frac{2857}{2}-\frac{5033}{18}n_{f}+\frac{325}{54}n_{f}^{2},
\end{equation}
and~\cite{the:beta_3 calculation MSbar}
\begin{multline}
\beta_{3}=\left(\frac{149753}{6}+3564\xi\left(3\right)\right)-\left(\frac{1078361}{162}+\frac{6508}{27}\xi\left(3\right)\right)n_{f}   \\
+ \left(\frac{50065}{162}+ \frac{6472}{81}\xi\left(3\right)\right)n_{f}^{2}  +\frac{1093}{729}n_{f}^{3},
\end{multline}
with $\xi\left(3\right)$ the Ap\'{e}ry constant, $\xi\left(3\right)\simeq1.20206$. 
Calculations are ongoing for $\beta_4$ \cite{the:beta_4 calculation MSbar}.
All the $\beta_{i}$ coefficients are gauge independent in the $\overline{MS}$
scheme~\cite{the:Caswell-Wilczek (1974)}. In the  momentum space subtraction (MOM) scheme 
\cite{the:Comparison alpha-g-gh alpha-3g alpha-4g, the: MOM} and Landau gauge, these
couplings are~\cite{the:beta_MOM}:
\begin{equation}
\beta_{2}=3040.48-625.387n_{f}+19.3833n_{f}^{2},
\end{equation}
and 
\begin{equation}
\beta_{3}=100541-24423.3n_{f}+1625.4n_{f}^{2}-27.493n_{f}^{3}.
\end{equation}  
They can be found for the minimal MOM scheme and Landau gauge in \cite{the:betas in mini MOM.}, as well as $\beta_1$ which is also gauge dependent in this scheme and different from Eq.~\ref{eq:beta_1}.
(The MOM scheme renormalization condition forces the virtual quark propagator to a free
 massless propagator form. Several MOM schemes exist and the above values of $\beta_2$ and $\beta_3$ are determined with the MOM scheme defined by subtracting the 3-gluon vertex to a point where one external momentum vanishes.) 
 A complication is that the coupling in the MOM scheme is gauge-dependent. The values of $\beta_2$ and $\beta_3$ given here are only valid in the Landau gauge. Expressions in the V-scheme can be found in ref.~\cite{Kataev:2015yha}. They are: $\beta_2 = 4224.18 - 746.01 n_f +20.88 n_f^2$ and $\beta_3 = 43175 - 12952 n_f +707.0 n_f^2$
Some important points must be made: first the $\beta_{i}$ are independent
of $\alpha_{s}$; they are expansions in orders of $\hbar$, see Section
\ref{sub:beta-coefficients calc.}. Second, the signs of the $\beta_{i}$
control 
 how $\alpha_{s}$ run. We have $\beta_{0}>0$ for $n_{f}\leq16$,
$\beta_{1}>0$ for $n_{f}\leq8$, $\beta_{2}>0$ for $n_{f}\leq5$,
and $\beta_{3}$ is always positive. Consequently, $\alpha_{s}$ decreases
at high momentum transfer, leading to the quark asymptotic freedom of pQCD. Finally,
the $\beta_{i}$ are sometimes defined with an additional multiplying
factor $1/\left(4\pi\right)^{i+1}$. 

The exact analytical solution to Eq. (\ref{eq:alpha_s beta series})
is known only to $\beta_{0}$ order. At  this order, Eq. (\ref{eq:alpha_s beta series})
is:
\begin{equation}
\frac{Q^{2}}{\alpha_{s}^{2}}\frac{\partial\alpha_{s}}{\partial Q^{2}}=-\frac{1}{4\pi}\beta_{0},
\end{equation}
that is
\begin{equation} \label{dEalpha}
-\frac{4\pi d\alpha_{s}}{\beta_{0}\alpha_{s}^{2}}=\frac{dQ^{2}}{Q^{2}}.
\end{equation}
Integrating (\ref{dEalpha}) between $Q^{2}$ and $\mu_{0}^{2}$ yields:
\begin{equation}
\frac{4\pi}{\alpha_{s}\left(\mu_{0}^{2}\right)}-\frac{4\pi}{\alpha_{s}\left(Q^{2}\right)}=\beta_{0}\mbox{ln}\left(\frac{\mu_{0}^{2}}{Q^{2}}\right).\label{eq:beta0 sol, 3}
\end{equation}
Eq. (\ref{eq:beta0 sol, 3}) expresses the general rule that pure
pQCD calculations only provide the evolution of an observable relative
to the value of this observable given at an arbitrary scale, here
 $\mu_0$. To convey this more conveniently, the QCD \emph{scale parameter}
 $\Lambda$ is introduced. At $\beta_{0}$ order, it is defined as:
\begin{equation}
\Lambda^{2} \equiv\mu^{2}e^{-\frac{4\pi}{\beta_{0}\alpha_{s}\left(\mu^{2}\right)}},
\end{equation}
which yields the familiar 1-\emph{loop} solution:
\begin{equation} \label{Eq.one loop}
\alpha_{s}(Q^{2})=\frac{4\pi}{\beta_{0}\mbox{ln}\left(Q^{2}/ \Lambda^{2}\right)}.
\end{equation}
We remark that the  scale parameter $\Lambda$ is RS-dependent,
see Table (\ref{Flo:Table of Lambda}). Consequently, although the expressions
of $\alpha_{s}$ at order $\beta_{0}$ or $\beta_{1}$ are universal,
their numerical values would still depend on the choice of the RS 
through $\Lambda$, unless the observable approximants in which $\alpha_s$
is used are also limited to their RS-independent orders.
The relation between $\Lambda_{1}$ in a scheme $1$ and $\Lambda_{2}$
in a scheme $2$ is, at the one-\emph{loop} order:
\begin{equation}
\Lambda_{2}=\Lambda_{1}e^{\frac{2v_{1}}{\beta_{0}}},
\end{equation}
where $v_{1}$ is the leading order difference between the $\alpha_{s}(Q^{2})$
in the two RS: $\alpha_{s}^{(2)}(Q^{2})=\alpha_{s}^{(1)}(Q^{2})[1+v_{1}\alpha_{s}^{(1)}(Q^{2})/\pi]$.
For example, the $\overline{MS}$ and MOM  scale parameters are related
by:
\begin{equation}
\Lambda_{MOM}=\Lambda_{\overline{MS}}e^{\frac{507-40n_{f}}{792-32n_{f}/3}},
\end{equation}
so that $\Lambda_{MOM}=1.847\Lambda_{\overline{MS}}$ for $n_{f}=2$,
$\Lambda_{MOM}=1.817\Lambda_{\overline{MS}}$ for $n_{f}=3$ and $\Lambda_{MOM}=1.783\Lambda_{\overline{MS}}$
for $n_{f}=4$. 

The value of $\Lambda$ is also dependent on the number
of active quark flavors, see Section \ref{sub:Quark-thresholds}. 

An exact analytical relation corresponding to Eq. (\ref{eq:beta0 sol, 3})
also exists at order $\beta_{1}$:
\begin{equation}
\frac{4\pi}{\alpha_{s}\left(Q^{2}\right)}-\frac{\beta_{1}}{\beta_{0}}\mbox{ln}\left(\frac{4\pi}{\alpha_{s}\left(Q^{2}\right)}+\frac{\beta_{1}}{\beta_{0}}\right)=\frac{4\pi}{\alpha_{s}\left(\mu^{2}\right)}-\frac{\beta_{1}}{\beta_{0}}\mbox{ln}\left(\frac{4\pi}{\alpha_{s}\left(\mu^{2}\right)}+\frac{\beta_{1}}{\beta_{0}}\right)+\beta_{0}\mbox{ln}\left(\frac{Q^{2}}{\mu^{2}}\right),
\end{equation}
and an exact, but non-analytical, $\alpha_{s}$ solution is given
by \cite{the: alpha_s exact beta_1}:
\begin{equation}
\alpha_{s}(Q^{2})=-\frac{4\pi\beta_{0}}{\beta_{1}}\frac{1}{1+W_{-1}\left(z\right)},\label{eq:GGK beta1}
\end{equation}
with $z\equiv-\frac{\beta_{0}}{e\beta_{1}}\left(\frac{\Lambda^{2}}{Q^{2}}\right)^{\beta_{0}^{2}/\beta_{1}}$
and $W_{-1}\left(z\right)$ is the lower branch of the real-valued
Lambert function solution of the equation $z=W(z)e^{W(z)}$. An accurate
approximation of $W_{-1}$ can be found in \cite{the:Lambert function aprox.}.
 The scale parameter at $\beta_{1}$ order is approximately (see Eq.
(\ref{eq:alpha_s}) truncated at $\beta_{1}$):
\begin{equation}
 \Lambda^2=Q^{2}\left(\frac{4\pi}{\beta_{0}\alpha_{s}\left(Q^{2}\right)}+\frac{\beta_{1}}{4\pi\beta_{0}}\right)^{\frac{\beta_{1}}{4\pi\beta_{0}}}e^{-\frac{4\pi}{\beta_{0}\alpha_{s}\left(Q^{2}\right)}}.
\end{equation}

Extending this approach to $\beta_{2}$ is difficult but the authors
of \cite{the: alpha_s exact beta_1} provided an approximate solution:
\begin{equation}
\alpha_{s}(Q^{2})=-\frac{4\pi\beta_{0}}{\beta_{1}}\frac{1}{1-\frac{\beta_{2}\beta_{0}}{\beta_{1}^{2}}W_{-1}\left(z\right)},\label{eq:GGK beta2}
\end{equation}
with $z\equiv-\frac{\beta_{0}}{\beta_{1}}\left(\frac{\Lambda^{2}}{Q^{2}}\right)^{\beta_{0}^{2}/\beta_{1}}e^{\frac{\beta_{2}\beta_{0}}{\beta_{1}^{2}}-1}$.
The exact and approximate $\alpha_{s}$ at order $\beta_{1}$ and
$\beta_{2}$, respectively, are shown in Fig. \ref{Flo:alpha_s from pQCD}.

\begin{figure}[ht]
\centering
\includegraphics[width=9.0cm]{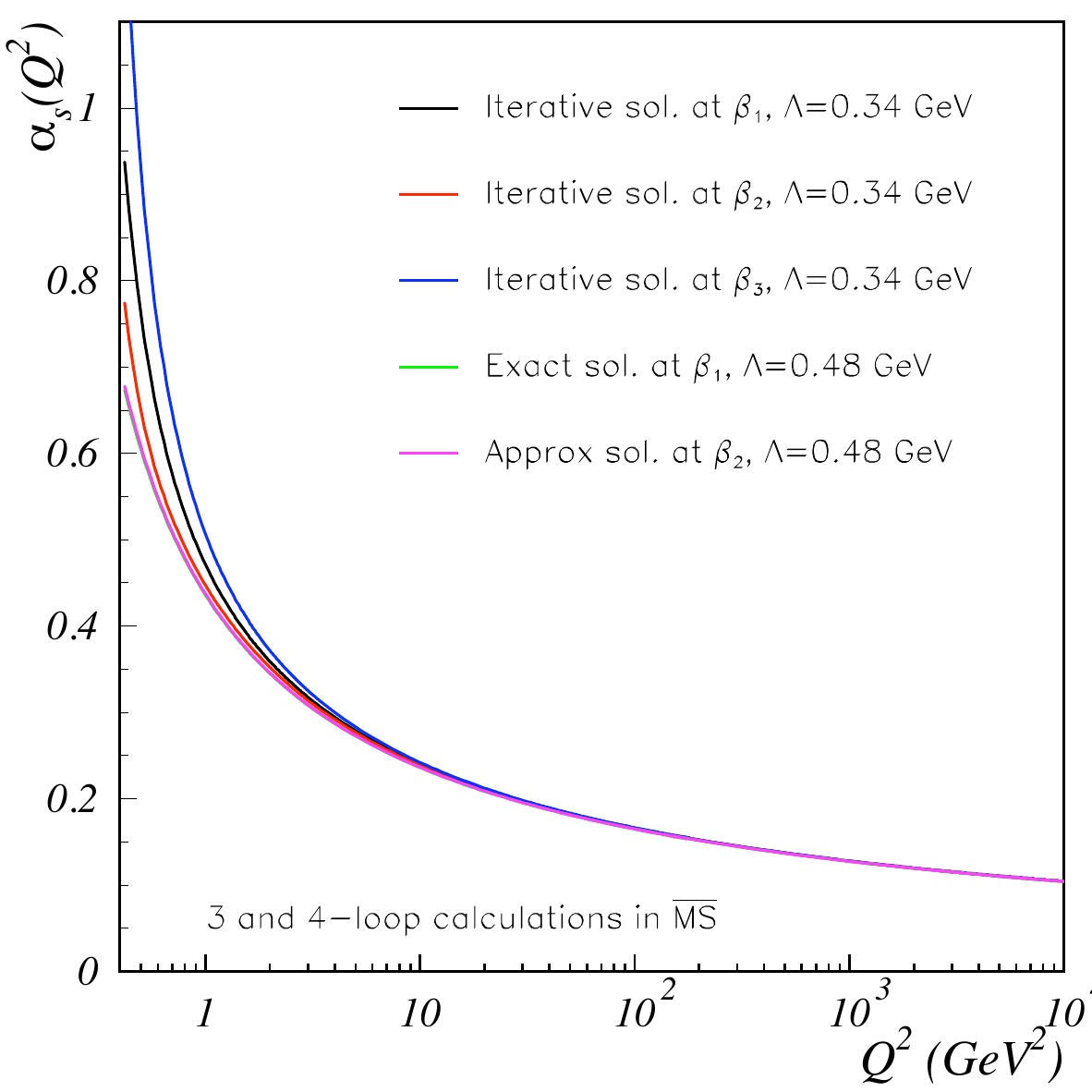}
\caption{\label{Flo:alpha_s from pQCD} \small The perturbative coupling $\alpha_{s}(Q^{2})$
computed at different orders in $\beta_{i}$ using the iterative method
(black, red and blue lines: Eq. (\ref{eq:alpha_s}) at orders $\beta_{1}$,
$\beta_{2}$ and $\beta_{3}$, respectively) and the exact and approximate
solutions, Eq. (\ref{eq:GGK beta1}) ($\beta_{1}$, green line) and
Eq. (\ref{eq:GGK beta2}) ($\beta_{2}$, magenta line), respectively.
The green and magenta lines are very close to each other. We use $n_{f}=3$
and $\Lambda=0.34$ GeV for the iterative method and adjusted $\Lambda$
to 0.48 GeV for the others so that all calculations match at large
$Q^{2}$. The coefficients $\beta_{2}$ and $\beta_{3}$ are calculated
in the $\overline{MS}$  renormalization scheme.}
\end{figure}

For orders up to $\beta_{3}$, an approximate analytical solution
to Eq. (\ref{eq:alpha_s beta series}) is obtained by an iterative
method \cite{the:alpha_s interative sol to beta func}: 
\begin{multline} 
\alpha_{s}(Q^2) = \frac{4\pi}{\beta_{0}\mbox{ln}\left(Q^{2}/\Lambda^{2}\right)} \biggl[ 1-\frac{\beta_{1}}{\beta_{0}^{2}}\frac{\mbox{ln}\left[\mbox{ln}(Q^{2}/\Lambda^{2})\right]}{\mbox{ln}(Q^{2}/\Lambda^{2})}  \\
+ \frac{\beta_{1}^{2}}{\beta_{0}^{4}\mbox{ln}^{2}(Q^{2}/\Lambda^{2})}\left(\left(\mbox{ln}\left[\mbox{ln}(Q^{2}/\Lambda^{2})\right]\right)^{2}-\mbox{ln}\left[\mbox{ln}(Q^{2}/\Lambda^{2})\right]-1+\frac{\beta_{2}\beta_{0}}{\beta_{1}^{2}}\right) \\
+ \frac{\beta_{1}^{3}}{\beta_{0}^{6}\mbox{ln}^{3}(Q^{2}/\Lambda^{2})}\biggl(-\left(\mbox{ln}\left[\mbox{ln}(Q^{2}/\Lambda^{2})\right]\right)^{3}+\frac{5}{2}\left(\mbox{ln}\left[\mbox{ln}(Q^{2}/\Lambda^{2})\right]\right)^{2} \\
+2 \mbox{ln}\left[\mbox{ln}(Q^{2}/\Lambda^{2})\right]-\frac{1}{2}-3\frac{\beta_{2}\beta_{0}}{\beta_{1}^{2}}\mbox{ln}\left[\mbox{ln}(Q^{2}/\Lambda^{2})\right]+\frac{\beta_{3}\beta_{0}^{2}}{2\beta_{1}^{3}}\biggr)\\
+\mathcal{O}(\frac{\left(\mbox{ln}\left[\mbox{ln}(Q^{2}/\Lambda^{2})\right]\right)^{4}}{\mbox{ln}(Q^{2}/\Lambda^{2})}) \biggr] .
\label{eq:alpha_s}
\end{multline}
The integration constant  $\Lambda$ is the only unknown parameter
in the monotonic   equation (\ref{eq:alpha_s}). Consequently, at a given
order and in a given RS, the perturbative  coupling $\alpha_{s}(Q^{2})$ can
be fully characterized either by giving the value of $\alpha_{s}$
at a conventional scale, usually, $Q^{2} = M_Z^2$, or by giving
the value of  $\Lambda$.

From Eq. (\ref{eq:alpha_s}), $\alpha_{s}$ can be extracted from
an observable at a given scale, evolved to the conventional scale
$M_Z^2$ and compared to $\alpha_{s}$ extracted using other
observables. The agreement of the values, see Section \ref{sub:Experimental-status},
demonstrates the universality of $\alpha_{s}$, the validity of Eq.
(\ref{eq:alpha_s beta series}), and constitutes an important consistency check
of QCD. To determine $\alpha_{s}(M_Z^{2})$, or equivalently $\Lambda$,
 actual measurements or nonperturbative calculations are necessary, see Section
\ref{sub:Experimental-status}.

\section{Quark thresholds \label{sub:Quark-thresholds}}

\begin{figure}[ht]
\centering
\includegraphics[width=9cm]{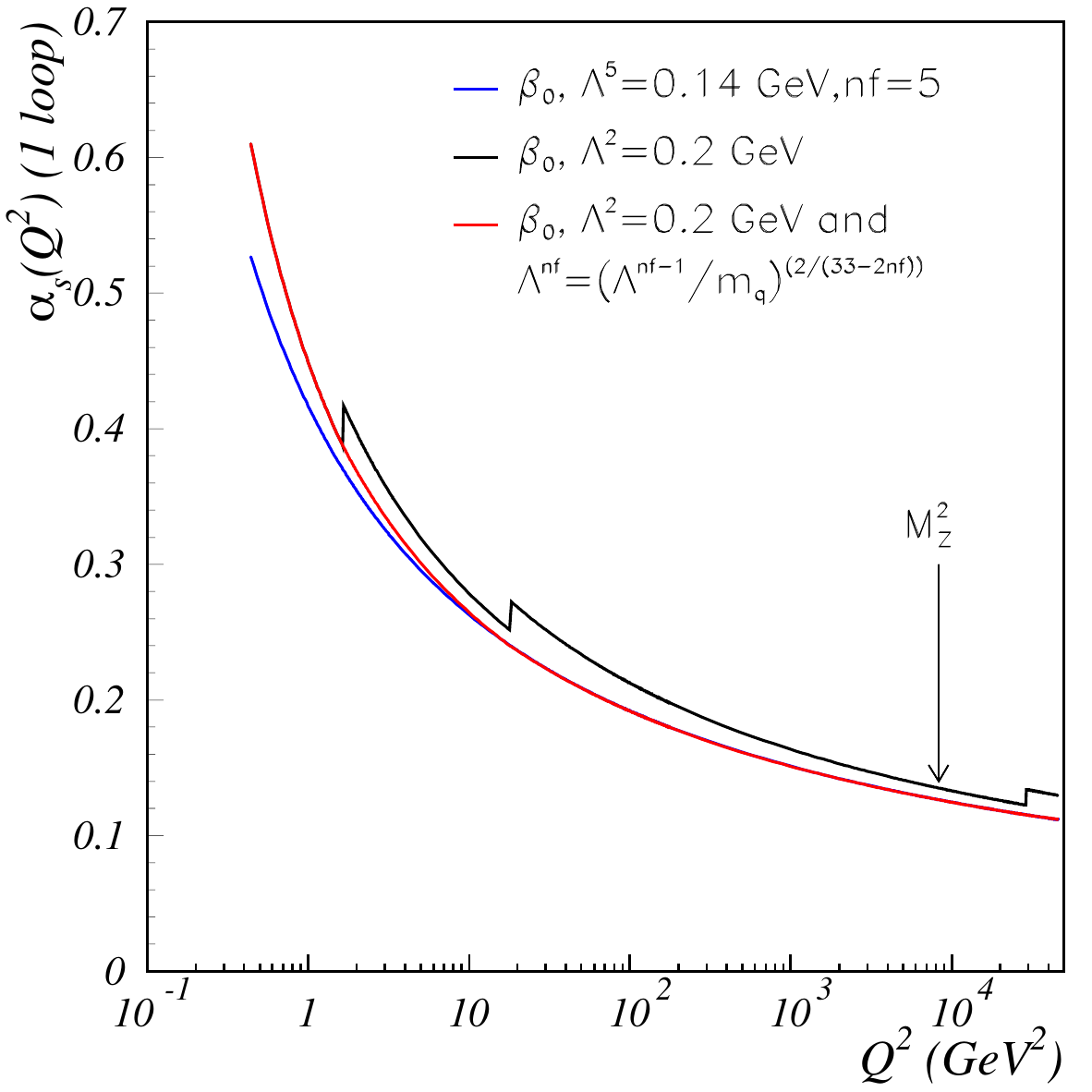}
\caption{\label{Flo:quark_thr} \small The running coupling $\alpha_{s}(Q^{2})$ at
order $\beta_{0}$ for a constant $n_{f}=5$ (blue line), with $\Lambda^{n_{f}=5}$
chosen so that it matches the experimental coupling value at  $Q^{2}=M_Z^{2}$
(shown by the arrow). The effect of quarks becoming active is shown
by the black line. The red line corrects for it by threshold matching.
We adjusted $\Lambda^{n_{f}=2}$ to match the data at $Q^{2}=M_Z^{2}$.}
\end{figure}

Quark masses create additional higher order perturbation terms that
have, for light quarks, a small influence. However, they do indirectly
and importantly affect $\alpha_{s}$ through $n_{f}$, the number
of active quark flavors at scale $Q^{2}$. A quark flavor is active
if its mass $m_{q}^{2}\ll Q^{2}$. For all other purposes, $m_{q}$
can be set to zero regardless whether it is active or not. The loose
requirement $m_{q}\ll Q$ implies that $\alpha_{s}$ varies smoothly
when passing a quark threshold, rather than in discrete steps  as depicted in
Fig. \ref{Flo:quark_thr}. The matching of the values of $\alpha_{s}$
below and above a quark threshold makes  $\Lambda$
 to depend on $n_{f}$.
For example, at leading order, requesting that 
\begin{equation}
\alpha_{s}^{n_{f}-1}\left(Q^{2}=m_{q}^{2}\right)=\alpha_{s}^{n_{f}}\left(Q^{2}=m_{q}^{2}\right),
\label{eq:LO matching}\end{equation}
 imposes:
\begin{equation}
\Lambda^{n_{f}}=\Lambda^{n_{f}-1}\left(\frac{\Lambda^{n_{f}-1}}{m_{q}}\right)^{2/(33-2n_{f})}.\label{eq:lambda(nf) LO}\end{equation}
The formula at next order, $\beta_{1}$, can be found in \cite{the:quark thr 2-loop match}.
The  four-loop matching in the $\overline{MS}$ RS is given in \cite{the:alpha_s interative sol to beta func}. 
Another procedure to treat the variation of active $n_f$ keeps the quark masses nonzero in the quark 
loops and treat $n_f$ analytically \cite{the:Binger-Brodsky}.

\section {Computation of the pQCD effective coupling \label{sub:beta-coefficients calc.}}

In this section, we illustrate how the coefficients $\beta_{i}$ of
the perturbative $\beta$-series are calculated by  giving the steps
leading to the expression of $\alpha_s$ to lowest order and thus of $\beta_{0}$. This allows us to understand
the physical origin of the processes contributing to the running of
the coupling in the UV domain. The $MS$ scheme is used and we neglect quark masses. 

\begin{figure}[ht]
\centering
\includegraphics[width=10cm]{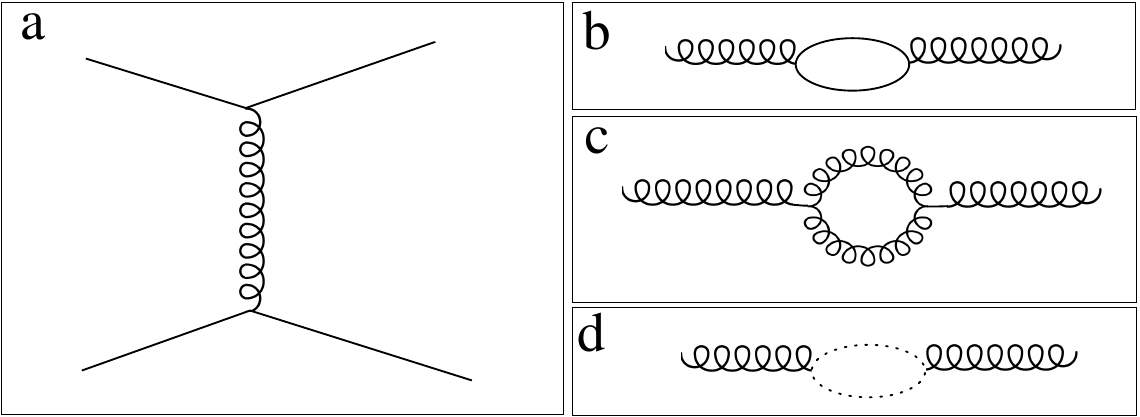}
\caption{\label{Flo:1st order corrections} \small Panel (a): leading order (Born) contribution
to quark--quark scattering. Panels (b), (c) and (d): quark, gluon and ghost
loop corrections to the gluon propagator, respectively.}
\end{figure}

As already discussed, $\alpha_{s}(Q^{2})$ can be defined using the
potential approach. The higher-order processes producing a deviation
of the quark--quark scattering amplitude from the leading  zero-order (Born
amplitude, see Fig. \ref{Flo:1st order corrections}(a) are folded
into the bare coupling, thereby defining the \emph{effective coupling}. The 
  processes contributing  to next order --first order-- are shown in 
 Figs.  \ref{QCD_vertex}(a), (b) and (c), and \ref{Flo:1st order corrections}(b), (c) and (d). We first consider the effect
of quark loops on the gluon propagator, see Fig. \ref{Flo:1st order corrections} (b).
Applying QCD's Feynman rules yields:
\begin{equation}
D_{q}^{\mu\nu}\left(q\right)=-\frac{\overline{\alpha_{s}}}{3\pi}\left(q^{\mu}q^{\nu}-\eta^{\mu\nu}q^{2}\right)\mbox{ln}\left(\frac{Q^{2}}{\mu^{2}}\right)n_{f}\frac{\delta_{ab}}{2},
\label{eq:quark loop}
\end{equation}
where $\overline{\alpha_{s}}$ is the bare coupling constant and $\mu^{2}$
the \emph{subtraction point}. The term in Eq. (\ref{eq:quark loop}) is analogous to the one causing the
QED \emph{effective coupling} to run. In addition, in QCD, gluon loops (Fig.
\ref{Flo:1st order corrections}(c) and tadpole graphs (Fig. \ref{QCD_vertex}(c)
may also contribute. In the $MS$ scheme, however, tadpoles are suppressed.
The calculation of the gluon loop term is similar to the quark loop
one and yields:
\begin{equation}
D_{g}^{\mu\nu}\left(q\right)=\frac{3\overline{\alpha_{s}}}{4\pi}\delta_{ab}\left[\frac{11}{6}q^{\mu}q^{\nu}-\frac{19}{12}\eta^{\mu\nu}q^{2}
+\frac{1-\xi}{2}\left(q^{\mu}q^{\nu}-\eta^{\mu\nu}q^{2}\right)\right]\mbox{ln}\left(\frac{Q^{2}}{\mu^{2}}\right),\label{eq:gluon loop}
\end{equation}
with $\xi$ a gauge fixing term. Contrary to the quark loop contribution,  the propagator
$D_{g}^{\mu\nu}(q)$ is not proportional to $\left(q^{\mu}q^{\nu}-\eta^{\mu\nu}q^{2}\right)$,
{\it  i.e.}, when gluon loops are accounted for, the gluon propagator
is not transverse anymore and current conservation is violated: $D_{g}^{\mu\nu}q_{\mu}\neq0$.
There are several ways to address this. Gauges where gluons are always
transverse, such as the axial gauge or the light-cone gauge, can be
chosen. Alternatively, fictitious particles of spin zero but obeying
the Fermi--Dirac statistics can be introduced, the Fadeev--Popov ghosts
\cite{the:Faddeev-Popov}. Their characteristics are chosen so that
they add a loop contribution to the gluon propagator, see Fig. \ref{Flo:1st order corrections}(d,
that complements the non-transverse term in Eq. (\ref{eq:gluon loop}),
to make it transverse. They contribute as:
\begin{equation}
D_{gh}^{\mu\nu}\left(q\right)=-\frac{3\overline{\alpha_{s}}}{4\pi}\delta_{ab}\left[\frac{1}{6}q^{\mu}q^{\nu}+\frac{1}{12}\eta^{\mu\nu}q^{2}\right]\mbox{ln}\left(\frac{Q^{2}}{\mu^{2}}\right).
\label{eq:ghosts}
\end{equation}
Contrary to QED, in QCD the fermion self-energy (Fig.
\ref{QCD_vertex}(a) and vertex corrections (Fig. \ref{QCD_vertex}(b)
do not cancel each other and need to be accounted for. At first order
the self-energy correction is:
\begin{equation}
G_{q}\left(p\right)=\frac{\not p}{p^{2}}\delta_{ab}\left[1-\xi\frac{\overline{\alpha_{s}}}{3\pi}\mbox{ln}\left(\frac{-p^{2}}{\mu^{2}}\right)\right],
\end{equation}
and the vertex correction yields:
\begin{equation}
\Gamma_{\mu}^{\alpha\beta;a}\left(q\right)=-i\sqrt{4\pi\overline{\alpha_{s}}}\frac{\lambda_{\alpha\beta}^{a}}{2}\gamma_{\mu}\left[1-\frac{\overline{\alpha_{s}}}{4\pi}\mbox{ln}\left(\frac{Q^{2}}{\mu^{2}}\right)\left\{ \frac{4}{3}\xi+3\left(1-\frac{1-\xi}{4}\right)\right\} \right],
\end{equation}
with $\lambda_{\alpha\beta}^{a}$ the Gell-Mann matrices, $a$ the
gluon color index and $\alpha, \beta$ the color indices for the two
quark lines. 

In renormalization schemes based on dimensional regularization, the sum of all the amplitudes is gauge invariant.
Summing them  yields the quark--quark scattering amplitude up to next-to-leading order (NLO)
\begin{equation}
\mathcal{M}=\mathcal{M}_{Born}\left[1+\frac{\overline{\alpha_{s}}}{4\pi}\mbox{ln}\left(\frac{Q^{2}}{\mu^{2}}\right)\left\{ \frac{2n_{f}}{3}-\frac{13}{2}-\frac{9}{2}\right\} \right].
\end{equation}
One can trace back the $2n_{f}/3$ term to the quark loop corrections
to the gluon propagator, the $-13/2$ term to the gluon and ghost
loop corrections to the gluon propagator, and the $-9/2$ term to
the vertex corrections. (The remaining graph --the quark self-energy,
Fig. \ref{QCD_vertex}(a)-- does not contribute in the Landau
gauge.) 

Folding the deviations from the Born term into the coupling constant
defines the \emph{effective coupling} $\alpha_{s}$, which is then at one
loop:
\begin{equation}
\alpha_{s}\left(Q^{2}\right)=\alpha_{s}\left(\mu^{2}\right)\left[1+\frac{\alpha_{s}\left(\mu^{2}\right)}{12\pi}\mbox{ln}\left(\frac{Q^{2}}{\mu^{2}}\right)\left\{ 2n_{f}-33\right\} \right].
\end{equation}
Inverting the expression and assuming that $\alpha_{s}(\mu^{2})$
is small  for $\mu^2 \gg Q^2$, yields:
\begin{equation}
\frac{4\pi}{\alpha_{s}\left(Q^{2}\right)}=\frac{4\pi}{\alpha_{s}\left(\mu^{2}\right)}+\frac{33-2n_{f}}{3}\mbox{ln}\left(\frac{Q^{2}}{\mu^{2}}\right).
\end{equation}
Using Eq. (\ref{eq:beta0 sol, 3}), we identify 
\begin{equation}
\beta_{0}=11-\frac{2n_{f}}{3}.\label{eq:beta0}
\end{equation}

In QED, only the photon propagator 
 contributes to the running of the coupling
constant since Ward--Takahashi identities  guarantee that the photon bremsstrahlung
cancels the vertex corrections. However, the Ward--Takahashi identities
do not hold for non-Abelian theories. A more restricted constraint
is given by the \emph{Slavnov--Taylor} identities \cite{the:Slavnov--Taylor id.},
which ensure that intermediate gauge-dependent quantities used in
calculations, such as  \emph{Green's function}s, yield final gauge-independent
results for observables. These constraints lead to relations between
the various   Green's functions characterizing the propagation and interaction
amplitudes and, as such, are especially important for solving the
Schwinger--Dyson   equations, see Section \ref{sub:Schwinger--Dyson-formalism}.

\section{Renormalization group \label{sub:Renormalization-group}}

In this paragraph, the  renormalization group (RG) approach is discussed. It provides a more
formal view of the mechanism inducing the running of $\alpha_{s}$.
This more formal derivation of $\alpha_{s}(Q^{2})$
is useful to understand how it is defined in some of the lattice and
Schwinger--Dyson studies in the IR, see Sections \ref{sub:Schwinger--Dyson-formalism}
and \ref{sub:Low Q Lattice-QCD-calculations}.

To regulate divergences, renormalization factors $Z_{i}(Q^{2})$ are
introduced to define renormalized fields: 
\begin{equation}
A_{a,\mu}={Z_{3}^{1/2}(Q^{2})}A_{a,\mu}^{R}(Q^{2}),  \quad
\psi={Z_{2}^{1/2}(Q^{2})}\psi^{R}(Q^{2}),  \quad C_{a}={\widetilde{Z}_{3}^{1/2}(Q^{2})}C_{a}^{R}(Q^{2}),
\end{equation}
where the superscript $R$ indicates the now scale-dependent renormalized
fields, and $C_{a}$ is the ghost field. $Z_{3}$ is the gluon propagator
renormalization factor, $Z_{2}$ is the quark self-energy renormalization factor and $\widetilde{Z}_{3}$
is the ghost propagator renormalization factor. The ghost propagator is obtained by definition from the
inverse of  the Fadeev--Popov operator $(-\partial+A)A$. The coupling constant
is likewise renormalized: $\overline{\alpha_{s}}=Q^{2\varepsilon}Z_{\alpha}(Q^{2})\alpha_{s}(Q^{2})$
where $\varepsilon$ is the dimensional parameter in the $MS$ scheme.
In such space, integrals are carried out in $4-2\varepsilon$ dimensions
and the UV divergences are regularized to $1/\varepsilon$ poles.
The $Z_{i}$ are constructed as functions of $1/\varepsilon$, such
that they  cancel the $1/\varepsilon$ poles. Since 
$\overline{\alpha_{s}}=Q^{2\varepsilon}Z_{\alpha}(Q^{2})\alpha_{s}(Q^{2})=\mu^{2\varepsilon}Z_{\alpha}(\mu^{2})\alpha_{s}(\mu^{2})$,
$Z_{\alpha}$ defines the scale dependence of the  effective coupling
such that $\alpha_{s}(Q^{2})=\mathcal{Z}_{\alpha}(Q^{2},\mu^{2})\alpha_{s}(\mu^{2})$,
with $\mathcal{Z}_{\alpha}(Q^{2},\mu^{2})\equiv(\mu^{2\varepsilon}/Q^{2\varepsilon})[Z_{\alpha}(Q^{2})/Z_{\alpha}(\mu^{2})]$.
The $\mathcal{Z}_{\alpha}$ form a group with a composition law: $\mathcal{Z}_{\alpha}(Q^{2},\mu^{2})=\mathcal{Z}_{\alpha}(Q^{2},\mu_{0}^{2})\mathcal{Z}_{\alpha}(\mu_{0}^{2},\mu^{2})$,
a unity element: $\mathcal{Z}_{\alpha}(Q^{2},Q^{2})=1$ and an inversion
law: $\mathcal{Z}_{\alpha}(Q^{2},\mu^{2})=\mathcal{Z}_{\alpha}^{-1}(\mu^{2},Q^{2})$.
The invariance under the infinitesimal transformation $Q\rightarrow Q+\delta Q$
leads to the RG Eq. (\ref{eq:alpha_s beta series}) with 
the leading order relation
\begin{equation}
Z_{\alpha}\left(Q^{2}\right)=1-\frac{\beta_{0}\alpha_{s}\left(Q^{2}\right)}{4\pi\varepsilon}\label{eq:z_alpha_beta0},
\end{equation}
necessary to cancel the $1/\varepsilon$ pole. This relation defines
the coefficient $\beta_{0}$. 

In addition to field and coupling renormalization factors, vertex
renormalization factors can also be introduced. They are straightforwardly
related to the $Z_{i}$ already introduced. For example, the quark--quark--gluon
vertex renormalization factor $Z_{1}$ is $Z_{1}=\left(Z_{2}Z_{\alpha}Z_{3}Z_{2}\right)^{1/2}$.
From this, $Z_{\alpha}$ can be extracted:
\begin{equation}
Z_{\alpha}=\frac{Z_{1}^{2}}{Z_{2}^{2}Z_{3}}\label{eq:z_alpha}.
\end{equation}
Clearly, $Z_{\alpha}$ can also be obtained from other graphs,
such as the ghost--gluon vertex. In the $MS$ scheme, the Feynman
rules yield at NLO $Z_{1}=1-\alpha_{s}(Q^{2})(3+4/3)/4\pi\varepsilon$,
$Z_{2}=1-\alpha_{s}(Q^{2})(4/3)/4\pi\varepsilon$ and $Z_{3}=1-\alpha_{s}(Q^{2})(-5+2n_{f}/3)/4\pi\varepsilon$.
Injecting these expressions in Eq. (\ref{eq:z_alpha}), assuming that
$\alpha_{s}(Q^{2})$ is small, and identifying with Eq. (\ref{eq:z_alpha_beta0})
yields:
\begin{equation}
\beta_{0}=11-\frac{2n_{f}}{3}.
\end{equation}

\section{The  Landau pole and the QCD parameter $\Lambda$ \label{sub:Landau-Pole}}

The \emph{Landau pole} is the point where the perturbative expression of
$\alpha_{s}$, Eq. (\ref{eq:alpha_s}), diverges. It occurs at the
value of the \emph{scale parameter}  $\Lambda$.  As we shall discuss, the divergence of the perturbative prediction for the 
running coupling does not usually appear in a nonperturbative approach such as AdS/QCD.

The value of  $\Lambda$ in pQCD
depends on the RS, on the order  of the $\beta$-series, $\beta_{i}$, on  the number of flavors $n_{f}$, and on the
approximation chosen to solve Eq. (\ref{eq:alpha_s beta series})
at orders higher than $\beta_{1}$.   At order $\beta_{0}$, the Landau
singularity is a simple pole whereas at higher order, it acquires
a more complicated structure.  The pole is located on the positive
real axis of the complex $Q^{2}$-plane,  and is thus unphysical: a pole
at  $\Lambda<0$ would correspond to production of on-shell particles.
If  $\Lambda>0$, a physical pole would correspond
to causality-violating tachyons. The existence of the  Landau pole thus implies that    the perturbative expression of
$\alpha_{s}$  is a non-observable quantity. 

The appearance of an
unphysical pole at  $\Lambda$ characterizes the scale at which pQCD
breaks down; {\it i.e.}, where $\alpha_{s}$ becomes large. Consequently,
the value of  $\Lambda$ is often associated with the confinement scale, or equivalently
the hadronic mass scale. An explicit relation between hadron masses
and  $\Lambda$ is known in the framework of holographic QCD; see Sec. \ref{sub:Holographic-QCD large Q}.
 $\Lambda$ can also
be related to  nonperturbative terms in the context of the  analytic
QCD approach, see Section \ref{sub:Handling of npert. terms}. 

We remark that  Landau poles were initially identified in the context
of Abelian QED. However, its value, $\Lambda\sim10^{30-40}$ GeV, is
well above the Planck scale \cite{the:QED pole}; it is thus
expected that new physics would suppress the occurrence of this unphysical divergence.

\section{Improvement of the perturbative series}

Naively, one may think that at leading order, the smaller the value
of  $\Lambda$; {\it i.e.}, the smaller the momentum scale at which the Landau
pole divergence occurs, the slower the increase of $\alpha_{s}(Q^{2})$
as $Q^{2}$ decreases. This would imply that for expansions at moderate
momentum scale, renormalization schemes with small  $\Lambda$ are preferable. Approximate
values of  $\Lambda$ in different schemes are given in Table \ref{Flo:Table of Lambda}.
However, such a simple criterium is foiled by the \emph{renormalon}
growth of the coefficients of the perturbative series.
Different growths in different renormalization schemes  may balance the difference in values
of  $\Lambda$.  In principle, at large enough order, the balancing
should be nearly exact since the observable described is RS-independent.
Some renormalization schemes  have inherently fast rising coefficients, independently of
the \emph{renormalon} problem, see {\it e.g.},  Ref. \cite{the:Brodsky 2001 BLM&squeletons}
in which it is shown that even for perturbative series
tailored to be free of  renormalons, the $\overline{MS}$ scheme leads
to intrinsically fast growing coefficients. The relations between
RS and their associated  $\Lambda$ are discussed in  Ref. \cite{the:Comparison alpha-g-gh alpha-3g alpha-4g}. 

\begin{table}[ht]
\centering
\begin{tabular}{|c|c|c|c|c|c|c|c|}
\hline 
RS/eff. charge & ${\overline{MS}}$  & $MS$ & $MOM$ & $V$ & $g_{1}$ & $R$ & $\tau$\tabularnewline
\hline
\hline 
 $\Lambda$ (GeV) & 0.34 & 0.30 & 0.62 & 0.48 & 0.92 & 0.48 & 1.10\tabularnewline
\hline
\end{tabular}
\caption{
\label{Flo:Table of Lambda} \small Examples of approximate values of  $\Lambda$
in different renormalization schemes  or  effective charge definitions, for $n_{f}=3$.}
\end{table}

In this section, we shall discuss several ways to improve perturbative series.  A principle of the renormalization group is that a prediction for an observable cannot depend on a theoretical convention such as the choice of the renormalization scheme.
The optimization of perturbative series can thus be tied to minimizing its 
RS-dependence since it brings the pQCD approximant closer to the RS-independent 
observable. Such an optimization and minimization of the RS-dependence clearly increases
the predictive power of pQCD.

\subsection{Effective charges and commensurate scale relations \label{sub:Effective charges and CSR}}

It was proposed by Grunberg that the QCD coupling can be defined
directly from any experimental observable \cite{the:Grunberg} which is predictable in pQCD. In such a
case, the coupling is called an ``effective charge''. The terminology
 ``effective charges''  is sometimes used with different definitions.
We will use it here in the sense defined by Grunberg.

The definition of the coupling from an observable obeys the RG Eq. (\ref{eq:alpha_s beta series}) 
and has important advantages:  an  effective charge is RS-independent, free of
divergences and analytic when crossing quark thresholds; it improves
perturbative expansions (suppressing in particular the \emph{renormalon}
problem), and its definition is extendible to the IR nonperturbative
domain. It also unifies QED and QCD coupling's definitions. In the case of QED,
the running coupling $\alpha(q^2)$ was defined by Gell-Mann and Low from the static  Coulomb
potential \cite{the:Gell-Mann--Low QED coupling}; {\it i.e.},  the strength of the Coulomb
interaction of heavy test charges at momentum transfer $t=q^2$. 
The apparent disadvantage of Grunberg's  effective charge approach is a process-dependence of the coupling.
However, different  effective charges obtained from different observables can be analytically related to each other
as we will see shortly.

The general prescription to form Grunberg's  effective charge is to truncate
the perturbative series of an observable to its first term in $\alpha_{s}$.
Since this term is RS-independent, so is the effective charge. For
example, in the $\overline{MS}$ RS, the Bjorken sum \cite{the Bj SR}
is known in the perturbative domain, up to $\alpha_{\overline{MS}}^{4}$
\cite{the: Kataev Bj SR coef.},  
\begin{eqnarray}
\label{eq:Bj SR, order alpha^4}
\int_{0}^{1^{-}}dx_{Bj} \biggl(g_{1}^{p}\left(x_{Bj},Q^{2}\right)-g_{1}^{n}\left(x_{Bj} ,Q^{2}\right)\biggr)= 
\frac{g_{A}}{6}\biggl[1-\frac{\alpha_{\overline{MS}}}{\pi}-3.58\left(\frac{\alpha_{\overline{MS}}}{\pi}\right)^{2}  \\ 
-20.21\left(\frac{\alpha_{\overline{MS}}}{\pi}\right)^{3}-175.7\left(\frac{\alpha_{\overline{MS}}}{\pi}\right)^{4}+\mathcal{O}\left(\alpha_{\overline{MS}}^{5}\right)\biggr] , \nonumber
\end{eqnarray}
where $\alpha_{\overline{MS}}$ is the coupling in the $\overline{MS}$
RS. The  effective charge $\alpha_{g_{1}}(Q^{2})$ is defined from
the truncation:
\begin{equation}
\int_{0}^{1^{-}}dx_{Bj} \biggl(g_{1}^{p}\left(x_{Bj} ,Q^{2}\right)-g_{1}^{n}\left(x_{Bj},Q^{2}\right)\biggr)\equiv\frac{g_{A}}{6}\left[1-\frac{\alpha_{g_{1}}\left(Q^{2}\right)}{\pi}\right].\label{eq:alpha_g_1 def.}
\end{equation}
In these equations, $g_{1}^{p}$ and $g_{1}^{n}$ are the spin-dependent
proton and neutron structure functions, respectively, $g_{A}$ is
the nucleon flavor-singlet axial charge, $x_{Bj} $ is the Bjorken scaling
variable. By definition,  the integration excludes the $x_{Bj} =1$ elastic contribution,
(which in any case is negligible at large $Q^{2}$). Such a definition amounts
to a particular choice of RS: In Eq. (\ref{eq:alpha_s beta series}),
the $\beta_{n}$ are RS-dependent for $n\geq2$, hence arbitrary,
and one is free to define them so that Eq. (\ref{eq:alpha_g_1 def.})
is realized. In a sense, this definition generalizes the introduction
of effective running discussed in  
 \ref{sec:Phenomenological-introduction:}:
all small distance quantum effects generating the higher order terms
in Eq. (\ref{eq:Bj SR, order alpha^4}) are folded into the definition
of $\alpha_{g_1}$. Another effective charge is $\alpha_D(Q^2)$, defined from the Adler function \cite{the:Adler function}, a spacelike continuation of $R_{e^+ e^-}(s)$. 
Another example of an  effective charge can be found
in  Ref.  \cite{the: Brodsky alpha_tau} where the observable is defined from the 
hadronic decay rate of the $\tau$ lepton.   

Clearly, effective charges depend on the choice of the observable. A natural
conventional choice would be to follow QED and use $\alpha_{V}$, the strong
coupling defined from the static heavy quark potential \cite{the:Appelquist, the:Brodsky alpha_v}.
At 3-\emph{loop} order and in  the $\overline{MS}$ scheme it is:
\begin{equation}
V\left(Q^{2}\right)=-\frac{1}{2\pi^{2}}\frac{4}{3}\frac{\alpha_{\overline{MS}}\left[1+a_{1}\left(\frac{\alpha_{\overline{MS}}}{4\pi}\right)+a_{2}\left(\frac{\alpha_{\overline{MS}}}{4\pi}\right)^{2}+a_{3}\left(\frac{\alpha_{\overline{MS}}}{4\pi}\right)^{3}\right]}{Q^{2}}
\end{equation}
with $a_{1}=31/3-10n_{f}/9$, $a_{2}=456.75-66.35n_{f}+1.23n_{f}^{2}$
and $a_{3}=13432.6-3289.91n_{f}+185.99n_{f}^{2}-1.37174n_{f}^{3}$.
Numerically, the $a_{i}$ coefficients are $a_{1}=5.88$, $a_{2}=211.03$
and $a_{3}=3161.00$ for $n_f=4$, displaying a typical factorial \emph{renormalon} growth.
The application of this formula to the spectroscopy and decays of $\mbox{Q--}\overline{\mbox{Q}}$
systems can be found in  Ref. \cite{the:Santia 2014}. 
The one-\emph{loop} corrections to $\alpha_V$ were
performed in the late 1970s and  early 1980s \cite{the:Appelquist, the:beta0 alpha_v}. 
The two-\emph{loop} calculations became available in the late 1990s and
early 2000s \cite{the:beta1 alpha_v, the:Badalian (2000)}.
Three-\emph{loop} calculations were made available recently, see \cite{the:beta2 alpha_v,
the:beta_n estimates alpha_V, the:V(r) pot.}
and the recent reviews \cite{the:Pineda review Q-Q potential, the:Garcia. Review alpha_s from stat. energy, Kataev:2015yha}. 
Alternatively, the determination of $\alpha_{V}$ from its $\overline{MS}$
(or other) RS can be done using \emph{Commensurate Scale Relations} --to be discussed next-- 
which eliminate the \emph{renormalon} growth problem of the high order
coefficients and thus improves the convergence of the pQCD series.

However, the definition of $\alpha_V$ as an effective 
charge must be made with care in a non-Abelian theory.  
In QED the Gell-Mann--Low coupling \emph{effective charge} $\alpha(q^2)$ is defined from the momentum space potential, 
$V(q^2) = [e_1 e_2 \alpha(q^2)] / q^2$, 
which generates the single-photon exchange scattering amplitude proportional to the charges of two infinitely heavy test charges $e_1$ and $e_2$.
In contrast to its QED equivalent,  in QCD multi-gluon exchange between the two static quarks cannot be separated from  single-gluon exchange due to the contributions of  multi-gluon diagrams. The exchanged gluons become connected by gluon exchange by the 3-gluon and 4-gluon couplings --the  ``H'' diagrams, see Fig. \ref{QCD_vertex}(d). 
In  QED, the analogous multi-photon exchange diagrams, such as those due to light-by-light scattering, can be suppressed by taking the formal limit of small external charges $e_1$ and $e_2$.
In pQCD, the multi-gluon exchange contributions are infrared divergent~\cite{the:Appelquist, the:gluon H graph divergence} as the test static quarks are separated; each additional horizontal rung that connects the two vertical gluons contributes an additional IR  divergence.   %
Thus a conventional \emph{effective charge} cannot be directly defined from the heavy quark scattering potential since it should be finite in the IR.
The IR divergence can be regularized by a parameter $\mu^{-1}$ which cuts off the loop  integration at large distances. For example, if one evaluates the QCD interaction between two heavy quarks in a quarkonium bound state, the value of $\mu$ corresponds to the (finite) size of the bound $\mbox{Q--}\overline{\mbox{Q}}$  system considered. 
The divergence occurs for $\mu=0$. Hence, the \emph{effective charge} $\alpha_V$ depends not only on the specific QCD process from which it is defined -- as any \emph{effective charge} does -- but it also depends on the hadron environment in which this process occurs. This is analogous to the IR-sensitive Bethe state-dependent  logarithm in the Lamb Shift in hydrogen atoms in QED. 

The physical IR divergence of the multi-gluon exchange H diagrams in pQCD signals that QCD can only be consistent if color is confined --a remarkable fact.
This can be contrasted with the fact that the \emph{Landau pole}, which also produces an IR divergence, does not imply confinement since the pole is unphysical.

Other measured or calculated effective
charges can be related to $\alpha_{V}$, or more generally, to any
other \emph{effective charge} by \emph{Commensurate Scale Relations} (\emph{CSR}) \cite{the:CSR}.
The \emph{CSR} are RS-independent. The relation between two schemes $A$
and $B$ is of the form: 
\begin{equation}
\label{eq:generic CSR}
\frac{\alpha_{A}\left(Q\right)}{\pi}=\frac{\alpha_{B}\left(Q^{*}\right)}{\pi}+a\left(\frac{\alpha_{B}\left(Q^{**}\right)}{\pi}\right)^{2}+b\left(\frac{\alpha_{B}\left(Q^{***}\right)}{\pi}\right)^{3} + \cdots .
\end{equation}
The commensurate scale, $Q^{*}$, gives the mean virtuality of the
exchanged gluon and thus the number of effective heavy quark flavors $n_f$.  One has 
$Q^{*}=Q$ for $\alpha_{V}$, just like the scale $\mu$ in the QED coupling is
set by the virtuality of the photon.
At leading order (LO), $Q=1.18~Q^{*}$ for $\alpha_{g_1}$, $Q=1.36~Q^{*}$
for $\alpha_{\tau}$ and $Q=0.435~Q^{*}$ for $\alpha_{\overline{MS}}$.
The first equality indicates that the $V$-scheme and $g_{1}$-scheme
are similar, $\alpha_{g_1}(Q)\sim\alpha_{V}(Q)$, while $\alpha_{\overline{MS}}(Q)<\alpha_{V}(Q)$
since having $Q^{*}>Q$ is equivalent to shifting the \emph{Landau pole}
to  a lower scale, see Fig. \ref{Flo:strong-coupling in diff schemes}.
Re-expressing the LO of Eq. (\ref{eq:beta0 sol, 3}) with $\mu=Q$
and $\mu=Q^{*}$ yields
\begin{equation}
\alpha_{s}\left(Q^{2}\right)=\frac{\alpha_{s}\left(Q^{*2}\right)}{1+\frac{\beta_{0}}{4\pi}\mbox{ln}\left(\frac{Q^{2}}{Q^{*2}}\right)}=\alpha_{s}\left(Q^{*2}\right)\sum_{n=0}^{\infty}\left[-\frac{\beta_{0}}{4\pi}\mbox{ln}\left(\frac{Q^{2}}{Q^{*2}}\right)\right]^{n},
\end{equation}
which shows that the shift of scales amounts to re-organizing the perturbative
expansion in $\alpha_{s}$ of an observable. The \emph{Commensurate Scale Relations} have been shown to
hold to any order of pQCD \cite{the:CSR hold at any pQCD order}.

\begin{figure}[ht]
\centering
\includegraphics[width=9.0cm]{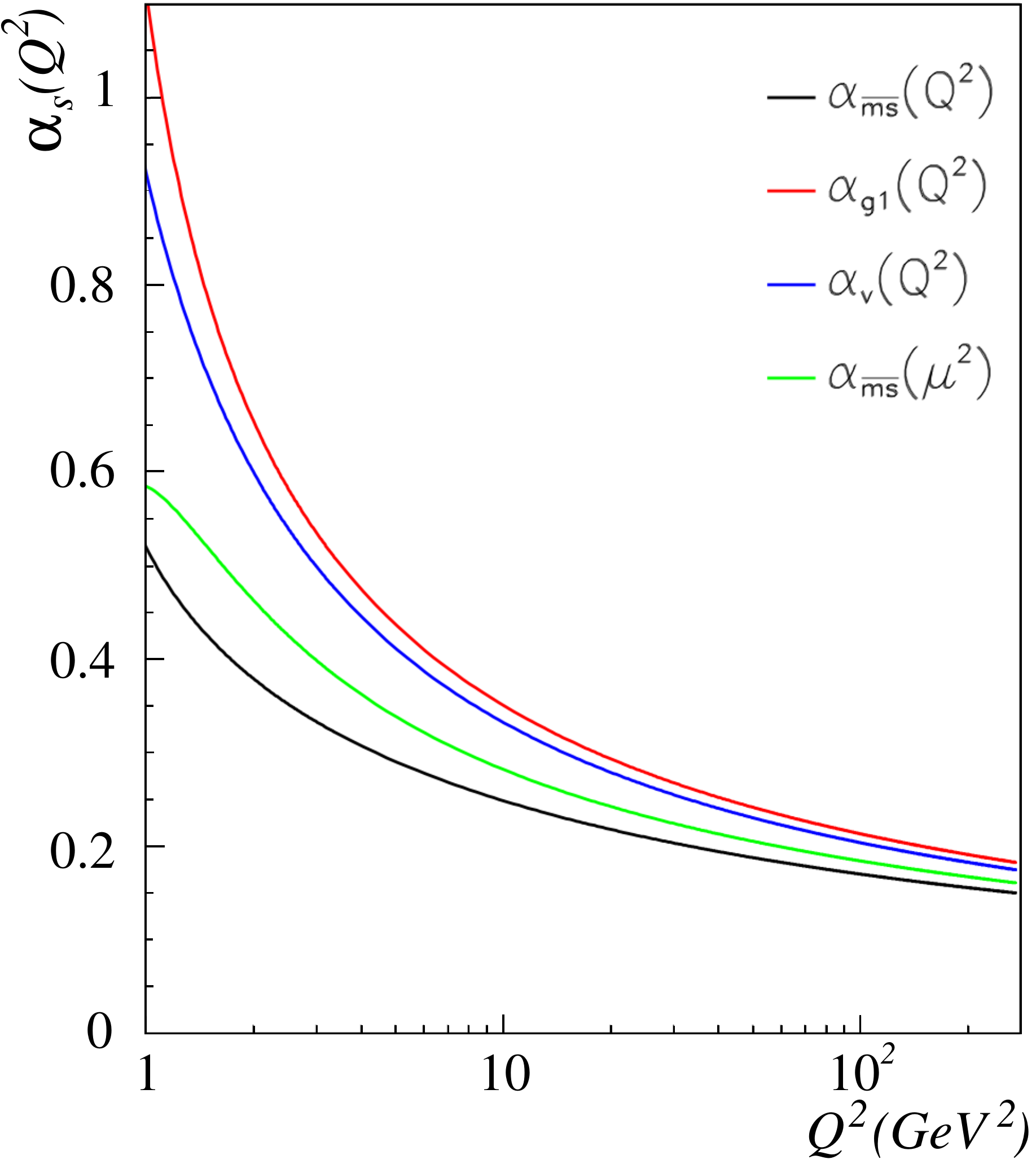}
\caption{\label{Flo:strong-coupling in diff schemes} \small The strong coupling $\alpha_{s}(Q^{2})$
expressed in different  renormalization schemes (black: $\overline{MS}$-scheme, red: $g_{1}$
effective charge, blue $V$-scheme) and using a scale $\mu\neq Q$
(green). We chose $\mu=0.708~Q$ which, in the \emph{CSR} context, is the
LO scale shift transforming $\alpha_{\overline{MS}}$ into $\alpha_{R}$
($\alpha_{R}$ is the \emph{effective charge} obtained from the ratio of
the $e^{+}+e^{-}\rightarrow$ hadrons rate to that for $e^{+}+e^{-}\rightarrow\mu^{+}+\mu^{-}$).
In this figure, $\alpha_{\overline{MS}}$ was computed with $n_{f}=3$,  $\Lambda=0.34$
GeV and to order $\beta_{2}$. }
\end{figure}

\subsection{The Brodsky, Lepage and Mackenzie procedure and its extensions}

The Brodsky, Lepage and Mackenzie (BLM) procedure \cite{the:BLM}
is a widely used method to optimize the perturbative series by setting the \emph{renormalization scales}  of  the running coupling $\alpha_s^n(Q^*_n)$ at each order $n$ to absorb all $\beta$ terms.   The coefficients of the pQCD series then matches the corresponding ``\emph{conformal}'' series with $\beta=0$.   The resulting predictions are then RS-invariant.

The BLM procedure originates from the observation
that in QED, the vacuum polarization contributions to 
the photon propagator are solely responsible
for the running of $\alpha$.   For example, in M\''oller scattering, the one-photon exchange amplitude depends on 
$\alpha(t) = {\alpha(0)\over 1-\Pi(t)}$  and $\alpha(u) = {\alpha(0)\over 1-\Pi(u)}$
 which sums all vacuum polarization contributions to all orders. 

For QCD, BLM showed that one can extend the QED scale-setting procedure
to NLO in pQCD by  identifying the $\beta_0$ and $\beta_1$ terms in the 
pQCD series  from their unique dependence on $n_f$ and absorbing them by 
shifting the  \emph{renormalization scale}.    As in QED, the resulting scales characterize 
the virtuality of the propagators in the amplitude  and the effective number of flavors.  
The BLM procedure reduces to the Gell-Mann--Low scale setting in the Abelian limit $N_C\to 0.$

More explicitly, the NLO expression of an observable
can be  rewritten in a form making its $n_{f}$-dependence explicit:
\begin{equation}
R=r_{0}\frac{\alpha_{s}^{n}\left(\mu\right)}{\pi}\left(1+\left[A+Bn_{f}\right]\frac{\alpha_{s}\left(\mu\right)}{\pi}\right).\label{eq:NLO nf-dep explicit}
\end{equation}
Using the RG Eqs. (\ref{eq:beta0 sol, 3}) and (\ref{eq:beta0}),
Eq. (\ref{eq:NLO nf-dep explicit}) becomes
\begin{equation}
R=r_{0}\frac{\alpha_{s}^{n}\left(\mu e^{3B/n}\right)}{\pi}\left(1+\left[A+\frac{33}{2}B\right]\frac{\alpha_{s}\left(\mu e^{3B/n}\right)}{\pi}\right),
\end{equation}
with the scale shifting $\mu\rightarrow\mu e^{3B/n}$. 

The  BLM procedure was extended to next-to-next-to leading order (NNLO) by Brodsky and Lu \cite{the:CSR}.
The renormalization scale $\mu^2(q^2_3, q^2_2, q^2_3) $ for the three-gluon 
vertex as a function of the Feynman virtualities of the three external gluons is given in
\cite{Binger:2006sj}
Extensions to higher orders are not as straightforward because of additional
$n_{f}$-dependences due to processes unrelated to the running of
$\alpha_{s}$ such as the UV finite corrections to the three and four-gluon vertices. 

Several extensions of the BLM procedure
have been proposed, for example by Grunberg and Kataev~\cite{Grunberg:1991ac}. An early extension of the BLM procedure was proposed
by Neubert in  Ref. \cite{the:Neubert BLM/renormalons}, along with an
explanation for the appearance of the $1/Q^{2}$ \emph{power corrections}
at low momentum transfer. The sequential extended BLM procedure 
proposed by Kataev and Mikhailov~\cite{the:seBLM} builds on Neubert's proposal. This
extension, which aims at improving the perturbative series convergence, is discussed 
further in a modified form by Ma \emph{et al.} in Ref.~\cite{the:MseBLM}.   
As we discuss in the next section, a rigorous procedure for identifying the $\beta$ terms  and setting the scales at any order in pQCD is given by the Principle of Maximum Conformality.

The BLM procedure and its extensions are physically motivated, 
minimizing the unphysical RS-dependence of the series. This is
achieved by shifting the scale $\mu$ so that at all orders of Eq.
(\ref{eq:pert exp of an observable with eff. alpha_s}), the RS-dependence
of $\alpha_{s}$ cancels the RS-dependence of $r_{n}$. Other procedures
have been developed in which the  physical basis is the non-relativistic
heavy quark phenomenology~\cite{the:Uraltsev 1997-1998 } or the
effects of gluon radiation~\cite{the:Catani et al. CMW scheme}.

\subsection{Principle of maximum conformality  \label{ss:PMC}}

The Principle of Maximum Conformality \cite{the:PMC, Brodsky:2012ik, the:CSR hold at any pQCD order} provides a rigorous 
generalization of the BLM procedure. In this method, one first generalizes the dimensional regularization by subtracting an extra constant $\delta$,
in addition to the $\mbox{ln} 4\pi - \gamma_E $ subtraction which defines the standard $\overline{MS}$ scheme.  This defines the $R_\delta$ scheme.
The coefficients of $\delta^n$ in  the resulting pQCD series uniquely  identify the  $\beta$ terms and the pattern of their occurrence at every order. 
The $R_\delta$ procedure thus systematically  identifies the non\emph{conformal} $\beta$ contributions to any perturbative QCD series.

All \emph{nonconformal} ({\it i.e.}, scale-dependent) terms
in the perturbative series describing an observable are  thus identified at each order 
and resummed in $\alpha^n_{s}$ by shifting its argument, 
thus allowing the automatic implementation of the BLM/\emph{PMC} procedure at all orders.
One thus obtains a perturbative
series that is maximally \emph{conformal} and thus scheme-independent.  The BLM/\emph{PMC} procedure restores the
original purpose of introducing a running coupling and resolves the
\emph{renormalon} and RS-dependence ambiguities.

The elimination of the \emph{renormalization scale} ambiguity greatly increases the precision, convergence, and reliability of pQCD predictions.
For example, \emph{PMC} scale-setting has been applied to the pQCD prediction for $t \bar t$ pair production at the LHC,  where subtle aspects of the \emph{renormalization scale} of the three-gluon vertex and multi-gluon amplitudes, as well as  large radiative corrections to heavy quarks at threshold, play a crucial role.  
The large discrepancy of pQCD predictions with  the $t \bar t$  forward--backward asymmetry measured at the Tevatron is significantly reduced from 
$3 \sigma$ to approximately $ 1 \sigma$.

In general, amplitudes of the same order may have different  \emph{renormalization scales}. 
For example, the $t$-channel and $u$-channel photon--exchange
amplitudes  appearing in M\''{o}ller scattering in QED at lowest order have scales 
$\mu^2 = t$ and $\mu^2 = u$, respectively, in the Gell-Mann
 Low scheme.  In addition, new scales will appear at  each higher order, reflecting 
 different momentum flow and gluon virtuality~\cite{Brodsky:1997vq}.
 The number of effective leptons $n_\ell$ in the QED $\beta$ function also changes. For
example, the  \emph{renormalization scales} of the two-photon exchange amplitudes in
M\''{o}ller scattering have a  smaller size than the scales of the
Born  amplitude since the two photons share the overall virtuality.

Distinctive  \emph{renormalization scales} have important phenomenological consequences
for pQCD, especially in processes which are sensitive to the interference
between contributing amplitudes such as the forward--backward heavy quark
asymmetries. In each case, the correct scales will be set automatically by
applying the PMC.

It should be emphasized that the conventional procedure of guessing a single
 \emph{renormalization scale} and its range gives pQCD results for physical
observables that depend on the renormalization scheme, contrary to the
principles of the renormalization group.  The conventional procedure is also
clearly incorrect for QED.   Varying the  \emph{renormalization scale} over a fixed
range does not give a reliable method for estimating uncertainties in a pQCD
expansion since the variation only exposes the nonconformal $\beta$-dependent
terms.

\section{Other optimization procedures \label{sub:Other optimization procedures}\label{sub:Optimized-Perturbation-Theory}}

We have discussed the BLM/\emph{PMC} prescription and \emph{Commensurate Scale Relations} in the previous
sections. Other optimization procedures exist. We briefly discuss
here the Fastest Apparent Convergence (FAC) principle, Principle of
Minimal Sensitivity (PMS) and Optimized Perturbation Theory (OPT).
The goal of these approaches is to devise an optimization procedure
that sets the value of the first ambiguously defined (RS-dependent)
parameter of the $\beta$-series ($\beta_{2}$). The 3-\emph{loop} perturbative
correction to a given observable (often unknown at the time when these
ideas were developed) is minimized by the optimization. This leads
to the removal of the \emph{Landau pole} and generally makes the coupling
observable dependent. 

The FAC principle \cite{the:Grunberg} is related
to \emph{effective charges} in the 
 sense that it fixes the scale $\mu$ so
that all perturbative coefficients beyond a given order are set to
zero.   However, the FAC procedure is not meant to be a scale-setting procedure.

The Principle of Minimal Sensitivity~\cite{the:OPT} assumes
that, since observables should be  RS-independent, their best approximations
obtained \emph{via} perturbation theory should be stable under small
RS variations. In practice, this is realized by imposing the independence
of $\alpha_{s}$ with respect to  $\Lambda$ and the other RS-dependent
parameters of a given perturbative series. However, as shown by Kramer 
and Lampe~\cite{Kramer 91}, the resulting PMS scales can be unphysical;  
for example, in the case of $e^+ e^- \to q \bar q g$ the PMS scale grows 
without bound when the gluon virtuality becomes soft.

Optimized Perturbation
Theory (OPT) is based on the PMS procedure \cite{the:OPT}. The convergence of the perturbative
expansion, Eq. (\ref{eq:pert. exp of an observable with coupling CONSTANT}),
truncated to a given order $n_{t}$, is enhanced by requesting its
independence from the choice of RS. The optimization is implemented by identifying
the RS-dependent parameters in the $n_{t}$-truncated series (the
$\beta_{n}$ for $2\leq n\leq n_{t}$ and  $\Lambda$), and requesting
that the partial derivative of the perturbative expansion of the observable
with respect to the RS-dependent parameters vanishes. Demanding RS-independence
modifies the series coefficients $r_{n}$ ($1\leq n\leq n_{t}$) and
the coupling $\alpha_{s}$ to {}``optimized'' values $\widetilde{r_{n}}$
and $\widetilde{\alpha_{s}}$.  At  first order the requirement of RS-independence
 imposes that Eq. (\ref{eq:pert. exp of an observable with coupling CONSTANT})
satisfies the Callan--Symanzik  equation (\ref{eq:Callan-Symanzik2}) with
$\beta$ truncated at order $n_{t}$. This implies that the perturbative
coefficients re-acquire a scale dependence: $r_{n}\left(Q^{2},\mu^{2},x_f \right)$.
This approach is based on a convergence criterion rather than physical
criteria as for \emph{effective charges} or the BLM procedure and its extensions,
or a simplicity criterion as for the $\overline{MS}$ scheme. In addition,
the OPT redefines $\alpha_{s}$ since part of the scale dependence
present in the usual definition is factored out of it and included
back in the $r_{n}\left(Q^{2},\mu^{2},x\right)$ coefficients.  In the context of 
the PMS/OPT, given the perturbative expansion of an observable
and of $\alpha_{s}$ at order $n_{t}$, one can assess whether their
combination is compatible with a small value of $\alpha_{s}$ at
low momentum transfer. If so, the perturbative calculations can then be improved
since the optimized coupling does not have a \emph{Landau pole} and so, its
growth at moderate momentum transfer is slower. This apparently allows one to extend the perturbative
series to lower energies \cite{the:OPT, Stevenson:2012ti}.  However, this procedure may 
hide situations where the physics of the subprocess requires consideration 
of nonperturbative dynamics \cite{Wu:2014iba}.  An example is $e^+ e^- \to b \bar b$,  where 
the scale of the final state gluon exchange is of order $v^2 m^2_b$ 
 and $v \to 0$ is the $b \bar b$ relative velocity. 
 
It should be noted that the BLM/\emph{PMC} scale-setting procedure automatically 
eliminates the dependence of the prediction on the choice of RS, 
so it automatically  achieves the goals set by the PMS and OPT \cite{Wu:2014iba}.  The resulting 
\emph{renormalization scales}  are always physical, reflecting the virtuality of the 
amplitude and, as in QED,  setting the number of active flavors $n_f$  appropriately  
at each order.  The dependence on the choice of  the initial scale $\mu_0$ is also 
minimized by the BLM/\emph{PMC} procedure.  

The \emph{PMC} method has now been applied to  many collider processes including 
multi-scale problems.  The results are independent of the renormalization scheme and the procedure
removes the problematic \emph{renormalon} growth of the  perturbative expansion, eliminates 
an unnecessary theoretical systematic error, and  gives increased  precision of the 
pQCD predictions.  This includes  pQCD predictions for LHC processes, such as  
Higgs and top quark production, thus greatly improving the sensitivity of LHC 
measurements to new physics.

\section{Determination of  the strong coupling $\alpha_{s}(M_Z^{2})$ or  the QCD scale  $\Lambda$\label{sub:Experimental-status}}

The coupling $\alpha_{s}$ in a convenient scheme, such as  the $\overline {MS}$ scheme,  can be extracted from a number of different measurements
involving hadronic reactions. Constraints  can also be obtained from lattice gauge theory calculations.
A comprehensive review of typical measures  is given in   Refs.
\cite{the:Workshop precis. meas., the:Workshop precis. meas. 2015}. The various determinations can
be compared with each other by either evolving them to a common scale,
typically the $Z^0$ mass $M_{Z}$, or by giving the value of  $\Lambda$
taking the appropriate  number of effective flavors $n_{f}$. There has been important progresses toward highly accurate
and precise determinations of $\alpha_{s}(M_Z^{2})$, as can be
judged from Fig. \ref{Flo:history}  which summarizes our knowledge
over the last 30 years as recorded by  the Particle Data Group \cite{the:PDG 2014}. The present
$Q^{2}$-range of the $\alpha_{s}$ determinations is approximately
$0.05<Q^{2}<10^{3}$ GeV$^{2}$, although only data above $Q^{2}$
greater than a few GeV$^{2}$ can be safely evolved with well-controlled 
perturbative equations. The large
range provides an essential check on the pQCD prediction for the running of the coupling,
Eq. (\ref{eq:alpha_s}),  a fundamental  check of   the theory.  In many cases,  the precision could be improved further by fixing the \emph{renormalization scale} using the \emph{PMC}  procedure.

\begin{figure}
\centering
\includegraphics[scale=0.45]{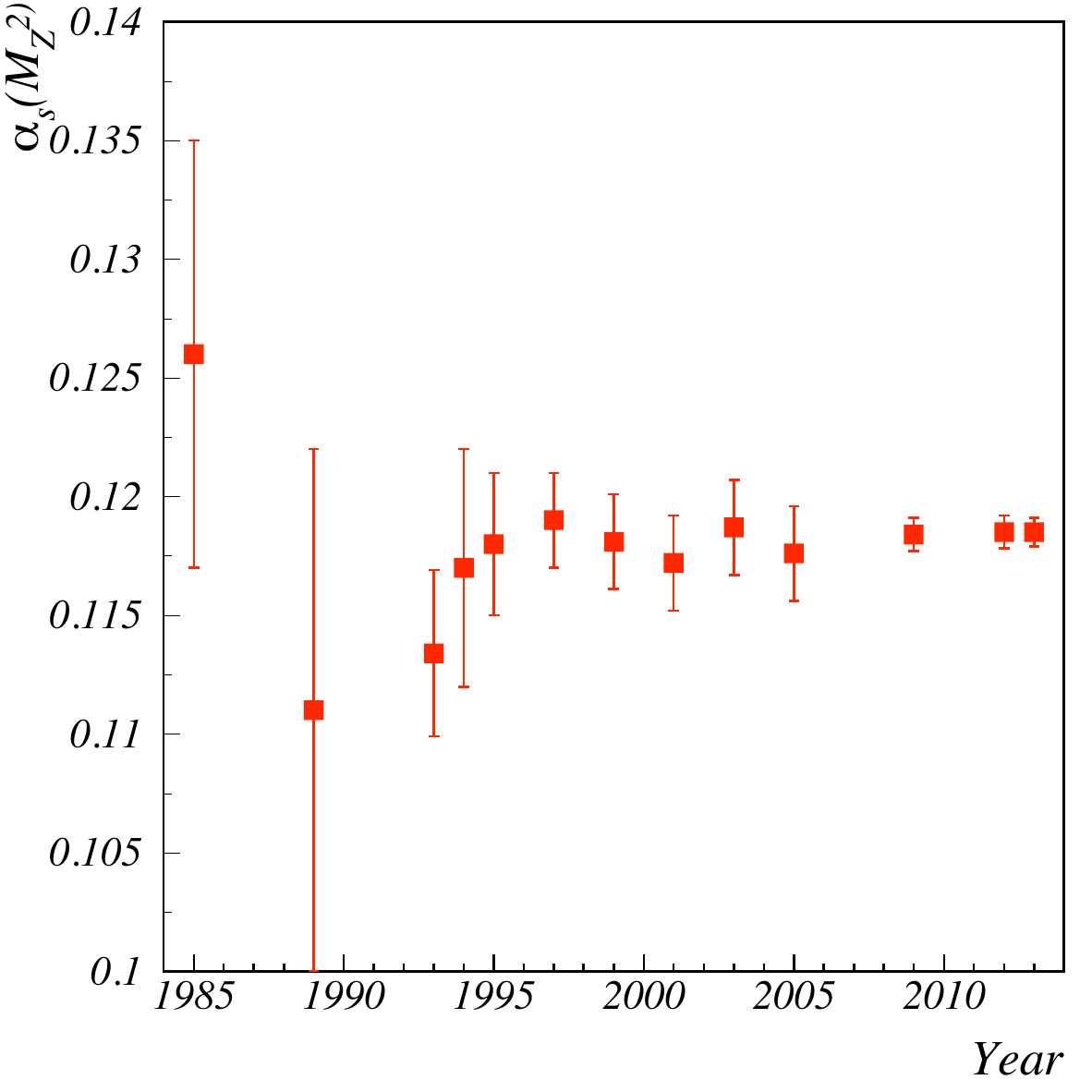}
\caption{\label{Flo:history} \small Evolution of the average world data for $\alpha_{s}(M_Z^{2})$
(Particle Data Group average \cite{the:PDG 2014}).}
\end{figure}

Several authors have compiled and compared the existing world data;  see, for example, the
recent Refs. \cite{the:PDG 2014, the:Altarelli
review, the:Pich, the:Erler}. The overall 
agreement between different determinations of $\alpha_{s}(M_Z^{2})$
within about 4\% provides an important consistency check of pQCD.
There are nevertheless tensions between various measurements which 
are presently attributed to the underestimation of uncertainties, rather than the
failure of QCD or effects from physics beyond the Standard Model. Typically, an
Unbiased Estimate is obtained by fitting the set of $\alpha_{s}(M_Z^{2})$
and scaling the individual uncertainties by an overall
factor, until the $\chi^{2}/ndf$ of the fit reaches unity
in order to account for underestimated uncertainties
and for possible correlations between different determinations of
$\alpha_{s}(M_Z^{2})$. For the most precise determinations,
theoretical uncertainties usually dominate.

We summarize in this section the most  effective methods for obtaining $\alpha_{s}(M_Z^{2})$.
In general, the conditions for a competitive extraction of $\alpha_{s}$ are a
precisely measured quantity, as inclusive as possible, with high
sensitivity to $\alpha_{s}$.  In the case of hadron production in
$e^{+}e^{-}$ collisions, QCD enters as a correction to the electroweak
process,  so the corresponding observables are only moderately sensitive to $\alpha_s$. In
contrast, hadronic decays of heavy mesons are directly proportional
to $\alpha_{s}^{3}$.   The perturbative expression of the quantity
must also be accurately known to high enough order, typically NNLO.   
The measurement must be done either at large $Q^{2}$ to suppress 
unknown nonperturbative $1/Q^{n}$ \emph{power corrections},
or at low $Q^2$ but with small nonperturbative corrections. 
The evolution
from low $Q^{2}$ to $M_{Z}^{2}$ then suppresses the overall uncertainties.

After describing the experimental status of $\alpha_{s}$,
we will discuss several non-ex\-per\-i\-men\-tal methods for predicting $\alpha_{s}$.
We then conclude by comparing some of the recent world data.

\subsection{Deep Inelastic Scattering 
 \label{sub:Deep-Inelastic-scattering}}

The basic process underlying deep inelastic lepton--hadron scattering  (DIS) is the elementary lepton--quark scattering process 
$\ell q \to \ell^\prime  q^\prime$.
The DIS (inclusive) data are sensitive to $\alpha_s$ through violations of Bjorken scaling  induced at LO by
gluon bremsstrahlung from the struck quark
as well as  the photon--gluon fusion and pair creation processes underlying 
the pQCD $Q^2$-evolution equations (Dokshitzer--Gribov--Lipatov--Altarelli--Parisi 
(DGLAP) equations \cite{the:DGLAP}).  At NLO,  the quark--photon vertex and the quark 
self-energy  corrections enter.
DIS data arguably provide the most robust way to obtain $\alpha_{s}(M_Z^{2})$ since 
the observables -- the nucleon unpolarized and polarized structure functions
$F_{2}(x_{Bj} ,Q^{2})$,  $g_{1}(x_{Bj} ,Q^{2})$ and $F_{3}(x_{Bj} ,Q^{2})$-- are fully
inclusive, and thus  have no uncertainties from final-state hadronic
corrections. 

The choice of the kinematic domain  where 
pQCD analyses have been  applied to DIS data is typically 
set at $Q^2 > $ 1 or 2 GeV$^2$ in order to  minimize \emph{higher-twist} power-law corrections arising from multiparton processes.   
One also limits the invariant mass $W>$ 2--4 GeV 
in order to exclude final-state high mass resonances, and one takes  $x_{Bj} >0.01$ to 
avoid low-$x_{Bj} $ resummation problems.  High-$x_{Bj} $ values are also  
excluded,  since \emph{higher-twist} subprocesses can become significant in this domain~\cite{the:LSS}.
The precision of the determinants of $\alpha_{s}(M_Z^{2})$ from DIS measurements
is at the percent level  due to the high precision of the measurements
of the unpolarized structure function $F_{2}(x_{Bj} ,Q^{2}$) --now
reaching 1\%  level over a wide kinematic  range. 
The precision of the strong coupling $\alpha_{s}(M_Z^{2})$
 in this case is limited by uncertainties 
of the gluon density distribution. 
The theoretical understanding underlying the evolution of  the polarized structure function $g_{1}(x_{Bj} ,Q^{2})$ is
 well developed, but the precision of the data needs further improvement. 

Precise measurements of $\alpha_{s}$ could in principle be obtained from the $Q^{2}$-dependences
of observables known to NNLO such as the Bjorken \cite{the Bj SR}
or the Gross--Llewellyn Smith (GLS)  sum rules~\cite{the:GLS sum rule}. The cleanest DIS observable for accessing $\alpha_{s}(M_Z^{2})$
is the Bjorken integral, $\int_{0}^{1}dx_{Bj} (g_{1}^{p}(x_{Bj} ,Q^2)-g_{1}^{n}(x_{Bj} ,Q^2))$.
As
such, the Bjorken integral $Q^{2}$-dependence is particularly simple
and known up to N$^{3}$LO, see Eq. (\ref{eq:Bj SR, order alpha^4}).
Since it is non-singlet in isospin, there is no gluon distribution
input,  and thus an absolute and rigorous pQCD prediction
exists (the Bjorken sum rule \cite{the Bj SR}). In practice, however,
the input data  lacks precision because one requires doubly polarized
measurements on both the neutron and proton. Furthermore, the most
precise measurements are at relatively low $Q^{2}$, $Q^{2}\leq5$
GeV$^{2}$, see \cite{Deur alpha_s from eg_1dvcs}. Finally,
there is an important uncertainty arising from the unmeasured low$-x_{Bj} $ part
of the integral. These caveats more than counterbalance the advantages
of the observable,  and thus the resulting $\alpha_{s}(M_Z^{2})$ is not
quite yet competitive with the best extractions of $\alpha_{s}(M_Z^{2})$.
The fit of the $Q^{2}$-dependence of the Bjorken integral was used
first in  Ref. \cite{the:Altarelli}. The latest measurements incorporate
recent DIS data from Jefferson Lab's Hall B,  yielding $\alpha_{s}(M_Z^{2})=0.1123\pm0.0061$
\cite{Deur alpha_s from eg_1dvcs}.  Since relatively low $Q^{2}$
data points are included, the nonperturbative \emph{higher-twist} correction
$\mu_{4}^{p-n}/Q^2$ to the $Q^{2}$-dependence of the Bjorken integral
has been accounted for with $\mu_{4}^{p-n}$ of the order of -0.02 GeV$^2$. 
The $\mu_{6}^{p-n}/Q^4$ and higher order \emph{power corrections} could be neglected.

The GLS sum rule \cite{the:GLS sum rule}, which relates the number of valence quarks
in the nucleon to its $F_{3}\left(x_{Bj} ,Q^{2}\right)$ structure function, offers the same 
advantages as the Bjorken sum rule. In the $\overline{MS}$ RS it reads:
\begin{multline}
\label{eq:GLS sum rule}
\int_{0}^{1}dx_{Bj} F_{3}(x_{Bj} ,Q^{2})=3\biggl[1-\frac{\alpha_{\overline{MS}}}{\pi}-3.58\left(\frac{\alpha_{\overline{MS}}}{\pi}\right)^{2} \\
-18.98\left(\frac{\alpha_{\overline{MS}}}{\pi}\right)^{3}+\mathcal{O}\left(\alpha_{\overline{MS}}^{4}\right) \biggr]+\mathcal{O}\left(1/Q^{2}\right)  ,
\end{multline}
with the same perturbative correction as the Bjorken sum rule, except
for a small contribution at order $\alpha_s^{3}$ 
from a light-by-light box-graph contribution. However,
the experimental difficulties in measuring the GLS sum are even greater
than for the Bjorken sum since it involves neutrino DIS measurements, 
and thus no precise determinations of $\alpha_{s}(M_Z^{2})$ from the GLS sum rule
exist at the moment. An example of extraction of $\alpha_s$ from a higher moment of $F_3$ can be found in \cite{Kataev:2001kk}.
The available data for the GLS sum rule come
from the CCFR measurement \cite{the:GLS CCFR}.

The $Q^2$-dependence of structure function moments can be used to  determine $\alpha_s$, even without 
a sum rule to anchor  its absolute magnitude.  Bernstein polynomials can be inserted
into the moments, to weight the integrand over the measured domain, thereby suppressing the uncertainties 
from the extrapolations to large and low $x_{Bj} $~\cite{the:Bernstein mom.}.

As we have noted, the most precise values of $\alpha_{s}(M_Z^{2})$ from DIS are at present obtained
from structure functions or parton distribution functions global fits.
Blumlein and collaborators \cite{the:Blumlein BBG and ABM} have
combined the world data for structure functions. The analysis at
NNLO, yields  $\alpha_{s}(M_Z^{2})=0.1133 \pm 0.0015$ (we have included
a theoretical uncertainty not quoted by the authors and estimated
from \cite{the:Jimenez-Delgado 2012}) or $\alpha_{s}(M_Z^{2})=0.1134\pm0.0020$,
if only non-singlet structure functions are included. Another analysis
from Jimenez-Delgado and Reya \cite{the:Jimenez-Delgado 2014},  which incorporates
most of the world data for parton distributions, leads to $\alpha_{s}(M_Z^{2})=0.1136\pm0.0014$.
The NNPDF collaboration has extracted $\alpha_{s}(M_Z^{2})=0.1191\pm0.0006$,
with a small systematic uncertainty estimated using neural
networks \cite{the:Workshop precis. meas., the:NNPDF}.

Jet production in DIS provides another observable for determining $\alpha_s$.  For example the production rate based on the subprocess $\gamma^* q \to q g$  is
directly proportional to $\alpha_s$. The phenomena underlying the production of a gluon jet are the same as 
the ones responsible for the DGLAP $Q^2$-dependence of the DIS structure functions: bremsstrahlung and
photon--gluon fusion. The appropriate kinematic domain is high $Q^2$ and large values of the mean
transverse energy of the two highest energy jets.   If one extracts $\alpha_s$ from the resulting scaling 
violations, the precision is limited by
the correlation between $\alpha_s$ and the gluon distribution.  An advantage of this
technique is that it allows one to directly measure the $Q^2$-dependence of $\alpha_s$.
A recent extraction from neutral-current DIS measurements of inclusive
jet, dijet and trijet cross-sections at HERA (H1 collaboration) compared
to the NLO expressions yields $\alpha_{s}(M_Z^{2})=0.1160\pm 0.0011$(exp) $\pm0.0032$(PDF, theory)
\cite{the:Andreev (H1) }.

\subsection{Observables from $e^{+}e^{-}$ collisions}

Observables from $e^{+}e^{-}$ collisions provide other inclusive processes to 
extract $\alpha_{s}(M_Z^{2})$.  At leading order, the  $e^{+}e^{-}$ pair annihilates into a virtual photon or a 
$Z^0$ which in turn decays into a quark--antiquark pair.  At the $Z^0$-pole
the observables are the $Z^0$ decay width
$\Gamma_{Z}$, the   ratio $R_{Z}$ of the $Z^0$ hadronic decay width normalized
to the leptonic width, and the hadronic and leptonic cross-sections.
The leptonic cross-section involves the $\Gamma_{Z}$ width and thus is sensitive
to $\alpha_{s}$.   Away from the pole, virtual photon production competes with $Z^0$ production. 
One can measure  the ratio of the hadron production to  lepton pair-production.
In each case, the sensitivity to $\alpha_{s}$ comes from $\Gamma_{Z}$ 
and its predominant hadronic decays.   The experimental
set-ups are distinct,  and thus yield results with largely independent
and uncorrelated experimental systematic uncertainties. Ratios
such as $R_{Z}$ and $R_{\tau}$ are experimentally
robust and thus have smaller systematic uncertainties. 
The sensitivity to $\alpha_s$ originates at LO from bremsstrahlung on the 
quark--antiquark lines into which the $Z^0$ or photon have decayed.  As  it is the case for DIS, vertex
and quark self-energy  corrections enter at NLO.
The pQCD corrections to the $e^{+}e^{-}$ collision observables
are usually known to NNLO or to N$^{3}$LO for $R_{Z}$ and $R_{\tau}$
\cite{the:Baikov R at NNNLO}. 
The uncertainties of the evolution equations
are small, and thus the  dominant uncertainty on the extraction of $\alpha_{s}$
is mostly experimental. 

The N$^{3}$LO expression for $R$ is:
\begin{multline} 
R(Q^{2})=R_{EW}\biggl(1+\frac{\alpha_{s}\left(Q^{2}\right)}{\pi}+
\left[1.9857-0.1153n_{f}\right]\frac{\alpha_{s}^{2}\left(Q^{2}\right)}{\pi}  \\+ 
\left[-6.6369- 1.2011 n_{f}-0.0052n_{f}^{2}-1.240\eta\right]\frac{\alpha_{s}^{3}\left(Q^{2}\right)}{\pi}  \\ + 
\left[-156.61-18.775n_{f}-0.7974n_{f}^{2}+0.0215n_{f}^{3}+\left(17.828-0.575n_{f}\right)\eta\right]
\frac{\alpha_{s}^{4}\left(Q^{2}\right)}{\pi} \\ 
+\mathcal{O}(\alpha_{s}^{5}\left(Q^{2}\right))+\delta_{NP}\biggr), \
\end{multline}
where $R_{EW}$ is the pure electroweak expectation, $\eta=(\Sigma e_{q})^{2}/(3\Sigma e_{q}^{2})$
 with $e_{q}$  the quark electric charges, and $\delta_{NP}$ includes
the  nonperturbative \emph{power corrections}.

A useful final state from $Z^0$ decay is a $\tau^+ \tau^-$ pair.  The $\tau$ can in turn undergo hadronic 
or leptonic decays, the ratio of which, $R_{\tau}$, allows the extraction of $\alpha_{s}$ 
at the low momentum scale $M_{\tau}^{2}=3.157$ GeV$^{2}$  \cite{the:Braaten a_s from tau decay}. This 
leads to an accurate $\alpha_{s}(M_Z^{2})$ determination, mostly because of the inclusiveness
of $R_{\tau}$ and because the evolution from $M_{\tau}^{2}$ to $M_{Z}^{2}$
suppresses the absolute experimental uncertainties by approximately
a factor $\alpha_{s}^{2}(M_{\tau}^{2})/\alpha_{s}^{2}(M_Z^{2})\simeq 7$.
However, low momentum transfer measurements require the control of 
higher order pQCD corrections as well as nonperturbative \emph{power corrections}.
For $R_{\tau}$, the nonperturbative corrections are suppressed as
$M_{\tau}^{6}$ and are estimated \cite{the:Le Diberder 1992}. However,
it is unclear that the perturbative corrections are determined properly
for the most precise extractions, since different approaches to the
perturbative series yield incompatible results, see {\it e.g.},   Refs. \cite{the:Workshop precis. meas.}
or \cite{the:Erler}. One pQCD expansion is done in the fixed order
perturbation theory (FOPT), while the other is done in the contour-improved
perturbation theory (CIPT). The pQCD expansion of $R_{\tau}$ is derived
from the vector and axial--vector current correlation functions. In
the $\overline{MS}$ scheme the pQCD approximant of $R_{\tau}$ is:
\begin{multline}
R_{\tau}=N_C\left|V_{ud}\right|^{2}S_{EW}(M_{\tau},M_{Z})\bigl[1+A_{1}+1.63982A_{2}+6.37101A_{3} \\
+49.07570A_{4}+\mathcal{O}(A_{5})+\delta_{NP}\bigr], 
\end{multline}
where $S_{EW}=1.01907\pm0.0003$ accounts for the electroweak radiative corrections,
$\delta_{NP}\simeq-0.006$ is the nonperturbative correction
and $A_{n}$ are contour integrals depending, at LO, on $\alpha_{s}$
only: 
\begin{equation}
A_{n}=\frac{1}{2\pi i}\oint_{\left|s\right|=M_{\tau}^{2}}\frac{ds}{s}\left(\frac{\alpha_{s}(s)}{\pi}\right)^{n}\left(1-2\frac{s}{M_{\tau}^{2}}+2\frac{s^{3}}{M_{\tau}^{6}}-2\frac{s^{4}}{M_{\tau}^{8}}\right).
\end{equation}
FOPT expands the $A_{n}$ in $\alpha_{s}$, $A_{n}=\alpha_{s}^{n}(M_{\tau}^{2})/\pi+\mathcal{O}(\alpha_{s}^{n+1}(M_{\tau}^{2})/\pi)$,
while CIPT uses a numerical approach to keep the $A_{n}$ unexpanded;  {\it i.e.}, 
the higher orders in $\alpha_{s}$ are resummed compared to the
FOPT perturbative series. The FOPT expansion yields:
\begin{multline}
R_{\tau}=N_C\left|V_{ud}\right|^{2}S_{EW}(M_{\tau},M_{Z})\biggl[1+\frac{\alpha_{s}(M_{\tau}^{2})}{\pi}+5.202\frac{\alpha_{s}^{2}(M_{\tau}^{2})}{\pi}+26.37\frac{\alpha_{s}^{3}(M_{\tau}^{2})}{\pi} \\ 
+127.1\frac{\alpha_{s}^{4}(M_{\tau}^{2})}{\pi}+\mathcal{O}(A_{5})+\delta_{NP}\biggr]. 
\end{multline}
The question of which expansion is preferable has not  yet been settled.
It has been argued that the FOPT uncertainties have been underestimated,
and if they are properly calculated, the two techniques will agree --see the contribution of S. Menke
to  Ref. \cite{the:Workshop precis. meas.}. The latest analysis
of $\tau$-decay data, using RG-improved FOPT expansion, yields $\alpha_{s}(M_{\tau}^{2})=0.3189_{-0.0151}^{+0.0173}$,
corresponding to $\alpha_{s}(M_Z^{2})=0.1184_{-0.0018}^{+0.0021}$ \cite{the:Abbas tau decay}.
The experimental data are from the ALEPH \cite{the:Aleph R_tau}
and OPAL \cite{the:OPAL R_tau} experiments at LEP (CERN) . 

Observables based on jet shapes can also be used  to access $\alpha_{s}(M_Z^{2})$.
These observables are less inclusive but are sensitive to $\alpha_s$ at 
leading order.  Event shapes measure the departure of the momentum flow
in an event from that of the 2-body $q\overline{q}$ configuration.
Extractions are carried out at NNLO.  Uncertainties from nonperturbative
hadronization processes dominate the total uncertainty. The shape
observable that has provided the most precise determination of the coupling
is the ``thrust'', $T$.  It measures the alignment of the produced particles with respect 
to the thrust axis, defined as the axis on which the projected momenta of the produced particles
is maximal. Thrust varies between $0.5<T<1$. The low values  characterize the 
3-jet region where pQCD is applicable, and high values  characterize the 2-jet region
where nonperturbative effects due to soft/collinear gluons are important.     
Recently, Gehrmann, Luisoni and Monni~\cite{the:Gehrmann 2013}
used thrust data from the TASSO experiment at PETRA (DESY) \cite{the:TASSO}
and the ALEPH \cite{the:ALEPH} and L3 \cite{the:L3} experiments
at LEP to obtain $\alpha_{Rgap}(M_Z^{2})=0.1131_{-0.0022}^{+0.0028}$
and $\alpha_{Rgap}(2\mbox{ GeV})=0.538_{-0.047}^{+0.102}$ (both in
the $Rgap-$scheme).  Another recent thrust result, from Abbate et
\emph{al}. \cite{the:Abbate}, includes additional experimental data from JADE (PETRA) \cite{the:JADE},
OPAL \cite{the:OPAL}, DELPHI \cite{the:DELPHI} (LEP) and AMY \cite{the:AMY}
(at TRISTAN). It yields $\alpha_{s}(M_Z^{2})=0.1135\pm0.0025$
($\overline{MS}$ scheme), in good agreement with an earlier more
precise extraction by the same group, $\alpha_{s}(M_Z^{2})=0.1135\pm0.0010$.
Another very recent  thrust
result~\cite{the:Hoang}  (still unpublished) yields $\alpha_{Rgap}(M_Z^{2})=0.1128\pm0.0012$, using ALEPH,
DELPHI, JADE, OPAL and SLD (SLAC) \cite{the:SLD} data. The same
group also provides another determination using a different event-shape
observable, the C-parameter distribution: $\alpha_{Rgap}(M_Z^{2})=0.1123\pm0.0015$
(both results are in the  $Rgap$ scheme). 

Another recent determination is from the analysis of the $Q^{2}$-evolution
of the average gluon and quark jet multiplicities using a recently
improved formalism. This analysis yields $\alpha_{s}(M_Z^{2})=0.1199\pm0.0026$
\cite{the:Bolzoni 2013}.  A global electroweak fit of weak decay
data by Erler and Freitas \cite{the:PDG 2014}, yields $\alpha_{s}(M_Z^{2})=0.1192\pm0.0027$
with, noticeably, a negligible theoretical contribution to the
uncertainty.

\subsection{Observables from $pp$ collisions}

Most of the determinations of $\alpha_{s}$ from hadronic collisions
have been  limited to NLO.  However, the CMS collaboration at  LHC has
recently determined $\alpha_{s}$ from the inclusive cross section for top-quark pair
$t\overline{t}$ production based on a  NNLO analysis that is constrained by 
PDF inputs. This analysis yields $\alpha_{s}(M_Z^{2})=0.1185_{-0.0042}^{+0.0063}$
\cite{the:CMS t-tbar}. 

Another important  result  of the high energy hadron
colliders, the LHC and the Tevatron, is the experimental verification
of the running of $\alpha_s(Q^2)$  at very large momentum transfers of order $1$
TeV. 

Inclusive jet production from $p\overline{p}$ collisions is 
proportional to $\alpha_s^2(Q^2)$ at LO,  and it provides 
constraints  on the  global PDF fits of DIS and hard scattering data. The
MSTW combined fit of DIS and jet data yields $\alpha_{s}(M_Z^{2})=0.1180\pm0.0014$ at NLO
\cite{the:MSTW}.

\subsection{Lattice QCD \label{sub: large Q Lattice-QCD}}

Lattice QCD can provide accurate determinations of $\alpha_{s}(M_Z^{2})$  at the 1\% level, although it is not certain that all uncertainties are understood,
see  {\it e.g.}, Refs.~\cite{the:Altarelli review} or \cite{the:Workshop precis. meas.}. 
Consequently, attention has focused recently on a better  understanding
and control of the lattice systematic uncertainties.  The systematic
uncertainties quoted in the older calculations (10 years ago or
earlier) are most likely underestimated and not well controlled. 
Recent \emph{unquenched} calculations explicitly include quark loops up to
2+1 or 2+1+1 quark flavors. This is adequate for the momentum transfer range
at which the calculations are performed (2+1 means that one of the
sea quark masses is set to the strange quark mass, while the two others
are taken as small as practically possible. For 2+1+1, the charm quark
mass is added).  

A recent exhaustive
review of the lattice results extracting $\alpha_{s}$ in the pQCD
domain can be found in \cite{the:Aoki review lattice alpha_s}. It
also provides a compilation of the lattice results, yielding an average value for the
coupling $\alpha_{s}(M_Z^{2})=0.1184\pm0.0012$. 
In this Section, we first outline the method, its benefits and its limitations. We will then report on the most 
recent and most accurate determinations of $\alpha_{s}$
in the UV domain.

\paragraph{Lattice calculation technique}

Lattice calculations use the path integral formalism \cite{Path integral}. Path integrals
provide the probability of evolving from an initial state $\left|x_{i}\right\rangle $
to a final state $\left|x_{f}\right\rangle $, summing over all possible
space-time trajectories. The integral is weighted by a factor depending
on the system's action, $S$. For example for a one--dimensional system
the propagator is given by:
\begin{equation}
\label{eq:1dprop}
\left\langle x_{f}\right|e^{-iHt}\left|x_{i}\right\rangle =\int Dx(t)e^{-iS\left[x(t)\right]/\hbar},
\end{equation}
where $\int Dx$ symbolizes the integration over all  paths for
which $x(t_{f})=x_{f}$ and $x(t_{i})=x_{i}$. In Eq.~(\ref{eq:1dprop}), we exhibit the dependence on
$\hbar$ explicitly in order to underline the link between path integrals (quantum
description) and the principles of Fermat/Maupertuis (the least action
principle), which specify that the classical path ($\hbar\rightarrow 0$)
must yield the smallest value of $S$. The fact that $\hbar\neq 0$ allows excursions
outside the classical path and is responsible for the quantum effects.

Path integrals are rarely solved analytically. They are also difficult to determine
numerically since, for a 4--dimensional space, one requires an $n$--dimensional integration,
with $n$ = 4$\times$(possible number of paths). Since the number of possible
paths is infinite, one must 
restrict the number of paths to a representative sample. This integration
is then carried on a finite sample. The most
efficient technique for such numerical integrations is to use the Monte
Carlo method. It is also more efficient to work in Euclidean 
rather than in Minkowski space.  After a Wick rotation
$it\rightarrow t$ \cite{Wick rot.}, the weighting factor becomes $e^{-S_{E}}$
which is easier to process than the oscillating function $e^{-iS}$.
Here, $S_{E}$ denotes the Euclidean action.  Using  these methods,  one 
can simply calculate correlation functions such as
\begin{equation}
\left\langle A_{1}\ldots A_{n}\right\rangle =\frac{\int Dx\mbox{ }A_{1}\ldots A_{n}e^{-S_{E}}}{\int Dx\mbox{ }e^{-S_{E}}},\label{eq:n-point function}
\end{equation}
where  $A_i$ is the gauge field at position $x_i$.  In particular, the two-point correlation function at $\left\langle x_{1}x_{2}\right\rangle $
represents the propagator of a boson.  As soon as interacting
fields are involved, no methods are known  to directly solve Eq. (\ref{eq:n-point function}) 
analytically.    However, if the strength of the interaction
is sufficiently weak, one can analytically evaluate the Gaussian
integrals by expanding the exponential involving the interaction term
({\it e.g.} pQCD or QED).  However, if the coupling is large, the integrals must be treated numerically.
The numerical technique used is as  follows: space is discretized (approximated
by a lattice) and the paths linking the different sites (nodes of discretized
space-time) are generated. The statistical precision depends on the
square root of the number of generated paths.  A path is generated according
to the probability $e^{-S_{E}}$, where $S_{E}$ is calculated for
that particular path. A correlation function can then be calculated by
summing the integrand over all paths.  Since paths are generated with the probability $e^{-S_{E}}$, this simple sum is equal to the weighted
sum $\sum_{path} x_{1} \ldots x_{n} e^{-S_E} \simeq \int Dx ~ x_1 \ldots x_n e^{-S_E}$.

The Monte Carlo technique can be used to generate paths with the appropriate
weight~\cite{metropolis}.  The procedure begins with a given path of 
action  $S_{1}$. The path is randomly changed to a new path of
action $S_{2}$ (several intermediate paths, which are not retained,
are  generated before producing  the $S_{2}$ path in order to avoid correlations between 
the $S_{1}$ and $S_{2}$ paths).  If $S_{2}<S_{1}$
the $S_{2}$ path is retained in the sample. Otherwise, it is retained
or rejected with probability $S_{2}-S_{1}$.

In order to ensure gauge invariance, a link involving the gauge field $A$
between the lattice sites must be introduced \cite{Wilson lattice}.
A link variable $U_{\overrightarrow{\mu}}=\mbox{exp}(-i\int_{x}^{x+a\overrightarrow{\mu}}dy\mbox{ }gA)$
is constructed, where $x$ is a lattice site, $a$ is the lattice
spacing, $\overrightarrow{\mu}$ is an elementary vector of the Euclidean
space, and $g$ the bare coupling.   The link variable $U_{\overrightarrow{\mu}}$ is explicitly
gauge-invariant. The action is then constructed using these 
gauge-invariant variables. The \emph{Wilson loop} $U_{1}\ldots U_{n}$
thus appears \cite{Wilson lattice}, where the closed path is given by the links $U_{i}$. It
is straightforward to show that the action can be expressed as a sum of
 \emph{Wilson loops}. In the continuum limit $a\rightarrow 0$, the simplest
loop (a square of side $a$) dominates. However, since $a\neq 0$ in the numerical
simulations, larger loops must be introduced as corrections to the
discretized expression of $S_{E}$. 

We have discussed only $A$ (i.e. gluons)
so far. The introduction of non-static quarks on the lattice is complicated
because of their fermionic nature. Including a fermion field
leads to the notorious fermion doubling problem, which multiplies the number of fermionic degrees of freedom, 
and introduces spurious particles. There are several possible methods 
which avoid this problem. One method is to break chiral symmetry (the Ginsparg--Wilson
approach \cite{Ginsparg Wilson}). Another method, called ``staggered fermions''  introduces 
non-local operators which respect chiral symmetry \cite{Staggered fermions}. Other methods also exist.
Each of these methods significantly slow down the lattice computations.  Once
fermions are included, the action becomes $S_{E}=S_{A}-\mbox{ln}\left(\mbox{Det}(K)\right)$
where $S_{A}$ comes from the gluon field and $K$ is similar to the
Dirac equation operator. Most early Lattice calculations, 
and some recent ones, simplify the calculations by taking $\mbox{Det}(K)=1$
(the \emph{quenched} approximation). This amounts to neglecting the dynamics
of fermions  which eliminates the effects of pair creation from the
QCD instant-time vacuum.

The lattice technique, although very powerful, has  its own limitations:

\paragraph{Critical slowing down} This phenomenon limits
the statistical precision. The problem, which is not specific to lattice
calculations, arises from the fact that, to keep discretization errors under
control,  the lattice spacing $a$ must be much smaller
than the characteristic sizes of the studied phenomena. The relevant physical measure 
is the correlation length $L_{c}$ defined by
$\left\langle x_{1}x_{2}\right\rangle \sim e^{-x/L_{c}}$. In general
$L_{c}$ is very small, except near a critical point. Therefore, calculations
need to be performed near such point. However, when $L_{c}$ is large,
many intermediate paths must be generated to obtain a path decorrelated
from the initial path. In the case of QCD, the statistical precision
varies as $\left(\frac{L_{R}}{a}\right)^{4}\left(\frac{1}{a}\frac{1}{m_{\pi}^{2}a}\right)$
where $L_{R}$ is the lattice size \cite{the:Lepage Lattice lectures}.  The first factor stems from the
number of sites and the second comes from the critical slowing down
(note the presence of the pion mass squared).

\paragraph{Extrapolation to the physical mass of the pion}

The lattice parameters can be chosen such that the mass of the pion is greater 
than its physical value, which minimizes the critical slowing down. 
This competes with the necessity to have 
calculations  with the pion mass as close as possible from the physical
mass. The lattice results are extrapolated to the physical pion mass
with guidance from Chiral Perturbation Theory~\cite{Bernard:2006gx}.  Nevertheless, an uncertainty
remains associated with this extrapolation.

\paragraph{Local operators} Local operators are well suited
for lattice calculations because selecting a given path involves calculating
the associated  difference of actions $S_{2}-S_{1}$. For a local action, this amounts
to computing $S_{2}-S_{1}$ only on one site and its neighbors
(since $S$ involves derivatives).  In four dimensions this represents 9
operations; in contrast if the action is not local, the calculation
needs  to be carried at each site of the lattice. This makes direct lattice
calculations of non-local operators impractical. For example, structure
functions are non-local, and thus they must be reconstructed on the lattice through 
their moments.

\paragraph{High momentum cut off} Momenta involved in lattice
calculations are automatically limited to $p\lesssim1/a$. This is not
a problem in practice if one can match to available pQCD calculations. 
The domain where
pQCD and lattice calculations are both valid allows one to establish the
renormalization procedure for the lattice calculations.

\paragraph{Finite lattice size} The lattice spacing $a$
must be chosen to be sufficiently small in order to reach the pQCD domain while keeping the
number of sites to a practical value for computation. This limits the total lattice
size.  On the other hand,  the lattice size must be taken large enough to encompass the physical system
and to minimize boundary effects.
$\\$

Severals approaches have been used to obtain $\alpha_{s}(M_Z^{2})$
using lattice gauge theory.
A first approach is to consider a short-distance quantity whose pQCD
prediction is known to high order, and compute it on the lattice  nonperturbatively.
The result is compared to the pQCD prediction in which the value of
$\alpha_{s}$ is adjusted so that there is a good match. The dominant
uncertainty is usually from the perturbative series truncation. 
Lattice finite size effects are suppressed when short-distance quantities
are computed. Space discretization errors are minimized by the use of improved
actions, such as \emph{tadpole} improved actions~\cite{the:Lepage Lattice lectures}.
A second lattice approach consists of direct calculations; {\it e.g.}, 
the computation of a QCD vertex.

\paragraph{The coupling from short-distance quantities}

A natural quantity to study on the lattice which determines $\alpha_s$ is
 the static $\mbox{Q--}\overline{\mbox{Q}}$ potential at short distances \cite{the:Michael Lattice stat. pot.}.
This quantity will be discussed in more detail in Section \ref{sub:Potential-approach}.
A recent review of the determination of $\alpha_{s}(M_Z^{2})$ using
this method is given in \cite{the:Garcia. Review alpha_s from stat. energy}.

The energy between a static quark and a static anti-quark separated
by a distance $r$ is calculated on the lattice. Unquenched results
at short distances are then compared to the pQCD one-gluon static potential $V(r)=-4\alpha_{R}(r)/3r$
allowing for an additional nonperturbative linear contribution $\sigma r$.   The coupling in coordinate space
$\alpha_{R}(r)$ is then Fourier transformed to $\alpha_{V}(Q^2)$.

At short distance, the 2-\emph{loop} expression is \cite{the:Michael Lattice stat. pot.}:
\begin{equation}
\label{eq:alpha_R(r)}
\alpha_R(r)=\frac{6 \pi}{\beta_0 \mbox{ln}\left(\frac{1}{r^2 \Lambda^2_R}\right)+\frac{4 \pi \beta_1}{\beta_0} \mbox{ln}\left(\mbox{ln}(\frac{1}{r^2 \Lambda^2_R})\right)}.
\end{equation}
The QCD parameter  $\Lambda_R$ in the $R$-scheme is given in Table \ref{Flo:Table of Lambda} for $n_f=3$. In the \emph{pure gauge} case, $n_f=0$, it is $\Lambda^0_R=0.70$ GeV. The
most recent calculation yields a coupling $\alpha_{s}=0.335_{-0.010}^{+0.012}$
at $Q^{2}=1.5$ GeV$^{2}$ for $n_{f}=2+1$, which is evolved  to 
$\alpha_{s}(M_Z^{2})=0.1166_{-0.0008}^{+0.0012}$ for $n_{f}=5$
\cite{the:Bazavov static pot}.  The  uncertainties
shrink by an order of magnitude when evolving from $Q^{2}=1.5$ GeV$^2$ to
$M_{Z}^{2}$. For the quoted results, the light quark masses are set
close to their physical values (5 MeV for $u$ and $d$) or at their
physical value (for $s$). Results are stable under variation of the value of the lattice
spacing $a$, indicating a negligible discretization error. Another recent calculation, from the ETMC collaboration,
was performed for $n_f=2$ and yields $\Lambda_{\overline{MS}}^{(2)}=0.332 \pm 0.0021$ \cite{the:ETMC1214}, in good agreement with
the world average.  

Another lattice approach is to evaluate the temporal n$^{\mbox{\small{th}}}-$moments
$G_{n}$ of current-current correlators $\left\langle 0\right|j_{5}(x,t)j_{5}(0,0)\left|0\right\rangle $
for the heavy quark pseudoscalar current $j_{5}=\overline{\psi}\gamma_{5}\psi$
\cite{the:McNeile Lattice c-c correlators}. As for the previous
method, the currents are calculated on the lattice and compared to
their pQCD expressions. The lattice results are fit using the pQCD
functional form with several fit parameters, including $\alpha_{\overline{MS}}$.
The pQCD expression is known to NNLO.  For example for $n=4$, $G_{4}=G_{4}^{LO}[1+0.7427\alpha_{\overline{MS}}(\mu)+0.0088\alpha_{\overline{MS}}^{2}(\mu)-0.0296\alpha_{\overline{MS}}^{3}(\mu)][1+\mathcal{O}(\Lambda^{4}/m^{4})]$,
where $m$ is the heavy quark mass. Higher pQCD orders are not known
and are treated  as free fit parameters up to $N^{14}LO$. This
constrains the uncertainty on the truncation of the pQCD series. The
 nonperturbative terms are assumed to be proportional to gluon and meson \emph{condensates}
which are suppressed as $(\Lambda/m)^{4}$.  Two results were
obtained by the HPQCD collaboration. In the first case, the light quarks $u$,
$d$ and $s$ were treated nonperturbatively,  taking  the heavy quarks
as perturbative effects. This $n_{f}=2+1$ calculation
yields $\alpha_{s}(25\mbox{ GeV}^{2})=0.2034\pm0.0021$. Evolved to
the $Z^0$ mass and corrected to $n_{f}=5$, it gives $\alpha_{s}(M_Z^{2})=0.1183\pm0.0007$.
A second calculation was done with the $c$ quark also treated nonperturbatively.
This analysis had improved statistics, as well as  improved determinations of the gluon action and
the lattice spacing. The $n_{f}=2+1+1$ calculation
yields $\alpha_{s}(25\mbox{ GeV}^{2})=0.2128\pm0.0025$, which leads
to  $\alpha_{s}(M_Z^{2})=0.11822\pm0.00074$
($n_{f}=5$), confirming the earlier result, including the assumption that heavy quarks can
be treated perturbatively. 
Calculations are done for several lattice
spacings $a$, varying between $0.06\leq a\leq0.12$ fm, and extrapolated
to the continuum case. The value of $a$ is measured by calculating
the dimensionless Wilson flow parameter $w_{0}/a$ and comparing it to
its dimensionful value $w_{0}$ known from an earlier simulation.
The light quark masses are chosen relatively close to their physical
values and corrections for the finite lattice size effects are included.  The JLQCD collaboration
also used current-current correlators, but  computed the vacuum polarization
function $\Pi(Q^{2})$ \cite{the:Shintani lattice} rather than $G_n$ moments. The vector and
axial--vector currents are both used. The N$^{3}$LO pQCD expression is
complemented by nonperturbative contributions up to $1/Q^{4}$.
The calculation yields $\alpha_{s}(M_Z^{2})=0.1181\pm0.0013$. The
chosen $u$ and $d$ quark masses range between 20 and 80 MeV. The
$s$ quark mass ranges between 95 and 125 MeV. Discretization effects are
estimated using lattice perturbation theory.   The effects of the finite lattice size are found to be
small.

The coupling $\alpha_{s}$ has also been extracted from the vacuum
expectation values of  \emph{Wilson loops} $W_{mn}$ --see  Ref. \cite{the:Mason Lattice wilson loop}
for a recent determination. The indices $m$ and $n$ characterize
the loop which forms a rectangle of size $ma\times na$. The loop definition
is $W_{mn}\equiv\left\langle 0\right|\mbox{Re Tr P}e^{-ig\oint_{nm}Adx}\left|0\right\rangle /3$,
with $g$ defined as $\overline{\alpha_{s}}=g^{2}/4\pi$ and P is the
path ordering operator.
The corresponding pQCD expression for a  flat $2a\times2a$ loop is:
$W_{22}^{pQCD}=\mbox{exp}[-9.20\alpha_{V}(2.582/a)+6.37\alpha_{V}^{2}(2.582/a)-17.11\alpha_{V}^{3}(2.582/a)+...]$.
In \cite{the:Mason Lattice wilson loop} calculations are done for
various loop sizes and flat and non-flat loops, for a total of 22
different loops. The value of $\alpha_{V}$ is adjusted so that 
the NNLO pQCD expectation matches the lattice result. The averaging
of the 22 different determinations yields $\alpha_{V}(56.25\mbox{ GeV}^{2})=0.2120\pm0.0028$.
The conversion from the $V$-scheme to $\overline{MS}$ and evolution
to the $Z^0$ mass lead to $\alpha_{s}(M_Z^{2})=0.1184\pm0.0008$.
The physical parameters of the simulation (the bare coupling constant
and bare quark masses) are tuned so that the calculation reproduces the
known experimental values of the $Y-Y'$ meson mass difference, of $m_{\pi}$, of
$2m_{K}^{2}-m_{\pi}^{2}$, of $m_{\eta_{c}}$ and of $m_{\Upsilon}$. Several
values of lattice spacing are used to extrapolate to the continuum
case.  In addition, the calculations are done for several values of the
light quark masses in order to reliably extrapolate to the physical case. The heavy quarks ($c$ and $b$) are treated
perturbatively. Finite lattice size effects are accounted for, as well as the
truncation error on the pQCD series. Earlier high precision calculations
using  \emph{Wilson loops} by Maltman \emph{et al.} \cite{the:Maltman Wilson loop}
and by the SESAM collaboration \cite{the:Spitz Wilson Loop} yield $\alpha_{s}(M_Z^{2})=0.1192\pm0.0011$
and $\alpha_{s}(M_Z^{2})=0.1118\pm0.0017$, respectively.

\paragraph{Vertex calculations of $\alpha_{s}$}

As will be discussed in more detail in Sections \ref{sub:Schwinger--Dyson-formalism}
and \ref{sub:Low Q Lattice-QCD-calculations}, the running QCD coupling $\alpha_{s}$ can
be computed from the ghost--gluon vertex, the quark--gluon
vertex or the multi-gluon vertices. Vertices are not observable and their calculations are
gauge-dependent. Most of the computations are done in the Landau gauge
and in a MOM RS. A recent calculation
by the ETM collaboration using the ghost--gluon vertex, see Eq. (\ref{eq:alpha_s SDE ghost--gluon}), yields
$\alpha_{s}(M_Z^{2})=0.1196\pm0.0014$~\cite{the:Blossier gh-gh-g vertex UV}. 

Another approach is to use the Schr\''{o}dinger functional \cite{the:Alpha lattice collaboration}.
There are no high precision ($\Delta\alpha_{s} \lesssim 0.002$) results or calculations with $n_{f}\geq3$ yet, although some should become available soon \cite{the:Sommer Lattice}.

\subsection{Heavy quarkonia}

Hadronic inclusive decay rates of heavy quarkonium systems
are very sensitive to the value of $\alpha_{s}$: for example,  the LO hadronic decay
rates of the $J^{PC} = 1^{--}$ bound states are proportional to $\alpha_{s}^{3}$ if the Zweig rule is operative. 
The dependence on the
not-well-known quarkonium wave function  is eliminated by considering the
ratio of hadronic to  leptonic decays. However, pQCD quarkonium decay rate expressions
are known only to NLO and thus, the extractions have significant theoretical
uncertainties. There is no  recent extraction of $\alpha_{s}$
using this method.   The latest one, from 2007, is $\alpha_{s}(M_{\Upsilon(1S)}^{2})=0.184{}_{-0.014}^{+0.015}$,
which is evolved to $\alpha_{s}(M_Z^{2})=0.119{}_{-0.005}^{+0.006}$
\cite{the:Brambilla heavy quarkonia decay (2007)}.  It is extracted
by comparing the measured $\Gamma(\Upsilon(1S)\rightarrow\gamma X)/\Gamma(\Upsilon(1S)\rightarrow X)$
decay ratio to non-relativistic QCD calculations (here, $X$ denotes final hadron states). 
The most precise rate measurement is provided by
CLEO~\cite{the:CLEO}.

\subsection{Holographic QCD\label{sub:Holographic-QCD large Q}}

Light-Front Holographic QCD~\cite{Brodsky:2006uqa, deTeramond:2008ht} originates from the direct connection of QCD quantized on
Dirac's light-front dynamics in our physical  {3+1}--dimensional space-time~\cite{Dirac},
to Einstein's gravity  in a 5--dimensional Anti-de Sitter.
(AdS) space-time. (An AdS space is the maximal symmetric space with 
constant negative curvature). The connection is based on the AdS/CFT correspondence
\cite{the:Maldacena duality}, where CFT stands for \emph{conformal}  field
theory, that is a theory without explicit scale dependence. 
The correspondence implies that a weakly interacting, gravity-like,
theory in $d+1$--dimensional AdS space can be mapped on the ($d$--dimensional) AdS space boundary
to a strongly interacting \emph{conformal} field theory in $d$--dimensions, 
thus also the name gauge/gravity correspondence.

 The holographic mapping to light-front physics  gives a relation between the fifth dimension holographic variable  of AdS space and the invariant impact light-front variable  in physical space-time~\cite{Brodsky:2006uqa, deTeramond:2008ht}.  Light-front holography  provides a precise relation between the boost-invariant light-front wavefunctions describing the internal structure of hadrons in physical space-time and the bound-state amplitudes in AdS space.  This connection also implies that the light-front  confining potential corresponds to an infrared distortion of AdS the  space,  which breaks the conformal invariance. The 
resulting valence Fock-state wavefunctions  of the light-front QCD Hamiltonian 
satisfy a relativistic equation of motion with an effective confining potential 
which incorporates the contribution from higher Fock-states. Holographic QCD gives 
a very good description of hadrons of arbitrary spin and incorporates many of
their observed spectroscopic and dynamical features~\cite{the:AdS/QCD review}.
The AdS/CFT correspondence offers new tools to analytically describe
the strong interaction at low-$Q^{2}$. This technique can then be used
to compute  $\Lambda$ and in turn $\alpha_s(M_Z)$.
 The light-front holographic approach to hadronic physics and its recent connection with superconformal quantum mechanics is described in more detail in Sec. \ref{sub:Holographic-QCD Lowq}.  We will also show in Sec. \ref{sub:Holographic-QCD Lowq} how the running of the strong coupling in the infrared can be obtained from
AdS gravity.

 Following Sec.\ref{sub:Holographic-QCD Lowq}, one can show how the mass scale underlying confinement and 
hadron masses  determines the scale controlling the evolution of the 
perturbative QCD coupling~\cite{the:Deur Lambda AdS/QCD}.  The 
relation between scales is obtained by matching the nonperturbative 
dynamics, as described by an effective  light-front theory embedded in AdS space, to the perturbative QCD  regime  
computed to four-\emph{loop} order.  While the AdS/QCD description is valid only in the  nonperturbative
QCD regime, one can actually match it to the perturbative regime thanks to the existence of an overlap between both regimes, 
called the parton--hadron duality~\cite{the:Bloom Gilman duality, 
the: H-P duality review}. The Holographic QCD predictions
for $\alpha_{s}$ and its \emph{$\beta$-function} are equated to their pQCD
counterparts at a transition scale, which is in turn determined by 
the matching procedure. One thus derives a running QCD coupling  
$\alpha_s(Q^2)$, defined at all momenta, which is consistent with the 
measured effective charge defined from the Bjorken sum rule and with
the measured perturbative scale $\Lambda_{\overline{MS}}$. At order $\beta_{0}$
and in the $\overline{MS}$ scheme, the relation is:
\begin{equation}
\Lambda_{\overline{MS}}^{(LO)}=M_{\rho}e^{-a}/\sqrt{a}\label{eq:Lambda M_rho relation in AdS/QCD},
\end{equation}
with $M_{\rho}$ the $\rho$ meson mass and $a=4\left(\sqrt{\mbox{ln}(2)^{2}+1+\beta_{0}/4}-\mbox{ln}(2)\right)/\beta_{0}\simeq0.55$.
The result at $\beta_{3}$ and $n_{f}=3$ is $\Lambda_{\overline{MS}} = 0.341\pm0.032$ 
GeV in good agreement with the combined
world data, $\Lambda_{\overline{MS}}^{(3)}=0.340\pm0.008$ GeV \cite{the:PDG 2014}.  The results are illustrated in Fig. \ref{Flo:matching_1}  in Sec.
\ref{sub:Holographic-QCD Lowq}, where the matching procedure and  the derivation of $\alpha_s$ using light-front holographic methods is discussed in more detail.

Eq. (\ref{eq:Lambda M_rho relation in AdS/QCD})  can be alternatively expressed
using the nucleon mass $M_{N}$:
\begin{equation}
\Lambda_{\overline{MS}}^{(LO)}=M_{N}e^{-a}/\sqrt{2a}.
\end{equation}
More generally one can use as input the mass of any meson 
or baryon composed of light quarks, since their masses are related to each other
within the holographic QCD framework or, conversely, meson and baryon
masses can be computed using only  $\Lambda$ as input ~\cite{the:Deur Lambda AdS/QCD}. 
Such a relation between the perturbative scale  $\Lambda$
and the hadron masses allows one to express the QCD fundamental mass scale
as a  function of a scheme-independent quantity rather than the scheme-dependent,
and thus  unphysical parameter,  $\Lambda$. Since QCD has no knowledge
of conventional units of mass such as GeV; only ratios are predicted.
Consequently any calculation necessarily can only yield ratios such as $\Lambda_{\overline{MS}}/M$,
with $M$ any hadron mass.

\subsection{Pion decay constant}

Recently, Kneur and Neveu have used Optimized Perturbation Theory
(see Section \ref{sub:Optimized-Perturbation-Theory}),  supplemented
by RG relations, to compute the dimensionless ratio of the pion decay
constant $f_{\pi}$ to  $\Lambda$ \cite{the:Kneur-Neveu}. The
modified optimized perturbative calculations allow to access the
strong $\alpha_{s}$ regime and to account for QCD's dynamical chiral
symmetry breaking characterized by $f_{\pi}$. The perturbative series
for  $f_{\pi}/\Lambda$ is known to fourth order: $f_{\pi}$ is
defined from the vacuum expectation of the autocorrelation function
of the axial current. It has been computed perturbatively up to fourth
order. Relating the mass scale in this perturbative series to  $\Lambda$
yields $f_{\pi}/\Lambda$, which reads at first order and in the
chiral limit:
\begin{equation}
\Lambda^{LO}=f_{\pi}/\sqrt{5/\left(8\pi^{2}\right)},
\end{equation}
that is  $f_{\pi}/\Lambda \simeq 0.25$ in the $\overline{MS}$ scheme.  The high order numerical calculations
depend on the assumed value of $n_{f}$. The results calculated at different
orders are stable. The fourth order result for $n_{f}=3$ yields
$\Lambda_{\overline{MS}}=0.317\pm0.013$ GeV, after correcting for
chiral symmetry breaking.  This correction produces a 10-20\% decrease
of  $\Lambda_{\overline{MS}}$ depending on $n_{f}$ as assessed
from lattice results. This yields a coupling $\alpha_{s}(M_Z^{2})=0.1174_{-0.0012}^{+0.0015}$.

As is the case of the holographic QCD approach discussed in the previous section,
it is expected that QCD predictions can only provide dimensionless
ratios $\Lambda_{\overline{MS}}/M$ or $\Lambda_{\overline{MS}}/f_{\pi}$
since the units of dimensionful quantities, meters or eV are arbitrary
(human  convention). 

\subsection{Grand unification}

Precise measurements of the fundamental force couplings offer a
way to investigate the physics beyond the Standard Model \cite{the:E158/Qweak/PVDIS}.
The merging of the couplings in supersymmetric extensions of the Standard
Model provides a theoretical prediction, $\alpha_{s}(M_Z^{2})=0.129\pm0.010$~\cite{the:P. Langacker 1995}, although due to threshold effects,
the runnings of the couplings have to be treated with care~\cite{the:Binger-Brodsky}.
A prediction from the minimal SU(5) supersymmetric extension of the
Standard Model in 5--dimensions yields a prediction closer to the measurements,
$\alpha_{s}(M_Z^{2})=0.118\pm0.005$ \cite{the:Hall-Nomura}.

\subsection{Comparison and discussion}

Recent world data compilations --together with the latest extractions of $\alpha_{s}(M_Z^{2})$ not included in
those compilations~\cite{the:recent alpha_s extractions}-- are shown in Fig. \ref{Flo:global_UV}.  
We also highlight the most accurate individual determinations. 
Recent $\alpha_s$ determinations that appeared during the final stage of writing this review could not be shown, such as the new result improving significantly the jet fragmentation function evolution shown in the figure \cite{d'Enterria:2015tha}.

\begin{figure}[ht]
\centering
\includegraphics[width=11cm]{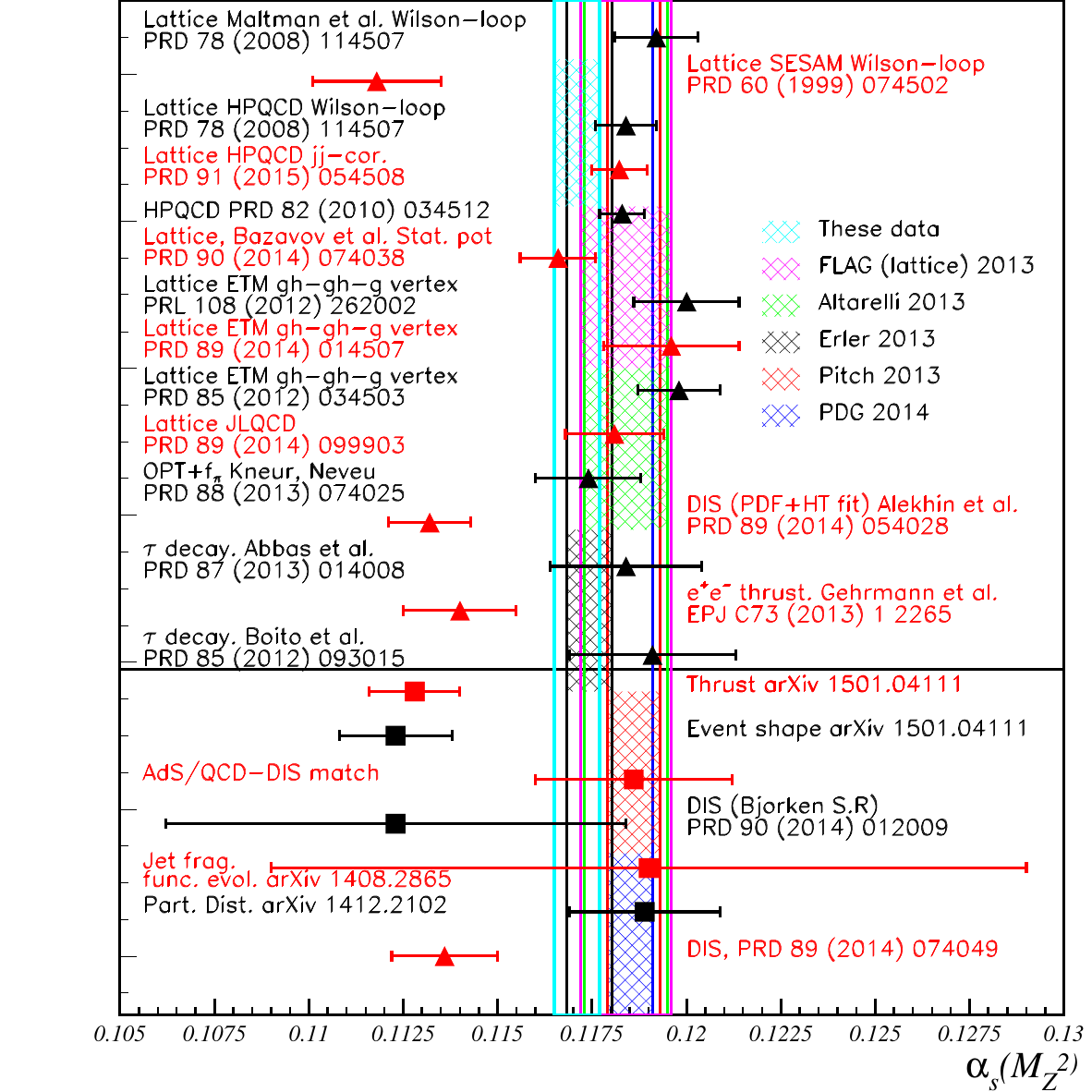}
\caption{\label{Flo:global_UV} \small The coupling $\alpha_{s}(M_Z^{2})$. Global averages
from the most recent compilations of the world data are shown by the vertical bands.
Also shown are recent determinations of $\alpha_{s}(M_Z^{2})$ not yet included in the
2013-1014 global averages (squares), and some of the most accurate available 
determinations of $\alpha_{s}(M_Z)^{2}$ (triangles).   Results
computed at $n_{f}<5$ are corrected to $n_{f}=5$.}
\end{figure}

This plot is only meant to be representative, rather than
exhaustive, of the recent and most accurate measurements of $\alpha_s$.
Fitting this choice of individual determinations (excluding one of
the two DIS highly correlated determinations) yields a $\chi^{2}/ndf$ of 4.3,
significantly larger than 1. This demonstrates the tensions between the different determinations.
Using the Unbiased Estimate, the average value of the coupling is $\alpha_{s}(M_Z^{2})=0.1171\pm0.0006$.
However, since we used only an arbitrarily selected sample of the available $\alpha_{s} $
determinations, this number should not be taken as a world average or compared to such averages.
The $\chi^{2}/ndf$ values from other compilations before applying the Unbiased Estimate procedure also underline
these tensions. In particular, there are clearly two classes of results differing by about 4\%. The lower
class of values comes from DIS measurements, $\tau-$decay and thrust measurements. 
The Unbiased Estimate is certainly not a flawless compilation procedure since it does not account
for possible correlations between results.  While the question of how to best combine all the data remains 
difficult, it is likely that the large $\chi^{2}/ndf$ values are due to a tendency to underestimate uncertainties rather than inadequate compilation procedures,  indications of physics beyond the Standard Model, or of a failing of QCD. 
Indeed, the consistency within a few percent between all determinations is a crucial verification of the validity of QCD.



\chapter{The strong coupling in the nonperturbative domain\label{sec:alpha_s in the IR}}

As mentioned in the Preamble, knowing the strong coupling in the  nonperturbative 
domain is necessary to understand both high energy  and hadronic phenomena.  An 
example is the calculation of the Sivers function \cite{the:Brodsky:2002cx}
encountered in single-spin pseudo-T-odd semi-inclusive DIS. In addition,
the \emph{renormalization scale} at arbitrarily small momenta $q^{2}\sim v^{2}S$
is required to evaluate heavy quark production as one approaches threshold
\cite{the:Brodsky:1995ds}. Finally, knowing the IR-behavior
of $\alpha_{s}$ is important to understand the mechanisms for  dynamical
chiral symmetry breaking \cite{the:dyn. chi. sym. breaking}. The following references
provide other examples of phenomena for which knowing the IR-behavior of $\alpha_s$
is required: \cite{the:Barnes 1982, the:Stern 1991,
the:Contopanagos 1994, the:Catani 1996, the:Badelek 1997,
the:Milton 2006, the:Courtoy 2011}. In addition, the review~\cite{the: Pennington 2011, Crewther:2012wd}
provides a concise description of the impact of $\alpha_{s}$  on our understanding of hadron dynamics. 
Additional reviews on $\alpha_{s}$ in the IR regime are given
in \cite{the:Prosperi alpha_s review, the:Stefanis alpha_s lowq review, the:Shirkov alpha_s low_Q review}.

Studying $\alpha_s$ in the IR is more challenging than in the UV domain since the perturbative 
formalism involving the fundamental QCD fields cannot be used.  Consequently,
a number of approaches have been explored.  Each approach has its benefits, justifications and limitations. In this second
part of this review, we will explore the insights brought by the various approaches. Since  there is still no consensus on how  $\alpha_s$ should be defined, such  discussion is necessary for  several reasons: 
1) the different definitions, and consequently meanings, of $\alpha_s$ need to be understood;
2) the approximations used and the arbitrary choices made ({\it e.g.}, the RS and gauge) need to be noted;
3) the connections between the different approaches need to be identified;
4) it illustrates how this challenging problem can be approached by a variety of methods;
5) it acknowledges the theoretical contributions, early or recent, toward solving the problem of the IR-behavior of $\alpha_s$;
6) finally, the variety of approaches to $\alpha_s$ in the IR domain can allow a cross-check of the conclusions.

Although there are important
results and constraints obtained from experiments, most investigations of the IR-behavior of $\alpha_s$ 
have been theoretical.  Some of the insightful theoretical approaches address theories which are closely related
to QCD, including pure gluonic Yang--Mills theory (without quark fields), spaces of different Minkowsky 
or Euclidean  dimensions, as well as SU($N_C$) with various numbers of  
colors $N_C$.  In that last context, it is useful to note 
that one recovers Abelian theory in the limit $N_C\to 0$ at fixed $\alpha_s 
C_F = \alpha$ where $C_F = {N^2_C-1\over 2 N_C}$ \cite{Brodsky:1997jk}.

 Perturbative QCD, via  Eq. (\ref{eq:alpha_s}), evidently predicts that  $\alpha_s(Q^2)$
diverges at the \emph{Landau pole}, when $Q^{2}\rightarrow\Lambda^{2}$.
However, this is not a meaningful prediction since it occurs, by definition, outside
the domain of pQCD's validity.  As we will discuss in 
Section \ref{sub:Handling of npert. terms}, the absence of  nonperturbative terms in the 
series (\ref{eq:alpha_s}) is responsible for this unphysical prediction. Consequently, the Landau
singularity cannot be cured by simply adding more perturbative terms to 
Eq. (\ref{eq:alpha_s}). In fact, they can worsen
the situation;  for example, a typical quantum field theory perturbative approximant
is a \emph{Poincar\'{e} series} ({\emph{asymptotic series}) which diverges beyond the order 
given approximately by the  inverse of the expansion coefficient \cite{the:Rivers 1988}. 
At $Q^{2}=1$ GeV$^{2}$, this is typically
$\alpha_{s}/\pi\simeq 0.2$. Thus, the perturbative approximant of an
observable will start diverging after $n\simeq 5$. 
Since the 
$\beta-$series is expanded in powers of $\hbar$, this discussion does not directly
concern its convergence, except that one traditionally chooses the $\beta_i$ order in Eq.  (\ref{eq:alpha_s}) 
to be the same as the $\alpha_s$ order in the approximant series. Thus,  the 
$\beta_i$ and $\alpha_s$ orders are linked.

It is often heard that the diverging behavior of the
perturbatively calculated $\alpha_{s}$ is responsible for quark confinement. This 
ignores the fact that the \emph{Landau pole} is unphysical and  that Eq. (\ref{eq:alpha_s}) does
not include important diagrams which connect quarks  via multi-gluon 
exchange such as the ``H'' diagrams, see Fig. \ref{QCD_vertex}(d). Such diagrams must be
included for large distances calculations.
Nevertheless, one can consider the possibility that $\alpha_{s}$
diverges as $1/Q^2$ in the $Q^2 \rightarrow 0$ limit, or  one can argue 
that a large value for the coupling is an ingredient
for confinement, but confinement does not require $\alpha_{s}$ to diverge.  For example, according
to lattice gauge theory simulations, the static $\mbox{Q--}\overline{\mbox{Q}}$ potential continues
to rise linearly well after the coupling has reached a maximum value, typically at distances larger than
0.4 fm. 
The predicted QCD potential
continues to rise linearly for  $\mbox{Q--}\overline{\mbox{Q}}$ separations of order 1.3 fm \cite{the:Bali 2006} 
or even larger distances for \emph{quenched} calculations \cite{the:Bolder 2001}.
Furthermore, the  IR value of  $\left\langle \alpha_{s}\right\rangle =\frac{1}{E}\int_{0}^{E} \, dQ\alpha_{s}(Q^2)$
obtained from analyzing jet shape observables in $e^{+}e^{-}$ annihilation
and DIS is finite and typically modest: $\left\langle \alpha_{s}\right\rangle = 0.47 \pm$0.07
for $E=2$ GeV \cite{the:Dokshitzer 1996-1998}. In the stochastic vacuum model approach to high-energy 
scattering \cite{Shoshi:2002rd}, it is found that $\alpha_{s} \simeq 0.81$ in the IR.
In another approach, Gribov showed that light quarks are super-critically bound if the averaged
coupling has a moderate value of 0.43 at large distances \cite{the:Gribov supercrit. binding}.

Most definitions of $\alpha_s$ in the IR  attempt to generalize the pQCD definition to include
QCD's confining effects. Thus, the question of the IR-behavior of $\alpha_s$ is intimately
linked to the topic of confinement.  Consequently, confinement 
 needs also to be 
addressed here. In the approaches that we will discuss, $\alpha_{s}$
is defined  from the behavior of quark, gluon, and/or ghost propagators which are believed to be directly
relevant to confinement.  Such definitions of $\alpha_s$ are used in the Schwinger--Dyson
framework (Section \ref{sub:Schwinger--Dyson-formalism}), the lattice
technique (Sections \ref{sub: large Q Lattice-QCD} and \ref{sub:Low Q Lattice-QCD-calculations}),
the Gribov--Zwanziger approach (Section \ref{sub:Gribov--Zwanziger-approach}),
the Functional Renormalization Group framework (Section \ref{sub:Functional renormalization group equations})
and Stochastic Quantization (Section \ref{sub:Stochastic-quantization}).

A popular picture of confinement, first put forth by Gribov \cite{the:Gribov 1978} 
and then developed by Zwanziger (Section \ref{sub:Gribov--Zwanziger-approach}), 
is supported by the different framework just mentioned. It is useful to have such confinement picture in mind 
in order to understand the implications of the IR-behavior of $\alpha_s$.  In the 
Gribov--Zwanziger scenario, the gluon propagator is IR-suppressed compared to an extrapolation of the $1/Q^2$
UV-behavior. Meanwhile, the ghost propagator is IR-enhanced.  Consequently, in this picture,
confinement results from the long-distance propagations of ghosts. Thus, the behavior of
the ghost and gluon propagators dictate the IR-evolution of $\alpha_{s}$. 

Since ghosts are unphysical artifacts reflecting gauge choices and introduced to force
the gluon propagator to be transverse (see Eqs.  (\ref{eq:gluon loop}) and (\ref{eq:ghosts}) 
and the discussion in between), a more intuitive but equivalent
view is provided by the Stochastic Quantization
framework in which there are no ghosts. Their role is played by
longitudinal gluons  in this formalism (see Section \ref{sub:Stochastic-quantization}).
At distances close to a color charge where
pQCD rules and where spherical symmetry is not yet disturbed by the presence 
of another color charge, short wavelength gluons can
be represented by plane waves (on-shell;  {\it i.e.}, transversely polarized  gluons).
At distances relevant to the large distance separation of the two color charges, 
the gluon field becomes highly asymmetric
and is postulated to collapse into a flux tube. Consequently, the plane waves are fully
distorted and longitudinally-polarized off-shell  gluons dominate.  Furthermore,
it was shown that the ghost propagator  (a manifestation of longitudinal gluons)  in 
Landau gauge, $\partial_\mu A^\mu=0$, is related to the instantaneous Coulomb propagator in
the Coulomb gauge,  $\nabla \cdot \mathbf A=0$. The Coulomb propagator is  thus enhanced at large
distance \cite{the:Cucchieri. 2002 proc.}, which leads to
a linear $\mbox{Q--}\overline{\mbox{Q}}$ potential \cite{the:Szczepaniak 2001,
the:Nakamura Coulomb pot from lattice, the:Cucchieri linear pot coulomb gauge}.
A dominating longitudinal gluon propagator indicates  off-shell gluons, and thus the 
concept of effectively massive gluons becomes relevant
(see Section \ref{sub:SDE massive gluon}).

This picture of confinement implies a specific IR-behavior of $\alpha_s$, namely that  
$\alpha_{s}$ loses its $Q^{2}-$dependence in the IR regime. In other
words, it effectively has a \emph{conformal behavior}, \emph{freezing}  at a given 
value: an \emph{infrared fixed point}. Such 
behavior appears to be supported by measurements and calculations 
using different techniques as we will see in the rest of the review. 

However, the Gribov--Zwanziger 
scenario described above  has been challenged,
as will be further discussed in Section \ref{sub:Classes-of-solutions in IF domain}.
In fact, as we already said, there is no agreement on  the IR-behavior of $\alpha_{s}$. 
The only certitude is that the \emph{Landau pole} is unphysical. 
Rather than validating the \emph{conformal} scenario, some studies indicate
that $\alpha_{s}(Q^{2})$ reaches its maximum in the IR-UV transition
region and then vanishes in the deep IR. Other investigations point to a
divergence of $\alpha_{s}(Q^{2})$ when $Q^{2}\rightarrow0$, where this
divergence is unrelated to the \emph{Landau pole}. 

There are several reasons for these different conclusions. First and
foremost, multiple definitions of $\alpha_{s}(Q^{2})$ in the IR exist.
Second, approximations are not always under control, and thus they may lead to
unphysical artifacts. Third, there is no agreement on the  more suitable  form (instant 
form or front form)~\cite{Dirac}, gauge, and RS to use. Finally, since some of the results are model-dependent,
some  assumptions on which the model rests may not be systematically valid.

In the next sections, we will review the  different
approaches used or developed to study the low IR-behavior of
$\alpha_{s}$, as listed by the theoretical techniques. Then we will compare the various results 
and discuss their differences.

First, we will briefly expand the discussion on the possible types of 
IR-behaviors for $\alpha_{s}$. 
The possibility that $\alpha_{s}(Q^{2})$ loses its $Q^{2}$-dependence
in the IR; {\it i.e.},  $\beta(Q^{2})\rightarrow0$
in the IR domain, was pointed out in the early days of QCD
\cite{the:Caswell beta_1 calculation -1, the:Sanda 1979, the:Banks-Zaks}.
A general argument, given in Refs. \cite{Brodsky:2007hb, the:Brodsky & Shrock}, provides a physical explanation:  color confinement implies
that long wavelengths of partons in hadrons are cut off. Consequently,
at this maximum wavelength corresponding to the typical hadron size,
the effects of loops in propagators and  vertex corrections disappear. 
Since these quantum effects are at the origin of the running of $\alpha_{s}$, it should
\emph{freeze} at the typical hadronic scale, provided that no other phenomena
are included in the IR-definition of $\alpha_s$:  an infrared fixed point becomes a natural consequence of confinement.
 In fact, a number of theories and models that include
confinement effects  produce this feature,
but their predictions of $\alpha_{s}$ do not
\emph{freeze} at a same value, and/or at the same momentum scale. 
The concept that $\alpha_{s}$ \emph{freezes}
is important since it  allows  one to compute quantities involving
integrals over the IR domain of the coupling constant. It also permits the use
of \emph{conformal} field theory for  nonperturbative QCD calculations. This connects to
the AdS/CFT approach (Sections \ref{sub:Holographic-QCD large Q}
and \ref{sub:Holographic-QCD Lowq}), including the extension of the
\emph{CSR} (Section \ref{sub:Effective charges and CSR} and Refs. \cite{the:alpha_g_1 from AdS}
and \cite{the:Brodsky CSR at low Q2}). In fact, the success of the AdS/CFT 
predictions may be  an indication that $\alpha_{s}$, with a definition relevant to
IR phenomenology, \emph{freezes} in the IR.  Experimental measurements
of \emph{effective charges},  results from lattice gauge theory, the
Schwinger--Dyson formalism, the phenomenology of the hadron mass spectrum, the Gribov--Zwanziger
confinement scenario and other approaches, also support this behavior.

Alternatively,  the QCD running coupling may diverge as $1/Q^2$, as suggested by the behavior of the $\mbox{Q--}\overline{\mbox{Q}}$ static
potential at large distances. 
Yet another possibility is a monotonic increase of $\alpha_{s}(Q^{2})$
as $Q^{2}$ becomes small, but without any divergence;  {\it i.e.} $\beta(Q^{2})$
remains non-zero and significantly negative. This is exemplified by
the `` analytic coupling'', see Section \ref{sub:Analytic approach}.
Finally, $\alpha_s$ may vanish in the IR.
Several models, see {\it e.g. } Refs. \cite{the:Dokshitzer 1996-1998, the:Arbuzov (2013), the:Boucaud Instantons 2003-2004} and some Schwinger--Dyson
and Lattice results indicate such behavior.  Some experimental results
\cite{the:Baldicchi alpha_an from quarkonium} are also suggestive of this behavior.

In the remaining part of this review, we will use $\alpha_s$ as a generic designation for the strong
coupling, and $\alpha_{***}$ for specific couplings, where ``$_{***}$'' indicates the method, authors or schemes,
defining $\alpha_s$.  We will start by discussing effective
charges and related definitions such as the coupling from holographic
QCD and Sudakov charges. Then we will discuss the coupling from the static $\mbox{Q--}\overline{\mbox{Q}}$
potential and the information that the hadron spectrum provides on the IR-behavior of $\alpha_s$.
The Schwinger--Dyson,  lattice, functional renormalization group, Gribov--Zwanziger and 
 stochastic quantization approaches will be discussed thereafter. These approaches use the same
definition of the coupling. Finally, we will describe other frameworks, the most developed being the
 analytic/dispersive approaches. 

Although the different approaches are organized in different sections for the sake of clarity, 
there are of course interrelations. For example, the effective 
coupling of Dokshitzer \emph{et al.} (Section \ref{sub:Dispersive-approach: Dok. Mar. Web.}) is related to the \emph{effective charge} 
approach of  Grunberg (Section \ref{sub:low Q Effective-charges}) and to  the Shirkov  \emph{et al.}  
analytic approach (Section \ref{sub:Analytic approach}). Dokshitzer's 
coupling can be interpreted in terms of effectively massive gluon fields, which ties it 
to the Schwinger--Dyson framework (Section \ref{sub:Schwinger--Dyson-formalism}), the lattice
results (Section \ref{sub:Low Q Lattice-QCD-calculations}) and 
other approaches addressing the IR-behavior of $\alpha_s$, since 
IR-regularizations are often associated with the emergence of a mass scale. 
The relevant mass scale can be interpreted as:
\begin{itemize} \label{IR mass scales}
\item the mass scale characterizing the harmonic oscillator potential in the light-front 
Schr\''odinger equation underlying  quark bound states. 
Equivalently, it is also the factor distorting the AdS space in AdS/QCD 
(Section \ref{sub:Holographic-QCD Lowq}); 
\item  the QCD string tension $\sigma$  (Section \ref{sub:Potential-approach});
\item an effective gluon mass (Sections \ref{sub:SDE massive gluon}, \ref{sub:Low Q Lattice-QCD-calculations} and \ref{sub:Analytic approach});
\item a dispersive variable ({}``dispersive mass'') (Section \ref{sub:Dispersive-approach: Dok. Mar. Web.});
\item the regulator which is introduced in the  functional renormalization group method (Section \ref{sub:Functional renormalization group equations});
\item the Gribov mass (Section \ref{sub:Gribov--Zwanziger-approach});
\item a glueball mass (Sections \ref{sub:Analytic approach} and \ref{sub:Background pert theo, Simonov}).
\end{itemize}
These scales are thus all related and  typically take values of order 0.5 GeV$\simeq 1.5 ~\Lambda_{\overline{MS}}$.   
In addition, all of these approaches have to connect to the phenomenological $\mbox{Q--}\overline{\mbox{Q}}$ linear potential.  While the existence of  such relations is physically suggestive, the explicit connection
between the various strong couplings $\alpha_s$ in the infrared is however often unknown.

\section{Effective charges} \label{sub:low Q Effective-charges}

As we have noted, Eq. (\ref{eq:alpha_s})  implies that $\alpha_s(Q^2)$, 
as derived from pQCD, diverges when $Q^{2}\rightarrow\Lambda^{2}$.   
However, observables measured across the domain extending from $Q^{2}\gg\Lambda^{2}$
to $Q^{2}<\Lambda^{2}$ display no sign of discontinuity or unusual behavior.
This is expected since the \emph{Landau pole} is unphysical and
  $\Lambda$ is an arbitrary quantity which  depends
on the choice of RS, see Table \ref{Flo:Table of Lambda}. In
contrast, observables must be RS-independent. This continuity of the observables,
along with Grunberg's \emph{effective charge} approach, see Section \ref{sub:Effective charges and CSR},
provides a definition of an effective coupling that behaves as
$\alpha_{pQCD}$ at large $Q^{2}$ but stays finite at small values of $Q^{2}$.

\emph{Effective charges} $\alpha^{eff}_s(Q^2)$ are defined directly from observables which are  calculable in the pQCD domain.  
A prominent example, which we will discuss in detail
in the following sections, is the \emph{effective charge} $\alpha_{g_1}(Q^2)$ defined from the Bjorken sum rule \cite{the Bj SR}.
This \emph{effective charge} definition of the running coupling 
incorporates QCD contributions inherent to deep inelastic scattering, 
QCD quantum corrections, and also the $Q^2$ -dependent perturbative effects 
responsible for the higher order terms of the observable's perturbative series, {\it e.g.}, 
gluon bremsstrahlung.

Since it extends to low $Q^2$, an effective charge $\alpha^{eff}_s(Q^2)$  also  incorporates  nonperturbative
QCD contributions. Those can be organized as power-law corrections and thus are often referred to as  
`` \emph{higher-twist}''  contributions as classified by the operator-product expansion.
For example, the \emph{effective charge} defined from the integral in the Bjorken sum rule,  
$\sum_{i=2,3...}^{\infty}\mu_{2i}^{p-n}(Q^{2})/Q^{2i-2}$,
incorporates the leading \emph{twist} series on the rhs of Eq. (\ref{eq:Bj SR, order alpha^4}).
The $\mu_{2i}^{p-n}(Q^{2})$ terms are themselves sums of  kinematical
or dynamical \emph{higher-twists} of \emph{twist} order up to $2i$. These terms
have the usual pQCD $Q^{2}$-logarithmic dependence associated with asymptotic freedom and also can be 
interpreted in terms of the confining force acting on the quarks \cite{the:Burkardt twists}.
Thus the loop effects responsible for the UV running, the pQCD effects responsible for the higher order
DGLAP evolution \cite{the:DGLAP}, and the  nonperturbative confining
forces are all incorporated into the effective QCD charge $\alpha^{eff}_s(Q^2)$.  In some cases
$\alpha^{eff}_s(Q^2)$ can take  values in the IR domain that would be unphysical for a quantity 
measuring the absolute strength of a force. It is e.g. the case for the
$\alpha_{D}\left(Q=0\right)=-\pi$ where $D(Q^{2})$ is the Adler function \cite{the:Adler function}. This is obviously
not a problem for an effective quantity.

A natural choice for an effective charge, close 
to the standard  Gell-Mann--Low definition of the QED running coupling
 \cite{the:Gell-Mann--Low QED coupling} is 
the potential scheme ($V$-scheme), where $\alpha_{V}(Q^2)$
is interpreted as the running coefficient of the $1/Q^2$ potential acting between heavy quarks
\cite{the:Appelquist, the:Brodsky alpha_v}.  However, as discussed in Section \ref{sub:Effective charges and CSR},
$\alpha_V$ is affected by infrared divergences at three loops and higher which prevent it from being calculated perturbatively, even at high $Q^2$. 

A related approach has been proposed by Dokshitzer \emph{et al.} 
\cite{the:Dokshitzer dispersive approach, the:Dokshitzer/webber: HT in event shapes.}.
As in the case of \emph{effective charges}, Dokshitzer's approach 
also generalizes the QED Gell-Mann--Low coupling. It  will be discussed in detail in Section 
\ref{sub:Dispersive-approach: Dok. Mar. Web.}. 
Dokshitzer's charge is intended to measure the effective 
interaction strength in the IR in order to extend the use of the fundamental QCD
fields outside the UV domain.  We will see that Grunberg's \emph{effective charge} defined
from the Bjorken sum rule, $\alpha_{g_1}$, may be interpreted in a similar manner.
In distinction to the standard effective charge,
Dokshitzer's effective coupling is built by  imposing dispersion 
relation constraints  on the coupling. The effective charge
$\alpha_{g_1}$ is also constrained by these restrictions since dispersion relations applied to photoproduction
cross-sections are the basis for deriving the 
Gerasimov--Drell--Hearn sum rule \cite{the:GDH},  and this in turn  constrains the 
IR-behavior of $\alpha_{g_1}$. Thus, although the respective formalisms of  
$\alpha_{g_1}$ and Dokshitzer's effective coupling seem to be different, the 
two couplings have some commonalities.

\subsection{Measurement of the  effective charge from the Bjorken sum rule.\label{sub: effective charge g1}}

The \emph{effective charge}  $\alpha_{g_1}(Q^2)$,   defined from the Bjorken sum rule,  has been discussed for large $Q^2$
in Sections \ref{sub:Effective charges and CSR} and \ref{sub:Deep-Inelastic-scattering},
where it was  shown that $\alpha_{g_1}(Q^2)$ has a simple 
perturbative series which is known to high orders in the $ \overline{MS}$ scheme.
An important advantage of $\alpha_{g_1}(Q^2)$  is that data exist at low, intermediate, and high
Q$^{2}$ \cite{Deur alpha_s from eg_1dvcs, the:Bj E142,the:Bj SMC, the:Bj. E143, the:Bj E154, the:Bj Hermes, the:E155/E155x, 
the:JLab  Bj SR, the:Bj COMPASS}.  In addition, 
rigorous sum rules  dictate the behavior of $\alpha_{g_{1}}$ in  the unmeasured $Q^{2}\rightarrow0$ and $Q^{2}\rightarrow\infty$
regions. 
A third advantage is that the Bjorken
sum is a non-singlet quantity, implying that some
resonance contributions to the sum, such as  the $\Delta$(1232) resonance,
cancel out. This has important consequences. For example, disconnected
diagrams  which are not easily  computed on the
lattice, do not contribute, thus  leading to more reliable lattice gauge theory estimates.
Similarly, the cancellation of the $\Delta$(1232)
makes chiral perturbation calculations more robust \cite{the:Burkert delta in BJ sum.}.
Finally, $\alpha_{g_{1}}$ is easier to interpret than other effective
charges due to similarities  between  the $g_{1}$ and $V-$schemes; 
see Fig. \ref{Flo:strong-coupling in diff schemes} and the associated
discussion. In effect, the \emph{effective charge} $\alpha_{g_{1}}$ can
be evaluated at any $Q^{2}$,  and its intuitive interpretation makes
it  easy to compare with theoretical expectations. 

The similarities between the $g_{1}$
and $V-$scheme have a physical origin: First, $g_{1}$ is extracted from inclusive reactions, 
and the implicit sum over all final states simplifies the theoretical expression
for the cross-section~\cite{the:Bloch-Nordsieck/TDL theorems.}.
Second, the partial suppression of  resonance contributions enhances 
the non-resonant background contribution. This allows  
DIS-like reactions to dominate the Bjorken integral
even at low $Q^{2}$.   In addition, the integral over $x_{Bj}$ and global
parton--hadron duality \cite{the:Bloom Gilman duality}  amplify
the dominance of non-resonant reactions.  In contrast,
coherent state contributions are not easily interpreted in terms of a QCD coupling: 
Indeed, if an elastic contribution would be added to the definition of $\alpha_{g_{1}}$, it would 
become significantly negative because the kinematic constraint
$\int_0^1 dx_{Bj}  ~  g_1(x_{Bj},Q^{2})\rightarrow 0$ as $Q^{2}\rightarrow 0$ (see
Eq. (\ref{eq:GDH limit of alpha_g_1}) below) would then not hold; 
this would yield $\alpha_{g_{1}+el}(Q^{2}=0)=-17.6$,  which includes the elastic contribution,
rather than $\alpha_{g_{1}}(Q^{2}=0)=\pi$.

Two rigorous sum rules constrain  $\alpha_{g_{1}}(Q^{2})$ in the limits
$Q^{2}\rightarrow0$ and $Q^{2}\rightarrow\infty$:  the
Gerasimov--Drell--Hearn (GDH) sum rule \cite{the:GDH, the:GDH*}
for $Q^{2}\rightarrow0$, and the Bjorken sum rule \cite{the Bj SR}
for $Q^{2}\rightarrow\infty$. Let us first consider the latter. At
large $Q^{2}$, the rhs of Eq. (\ref{eq:Bj SR, order alpha^4}) can
be computed.  Equating it to the rhs of 
Eq. (\ref{eq:alpha_g_1 def.}) yields the result:
\begin{equation}
\alpha_{g_{1}}=\alpha_{\overline{MS}}+3.58\frac{\alpha_{\overline{MS}}^{2}}{\pi}+20.21\frac{\alpha_{\overline{MS}}^{3}}{\pi^{2}}+175.7\frac{\alpha_{\overline{MS}}^{4}}{\pi^{3}}+\mathcal{O}\left(\alpha_{\overline{MS}}^{5}\right)\label{eq:msbar to g_1},
\end{equation}
which is indicated by the blue band shown in Fig. \ref{Flo:lowq alpha. exp. }.
The width of the band represents the uncertainties due  to the value of $\Lambda_{\overline{MS}}$,
the truncation of the Bjorken series in Eq. (\ref{eq:msbar to g_1}),
and the truncation of the $\beta$ series used to compute $\alpha_{\overline{MS}}$
in Eq. (\ref{eq:msbar to g_1}).  At the smallest $Q^{2}$  typically considered for the applicability of pQCD,
$Q_{min}^{2}\gtrsim1$ GeV$^{2}$, the \emph{asymptotic series} 
(\ref{eq:msbar to g_1}) converges up to order $n\sim\pi/\alpha_{\overline{MS}}(Q_{min}^{2})\simeq4$
so one should stop at this order, lest 
$175.7\alpha_{\overline{MS}}^{4}/\pi^3$ becomes comparable to $ \mathcal{O}\left(\alpha_{\overline{MS}}^{5}\right)$.

\begin{figure}[ht]
\centering
\includegraphics[scale=0.45]{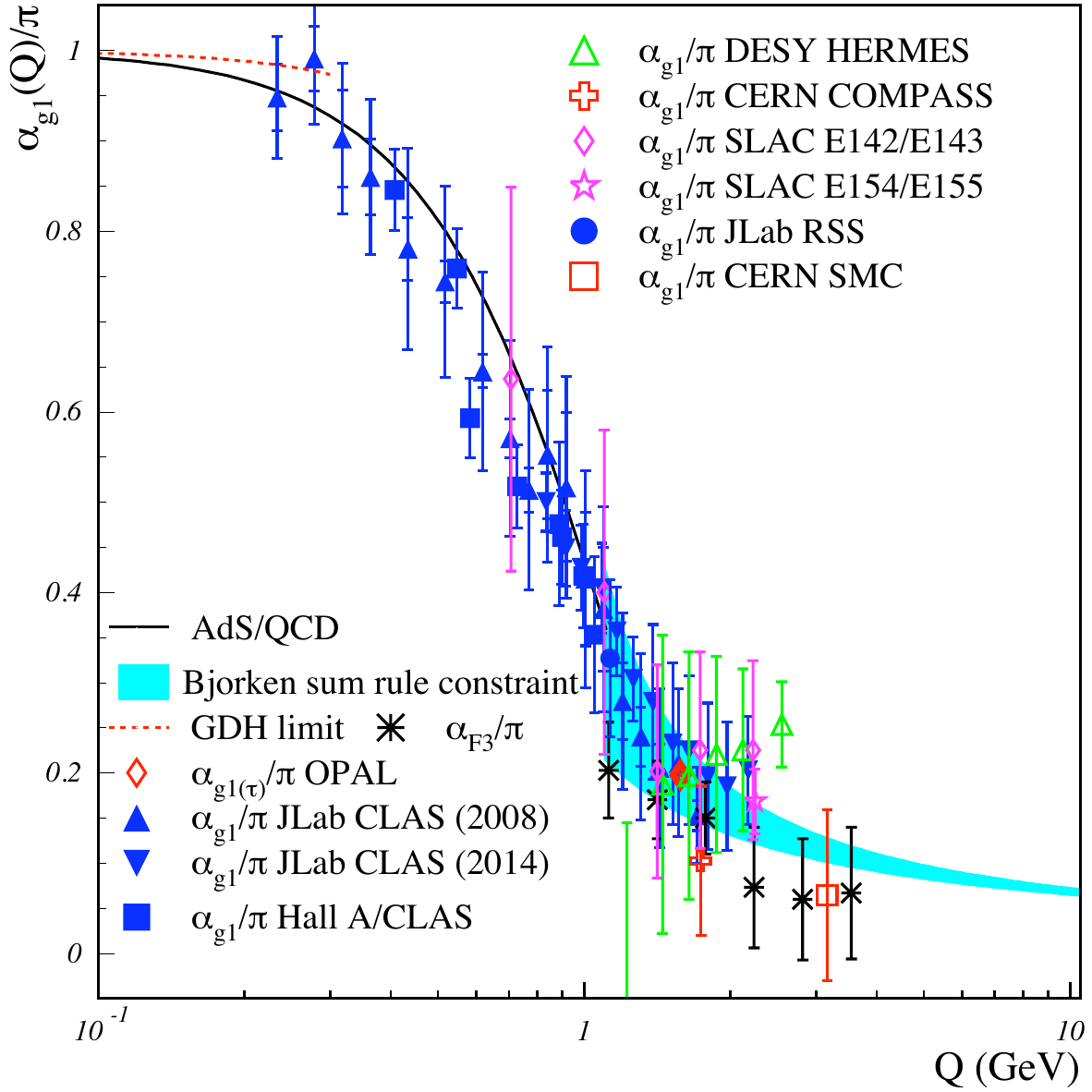}
\caption{\label{Flo:lowq alpha. exp. } \small Experimental data and sum rule constraints
for the \emph{effective charges} $\alpha_{g_{1}}(Q)/\pi$ and $\alpha_{F_{3}}(Q)/\pi$.
The blue data points are from Jlab \cite{the:Deur lowq alpha_s},
 the green points are from Hermes \cite{the:Bj Hermes},
 the black points are from Fermilab \cite{the:GLS CCFR}, 
the red points are from CERN \cite{the:Bj COMPASS}
and the magenta points are from SLAC \cite{the:Bj E142, the:Bj. E143, the:Bj E154, the:E155/E155x}.}
\end{figure}

As an alternative to Eq. (\ref{eq:msbar to g_1}), one can use the BLM/\emph{PMC} expression 
(see Section \ref{sub:Effective charges and CSR} and Eq. (\ref{eq:generic CSR})) for  
$\alpha_{g_{1}}(Q^{2})$ in  terms of $\alpha_{\overline{MS}}(Q^{2})$ \cite{the:CSR}.  
In this case all nonzero $\beta$ terms are shifted into the scales $Q^*$, $Q^{**}$, $\cdots$
of the $\alpha_{\overline{MS}}$ coupling thus matching a \emph{conformal} expansion.  
However, the domain of validity of  this particular relation --a \emph{CSR}--  
may not be useful due to the large
decrease in the  scales $Q^*_{\overline{MS}}=0.37Q_{g_1}^{*}$
and $Q^{***}_{\overline{MS}}=0.25Q_{g_1}^{***}$. This decrease renders the
computation of $\alpha_{\overline{MS}}$ problematic.

Let us consider now the GDH sum rule  \cite{the:GDH} which can be expressed as the limit: 
\begin{equation}
\frac{8}{Q^{2}}\int_{0}^{1^{-}}dx_{Bj} ~g_{1}\left(x_{Bj},Q^{2}\right)~ \overrightarrow{_{Q^2 \to 0}}~ \frac{-\kappa_N^{2}}{M_N^{2}}, \label{eq:GDH limit of alpha_g_1}
\end{equation}
where $\kappa_N$ is the anomalous magnetic moment of the target particle
(here a nucleon) and $M_N$ its mass. The elastic scattering contribution
to $g_{1}$, $x_{Bj} =1$ is excluded from the integral in Eq. (\ref{eq:GDH limit of alpha_g_1}).
At $Q^2=0$, there is no elastic reaction and at large $Q^2$ it is negligible,
 thus its exclusion is unimportant. At intermediate $Q^2$, the elastic contribution
 would be sizable if one chooses to include it to Eq. (\ref{eq:alpha_g_1 def.}) 
(in that case, Eq. 
(\ref{eq:GDH limit of alpha_g_1}) becomes invalid). 
However, as already 
discussed, this contribution should be excluded for $\alpha_{g_{1}}$. 

Eqs. (\ref{eq:GDH limit of alpha_g_1}) and (\ref{eq:alpha_g_1 def.})
provide an important constraint on $\alpha_{g_{1}}$ at $Q^{2}=0$:
\begin{equation}
\frac{d\alpha_{g_{1}}\left(Q^{2}\right)}{dQ^{2}}|_{Q^{2}=0}=\frac{3\pi}{4g_{A}}\left(\frac{\kappa_{p}^{2}}{M_{p}^{2}}-\frac{\kappa_{n}^{2}}{M_{n}^{2}}\right).\label{eq:GDH slope for alpha_g_1}
\end{equation}
Furthermore, there is an additional kinematic constraint, $\int_{0}^{1^{-}}dx_{Bj} g_{1}\left(x_{Bj},Q^{2}=0\right)=0$,
because when $Q^{2}$ is small, only the elastic scattering contribution
is permitted. However, the exclusive reaction is excluded from Eq. (\ref{eq:alpha_g_1 def.})
leading to a vanishing integral at $Q^{2}=0$. This result also follows from
Eq. (\ref{eq:GDH limit of alpha_g_1}).  Consequently, the infrared fixed-point value
\begin{equation}
\alpha_{g_{1}}\left(0\right)=\pi.\label{eq:alpha_g_1 IR value}
\end{equation}
follows from  (\ref{eq:alpha_g_1 def.}).

The two constraints given by Eqs. (\ref{eq:GDH slope for alpha_g_1})
and (\ref{eq:alpha_g_1 IR value}) are shown in Fig. \ref{Flo:lowq alpha. exp. }
by the dashed red line. The range between the low and high $Q^{2}$
constraints is densely filled by experimental data \cite{the:Deur lowq alpha_s},
mostly coming from the Bjorken integral data \cite{Deur alpha_s from eg_1dvcs, the:Bj E142,the:Bj SMC, the:Bj. E143, the:Bj E154, the:Bj Hermes, the:E155/E155x, 
the:JLab  Bj SR, the:Bj COMPASS}.
Other data come from the CCFR measurement \cite{the:GLS CCFR} of
the Gross--Llewellyn Smith sum rule \cite{the:GLS sum rule}, Eq.
(\ref{eq:GLS sum rule}).
At leading \emph{twist}, the GLS sum rule has the same $Q^{2}$-dependence
as the Bjorken sum rule, except for a small difference at order
$\alpha_{\overline{MS}}^{3}$ coming from the light-by-light contribution
to the GLS sum rule.  Consequently, we expect in the perturbative domain that 
$\alpha_{g_1}\left(Q^{2}\right)=\alpha_{F_{3}}\left(Q^{2}\right)$
up to $\mathcal{O}(\alpha_{\overline{MS}}^{3})$, with $\alpha_{F_{3}}$ defined
from $\int^{1^-}_0 dx~F_3(x,Q^2) \equiv 3[1-\alpha_{F_{3}}/\pi]$.  
In addition, the kinematic
constraint leading to Eq.  (\ref{eq:alpha_g_1 IR value})  also applies to the GLS integral
and thus $\alpha_{F_{3}}\left(0\right)=\pi=\alpha_{g_{1}}\left(0\right)$.
In addition, as shown in Fig. \ref{Flo:lowq alpha. exp. }, one can also use the value of 
$\alpha_{g_{1}(\tau)}$,  which is obtained using  a  \emph{CSR},
with the effective charge $\alpha_{\tau}$ defined from hadronic $\tau$ lepton decay, as reported in  Ref.~\cite{the: Brodsky alpha_tau}.
One also notes that the  JLab data  at low-$Q^{2}$ tend toward $\alpha_{g_{1}}\simeq \pi$, suggesting
that $\alpha_{g_1}$ becomes  nearly flat and thus ``\emph{conformal}'' at $Q\lesssim0.03$ GeV.

\paragraph{A prediction of QCD's \emph{conformality} from the GDH sum rule}

The approximate isospin symmetry between the proton and neutron implies
that $M_{p}\simeq M_{n}$.   Although $\kappa_p$ and $\kappa_n$ differ in sign, the squares are similar in magnitude: $\kappa_{p}^{2}=3.21\simeq\kappa_{n}^{2}=3.66$.
Consequently, the slope of $\alpha_{g_{1}}\left(Q^{2}\right)$ near
$Q^{2}=0$ as given by Eq. (\ref{eq:GDH slope for alpha_g_1}) is suppressed.
For example, in the deep IR region, $\alpha_{g_{1}}$ is
expected to decrease by  only  $\simeq  1\% $ over the $Q^{2}$-span of 0.1 GeV$^{2}$,
consistent with near-\emph{conformal} behavior in the low-$Q^{2}$ region.

\subsection{Measurement of the effective charge defined  from $e^{+}e^{-}$ annihilation.}

The ratio $R_{e^+ e^-}(s)  $ of cross sections for  $e^{+}e^{-}$ annihilation to hadrons divided by muon pair
production, and the ratio of hadronic over muonic $\tau-$decay widths, as well as 
the Adler function $D(Q^{2})$ \cite{the:Adler function}, have each been
used to define various \emph{effective charges} \cite{the:CSR, the:BLM}. 
They are closely related to each other by \emph{CSRs}. For example,
in the purely perturbative domain, the couplings from $\tau$-decay
and $R_{e^{+}e^{-}}$ ratios,   $\alpha_\tau$ and $\alpha_R$ respectively, are predicted by the CSRs to obey: 
\begin{equation}
\alpha_{\tau}(s)=2\int_{0}^{s}\frac{dt}{s}\left(1-\frac{t}{s}\right)^{2}\left(1+\frac{2t}{s}\right)\alpha_{R}(t),
\end{equation}
where $s$ and $t$ are the Mandelstam variables. The coupling $\alpha_{\tau}(s)$
has been extracted from the OPAL experimental data \cite{the:OPAL R_tau}
by Brodsky and collaborators \cite{the: Brodsky alpha_tau}, 
with the result that  $\alpha_{\tau}(0) \simeq 7.0$.

\subsection{Sudakov effective charges \label{sub:Sudakov-Effective-Charges}}

As is the case for the \emph{effective charges} just discussed  above, the Sudakov effective
charges introduced by Gardi and Grunberg \cite{the:Grunberg Sudakov charge}
essentially fold perturbative corrections into the definition
of the \emph{effective coupling}. In the large-$x_{Bj}$ limit of inclusive structure functions,
collinear soft multi-gluon emissions dominate the
perturbative series. 
These effects induce large logarithmic corrections (Sudakov double
logarithms) which increase with the order of the perturbative series. These
corrections can be incorporated  by exponentiating (resumming) the Mellin transform
of the structure function, leading to an exponential  $e^{-S}$ damping factor,
where $S$ is the Sudakov factor proportional to $\alpha_{s}\mbox{ln}^{2}$.
In principle,  this analysis  includes all  the relevant perturbative contributions.  Nonperturbative
\emph{power corrections} are not explicitly included in the Sudakov factor.
However, $S$ arguably does depend on such corrections since they influence the perturbative
series through IR \emph{renormalons}. 
The sum of the Sudakov anomalous dimensions,  which appear in the
exponentiation, defines the Sudakov effective charge.  Such charges generalize
the coupling definition of   Ref.~\cite{the:Catani et al. CMW scheme}. Most important, the Sudakov \emph{effective charges}
have a universal (scheme/observable-independent) \emph{freezing}  value. 

The  definition of the  Sudakov effective charge is
illustrated by an example provided in    Ref.~\cite{the:Grunberg Sudakov charge}. The Mellin
transform of the DIS structure function $F_{2}(Q^{2},x_{Bj} )$ is:
\begin{equation}
\hat{F_{2}}\left(Q^{2},N\right)=\int_{0}^{\infty}dx_{Bj}\, x_{Bj} ^{N-1}F_{2}\left(Q^{2},x_{Bj}\right),
\end{equation}
where $\hat{F_{2}}$ can be expressed, as usual, as a perturbative
series in powers of $\mbox{ln}(Q^{2})$ which can be exponentiated. The argument
of the exponentiated series is $\mbox{ln}(\hat{F_{2}})$.
Differentiating it with respect to $\mbox{ln}(Q^{2})$ yields:
\begin{equation}
\frac{d~\mbox{ln}~(\hat{F_{2}})}{d~\mbox{ln}\left(Q^{2}\right)}=\frac{16}{3}\left[H\left(Q^{2}\right)+\int_{0}^{1}dz\,\frac{z^{N-1}-1}{1-z}\mathcal{A_{S}}\left[\left(1-z\right)Q^{2}\right]+\mathcal{O}\left(1/N\right)\right],
\end{equation}
where $H(Q^{2})$ and $\mathcal{A_{S}}(Q^{2})$ are given as perturbative series
in powers of $\alpha_{s}$.  In particular, $\mathcal{A_{S}}(Q^{2})=\frac{\alpha_{s}(Q^{2})}{4\pi}\left[1+A_{1}\alpha_{s}(Q^{2})+ \cdots \right]$
defines the Sudakov effective charge. It is connected to the two Sudakov
anomalous dimensions $A$ and $B$:
\begin{equation}
\mathcal{A_{S}}\left(Q^{2}\right)=\frac{3}{16}\left[A\left(\alpha_{s}\left(Q^{2}\right)\right)+\frac{dB\left(\alpha_{s}\left(Q^{2}\right)\right)}{d~\mbox{ln}\left(Q^{2}\right)}\right].
\end{equation}
In the large $Q^{2}$-limit, one has:
\begin{equation}
\mathcal{A_{S}}\left(Q^{2}\right)=\alpha_{s}\left(Q^{2}\right)+\frac{3}{4}\beta_{0}\left(\alpha_{s}\left(Q^{2}\right)\right)^{2}+\left(\frac{5}{4}-\frac{\pi^{2}}{3}\right)\beta_{0}^{2}\left(\alpha_{s}\left(Q^{2}\right)\right)^{3}+  \cdots .
\end{equation}
The IR limit is:
\begin{equation}
\mathcal{A_{S}}\left(Q^{2}\right)=\frac{1}{2\beta_{0}}\left[1-\frac{2}{\pi}arctan\left(\frac{\mbox{ln}\left(Q^{2}/\Lambda^{2}\right)}{\pi}\right)+\frac{\Lambda^{2}}{Q^{2}}-\frac{\Lambda^{4}}{Q^{4}}\right],
\end{equation}
which has no Landau singularity,  but it diverges toward negative values
as $Q^{2}\rightarrow0$.   Nonperturbative \emph{power corrections}
may regularize this unphysical behavior. An example of a simple ansatz for  the nonperturbative
contributions,  which can cancel divergences and lead to a $1/\beta_{0}$
\emph{fixed point}, is given in \cite{the:Grunberg Sudakov charge 2}. 
$\mathcal{A_{S}}\left(Q^{2}\right)$ does not \emph{freeze} in the IR, 
increasing from  $1/\beta_{0} \simeq 0.10$ at $Q^2$=0 toward a maximum at $Q^2/\Lambda^2 \simeq 0.84$.

It was already noticed in \cite{theSudakov damping in FF} that the \emph{Landau pole} can be suppressed by the Sudakov damping factor $e^{-S}$, although this suppression was not incorporated in
the coupling discussed in Ref.~\cite{theSudakov damping in FF} which retained the standard pQCD definition.

\section{AdS/CFT and Holographic QCD \label{sub:Holographic-QCD Lowq}}

The behavior of the QCD coupling in the IR can be obtained by enforcing the underlying \emph{conformal} symmetry of QCD.  In fact,  conformal symmetry even allows one to identify the form of the long-range color-confining potential by incorporating a method due to  de Alfaro, Fubini and Furlan (dAFF), as  discussed below. Its scale-dependence then leads to the running of the coupling in the IR as well as a constraint in the pQCD domain.

As we discussed in  Section \ref{sub:Holographic-QCD large Q},  light-front holographic QCD is based on the AdS/CFT correspondence \cite{the:Maldacena duality} and the light-front quantization procedure based on 
the frame-in\-dep\-end\-ent front form devised by Dirac~\cite{Dirac}.  In the   front form,  the time evolution variable is $\tau = t+z/c$; {\it i.e.}, the time along the
light-front   (LF). The front-form results are independent of the observer's Lorentz frame. 

In this framework, one can reduce the full QCD light-front Hamiltonian  equation  to obtain a one--dimensional light-front 
Schr\''odinger Equation  (LFSE) acting on the valence Fock state. The eigenvalues  of the LFSE determine the  hadron spectrum consisting of light quarks and the eigensolutions determine their LF wavefunctions.  The LFSE is analogous to the
Schr\''odinger equation describing hydrogenic atoms in QED~\cite{deTeramond:2008ht} but  it is
relativistic and frame-independent. The radial variable for the LFSE is identified as the invariant distance
between the $q$ and $\bar q$:   $\zeta = b_\perp \sqrt{x(1-x)}$, where 
$\zeta^2 $ is conjugate to the light-front kinetic energy 
$k^2_\perp / [x(1-x)]$.   Here, $x$ is the light-front momentum fraction 
$x= k^+ / P^+ = (k^0 +k^z)/ (P^0 + P^z)$ and $b_\perp$ is the transverse 
impact parameter.

If one requires the effective action  which underlies the  QCD Lagrangian 
to remain \emph{conformally} invariant and extends the dAFF formalism 
to light-front Hamiltonian theory~\cite{deAlfaro:1976je, Brodsky:2013ar}, 
the light-quark antiquark  potential has the unique form of a harmonic oscillator 
potential $V(\zeta^2) = \kappa^4 \zeta^2$, and a  mass gap arises.   The dAFF mechanism,  originally derived in the  context of $1+1$  quantum mechanics, enables the emergence of a mass scale  $\kappa$ while the action remains conformal.  The result is a nonperturbative relativistic light-front wave equation  which 
incorporates confinement and other essential spectroscopic and dynamical 
features of hadron physics, including linear Regge trajectories with the same slope in the 
radial quantum number $n$ and orbital angular momentum $L$.   

\begin{figure}[ht]
\begin{center}
\includegraphics[width=8.6cm]{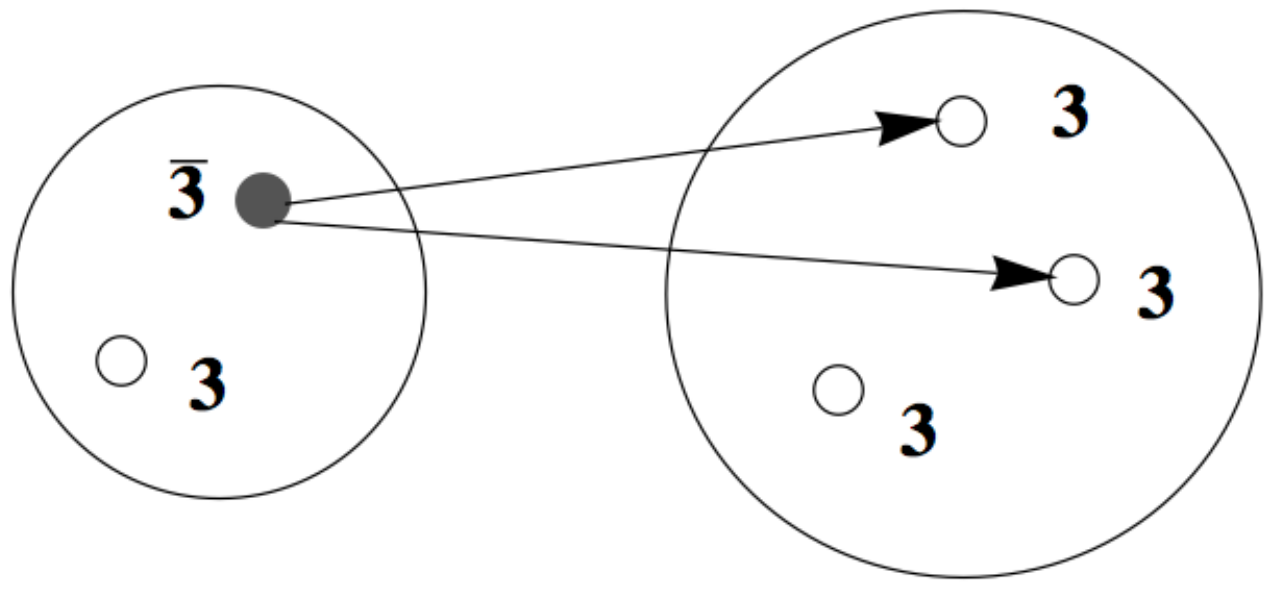}
\end{center}
\caption{\label{MNSUSY} \small In QCD, hadronic dynamical supersymmetry is rooted in the dynamics of color $SU(3)$ where $\bf \bar 3 \sim \bf 3 \times \bf 3$.}
\end{figure}

It has been shown recently  how the main features of nonperturbative QCD dynamics are well captured in a semiclassical effective theory based on superconformal quantum mechanics~\cite{Akulov:1984uh, Fubini:1984hf} and its extension to light-front physics~\cite{deTeramond:2014asa, Dosch:2015nwa}. This new approach to hadron physics incorporates several basic aspects one expects from QCD, namely confinement and the appearance of a massless pion in the limit of zero quark masses. 
The specific breaking of \emph{conformal} invariance uniquely determines the 
effective confinement potential. The generalized supercharges connect 
the baryon and meson spectra to each other in a remarkable manner.   
The partner mesons and baryons have the same mass provided one identifies the relative 
orbital angular momentum $L_M$ of the $q \bar q$ mesons with  the 
relative internal orbital angular momentum $L_B$ of the quark--diquark baryons in a cluster decomposition, where $L_M = L_B+1$~\cite{deTeramond:2014asa,Dosch:2015nwa}.
Furthermore, this framework gives remarkable connections  across the full light~\cite{Brodsky:2016yod}  and the full heavy-light hadron spectra, where heavy quark masses break the conformal invariance, but the underlying supersymmetry  still holds~\cite{Dosch:2015bca}. It is important to recall that  in the context of hadronic physics the supersymmetric relations  are not a consequence of supersymmetric QCD, at the level of fundamental fields,  but an emergent dynamical supersymmetry from color $SU(3)_C$. This relies on the fact that in $SU(3)_C$ a diquark can be in the same color representation as an antiquark, namely a $\bf \bar 3 \sim \bf 3 \times \bf 3$~\cite{Miyazawa:1966mfa, Catto:1984wi, Lichtenberg:1999sc} as illustrated in Fig. \ref{MNSUSY}.

The light-front harmonic oscillator  potential corresponds
to a linear potential for bound states of nonrelativistic heavy quarks in
the usually employed instant-form of relativistic dynamics~\cite{the:Trawinski 2014}.  This  links  light-front QCD  to   lattice gauge theory as 
well as heavy-quark effective theory. 
The  mass parameter $\kappa$ can be determined  from a hadron mass; {\it e.g. } 
$\kappa= M_\rho/ \sqrt{2}$ or $\kappa= M_p/ 2$~\cite{the:AdS/QCD review}, from  the scale $\Lambda$ controlling pQCD evolution, as discussed in Section \ref{sub:Holographic-QCD large Q}, or from other experimental inputs such as 
the Bjorken sum $Q^2$-evolutions.  Remarkably, all of these determinations are in good agreement.

The underlying \emph{conformal} symmetry of QCD allows one to perform equivalent calculations in
a higher dimensional curved space-time where the dual theory is weakly coupled and described by gravity.
 Mathematically, this dual representations is based on the isomorphism of the  $SO(4,2)$ group of conformal transformations in physical space-time  with the group of isometries of AdS$_5$ space. Because of this correspondence, one can identify the fifth--dimension coordinate $z$ of AdS$_5$ space with the light-front coordinate $\zeta$.    Light-front holography also implies that the potential $V(\zeta^2) = \kappa^4 \zeta^2$ appearing in the 
meson LFSE can be obtained by modifying  AdS$_5$ space using a specific ``dilaton profile'' $e^{\pm \kappa^2 z^2}$~\cite{the:AdS/QCD review, Karch:2006pv} (The plus sign dilaton profile is required in light-front holographic QCD~\cite{the:AdS/QCD review}).  
The quark--antiquark binding interaction is thus related  in AdS/QCD to the modification of the AdS space curvature ~\cite{Brodsky:2013ar}.  The same harmonic oscillator potential also emerges from  the application of the dAFF 
procedure to LF theory \cite{deAlfaro:1976je, the:AdS/QCD review}. 

Remarkably, one also finds that the same modification of the AdS$_5$ action  
also determines the IR-behavior of  $\alpha_s \propto \exp{\left(-Q^2\over 4 \kappa^2\right)}.$
Recall that in pQCD, the \emph{effective coupling} $\alpha_{pQCD}(Q^2)$ 
is defined  by folding short-distance quantum effects into its evolution.  
Analogously, the $Q^2$-dependence of the AdS/QCD effective coupling stems from the effects of the 
large-distance forces folded into the coupling constant~\cite{the:alpha_g_1 from AdS}.   To obtain this result, 
one notes that the AdS$_5$ action is similar 
to the action of  general relativity:  $S_{GR}\propto\int d^{4}x\sqrt{\mbox{det}(g_{\mu\nu})}R/G_{N}$,
where $G_{N}$ is Newton's constant and $R$ is the Ricci scalar.  In 
AdS/QCD, $\sqrt{R}$ is replaced by the gluon field $F$, $\sqrt{G_{N}}$ is replaced  by
the gauge coupling $g_{AdS}$, and the metric determinant becomes $\sqrt{\mbox{det}(g_{AdS})}e^{\kappa^{2}z^{2}}$,
which includes the $e^{\kappa^{2}z^{2}}$ distortion  of AdS space. 
This analysis yields the
AdS action:
\begin{equation} 
S_{AdS}=  - \frac{1}{4}\intop d^{4}x \, dz \sqrt{\mbox{det}(g_{AdS})}e^{\kappa^{2}z^{2}}\frac{1}{g_{AdS}^{2}}F^{2}.
\end{equation}
Just as $\alpha_{s}\equiv g^{2}/4\pi$ acquires its effective $Q^{2}$ dependence
from short-distance QCD quantum effects, the initially constant AdS
coupling $\alpha_{AdS}\equiv g_{AdS}^{2}/4\pi$ is redefined to
absorb the long distance confining forces;  {\it i.e.} the effects of the modification of the AdS space curvature from \emph{nonconformal} confinement dynamics~\cite{the:alpha_g_1 from AdS}:
$g_{AdS}^{2}\rightarrow g_{AdS}^{2}e^{ -\kappa^{2}z^{2}}$.   The five--dimensional coupling $g_{AdS}(z)$
is mapped,  modulo a  constant, into the QCD coupling $g_{s}$ of the confining theory in physical space-time using light-front holography. Thus, by identifying $z$ with the invariant impact separation variable $\zeta$,  
$g_{AdS}(z) \to g_{s}(\zeta)$, one predicts 
\begin{equation}  \label{eq:gYM}
\alpha_s^{AdS}(\zeta) \equiv g_{s}^2(\zeta)/4 \pi \propto  e^{-\kappa^2 \zeta^2} . 
\end{equation}

The physical coupling measured at the scale $Q$ is the light-front transverse Fourier transform
of the  LF  coupling $\alpha_s^{AdS}(\zeta)$  (\ref{eq:gYM}):
 \begin{equation} \label{eq:2dimFT}
\alpha_s^{AdS}(Q^2) \sim \int^\infty_0 \! \zeta d\zeta \,  J_0(\zeta Q) \, \alpha_s^{AdS}(\zeta),
\end{equation}
in the $q^+ = 0$ light-front frame where $Q^2 = -q^2 = - \mathbf{q}_\perp^2 > 0$ and $J_0$
is a Bessel function. 
Using this ansatz we then have from  Eq.  (\ref{eq:2dimFT})
\begin{equation}
\label{eq:alpha_g_1 from AdS/QCD}
\alpha_s^{AdS}(Q^2) = \alpha_s^{AdS}(0) \, e^{- Q^2 /4 \kappa^2},
\end{equation}
where  $\alpha_s^{AdS}(0) = \pi$ in the $g_{1}$ scheme.
The identification of $\alpha_s^{AdS}(Q^2)$ with the physical QCD running coupling
in its nonperturbative domain   determines the space-time dependence at large distances of the physical four--dimensional coupling $\alpha_s$.

The harmonic oscillator form of the LF potential $V(\zeta^2) = \kappa^4 \zeta^2$, and thus the Gaussian form for the running coupling, also 
follows from the dAFF procedure; {\it i.e.},  the requirement
that the  action remains \emph{conformal} \cite{deAlfaro:1976je, the:AdS/QCD review}, even though a mass scale appears in the confining potential
of the light-front Hamiltonian.

Eq. (\ref{eq:alpha_g_1 from AdS/QCD}) is valid only
at the domain of $Q^{2}$ where QCD is a strongly coupled theory and
the AdS/CFT correspondence can be applied. 
Furthermore, since the semi-classical approximation used in the light-front holographic approach
neglects quantum effects,  vacuum  polarization effects  are not  incorporated in $\alpha_{AdS}$,  and consequently it is not valid at large $Q^2$. 
Nonetheless, $\alpha_{AdS}$ can be supplemented
at large $Q^{2}$ by either parameterizing the well-known pQCD effects
at the origin of the large $Q^{2}$ dependence, as done in \cite{the:alpha_g_1 from AdS},
or by matching Eq. (\ref{eq:alpha_g_1 from AdS/QCD}) to Eq. (\ref{eq:alpha_s}),
as done in \cite{the:Deur Lambda AdS/QCD} and discussed in Section \ref{sub:Holographic-QCD large Q}.
This matching procedure is illustrated in Fig. \ref{Flo:matching_1}.

\begin{figure}[ht]
\centering
\includegraphics[scale=0.45]{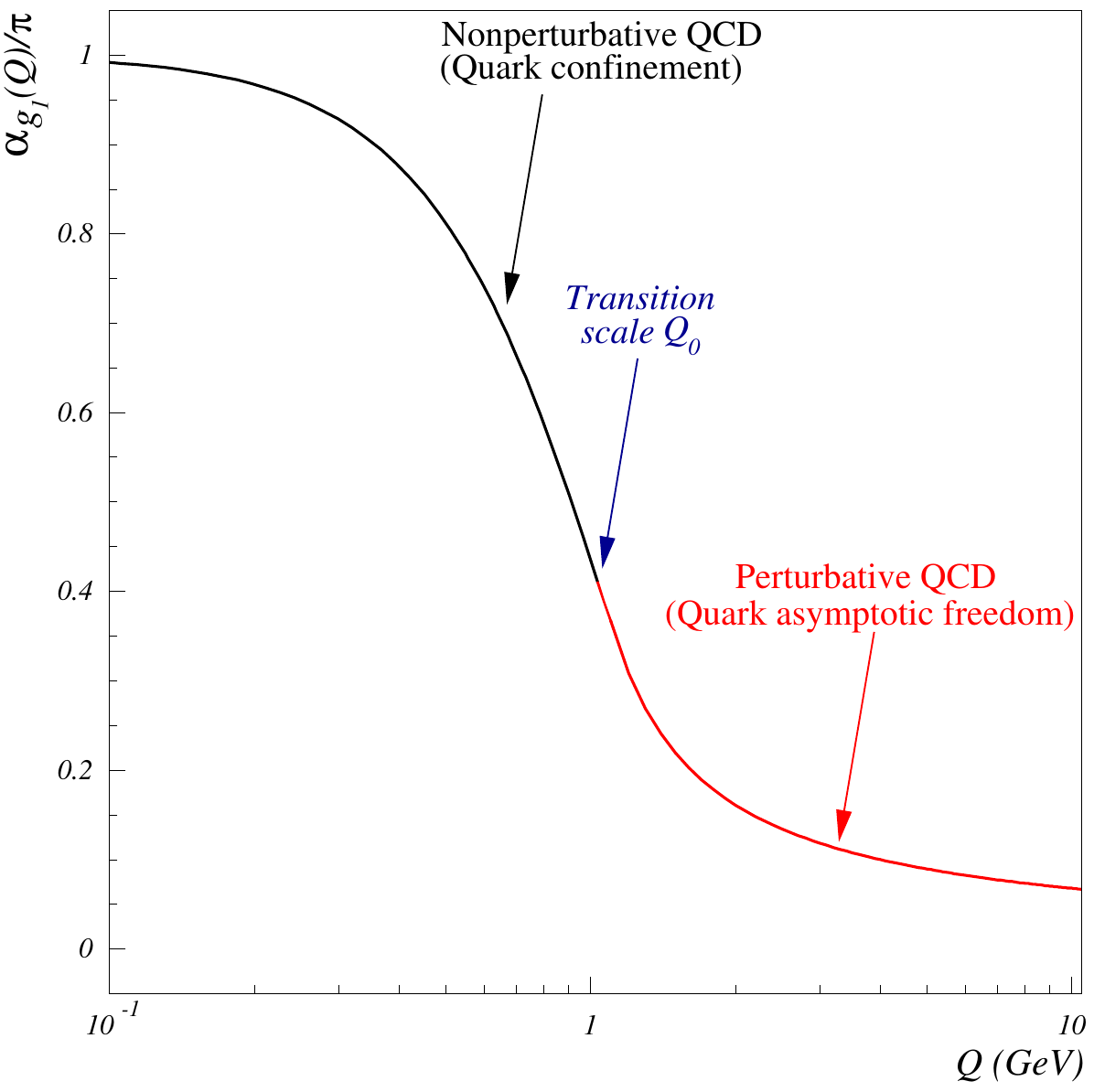}
\caption{\small \label{Flo:matching_1} Unified coupling obtained from the analytic matching of nonperturbative  and perturbative QCD regimes. The procedure determines the relation between $\Lambda_{\overline {MS}}$ and $\kappa$ or equivalently hadron masses. The transition scale $Q_0$ between the large and short-distance regimes of QCD is determined as well.}
\end{figure}

The Gaussian form $\alpha_{s}(Q^{2})=\alpha_{s}(0)e^{-Q^{2}/4\kappa^{2}}$
predicted by
 the light-front holographic approach describes  the available experimental data for the effective
charge $\alpha_{g_1}(Q^{2})$ very well  over  a large range of moment transfer; see Fig. \ref{Flo:lowq alpha. exp. }. 
Conversely, a fit of the experimental data  shown in Fig.~\ref{Flo:lowq alpha. exp. } with $\kappa$ as the free fit 
parameter yields $\kappa=0.496 \pm 0.007$, in good agreement with the 
determination from the value $\kappa=0.50 \pm 0.04$ obtained from 
the relation between $\kappa$ and hadron masses: 
$\kappa= M_\rho/ \sqrt{2}$ or $\kappa= M_p/ 2$~\cite{the:AdS/QCD review}. 
It is remarkable that the parameter $\kappa$ which fits the Bjorken
sum rule data is set independently by a hadron mass.

 Additionally, it has been shown using
the Schwinger--Dyson equations for the quark propagator that if one
assumes a Gaussian form for the QCD running coupling, the quark propagator
has poles above and below the real axis, 
consistent with quark confinement \cite{the:Cloet 2014}.
Finally, the \emph{scaling solution} of the Schwinger--Dyson equations yields a coupling in close agreement 
with $\alpha_{AdS}$, with a \emph{freezing}  value of 2.97 (Section \ref{sub:Schwinger--Dyson-formalism}).

\section{The  effective potential approach\label{sub:Potential-approach}}

\subsection{The static $\mbox{Q--}\overline{\mbox{Q}}$ potential}

A remarkable feature of hadron physics is that mesons and baryons composed of light quarks all  lie on Regge trajectories. 
The square of the hadron masses are linear in both the internal orbital angular momentum $L$ and the principal quantum number $n$ which is associated with the number of nodes in its wavefunction. Moreover the slope in $L$ and $n$ are identical phenomenologically.
These universal features in fact follow from the eigenvalues of LFSE with the harmonic oscillator potential $V(\zeta^2) = \kappa^4 \zeta^2$~\cite{the:AdS/QCD review}. 
The  lightest  eigenvalue, corresponding to the pion, has zero mass in the  chiral limit $m_q \to 0.$

In the case of heavy quark $\mbox{Q--}\overline{\mbox{Q}}$ states,  the effective nonrelativistic 
static potential between two heavy quarks ($\mbox{Q--}\overline{\mbox{Q}}$) of mass
$m_{Q}\gg\Lambda$ separated by a distance $r$  is well described by:
\begin{equation}
V(r)=-\frac{4}{3}\frac{\alpha_{V}\left(r\right)}{r}+\sigma r. \label{eq:Q-Q stat pot.}
\end{equation}

The first term is a perturbative, Coulomb-type, one-gluon exchange contribution.
The hadronic wavefunctions and the fine structure of the hadron spectrum are sensitive to this term.
When $\alpha_{V}(r)$ is approximated as an averaged 
 coupling $\left\langle \alpha_{V}\right\rangle $,
Eq. (\ref{eq:Q-Q stat pot.}) is known as the Cornell potential \cite{the:Eichten. Cornell pot.}.
The mean value of the 
 coupling $\left\langle \alpha_{V}\right\rangle $ depends
on the size of the hadrons considered. Its value in GeV units spans from 0.19
to 0.4, with an optimal value of 0.22 \cite{the:Eichten. Cornell pot., the: Celmaster et al. 1979-1981, the:More Cornell Pot.}.
In the context of an IR-\emph{freezing}  behavior of $\alpha_{V}(Q^{2})$,
this implies a \emph{freezing}  value of $\alpha_{V}(0)=0.42 \pm 0.03$ \cite{the:Eichten. Cornell pot., the: Celmaster et al. 1979-1981,
the:More Cornell Pot.}.

The second term in Eq. (\ref{eq:Q-Q stat pot.}) is a linear confining
potential which  can be interpreted as the string tension between the $\mbox{Q--}\overline{\mbox{Q}}$ pair, with
$\sigma\sim0.18$ GeV$^{2}$. This term controls the slopes and 
intercepts of the Regge trajectories. The QCD string picture was  first postulated 
by Nambu, Nielsen and Susskind
to interpret the Regge trajectories \cite{the:Nambu}. A common interpretation is that the string is
formed by a chromoelectric flux tube and is responsible for quark
confinement. 
Regge trajectories 
stem straightforwardly from the string picture:
the faster a hadron spins, the larger the string tension must be
to compensate for the centrifugal force, and hence the larger its
mass \cite{the: Greensite conf.}.

In fact, the AdS/QCD holographic QCD framework which 
yields the harmonic oscillator potential $V(\zeta^2) = \kappa^4 \zeta^2$ in the light-front  
transverse coordinate $\zeta$  for light quark pairs, corresponds to a linear potential for heavy quarks 
in the non-relativistic
 instant-form radial coordinate $r$ for large distances~\cite{the:Trawinski 2014}.
The string tension for heavy quarks can then be predicted from holographic QCD to be:
\begin{equation}
\sigma_{AdS} = 2\kappa^{2}/\pi\simeq0.18~\mbox{GeV}^2,
\end{equation}
which is remarkably consistent with fits to the quarkonium data.

A more accurate potential than that provided by Eq. (\ref{eq:Q-Q stat pot.}) 
includes velocity-dependent and spin-dependent 
corrections, see Refs. \cite{the:Brambilla Hugs lectures}
and \cite{the:Pineda review Q-Q potential} for reviews.

The validity of Eq. (\ref{eq:Q-Q stat pot.}) is confirmed for heavy quarks by
lattice QCD simulations. However, in  (2+1) lattice simulations, $\alpha_{V}(0)\sim0.3$
\cite{the:Bali 1995 et al.}, somewhat smaller than  the value determined
from the Cornell potential. Earlier \emph{quenched} calculations resulted
in an even smaller value of the coupling, $\alpha_{V}(0)\sim0.22$
\cite{the:Bali 1995 et al.}.

In addition to lattice QCD,  other  nonperturbative approaches also suggest the form of Eq. (\ref{eq:Q-Q stat pot.}).
In the Coulomb gauge, the instantaneous gluon propagator leads to
a linear static $\mbox{Q--}\overline{\mbox{Q}}$ potential \cite{the:Cucchieri. 2002 proc., the:Szczepaniak 2001,
the:Nakamura Coulomb pot from lattice,
the:Cucchieri linear pot coulomb gauge}, as was advocated
by Gribov and Zwanziger~\cite{the:Gribov 1978, the:Zwanziger 1981-1993};  
see Section \ref{sub:Gribov--Zwanziger-approach}.
Generally, it is expected that 
in the IR, the gluon propagator is modified by  nonperturbative effects
and displays a $1/\left(Q^{2}+f(Q^2) \right)^2$ dipole form  that would lead
to the linear IR potential  \cite{the:Gracey stat pot 2009-2010}. (The function  $f(Q^2)$ is  unimportant 
here. It is typically related to an effective gluon mass or a glueball mass.)
This is studied in the Schwinger--Dyson formalism, see Section
\ref{sub:Schwinger--Dyson-formalism} and in particular Refs. \cite{the:Linear pot. in SDE, the:Epple 2007},
and in the related approach of the   functional renormalization group
\cite{the:Ellwanger linear pot in FRG}, see Section \ref{sub:Functional renormalization group equations}.
A linear potential is also recovered using   background perturbation theory~\cite{the:Simonov 1993}, 
see Section \ref{sub:Background pert theo, Simonov}
for details.

 To summarize, Eq.  (\ref{eq:Q-Q stat pot.}) is well established. Nevertheless, there is no agreement
on the IR-behavior of $\alpha_s$, owing to its various possible definitions. Still, within a specific definition,
Eq.  (\ref{eq:Q-Q stat pot.}) -or more generally hadron spectroscopy- does constrain $\alpha_s$ as we 
will now discuss.

\subsection{Constraints on the running coupling from the hadron spectrum\label{Sub: hadron spectrum}}

The hadron spectrum is sensitive to the IR-behavior of $\alpha_s$ when 
one identifies the strong coupling as the \emph{effective charge} 
underlying the heavy $\mbox{Q--}\overline{\mbox{Q}}$ potential.
Hence, the heavy quark potential can provide a useful information on
$\alpha_s(Q^2)$ in the IR. 
For example in the Godfrey--Isgur quark/flux-tube model 
\cite{the:Godfrey-Isgur}, $\alpha_{s}$
must  \emph{freeze} in the IR at $\alpha_{s}(0)=0.6$.
Other examples are the model of Zhang and Koniuk
\cite{the:Zhang Koniuk quark model}, or the work
of Badalian and collaborators using the framework of  background perturbation
theory.  Badalian \emph{et al.} obtain $\alpha_{s}(0)=0.58$ from the analysis
of the bottomonium, see Section \ref{sub:Background pert theo, Simonov}
and  Ref. \cite{the:Badalian Q-Q static force (2005)}. Likewise,
Andreev \cite{the:Andreev 2011-2013} uses several 
of the forms for  $\alpha_{s}(Q^2)$
suggested in the literature and constrains them with
a Poincar\'{e} covariant quark model to reproduce meson characteristics
and experimental data, including the Bjorken sum rule, 
Eq. (\ref{eq:Bj SR, order alpha^4}), and the GLS sum rule, Eq. (\ref{eq:GLS sum rule}). 
Andreev obtains a consistent set of behaviors
by using $\alpha_{s}(Q^2)$  defined in the $\overline{\mbox{MS}}$ 
scheme, but regularized with a constant effective
gluon mass using the massive analytic perturbation theory
approach (Section \ref{sub:Analytic approach}), or
with Cornwall's coupling (see Section \ref{sub:SDE massive gluon}) or
with the ``synthetic coupling'' (also
discussed in Section \ref{sub:Analytic approach}).
This procedure results in a  \emph{freezing}  value $\alpha_{s}(0) \simeq 0.71 $ in the $\overline{\mbox{MS}}$
renormalization scheme.

Baldicchi and Prosperi also tested
several different expressions of $\alpha_{s}(Q^{2})$ to check their
compatibility with  the meson spectrum \cite{the:Baldicchi 2002}. One of the expressions
studied is a crude \emph{freezing}  implementation: below a given $Q_{0}$, 
$\alpha_{s}$ is constant, and above $Q_0$ it follows the pQCD form. 
Another expression of $\alpha_{s}$ is given by the
``analytic''  approach  described in Section \ref{sub:Analytic approach}
and given by Eq. (\ref{eq:alpha_s from analytic QCD}). This $\alpha_{an}$ is finite and decreases
monotonously with $Q^2$. A  related expression
has been give  by Dokshitzer \emph{et al.}, and is given by Eq. (\ref{eq:alpha_dmw}). 
Baldicchi and Prosperi  conclude that the meson spectrum appears to be rather insensitive
to the type of coupling used \cite{the:Baldicchi 2002}. 

Historically, these approaches follow from the pioneering works of Richardson and of Buchmuller \emph{et al.}.  These authors
suggested that $\alpha_{s}$ should be defined to incorporate
a  long range force in addition to the usual pQCD effects responsible
for the running of the coupling \cite{the:Richardson 1979, the:Buchmuller Grunberg}. 
In the non-relativistic (static) approximation, Eq. (\ref{eq:Q-Q stat pot.}) becomes in momentum space:
\begin{equation}
V(Q)=\frac{1}{\left(2\pi\right)^{3}}\int d^{3}re^{-ir.Q}V(r)=-\frac{1}{\pi^{3}}\frac{2}{3}\frac{\alpha_{V}\left(Q^{2}\right)}{Q^{2}}-\frac{\sigma}{\pi^{3}Q^{4}}.
\end{equation}
Incorporating in the coupling the long distance  $\sigma$-term  leads to the definition of $\alpha_{Rich}(Q^{2})$:
\begin{equation}
V(Q^{2})=-\frac{1}{\pi^{3}}\frac{2}{3}\frac{\alpha_{Rich}(Q^{2})}{Q^{2}}.\label{eq:Q-Q stat pot. Mom space.}
\end{equation}
Richardson  \cite{the:Richardson 1979} proposed that:
\begin{equation}
\label{eq:Richardson}
\alpha_{Rich}\left(Q^{2}\right)=\frac{4\pi}{\beta_{0}\mbox{ln}\left(1+Q^{2}/\Lambda_{V}^{2}\right)},
\end{equation}
where $\Lambda_V$ should be used since $\alpha_{Rich}$ identifies to $\alpha_V$ in 
the UV. The corresponding \emph{$\beta$-function} is:
\begin{equation}
\beta=-\beta_{0}\alpha_{Rich}^{2}\left(1-e^{-1/(\beta_{0}\alpha_{Rich})}\right).
\end{equation}
This expression anticipated results from the Schwinger--Dyson  formalism;  see Eq.
(\ref{eq:alpha_s cornwall}), in which the coupling is regulated by
an effective dynamical gluon mass $m_g(Q^2)$.  Using Richardson's Eq. (\ref{eq:Richardson}), the gluon mass is
assumed to be
constant and equal to $\Lambda_V$, which is close to the $m_g(0)$ values suggested by Schwinger--Dyson
and lattice studies. The $Q^{2}\rightarrow 0$ limit of Eq. (\ref{eq:Richardson}) is:
\begin{equation}
\alpha_{Rich}\left(Q^{2}\right)=\frac{4\pi\Lambda_{V}^{2}}{\beta_{0}Q^{2}}, \end{equation}
 and the string tension
\begin{equation}
\label{Eq:Richardson string tension}
\sigma=\frac{8\pi\Lambda_V^{2}}{3\beta_{0}}.
\end{equation}
Using the phenomenological value $\sigma=0.18$ GeV$^2$, one obtains $\Lambda_{V}=0.46$ GeV for $n_{f}=2$
in agreement with its high-$Q^{2}$ pQCD determination $\Lambda_{V}=0.48$
GeV, see Table \ref{Flo:Table of Lambda}.  Thus, if all the mechanisms leading to the
long-distance linear potential are included in the definition of $\alpha_{V}$,
then the $1/Q^4$ divergence of $V$ in the instant-form coordinate system 
implies that $\alpha_V$ should diverge 
as $3\sigma/4Q^{2}$  in the limit $Q^2 \rightarrow 0$.

The fact that the linear $\mbox{Q--}\overline{\mbox{Q}}$ potential leads to $\alpha_V \propto 1/Q^2$ 
has inspired several phenomenological models. Belyakova
and Nesterenko \cite{the:Belyakova 2011} have constructed a coupling
with the constraints that it follows $\alpha_{\overline{MS}}(Q^{2})$
in the UV and $\alpha_{s}(Q^{2})\simeq\frac{4\pi}{\beta_{0}}\frac{\Lambda^{2}}{Q^{2}}$
in the IR. The Belyakova--Nesterenko coupling 
is defined by the \emph{$\beta$-function}:
\begin{equation}
\beta\left(\alpha_{BN}\right)=\beta_{\overline{MS}}^{(l)}\frac{1-e^{-8\pi/\alpha_{BN}\beta_{0}}\left(1-l^{2}\beta_{0}^{l}/\beta_{l}\right)}{1+l^{2}\left(\alpha_{BN}\beta_{0}/\left(4\pi\right)\right)^{l}}.
\end{equation}
It resembles the \emph{$\beta$-function}s  proposed
in Refs. \cite{the:Richardson 1979, the: Celmaster et al. 1979-1981}.
At leading order, it yields:
\begin{equation}
\alpha_{BN}\left(Q^{2}\right)=\frac{4\pi}{\beta_{0}}\frac{1}{W_{0}\left(Q^{2}/\Lambda^{2}\right)},
\end{equation}
where $W_{0}$ is the principal branch of the real-valued Lambert function;
see Section \ref{sub:pQCD evolution-equation}. The static $\mbox{Q--}\overline{\mbox{Q}}$
potential obtained for $\Lambda^{(3)}=375\pm40$ MeV agrees with both
lattice QCD and the Cornell potential, Eq.
(\ref{eq:Q-Q stat pot.}). The model provides the same  relation between  $\Lambda$ and
$\sigma$  found by Richardson, see Eq. (\ref{Eq:Richardson string tension}).

\section{The Schwinger--Dyson formalism \label{sub:Schwinger--Dyson-formalism}}

Studies of the Schwinger--Dyson equation (SDE) for QCD have also provided  important
insights into the  IR-behavior of $\alpha_s(Q^2)$. 
In effect, the SDE (and the corresponding  Bethe-Salpeter formalism  for bound 
states)  incorporates an infinite set of coupled non-linear
equations relating  \emph{Green's function}s as well as other constraints from the QCD equations of motion. 
In this formalism,  the equation for the
$n$-point function depends on the equation  for the $(n+1)$-point function. 
Although the SDE formalism can be described  using perturbative
Feynman graphs, infinite sets of diagrams are resummed into one-particle
irreducible graphs, and thus  the SDE can account for the  nonperturbative
content of the theory. Alternatively, the SDE can be derived using the generating
functional of the  \emph{Green's function}s without reference to perturbative
graphs. The SDE formalism thus stems from the fundamental equations of  QCD, 
encompasses its  nonperturbative effects, and is able in principle
to provide analytical  results. As such, it provides   a powerful
tool. Furthermore, unlike Lattice QCD, the SDE formalism is not subject to unphysical artifacts
resulting from space-time discretization, nor unphysically large quark
masses.  These effects can complicate studies of important hadronic phenomena. 

However, the SDE framework has its
own difficulties: the infinite system of equations must be truncated
to be amenable to solutions.  In fact, any truncation of the kernel conflicts with crossing symmetry. 
In the case of QED, one must include an infinite number of crossed graph kernels  in the Bethe-Salpeter equation in order  to derive the Dirac Coulomb equation for muonium $\mu^+ e^-$ in the limit of $m_{\mu}\to \infty.$
This implies that the SDE approach contains approximations which may  lead to effects that
are not easily understood and controlled. Also, the equations are
difficult to solve and analytical answers can be hard to obtain without
further approximations. Consequently, the SDE quantitative results
are often obtained numerically. 
The SDE truncations are typically chosen to ensure 
that cut off propagators are positive (to retain unitarity), that the high-$Q^{2}$ behavior of
the QCD coupling matches RGE expectations,
and that the  \emph{Green's function}s retain their  multiplicative
renormalizability properties. The last property implies that a full
propagator is proportional to its corresponding bare propagator via a {}``dressed
function''.   

Despite the difficulties inherent to the SDE, this approach provides
an important method for furthering our understanding of QCD in general and
of $\alpha_{s}$ in particular. It is especially powerful when studied
alongside other techniques which will be discussed in the next 
sections, noticeably lattice QCD, so that these methods can supplement
each other. In these different approaches, couplings have
the same definition and the RS is usually the same (the MOM RS or
derivatives). Consequently, comparing quantities in SDE 
to other techniques is possible. As an example of the complementarity
between the SDE framework and lattice approaches, the SDE can be used to understand the effects
of space-time discretization and  of choice of boundary conditions for
the lattice approach, whereas the lattice can be used to understand the effect
of truncating the infinite system of equations of the Schwinger--Dyson
formalism.  For general reviews on SDE see Ref. \cite{the:SDE review}.
Reviews of SDE results discussing the IR-behavior
of $\alpha_{s}$ can be found in \cite{the:Alkofer von smekal SDE review, the:Fischer Review SDE in IR}.

\subsection{The QCD coupling defined from Schwinger--Dyson equations  \label{sub:coupling from SDE}}

The behavior of the gluon and ghost Green's  functions is directly related to the running of $\alpha_{s}$. 
The required propagators 
and vertices which define these  \emph{Green's function}s have been extensively studied using  the Schwinger--Dyson formalism,  on the lattice,
and by using the Gribov--Zwanziger approach \cite{the:Gribov 1978, the:Zwanziger 1981-1993}, as well as other methods.
See \cite{the:Boucaud et al review Yang Mills in IR, the:Mass report on correl func. 2013}
for recent reviews.   For example, the strong coupling derived  from the
ghost--gluon vertex $\alpha_{s}^{gh}(Q^{2})$ is given in the
Landau gauge and the MOM RS by 
\begin{equation}
\alpha_{s}^{gh}\left(Q^{2}\right)= \alpha_{s}^{gh} \left(\mu\right)G^{2}
\left(Q^{2},\mu\right)Z\left(Q^{2},\mu\right),\label{eq:alpha_s SDE ghost--gluon}
\end{equation}
where $\alpha^{gh}_{s}(\mu)$ is the coupling determined at the  \emph{renormalization scale}
$\mu$,  $Z(Q^{2},\mu)$ is the gluon propagator dressing function, and $G(Q^{2},\mu)$ 
is the ghost propagator dressing function (the ghost propagator being the
inverse of the Fadeev--Popov operator). The RG invariance of Eq. (\ref{eq:alpha_s SDE ghost--gluon})
is evident: the quantities forming the right-hand side depend on $\mu$,
but on the left hand side, $\alpha_{s}^{gh}$ does not.  The dressing functions
are defined as 
$\delta^{bc} G(Q^{2},\mu)/Q^{2}=G^{bc}(Q^{2},\mu)$ 
and 
$-\delta^{bc}Z(Q^{2},\mu)(\delta_{\mu\nu}-q_{\mu}q_{\nu}/q^{2})/Q^{2}=D^{bc}_{\mu\nu}(Q^{2},\mu)$
where $b,c$ are color indices, $G^{bc}(Q^{2},\mu)$ is the ghost propagator, and $D^{bc}_{\mu\nu}(Q^{2},\mu)$
is the gluon propagator.  Since the dressing functions relate the first-order
({\it i.e.} bare)  \emph{Green's function}s to their fully ({\it i.e.} dressed) expressions,
they are often called ``form factors'' in analogy with the functions
modifying the point-like expression for  elastic scattering
due to hadron structure. There is no contribution from
the vertex dressing function in Eq. (\ref{eq:alpha_s SDE ghost--gluon}) owing to the \emph{Slavnov--Taylor} identity \cite{the:Slavnov--Taylor id.}.

One sees from Eq. (\ref{eq:alpha_s SDE ghost--gluon}) that the analytic form of the  
propagators  provides a  definition of the strong coupling.
Other definitions are possible using  the three-gluon vertex (defining
the coupling $\alpha_{s}^{3g}$), the four-gluon vertex (defining
$\alpha_{s}^{4g}$) or the
quark--gluon vertex (defining $\alpha_{s}^{qg}$). 
Since ghosts do
not couple to themselves or to quarks, no further coupling definitions are possible.

The  $\alpha_{s}^{3g}$,  $\alpha_{s}^{4g}$ and $\alpha_s^{qg}$ definitions involve 
vertex dressing functions which are more difficult to study  
\cite{the:Comparison alpha-g-gh alpha-3g alpha-4g}. In particular,
there are a large number of 
contributing Feynman graphs. 
Thus $\alpha_{s}^{gh}$
has been the focus of most of the initial SDE studies, although several
groups have calculated the other couplings,
The quantities $\alpha_{s}^{gh}$, $\alpha_{s}^{3g}$, $\alpha_{s}^{4g}$,
$\alpha_{s}^{qg}$ are the same in a massless RS, such as $\overline{MS}$, but 
not for a massive RS such as the MOM RS \cite{the:Shirkov alpha_s low_Q review}.
Nevertheless, these couplings can be related to each other
\cite{the:Comparison alpha-g-gh alpha-3g alpha-4g, the:betas in mini MOM., the:Shirkov alpha_s low_Q review, Gracey:2011vw}, 
and the differences vanish
at large $Q^{2}$.   Another coupling used in the SDE is $\alpha_{gse}$, defined
from the gluon self-energy \cite{the:Cornwall alpha_s}. It has been shown to be 
closely related  to $\alpha_s^{gh}$ in the Landau gauge \cite{the:Aguilar (2009)}.

\subsection{Gauge choices \label{sub:SDE Gauge-choices}}

 \emph{Green's function}s are not directly observable, and they become gauge-dependent if one utilizes 
the MOM RS.  This also makes the derived results for  $\alpha_{s}^{gh}$, $\alpha_{s}^{3g}$, 
$\alpha_{s}^{4g}$ and $\alpha_{s}^{qg}$ gauge dependent.   
An alternative procedure is to use  Cornwall's \emph{Pinch} scheme to define 
a gauge-independent three-gluon coupling \cite{Binger:2006sj,  the:Cornwall alpha_s PT}.

Most SDE
results relevant to $\alpha_{s}$ are obtained in the Landau gauge,  $\partial_\mu A^\mu =0$,
since the relatively small numbers of fields 
and vertices make computations  of the SDE simpler relative to many other
gauges. It is also a Lorentz-covariant gauge, which, beside making the
\emph{Kugo--Ojima confinement criterion} applicable, further simplifies 
solving the SDE.  For example, of all the possible tensors (from bosons)
or spinor (from fermions) terms which form the 2-point  \emph{Green's function},
only two survive generic symmetry requirements.  They lead to a boson
propagator that can be written as $A(Q^{2})q^{2}\eta^{\mu\nu}-B(Q^{2})q^{\mu}q^{\nu}$. 
These  two functions,  which are \emph{a priori} independent,
become evidently equal in Lorentz-covariant gauges such as the 
Landau gauge:  $A(Q^{2})=B(Q^{2})$, yielding a propagator that depends
only on $Q^{2}$.  In contrast, gauges which are not covariant have
propagators -- and consequently  predictions for $\alpha_{s}$ -- which can depend on
another variable in addition to $Q^{2}$.   For example, in the first calculations of the
gluon propagator done in the (non-covariant) axial gauge \cite{the:Baker gluon prop SDE 1981.},
$B(Q^{2})$ was  set to zero in order to force the gluon propagator to
only depend on $Q^{2}$, as is the case in the perturbative domain. 
The assumed vanishing of  $B(Q^{2})$ leads
to a $1/Q^{4}$ behavior of the gluon propagator.  Since there are
no ghosts in this gauge, one predicts a singular behavior 
$\alpha_{s}\propto1/Q^{2}$ in the IR. This behavior seems satisfactory,  given the
expectation that it leads to a  linear $\mbox{Q--}\overline{\mbox{Q}}$ potential; see Section 
 \ref{sub:Potential-approach}. However, more complete SDE
calculations in covariant gauges and subsequent lattice calculations have shown
this behavior to be an artifact of the assumption $B(Q^{2})=0$.

Another advantage of the Landau gauge is that the equations involving
the transverse and longitudinal propagators decouple. The on-shell gluon
propagator should be transverse, and thus it can be studied independently.
In addition, in the Landau gauge the ghost--gluon vertex has a finite renormalization
constant \cite{the:Slavnov--Taylor id., the:Bloch lattice SU2 (2004)},
which allows the coupling to be defined and extracted from Eq. (\ref{eq:alpha_s SDE ghost--gluon}); 
using  calculations involving only the two-point  \emph{Green's functions}
$G(Q^{2})$ and $Z(Q^{2})$.  The Landau gauge is also symmetric under
the exchange of ghost and anti-ghost.  
The gauge-dependence of the running coupling using this procedure has been
argued to be weak for couplings computed in gauges close to the
Landau gauge. This was checked by using various linear covariant 
gauges \cite{the:Fischer Review SDE in IR}. Notwithstanding the preeminence
of the Landau gauge, other gauges, such as the Coulomb, Feynman 
\cite{the:Aguilar ghost--gluon vert. in feynman gauge 2008, 
the:Aguilar SDE 2008, the:Aguilar Papavassiliou SDE (2006)}, axial \cite{the:Baker gluon prop SDE 1981.}, 
light-cone \cite{the:Cornwall alpha_s},
maximal Abelian gauge, \cite{the:Capri. SDE props in Max Abelian gauge}
and linear covariant gauges (which interpolate between the Landau
and the Feynman gauges) have also been used \cite{the:Fischer Review SDE in IR, the:Schleifenbaum 2006, 
the:Aguilar m_g gauge dependence}.

\subsection{Classes of solutions in the IR domain \label{sub:Classes-of-solutions in IF domain}}

There is presently a consensus that two types of solutions are possible for  the IR behavior of
the gluon propagator, and thus for  the IR behavior of $\alpha_s$. The first solution, called {}``scaling
scenario'' (or also {}``\emph{conformal} scenario'', {}``critical solution''
or {}``ghost dominance scenario''), was first obtained
and studied using the SDE formalism.   Many results obtained using the SDE (see {\it e.g.}
\cite{the:Zwanziger Stoch. quant. 2002}-\cite{the:Bloch S-D alpha_s 2003}), 
Lattice (see Section \ref{sub:Low Q Lattice-QCD-calculations}),
variational approach to the Yang--Mills Schr\''odinger equation of the
vacuum \cite{the:Schleifenbaum 2006} and the  exact renormalization
group equations method (Section \ref{sub:Functional renormalization group equations}),
show that in the IR, $G(Q^{2},\mu^{2})$ and $Z(Q^{2},\mu^{2})$ scale with related power
laws: $G(Q^{2},\mu^{2})\propto(Q^{2})^{-\kappa'}$ and $Z(Q^{2},\mu^{2})\propto(Q^{2})^{2\kappa'}$,
where $\kappa'$ is called the \emph{IR exponent}. (This coefficient is
generally labeled $\kappa$, or sometimes $\alpha$, in the 
literature. We write it here $\kappa'$ to differentiate it
from the mass scale $\kappa$ used in  holographic QCD in Sections
\ref{sub:Holographic-QCD large Q} and \ref{sub:Holographic-QCD Lowq}.).
Such relations imply an IR \emph{freezing}  of $\alpha_{s}^{gh}$, see Eq.
(\ref{eq:alpha_s SDE ghost--gluon});  thus  the alternative name of {}``\emph{conformal} scenario''. 
These relations  also imply that ghosts dominate in the IR over gluons.
It was shown that the scaling scenario in the Landau gauge  leads to
a well defined running of the coupling \cite{the:Lerche von Smekal 2002,
the:Fischer against decoupling 2007, the:Alkofer 2010,
the:Leder 2011 FRG}. In this gauge, the \emph{IR exponent} was calculated
to be $\kappa'=(93-\sqrt{1201})/98\simeq0.595$, using the SDE
\cite{the:Schleifenbaum 2006, the:Zwanziger Stoch. quant. 2002, the:Lerche von Smekal 2002}
or the  functional renormalization group method \cite{the:Fischer  IR exp from FRG, 
the:Pawlowski IR exp from FRG}. This value leads to an IR \emph{freezing} 
value  $\alpha_{s}^{gh}\left(0\right)\simeq2.972$.

It should be noted that the SDE computations of $\kappa'$ can depend on
the approximations used to solve the SDE.  This has resulted in a range
of values of $0.3\lesssim\kappa'\lesssim1$. Lattice calculations
indicate that $0.5\lesssim\kappa'\lesssim0.6$ with $\kappa'\simeq0.5$
favored, see \emph{e.g.}~\cite{the:Furui Nakajima}.  Lerche and von Smekal
\cite{the:Lerche von Smekal 2002}, and then Alkofer and collaborators
\cite{the:Alkofer 2003} have studied $\alpha_{s}^{gh}$ within this range
using the Landau gauge and have obtained $2.5\lesssim\alpha_{s}^{gh}\left(0\right)\lesssim2.972$.
The \emph{scaling solution} also appears in other gauges. It has been studied in
the Coulomb gauge by Burgio \emph{et al.} \cite{the:Burgio 2010} using the lattice, and by 
Epple \emph{et al.} \cite{the:Epple 2008 SDE scaling sol. Coul. Gauge} 
using the SDE. The \emph{IR exponent} 
$\kappa'$ is computed in the Landau,  interpolating and Coulomb
gauges in Ref.~\cite{the:Schleifenbaum 2006}, confirming $\kappa'\simeq0.595$ in the Landau and  interpolating gauges.
In the Coulomb gauge, two possible values are found: $\kappa'\simeq 0.398$
and  $\kappa'=0.5$. These values are also found
using the ``functional renormalization group'' framework \cite{the:Leder 2011 FRG} --  see Section
\ref{sub:Functional renormalization group equations}   and by using
a variational approach  in Refs. \cite{the:Epple 2007, the:Epple 2008 SDE scaling sol. Coul. Gauge}.
However, a solution with $\kappa'=0.5$ seems not to be favored \cite{the:Leder 2011 FRG}.

Fisher and collaborators have studied the effects of  a finite volume in lattice simulations
and the corresponding effects of boundary conditions on the value of $\kappa'$ \cite{the:Fischer lattice artifact study with SDE}
by performing SDE calculations both on a torus and for infinite space;  one 
finds $\kappa'\simeq0.5$ and $\kappa'\simeq0.6$, respectively.
The  coupling $\alpha_{s}^{gh}$ \emph{freezes} in the IR, irrespective of the
value of $\kappa'$.  The work of Fisher \emph{et al.} also indicates an IR-vanishing gluon propagator
in the case of infinite space, whereas it stays finite in the finite-volume
case. The dependence of $\kappa'$ with variable spatial dimensions was studied
in \cite{the:Huber kappa' in 2 3 4D} for 2, 3 and 4 dimensions. The exponent
$\kappa'$ was found to not vary dramatically, suggesting similar
behavior of the underlying \emph{Green's function}s in these spaces. 

A second  possible solution for the IR behavior of $\alpha_s$, called the {}``\emph{decoupling solution}'' (or {}``massive
solution'', or {}``subcritical solution''),
was discovered by Boucaud and collaborators using lattice
calculations \cite{the:Boucaud decoupling sol. discovery}.  In this
case, the ghost propagator dressing function $G(Q^{2},\mu^{2})$ is constant in the IR
and the gluon propagator dressing function scales as $Z(Q^{2},\mu^{2})\propto Q^{2}$.
The predicted \emph{IR exponents} are different:  0 for the ghosts
and 1 for the gluons.  Thus according to Eq. (\ref{eq:alpha_s SDE ghost--gluon})
this solution produces an IR-vanishing $\alpha_{s}^{gh} \propto Q^2$.  The name  \emph{decoupling solution} 
originates from the vanishing of $Z(Q^{2},\mu^{2})$ as $Q^2 \rightarrow 0$, 
leading to free ghosts in the IR limit. That is, the ghosts decouple from an effectively 
massive longitudinal gluon field.  In the decoupling scenario, the IR regime is still dominated by ghosts, although less
strongly than in the scaling scenario where they diverge. 

The suppression of the  gluon field in the \emph{decoupling solution} can be 
interpreted as due to a dynamically generated gluon mass. The \emph{decoupling solution} is
thereby consistent with a Yukawa-like gluon propagator in the IR.
The massive solution is thus sometimes associated with massive scenarios,
such as Cornwall's \cite{the:Cornwall alpha_s, the:Cornwall 1975 m_g} massive gluon which 
will be discussed in the next  section. However, Cornwall's pioneering
work is not in the Landau gauge;  it does not require ghosts,  and it
uses a different coupling, $\alpha_{gse}$ (although it has been shown to be 
similar  to $\alpha_s^{gh}$ \cite{the:Aguilar (2009)}).  Cornwall's results produce a non-vanishing \emph{freezing} 
$\alpha_{gse}$.  

Although it was first found on the lattice, the \emph{decoupling solution} is also present in the SDE calculations
\cite{the:Alkofer 2010,  the:Boucaud decoupling sol. discovery, the:Aguilar-Natale SDE 2005, the:Aguilar 2008,
the:Fischer 2009 against decoupling}, the  \emph{pinch} technique background field method \cite{the:PT bg fd}, an
extended Gribov--Zwanziger approach~\cite{the:Dudal 2010, the:Dudal 2008-2011},
and from the  stochastic quantization equations \cite{the:Llanes-Estrada 2012}.
This last work suggests, within the approximations used to truncate
the equations, that the \emph{decoupling solution} is a general solution
of the theory. In contrast, the \emph{scaling solution} would appear in specific gauges only,
and represent a particular solution stemming from the existence
of a Gribov horizon ({\it i.e.}, the limit of the finite region of the field configuration
space where the ghost propagator remains positive and finite, see
Section \ref{sub:Gribov--Zwanziger-approach}). This work also indicates
that the \emph{decoupling solution} has less effective action, which explains why it seems
favored by the lattice technique. 

It is still unclear which solution is realized in the physical world.
Certainly, the description  given above of the two solutions is gauge-dependent, 
since some gauges, 
some relativistic form of dynamics, {\it e.g.}, the light-front, and some quantization methods do not require ghosts. 
In such a case, the ghosts are replaced by other degrees of freedom, such as the longitudinal 
component of the gluon propagator, see Section \ref{sub:Stochastic-quantization}.
Thus the two solutions can still exist in a gauge without ghosts. 

Some 
 authors~\cite{the:Fischer lattice artifact study with SDE}  have indicated 
that some  decoupling-like results (that is calculations leading to a vanishing coupling)
may be  due to a finite lattice-size effect. However,  an unphysical decoupling originating 
solely from finite size effects seems to be now ruled out~\cite{the:Fischer 2007},
and it would not explain why the solution is also present in analytical methods.
Fischer, Maas and Pawlowski have argued that only the \emph{scaling solution}
is compatible both with the
equations respecting the BRST symmetry \cite{the:BRST}, and with
the color-confining  gluons  \cite{the:Fischer 2009 against decoupling}.
Fischer and Pawlowski have also showed that the \emph{scaling solution}
is unique for the SDE and the  exact renormalization group equations. 
In particular,
they showed that if the ghost and gluon external momenta vanish, then
one must realize the \emph{scaling solution} \cite{the:Fischer against decoupling 2007}.
The study of Alkofer, Huber and Schwenzer also disfavors the decoupling
solution, at least in the simplest realizations of QCD's confinement \cite{the:Alkofer 2010}.
Furthermore, the \emph{decoupling solution} cannot satisfy the \emph{Kugo--Ojima
confinement criterion} \cite{the:Kugo--Ojima}.  On the other hand, there are
models supporting the \emph{decoupling solution}:  Boucaud \emph{et al.}  have
interpreted this behavior in the context
of instantons \cite{the:Boucaud Instantons 2003-2004}: the initial
$\alpha_{s}(Q^{2})\propto Q^{4}$ IR-behavior is predicted by
the  instanton liquid model, which represents the QCD vacuum as an
instanton liquid.

It has been shown in \cite{the:Baldicchi 2002}
and discussed in \cite{the: Pennington 2011} that the two solutions lead to sensibly
similar descriptions of Nature in the IR domain, corresponding to hadron
mass/size scales.  However, it is important to determine which
solution is being realized; for example, the \emph{scaling solution} justifies the approach
based on \emph{conformal} field theories, see Sections \ref{sub:Holographic-QCD large Q} and
\ref{sub:Holographic-QCD Lowq}. In this regard, the excellent description
of QCD phenomenology by this approach  and the physical arguments for IR freezing of the QCD coupling given in Refs. \cite{Brodsky:2007hb, the:Brodsky & Shrock}, would suggest that the \emph{scaling solution}
is the one realized in Nature.

\subsubsection{Typical forms of the gluon propagator for the two solutions}

The \emph{scaling solution},   which leads to a freezing infrared fixed point value for the QCD coupling, is consistent with the original Gribov--Zwanziger
approach of confinement (Section \ref{sub:Gribov--Zwanziger-approach}). In this case the
gluon propagator behaves as
\begin{equation}
D_{\mu\nu}^{bc}(q)=\delta^{bc}\left(\eta_{\mu\nu}-\frac{q_{\mu}q_{\nu}}{q^{2}}\right)\frac{q^{2}}{q^{4}+m_{gr}^{4}}.\label{eq:Gribov gluon propagator 2}
\end{equation}
Here, $m_{gr}$ is the Gribov mass. 
In addition, if $Q^2 \gg m^2_{gr}$ one recovers the
perturbative gluon propagator $\delta^{bc}(\eta_{\mu\nu}-q_{\mu}q_{\nu}/q^{2})/q^{2}$.

The \emph{decoupling solution},    which leads to an IR-vanishing coupling when $Q^2 \to 0$, is itself consistent with an effective gluon propagator obeying
a Yukawa form \cite{the:Boucaud SDE massive gluon prop.},
\begin{equation}
D_{\mu\nu}^{bc}(q)_{q\rightarrow0}
 \propto\delta^{bc}\left(\eta_{\mu\nu}-\frac{q_{\mu}q_{\nu}}{q^{2}}\right)\frac{1}{q^{2}-m_{g}^{2}},\label{eq:Yukawa gluon propagator 2}
\end{equation}
which can also  be consistent with the generalized Gribov--Zwanziger scenario. Nevertheless,
in this refined Gribov--Zwanziger scenario,  the \emph{scaling solution} still seems
preferred \cite{Thelan 2014}. 

We will now discuss the topic of the effective gluon mass. As we have mentioned earlier, the emergence of such a
mass is equivalent to the regularization of the behavior of $\alpha_s$ in the IR.

\subsection{The massive gluon propagator \label{sub:SDE massive gluon}}

The first  studies of the gluon propagator in the IR regime using the
Schwinger--Dyson formalism began in the 1970s~\cite{the:Cornwall 1975 m_g,
the:Ball 1978, the:Mandelstam SDE}. 
Using the Landau gauge,  these 
 early results indicated  that $\alpha_{s}$ has
no \emph{Landau pole}, and that in the IR, $\alpha_{s}(Q^{2})\propto1/Q^{2}$
\cite{the:Ball 1978, the:Bar-Gadda}. 
A calculation  in the axial gauge \cite{the:Baker gluon prop SDE 1981.} reached
the same  conclusion.  The singular dependence in the IR agrees naively with the  expectation
of a linear potential in the static limit; see Section \ref{sub:Potential-approach}.  However, in spite
of this apparent consistency, Cornwall \cite{the:Cornwall alpha_s}
has argued that the $1/Q^{2}$ behavior is actually erroneous due to the explicit
exclusion of solutions which generate a dynamical gluon mass that  regulates
the \emph{Landau pole}. Earlier
suggestions that a gluon mass should suppress the unphysical Landau
pole were made in \cite{the:Cornwall 1975 m_g, the:Parisi m_g,
the:Poggio m_g}.  

In Cornwall's work an analytic form for the gluon
propagator is  proposed for solving
the SDE  in the \emph{pure gauge} sector and in the light-cone
gauge where ghosts are unnecessary. Feynman graphs are resummed to form
a gauge-independent gluon propagator dressing function that modifies the
free gluon propagator which carries the gauge dependence. The resulting
dynamically generated gluon effective mass is $Q^{2}$-dependent and
vanishes at large-$Q^{2}$. This feature ensures compatibility with 
pQCD expectations,  and it accommodates the fact that
a physical gluon mass is forbidden by gauge invariance. 
Cornwall's SDE solution to the gluon propagator suggests that the
perturbative form of $\alpha_s$, defined from the gluon self-energy, 
should be modified in the IR as : 
\begin{equation}
\alpha_{gse}\left(Q^{2}\right)=\frac{4\pi}{\beta_{0}\mbox{ln}\left(\left[Q^{2}+\varsigma m_{g}^{2}\left(Q^{2}\right)\right]/\Lambda^{2}\right)},\label{eq:alpha_s cornwall}
\end{equation}
with $\varsigma=4$ and $m_g(Q^2)$ is an effective gluon mass: 
\begin{equation}
m_{g}^{2}\left(Q^{2}\right)=\frac{m^{2}}{\left[\mbox{ln}\left(\frac{Q^{2}+\xi m^{2}}{\Lambda^{2}}\right)/\mbox{ln}\left(\frac{\varsigma m^{2}}{\Lambda^{2}}\right)\right]^{12/11}},\label{eq:gluon mass Cornwall}
\end{equation}
with $m=0.5\pm0.2$ GeV and $\Lambda=0.26\pm0.05$
GeV.   The \emph{freezing}  value, $\alpha_{gse}(0)=0.42 \pm 0.20$
agrees with the characteristic value of $\alpha_{s}$
estimated using magnetic and color--magnetic spin--spin interactions
\cite{the:Franklin 1982} and other determinations done in the context of hadron spectroscopy, 
see Section \ref{sub:Potential-approach}.    The limit $m = \Lambda$ is consistent with  Richardson's form, 
Eq. (\ref{eq:Richardson}). 
Renormalization
group equations predict that at large $Q^{2}$, $m_{g}^{2}(Q^{2})\propto \mbox{ln}(Q^{2})^{-12/11}$
where the exponent is related to the strength of the static force
in the \emph{pure gauge} sector; $4\pi(\beta_{0})^{-1}=12\pi/11$.

The  gluon mass must satisfy the condition $m_g \geq \Lambda$ in order to eliminate the \emph{Landau pole}.  
Cornwall showed that using the  \emph{pinch} technique $m \simeq1.2 \, \Lambda$ \cite{the:Cornwall alpha_s PT}
(a review of the  \emph{pinch} technique, including its application
to the SDE and the computation of $\alpha_{s}$, can be found in \cite{the:Binosi Pinch technic review}). 
Values of $m$ and $\Lambda$ can also be obtained by assuming
Cornwall's massive form for the gluon propagator and constraining it 
with hadron phenomenology.  For example, the isovector GDH sum rule for the nucleon
can be used to constrain $m_g$ by determining $\alpha_{\overline{MS}}$ at $Q^{2}=0$.
Solving Eqs. (\ref{eq:msbar to g_1}) and (\ref{eq:alpha_g_1 IR value}),
one finds  $m=0.350$ GeV and $\Lambda=0.268$ GeV ($\alpha_{gse}(0)=0.59$). 
The ratio $m/\Lambda=1.31$ validates the pinch technique expectation that $m/\Lambda\simeq1.2$   \cite{the:Cornwall alpha_s PT}.
The GDH-based result is gauge
independent,  but this analysis depends on the truncation order of Eq. (\ref{eq:Bj SR, order alpha^4}).
Since an \emph{asymptotic series} will reach its optimal order at $n\sim(1/\mbox{coupling)}$,
the value of $\alpha_{s}(Q^{2}\simeq1\mbox{ GeV}^{2})$
suggests that truncating Eq. (\ref{eq:Bj SR, order alpha^4}) at $n\sim 4$ is
adequate.

Another way to use phenomenological information is
to constrain Eqs. (\ref{eq:alpha_s cornwall}) and (\ref{eq:gluon mass Cornwall})
using data for  hadron form factors.   A good description of the experimental
pion and kaon form factors is obtained for $m=0.54$ GeV 
by Ji and collaborators \cite{the:Ji 1987-1990}. It yields 
$\alpha_{gse}(0)=0.4$ for $\Lambda=0.26$ GeV.   Brodsky
and collaborators have combined the dimensional counting rules for the 
form factors in the UV with Cornwall's \emph{freezing}  coupling for $m=0.44$
GeV ({\it i.e.} $\alpha_{gse}(0)=0.47$ for $\Lambda=0.26$ GeV) to analyze the 
photon--to--pion transition form factor and $\gamma\gamma\rightarrow\pi^{+}\pi^{-}$. 
The data is well reproduced, but the predicted normalization of the space-like pion 
form factor is higher than measurements~\cite{the:Brodsky 1998 }.   On
the other hand, Aguilar and collaborators find that $m=0.300$ GeV
\cite{the:Aguilar SDE 2001} describes  the pion form factor
data well, which implies a \emph{freezing}  value $\alpha_{gse}(0)=0.68$ ($\Lambda=0.26$ GeV). 

In  Ref. \cite{the:Aguilar 2002}, the value of $m$ is constrained using measurements
of the proton--proton elastic cross section. This  yields $m=0.370$ GeV 
($\alpha_{gse}(0)=0.55$ for $\Lambda=0.26$ GeV) and
in  Refs.  \cite{the:Binosi eq. for mg(Q^2), the:Binosi eq. for mg(Q^2) from RGE}
the value $m=0.374$ GeV ($\alpha_{gse}(0)=0.54$ for $\Lambda=0.26$ GeV) 
is obtained.  
The SDE are used in Ref.~\cite{the:Binosi (2015)} to obtain $m=0.68$ GeV 
($\alpha_{gse}(0)=0.35$ for $\Lambda=0.26$ GeV).  In this analysis,  the predicted SDE coupling 
\emph{freezes} at $\alpha_s(0)=2.8$, a much larger value than the $\alpha_{gse}(0)$ 
expected from Cornwall's form. 
In Ref. \cite{the:Binosi (2015)}, the gluon mass and coupling are well 
parameterized in the UV and IR-UV transition region $1<Q^2<18.5$ GeV$^2$ by 
\begin{equation}
m_g^2\left(Q^2\right)=\frac{0.22+0.019Q^2}{1+1.76Q^2}
\end{equation}
and 
\begin{equation}
\alpha_s\left(Q^2\right)=12.9m_g^2\left(Q^2\right).
\end{equation}
All of these values of $m$ are within Cornwall's proposed
range.  Lattice QCD estimates of $m$
also fall within this range \cite{the:Bernard et al early mg lattice estimates}.
An analysis of the data for inclusive radiative decays of the $J/\psi$
and $\Upsilon$ yields a somewhat higher value, $m\sim1$ GeV 
($\alpha_{gse}(0) \sim 0.28$ for $\Lambda=0.26$ GeV) \cite{the:Field m_g 2002}.

Aguilar and Papavassiliou have summarized the studies of $m_{g}(Q^{2})$
in  Ref.~\cite{the:Aguilar SDE 2008}. These authors also show that a
massive propagator effectively describes the solution of the SDE in
the gauge-independent  \emph{pinch} technique in the \emph{pure gauge} sector,  but with two
possible behaviors at large $Q^{2}$. One solution is similar to Eq. (\ref{eq:gluon mass Cornwall}),
with $m_{g}(Q^{2})\propto \mbox{ln}(Q^{2})^{-\gamma}$ (taking a typical value $ \gamma=0.86$), as expected from renormalization group equations.  The second solution
falls with a power law: $m_g(Q^{2})\propto \mbox{ln}(Q^{2})^{\gamma}/Q^{2}$
(with a typical value  $\gamma=1.12$), similarly to the power
corrections obtained from the \emph{OPE}.   The  behavior of the gluon mass in Eq. (\ref{eq:gluon mass Cornwall}) is consistent 
with the logarithmic running of $m_{g}(Q^{2})$, as predicted by the first solution,
with the values $\varsigma=1.007$, $m=0.3$ GeV, $\Lambda=0.3$ GeV.
This leads to a  \emph{freezing}  value for $\alpha_{gse}(0)\simeq 1.2$.
For the second power-law  solution for $m_{g}(Q^{2})$, the effective
mass is well described by 
\begin{equation}
m_{g}^{2}\left(Q^{2}\right)=\frac{m^{4}}{\left[Q^{2}+m^{2}\right]\left[\mbox{ln}\left(\frac{Q^{2}+\epsilon m^{2}}{\Lambda^{2}}\right)/\mbox{ln}\left(\frac{\epsilon m^{2}}{\Lambda^{2}}\right)\right]^{\gamma}},
\end{equation}
with  $\epsilon=1.046$, $m=0.5$ GeV, $\Lambda=0.3$ GeV
and $\gamma=1.12$. This leads to the \emph{freezing}  value $\alpha_{gse}(0)\simeq1.0$.

The results of Aguilar and Papavassiliou are gauge-invariant,  and the effective gluon mass can be interpreted as 
a gauge-invariant gluon vacuum \emph{condensate} or nonperturbative correction of dimension four.
In subsequent work,  \cite{the:Aguilar Papavassiliou SDE (2006)}, Aguilar and Papavassiliou 
used the  Feynman gauge,  the  \emph{pinch} technique,  and 
gauge invariant truncations of the gluon propagator to obtain forms for $m_{g}(Q^{2})$
and $\alpha_{gse}(Q^{2})$. Effects from ghost loops were neglected. Although the authors did not
specify the \emph{freezing}  value, they provide a numerical example with a \emph{freezing}  
value of $\alpha_{gse}\left(0\right)=0.91$ and a gluon mass normalization, $m_{g}(0)=0.45$
GeV. 
An alternative nonperturbative approach to the gluon mass 
has been developed by Aguilar and collaborators based on the SDE in the
Landau gauge \cite{the:Binosi eq. for mg(Q^2)} and the 
RGE \cite{the:Binosi eq. for mg(Q^2) from RGE}. The
authors provide a numerical fit of their results:
\begin{equation}
m_{g}^{2}\left(Q^{2}\right)=\frac{m^{2}}{1+\left(Q^{2}/M^{2}\right)^{\gamma}}
\label{eq:Aguilar m_g(Q2)},
\end{equation}
with $M=0.557$ GeV, $m=0.375$ GeV, and $\gamma=1.08$. 
Cornwall's
solution (\ref{eq:gluon mass Cornwall}), agrees well with these results.
The gauge-dependence of $m_{g}(Q^{2})$  in these formulae has
been investigated using linear covariant gauges in which a gauge parameter
$\xi$ continuously transitions between the Landau gauge ($\xi=0$)
and the Feynman gauge ($\xi=1$).   The resulting gluon mass normalization is typically
 $m \simeq0.37 \pm 0.1$ GeV \cite{the:Aguilar m_g gauge dependence}. 
This implies $\alpha_{gse}(0)=0.55 \pm 0.03$ for $\Lambda=0.26$ GeV.

A closely related model has been recently proposed 
by Ayala \emph{et al.} \cite{the:Ayala 2015}. In this analysis the \emph{effective charge} $\alpha_V$ has the form proposed
by Cornwall, Eq. (\ref{eq:alpha_s cornwall}),  assuming the Aguilar \emph{et al.} 
$Q^2$-dependence of the effective gluon mass, Eq. (\ref{eq:Aguilar m_g(Q2)}), with $\gamma=1.15$ and
a normalization  $m=\Lambda$.  Thus Richardson's limit, Eq. (\ref{eq:Richardson}), of Cornwall's coupling is
recovered at $Q^2=0$.  Since the resulting coupling diverges as $1/Q^2$ when $Q^2 \to 0$,  this also leads to a linear potential for static quarks
consistent with quarkonium spectroscopy.

~

Cornwall's coupling, Eq. (\ref{eq:alpha_s cornwall}), takes the
form of the pQCD coupling at $\beta_{0}$ order with an additional
dynamical effective gluon mass term.  In effect, the  nonperturbative
effects which regularize $\alpha_{s}$ in the IR are effectively incorporated
into $m_{g}(Q^{2})$, together with higher order pQCD loops. This is similar to
the case of the \emph{effective charges} defined by Grunberg, see Section
\ref{sub:low Q Effective-charges}: both  nonperturbative and perturbative
contributions are incorporated into $\alpha_s(Q^2)$.  However, in Cornwall's case, observable-independent
higher loops are folded in, whereas in Grunberg's case,  observable-dependent gluonic  bremsstrahlung 
is folded in. Nevertheless, the two results can agree, once the scheme 
dependence is accounted for, see Section
\ref{sub:alpha_s IR: Comparison-and-discussion}.

Eqs. (\ref{eq:alpha_s cornwall}) and (\ref{eq:gluon mass Cornwall})
have been augmented by further work by Cornwall and Papavassiliou \cite{theCornwall alpha_s 2}
using the  \emph{pinch} technique} and the SDE.
One finds the same form for the running coupling except that quark loops are now
included:
\begin{equation}
\alpha_{gse}\left(Q^{2}\right)=\frac{12\pi}{33\mbox{ln}\left[\left(Q^{2}+
\epsilon m_{g}^{2}\left(Q^{2}\right)\right)/\Lambda^{2}\right]-2n_{f}
\mbox{ln}\left[\left(Q^{2}+\epsilon M^2\right)/\Lambda^{2}\right]}\label{eq:alpha_s cornwall 2}.
\end{equation}
In this analysis, $\epsilon\simeq4.8$, 
$M$ is identified with the string
tension: $M\simeq\sqrt{\sigma}\simeq0.42$ GeV, and  Eq. (\ref{eq:gluon mass Cornwall}) is  used
for $m_g(Q^2)$. The screening 
role of the quark loops can be recognized in Eq. (\ref{eq:alpha_s cornwall 2}). 
With $n_f=2$, $\Lambda=0.26$ GeV and $M=0.42$, $\alpha_{gse}(0)=0.44$.

Other forms for the massive  gluon propagator have been proposed using 
the  Gribov--Zwanziger approach;  see Eq. (\ref{eq:Gribov gluon propagator 2}). If one uses 
the SDE solution of Stingl and of Habel \emph{et al.}~\cite{the:Stingl},
the gluon propagator  has the form
\begin{equation}
D_{\mu\nu}^{bc}(q)=\delta^{bc}\left(\eta_{\mu\nu}-
\frac{q_{\mu}q_{\nu}}{q^{2}}\right)\frac{q^{2}}{q^{4}+2aq^{2}+m_{g}^{4}},
\end{equation}
where  $a$ is a parameter constrained by phenomenology.

Lattice results can also be interpreted
in terms of a massive gluon propagator \cite{the:Leinweber 1998-2000}. 
For example, the
results of Marenzoni and collaborators~\cite{the:Marenzoni} can be fit by the form
\begin{equation}
D(Q^{2})=\frac{1}{bQ^{2}\left(Q^{2}/\Lambda^{2}\right)^{\eta}+m_{g}^{2}},
\end{equation}
with $b=0.102(1)$, $\eta=0.532(12)$ and $m_{g}\simeq0.4$ GeV. 
Similarly, Oliveira and Bicudo have interpreted their lattice data
in terms of a massive gluon propagator. Their result
for $Q^{2}\lesssim0.25$ GeV$^{2}$ can be described assuming a constant
IR gluon mass of about $m_{g}\simeq 0.7$ GeV \cite{the:Oliveira m_g (2011)}.
Their calculations are performed in the Landau gauge. A similar value
is obtained in Ref.  \cite{the:Dudal 2010}. The 
results are consistent with the \emph{decoupling solution}, yielding
an IR-vanishing $\alpha_s$.  In
recent work~\cite{the:Binosi (2015)},  
the SDE and lattice data  are also combined in the form of 
a  massive gluon propagator.  This leads 
to an IR value of $\alpha_{s}(0)\simeq 0.9 \, \pi$.
Thus lattice studies can in effect  be interpreted as the emergence of an effective gluon mass.

Massive gluons have also been considered in approaches other than SDE and
lattice QCD; {\it e.g.}, the Gribov--Zwanziger approach~\cite{the:Capri 2007}, the
 background perturbation theory (Section \ref{sub:Background pert theo, Simonov}), 
or the variational approach of Szczepaniak and Swanson \cite{the:Szczepaniak 2001} 
which yields a constituent gluon mass of about 0.6 GeV. 

Finally, we notice that an effective gluon mass can be used to generate 
$1/Q^{2}$ \emph{power corrections};  this is in spite of the fact that in  the \emph{OPE} there is 
no \emph{condensate} of dimension-2
arising from a local gauge-invariant operator \cite{the:Chetyrkin 1999}. 
However, in this case, the resulting  effective gluon mass must be imaginary (tachyonic gluon mass).

\subsection{The coupling defined from the ghost--gluon vertex}

The ghost--gluon coupling $\alpha_{s}^{gh}$ from Eq. (\ref{eq:alpha_s SDE ghost--gluon}) 
was first obtained using  the SDE  formalism by von Smekal, Hauck and Alkofer \cite{the:von-Smekal SDE alpha_gh}.
Their  Landau gauge analysis  is based on a self-consistent truncation scheme, 
where the truncation implements the \emph{Slavnov--Taylor} identities
\cite{the:Slavnov--Taylor id.} for the three-point vertex
function. The numerical results lead  to a coupling agreeing with
pQCD evolution at large $Q^{2}$,  while \emph{freezing} to a value of $\alpha_{s}^{gh}\left(0\right)\simeq9.5$.
The following forms match well the numerical solution \cite{the:Aguilar SDE 2001}:
\begin{eqnarray}
\alpha_{s}^{gh}&=&0.261+9.2621e^{-2\frac{\left(Q^{2}-0.0297\right)^{2}}{0.4689}};\mbox{ for }Q^{2}<0.31\mbox{ GeV}^{2}, \nonumber\\
\alpha_{s}^{gh} &=&1.4741+8.6072e^{-\frac{Q^{2}-0.1626}{0.3197}};\mbox{ for }0.31<Q^{2}<1.3\mbox{ GeV}^{2}, \nonumber\\
\alpha_{s}^{gh}&=&\frac{1.4978}{\mbox{ln}(1.8488Q^{2})};\mbox{ for }Q^{2}>1.3\mbox{ GeV}^{2}.
\end{eqnarray}
The numerical techniques have been subsequently  improved, yielding smaller \emph{freezing}  values.  Fischer \emph{et al.}
\cite{the:Fisher S-D alpha_s} obtained a numerical coupling (Landau gauge, \emph{pure gauge} sector) which is well fitted by:
\begin{equation}
\alpha_{s}^{gh}\left(Q\right)=\frac{2.972}{\mbox{ln}\left(e+5.292Q^{2.324}+0.034Q^{3.169}\right)},
\end{equation}
where the numerical coefficients are determined by matching to the value of the pQCD coupling at the $Z^0$ mass: $\alpha_{s}^{gh}(M_Z^{2})=\alpha_{s}^{pQCD}(M_Z^{2})$.
This implies, in the $MOM$ scheme, $\Lambda_{MOM}=0.715$ GeV and the
\emph{freezing}  value $\alpha_{s}^{gh}\left(0\right)\simeq2.972$; see
Section \ref{sub:Classes-of-solutions in IF domain}. 

An alternative truncation prescription was used by Bloch \cite{the:Bloch S-D alpha_s 2002}. The 
result in the \emph{pure gauge} theory is fitted by: 
\begin{equation}
\alpha_{s}^{gh}\left(Q^{2}\right)=\frac{1}{15+Q^{2}/\Lambda^{2}}\left(15\times2.6+\frac{4\pi}{\beta_{0}}\left(\frac{1}{\mbox{ln}(Q^{2}/\Lambda^{2})}-\frac{1}{Q^{2}/\Lambda^{2}-1}\right)\frac{Q^{2}}{\Lambda^{2}}\right),\label{eq:SDE Bloch-1}
\end{equation}
where $\Lambda_{\overline{MS}}=0.33$ GeV in the $\overline{MS}$ scheme. In this Landau gauge analysis, the \emph{freezing} 
value for $\alpha_{s}^{gh}$  is 2.6, and the dynamically generated gluon mass is $m_g\simeq 0.35$
in the IR domain.   The form for the running  ghost--gluon coupling is similar to the behavior of $\alpha_{an}$
obtained in the analytic approach, Eq. (\ref{eq:alpha_s from analytic QCD}), see Section \ref{sub:Analytic approach}.
The truncation prescription was subsequently improved in  Ref.~\cite{the:Bloch S-D alpha_s 2003},
leading to numerical results parameterized as

\begin{multline} 
\alpha_{s}^{gh}\left(Q^{2}\right) = \frac{1}{1.16+Q^{4}/\Lambda^{4}}
\biggl(3.49\left[1.16-0.070\left(Q^{2}/\Lambda^{2}\right)^{0.66}\right]\\  
+ \left[Q^{4}/\Lambda^{4}\right]\left(Q^{2}/\Lambda^{2}+2\right) 
\alpha_{s}^{pQCD,\beta_{1}}\biggr), 
\end{multline}
where $\alpha_{s}^{pQCD,\beta_{1}}$ is the perturbative coupling at  order $\beta_{1}$
obtained from Eq. (\ref{eq:alpha_s}) and $\Lambda=0.856$ GeV.
This gives a \emph{freezing}  value of 3.49. 
Comparisons with the earlier results 
illustrate the effects of modifying the truncation prescription
as well as choosing a different RS. 
To reinforce this point, we note that the earlier work by Atkinson and Bloch with a bare truncation that leads to a \emph{freezing}  value
of 11.47 by averaging the SDE angular integrals, and to a \emph{freezing} 
at $4 \pi / 3$ with the angular integrals done exactly \cite{the:Bloch S-D alpha_s 1998}.
Bloch systematically studied the effect of varying the  truncation
scheme and found that $2\pi/3<\alpha_{s}^{gh}\left(0\right)<8\pi/3$
in SU(3) \cite{the:Bloch (2001)}.

Aguilar and Natale \cite{the:Aguilar-Natale SDE 2005}  
obtained an IR-vanishing coupling $\alpha_{s}^{gh}$. These authors 
assumed a massless gluon propagator behaving as $1/Q^{2}$.  
The calculations were done in the Landau gauge
and pure Yang--Mills using the renormalization prescription of 
\cite{the:Cornwall alpha_s, theCornwall alpha_s 2}. In subsequent works
Aguilar \emph{et al.} employed a massive gluon
propagator, rather than a massless one,  leading to a \emph{freezing}  behavior
of the coupling. Furthermore, in  \cite{the:Aguilar (2009)}, Aguilar \emph{et al.} computed both 
$\alpha_{gse}$ and  $\alpha_s^{gh}$ and found that the two couplings have the same \emph{freezing}  value
of  order 0.6 for  $m_{g}(0)=0.5$ GeV.

Schleifenbaum and collaborators~\cite{the:Schleifenbaum 2006} have computed the gluon and ghost propagators
in both Landau and Coulomb gauges, finding that  the gluon propagator in the IR
is suppressed, whereas the ghost propagator is singular.
These features yield an approximately linear potential and a ghost--gluon
coupling \emph{freezing}  at $\alpha_{s}^{gh}\left(0\right)\simeq3.65$ for the Landau gauge.   They also find
two solutions in the Coulomb gauge $\alpha_{s}^{gh}(0) \simeq 14.21/N_C=4.74$ 
or $\alpha_{s}^{gh}(0)=16\pi/N_C\simeq5.59$, with the value 4.74 favored~\cite{the:Leder 2011 FRG}.

\subsection{Couplings defined from the 3-gluon and 4-gluon vertices}

In addition to the definition from the ghost--gluon vertex, Eq.
(\ref{eq:alpha_s SDE ghost--gluon}), couplings can also be defined
from the 3-gluon and 4-gluon vertices:
\begin{equation}
\alpha_{s}^{3g}\left(Q^{2}\right)=\alpha_{s}^{3g}\left(\mu\right)\left[\Gamma^{3g}(Q^{2},\mu)\right]^{2}Z^{3}(Q^{2},\mu),\label{eq:alpha_s 3g vertex}
\end{equation}
\begin{equation}
\alpha_{s}^{4g}\left(Q^{2}\right)=\alpha_{s}^{4g}\left(\mu\right)\left[\Gamma^{4g}(Q^{2},\mu)\right]Z^{2}(Q^{2},\mu),\label{eq:alpha_s 4g vertex}
\end{equation}
where $\Gamma^{3g}$ and $\Gamma^{4g}$ are the dressing functions
of the tree-level 3-gluon and 4-gluon vertices.
In contrast  to the case of $\alpha_{s}^{gh}$,  dressing functions appear since there are no
\emph{Slavnov--Taylor} identities for these vertices. 
The three couplings
$\alpha_{s}^{gh}(Q^{2})$, $\alpha_{s}^{3g}(Q^{2})$ and $\alpha_{s}^{4g}(Q^{2})$
are different in the MOM RS but can be related to each other 
\cite{the:Comparison alpha-g-gh alpha-3g alpha-4g, the:Shirkov alpha_s low_Q review}. 

The 3-gluon
and 4-gluon vertices have been  studied in the pure
gauge sector by Baker and Lee \cite{the:Baker-and Lee SDE 4g vertex},
Celmaster and Gonsalves \cite{the:Comparison alpha-g-gh alpha-3g alpha-4g}, by 
Brandt and Frenkel \cite{the:Brandt SDE 4g vertex}, as well as by Papavassiliou
using the   \emph{pinch} technique~\cite{the:Papavassiliou SDE 4g vertex 1993}.
The resulting couplings $\alpha_{s}^{3g}(Q^{2})$ and $\alpha_{s}^{4g}(Q^{2})$
were first
investigated by Alkofer, Fischer, and Llanes-Estrada~\cite{the:Alkofer Definition of couplings in SDE}.
They concluded that these  couplings
must \emph{freeze} in the IR, but they did not 
provide \emph{freezing}  values. 
A recent perturbative analytical calculation of the IR and UV 
behaviors
of the 3-gluon vertex (Landau gauge and \emph{pure gauge} sector) is reported
in \cite{the:Pelaez 3-pt cor func. SDE}.  

\begin{figure}[ht]
\centering
\includegraphics[scale=0.45]{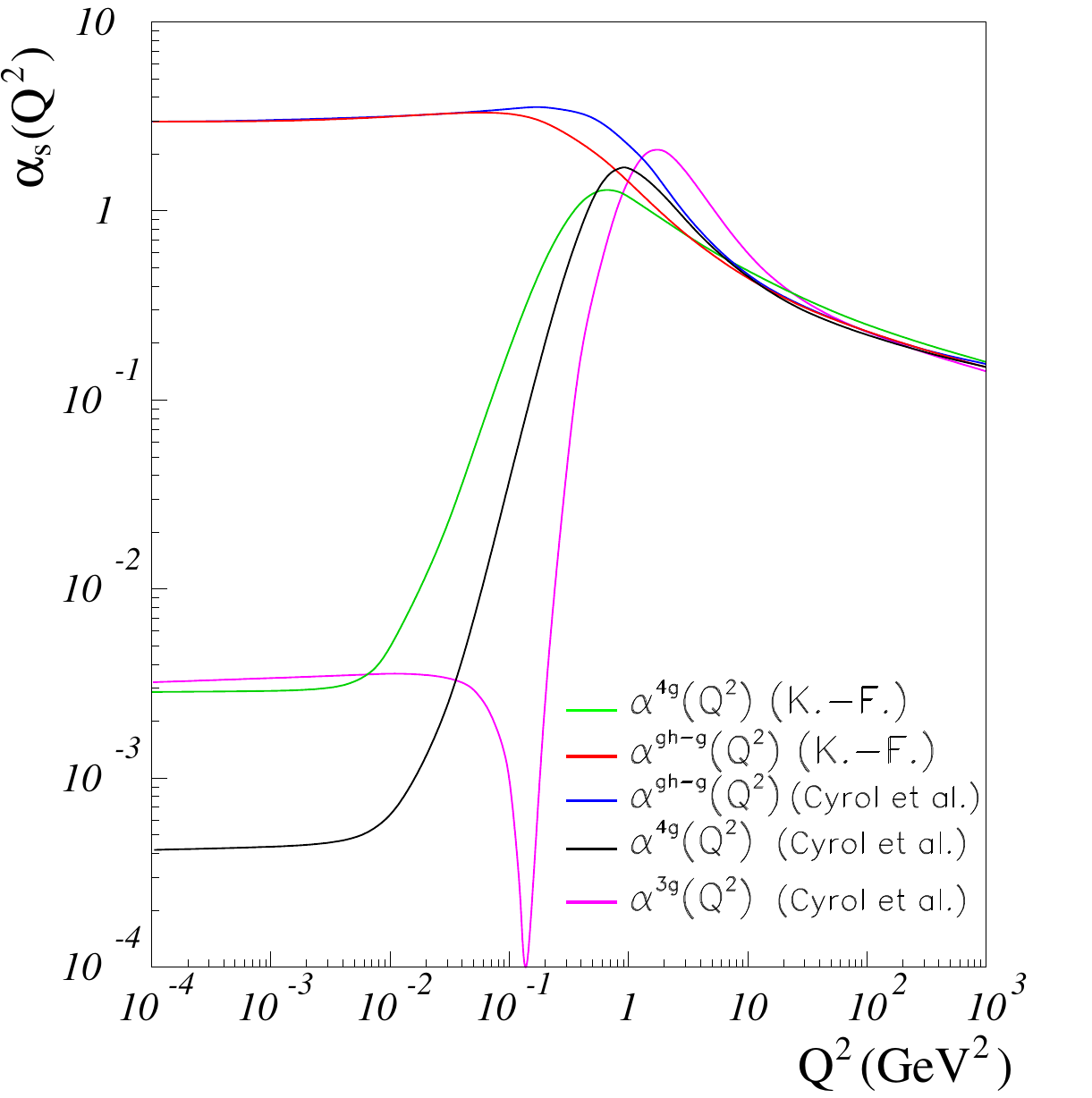}
\caption{  \small \label{Flo:lowq alpha. SDE} SDE computations of $\alpha_{s}$ defined
using the 4-gluon vertex (the green line is from Ref. \cite{the:Kellermann-Fisher alpha_4g (2008)}
and black line is from Ref. \cite{the:Cyrol 4g alpha (2014)}). 
The red line, using  the ghost--gluon vertex definition,  is from Ref.~\cite{the:Kellermann-Fisher alpha_4g (2008)}
and  the blue  line is  from Ref. \cite{the:Cyrol 4g alpha (2014)}.
The magenta line for the 3-gluon vertex definition is from Ref.~\cite{the:Cyrol 4g alpha (2014)}.
Quark degrees of freedom are not included and the calculations are
done in the Landau gauge and MOM scheme.}
\end{figure}

The 3-gluon vertex was
calculated  nonperturbatively using the SDE by Eichmann\emph{
et al.}~\cite{the:Eichmann SDE 3g vertex} 
using the Landau gauge.
Both \emph{scaling} and  \emph{decoupling solutions} were obtained.  In the
decoupling case, $\alpha_{s}^{gh}$ and $\alpha_{s}^{3g}$ vanish in the IR, whereas
in the scaling case $\alpha_{s}^{gh}(0)=2.972$ and $\alpha_{s}^{3g}(0)=1.6\times10^{-3}$;
see Fig. \ref{Flo:lowq alpha. SDE}. 
A self-consistent SDE calculation
of $\alpha_{s}^{3g}(Q^{2})$, also in the Landau gauge, yields vanishing
$\alpha_{s}^{gh}(Q^{2})$ and $\alpha_{s}^{3g}(Q^{2})$ in the IR
\cite{the:Blum SDE 3g vertex}. 
Huber, Campagnari, and Reinhardt
solved numerically, in the Coulomb gauge and \emph{pure gauge} sector,
the canonical recursive SDE for the three-gluon and ghost--gluon vertices.
The resulting ghost--gluon vertex is IR finite,  whereas the three-gluon vertex
is IR singular \cite{the: Huber 3g SDE vertex}. 

The 4-gluon vertex
was first studied  nonperturbatively by Stingl, Driesen and collaborators~\cite{the:Stingl SDE 4g vertex}. 
The calculation for the 4-gluon coupling
$\alpha_{s}^{4g}$  in the \emph{pure gauge} sector,  assuming Landau gauge and the MOM RS,  by  Kellermann
and Fischer~\cite{the:Kellermann-Fisher alpha_4g (2008)}  yields a very small \emph{freezing}  value, $\alpha_{s}^{4g}(0)\simeq2.77\times10^{-3}$. 
The resulting behavior of $\alpha_{s}^{4g}(Q^{2})$ follows the pQCD expectation at large $Q^2$ and increases
with decreasing $Q^{2}$ to about $Q^{2}=0.5$ GeV$^{2}$. It
then drops to its \emph{freezing}  value, which is reached at $Q^{2} \sim 10^{-2}$
GeV$^{2};$ see Fig.~\ref{Flo:lowq alpha. SDE}. 

Cyrol, Huber and von Smekal~\cite{the:Cyrol 4g alpha (2014)} have recently computed self-consistent values for the couplings controlling 
the 3-gluon and 4-gluon vertices.  Again this calculation is  in the \emph{pure gauge} sector, assumes Landau gauge
and the MOM RS.   In the \emph{scaling solution},
the 3-gluon and 4-gluon vertex couplings \emph{freeze} at values that are
very small compared to $\alpha_{s}^{gh}(0)=2.972$; $\alpha_{s}^{3g}(0)=3.2\times10^{-3}$
and $\alpha_{s}^{4g}(0)=4.2\times10^{-4}$, confirming qualitatively
the finding in  Refs. \cite{the:Eichmann SDE 3g vertex, the:Kellermann-Fisher alpha_4g (2008)}, namely  
$\alpha_{s}^{3,4g}(0) \ll \alpha_{s}^{gh}(0)$. These running couplings are shown in Fig. \ref{Flo:lowq alpha. SDE}.
In the \emph{decoupling solution} case, all three definitions of $\alpha_{s}$
lead to IR-vanishing couplings.

\subsection{Including quarks\label{sub:SDE: Including quarks}}

Quark loops contribute to the running of $\alpha_s$ in the UV in the opposite 
  direction to the gluons, making $\alpha_s$ increase with $Q^2$.
Thus one  expects 
that their inclusion will be to decrease $\alpha_s(0)$ compared
to the results obtained in the \emph{pure gauge} sector. 
This comparison is done with $\Lambda$
adjusted so that $\alpha^{n_f=0}_s(M_Z^2) = \alpha^{n_f =5}_s(M_Z^2)$.  \emph{A contrario}, if the
value of $\Lambda$ is kept the same for any $n_f$, the quark loops would increase the value of  
$\alpha^{n_f > 0}_s(0)$ compared to $\alpha^{n_f=0}_s(0)$ since 
$\alpha^{n_f > 0}_{pQCD}(Q^2)>\alpha^{n_f = 0}_{pQCD}(Q^2)$
and the \emph{Landau pole} divergence of the perturbative coupling is steeper. 
In any case, since effects other than vacuum polarization are included in the IR definition of the coupling, 
the effect of quarks is not obvious, as illustrated by the following example.  

The influence of quark loops on the coupling has been investigated
in ref.~\cite{theCornwall alpha_s 2};  see Eq. (\ref{eq:alpha_s cornwall 2}). The 
quark loops increase the \emph{freezing}  value: 
$\alpha^{n_f=2}_{gse}(0)=0.81$ in comparison with  $\alpha^{n_f=0}_{gse}(0)=0.60$. The values 
$\Lambda^{(5)}=0.09$ GeV and $m=1.2 \, \Lambda^{(5)}=0.11$ GeV are used for the $n_f=2$ calculation, and 
 $\Lambda^{(0)}=0.65$ GeV and $m=1.2 \, \Lambda^{(0)}=0.78$ GeV are used for the $n_f=0$ calculation  --so that
 $\alpha^{n_f=2}_{gse}(M_Z^2)=\alpha^{n_f=0}_{gse}(M{_Z}^2)=0.119$ at the $Z^0$ pole.
The quark loop effect is thus non-negligible according to \cite{theCornwall alpha_s 2}.  However, 
Fisher and collaborators have recently reported  that  quark loops are suppressed in the IR and thus 
have only a small effect for a physical numbers of quark flavors, $n_{f}<6$ \cite{the:Fischer SDE unquenching}. 
This result is apparent from the function that fits their results (apart for a bump at $Q^{2}\simeq0.5$ GeV$^2$,  which 
is believed to be unphysical and is omitted):
\begin{equation}
\alpha_{s}^{gh}\left(Q^{2}\right)=\frac{1}{1+\frac{Q^{2}}{{\Lambda^{n_f}}^{2}}}
\left(2.972+\frac{4\pi}{\beta_{0}}\left(\frac{1}{\mbox{ln}(\frac{Q^{2}}{{\Lambda^{n_f}}^{2}})}-\frac{1}{\frac{Q^2}{{\Lambda^{n_f}}^2}-1}\right)\frac{Q^{2}}{{\Lambda^{n_f}}^{2}}\right),\label{eq:Fisher SDE -2}
\end{equation}
with a \emph{freezing}  value which is independent of $n_f$.  The quark effects are encoded in the \emph{scale parameter}: 
$\Lambda^{n_f=0}_{MOM}=0.71$ GeV and $\Lambda^{n_f=3}_{MOM}=0.51$ GeV.
(We note that the fit form is similar to Eq. (\ref{eq:SDE Bloch-1}) and to the coupling found in the 
  analytical approach, Eq. (\ref{eq:alpha_s from analytic QCD})).
Eq. (\ref{eq:Fisher SDE -2}) was used to study analytic properties
of the gluon and quark propagators in the Landau gauge \cite{the:Alkofer gluon & q prop 2008}. 

Related to the inclusion of quarks, Alkofer, Fisher, and Llanes-Estrada
have studied the influence of chiral-symmetry breaking on the running
coupling $\alpha_{s}^{qg}(Q^{2})$, now defined by the quark--gluon
vertex \cite{the:Alkofer SDE & chiral sym}:
\begin{equation}
\alpha_{s}^{qg}\left(Q^{2}\right)=\alpha_s^{qg}\left( \mu \right) \xi^2_1\left( \mu,Q^2 \right)P^2\left( \mu,Q^2 \right)Z\left( \mu,Q^2 \right),
\label{eq:alpha_s q-g vertex}
\end{equation}
where $\xi_1$ is the vertex dressing function and $P$ the quark propagator dressing function.
It was found that the
reduction of the number of linearly independent Dirac tensors using 
chiral symmetry leads to different behavior of the coupling. In the
chiral symmetry case,   which approximates well the light quark sector, 
the coupling is IR-finite: $\alpha_{s}^{qg}(Q^{2}) \simeq 2.5$. In contrast,
in the broken chiral symmetric case (as for the heavy quark sector where quark loops are \emph{quenched}),
the additional tensors induce a coupling diverging in the IR: $\alpha_{s}^{qg}(Q^{2})\propto1/Q^{2}$.
As discussed in Section \ref{sub:Potential-approach} the $1/Q^{2}$
behavior of the coupling gives rise to a  linear $\mbox{Q--}\overline{\mbox{Q}}$ potential for heavy static quarks.
This behavior can be compared to the expectation that QCD strings
develop only in the heavy-quark sector \cite{the:Light quarks no-strings}.

\subsection{Gauge dependence\label{sub:SDE Gauge-dependence}}

Many calculations in the SDE formalism and other frameworks yield a gauge-de\-pen\-dent  coupling $\alpha_{s}$.  The
question of the best choice of gauge and its influence on the IR-behavior of $\alpha_{s}$ is open.
(The gauge-dependence subsides in the UV regime). 
Fisher has reported that couplings computed in gauges close to the
Landau gauge have a weak gauge-dependence \cite{the:Fischer Review SDE in IR}.
Fischer and Zwanziger \cite{the:Fischer-Zwanziger SDE gauge-dep study}
have studied the gauge-dependence of the \emph{scaling solution} of $\alpha_{s}^{gh}$
using a class of transverse gauges which interpolate between the Landau
and Coulomb gauges. The IR value $\alpha_{s}^{gh}\left(0\right)\simeq2.972$,
found for  the \emph{IR exponent}  $\kappa'=0.595$, holds across the gauge range,
except the Coulomb case which has a singular limit.  This particular
case is studied in detail in   Ref. \cite{the:Alkofer-Maas-Zwanziger 2010}. The \emph{freezing}  value is found
to be slightly lower: $\alpha_{s}^{gh}\left(0\right)\simeq2.333$.
However, the already mentioned calculations of the same quantity in  Ref. \cite{the:Schleifenbaum 2006},
in both Landau and Coulomb gauges, lead to the opposite trend: $\alpha_{s}^{gh}\left(0\right)\simeq3.65$
in the Landau gauge and  $\alpha_{s}^{gh}\left(0\right)\simeq4.74$ in the Coulomb gauge. (Two 
solutions are actually found in the Coulomb gauge: $\alpha_{s}^{gh}\left(0\right)\simeq4.74$
or $\alpha_{s}^{gh}\left(0\right)\simeq5.59$ for $\kappa'\simeq0.398$
or $\kappa'=0.5$, respectively, with $\alpha_{s}^{gh}\left(0\right)\simeq4.74$ 
preferred).   Recently, Huber has numerically obtained the decoupling
solutions of the SDE for linear covariant gauges, varying its gauge
fixing parameter between $0\leq\xi\leq0.2$ \cite{the:Huber SDE 2015}.
In each case,  the couplings extracted from the ghost--gluon and 3-gluon
vertices vanish as $Q^{2}\rightarrow0$, albeit with a large gauge
dependence in the intermediate $Q^{2}$ domain of a few GeV$^{2}$.

\section{Lattice gauge theory \label{sub:Low Q Lattice-QCD-calculations}}

Lattice calculations provide another important method for studying  the behavior of the QCD coupling $\alpha_{s}$
in the IR regime. These studies connect with the SDE framework.
Both approaches have greatly benefited from their complementarity.
The basics of lattice QCD have been given in Section \ref{sub: large Q Lattice-QCD}.
Here, we will first describe the steps for typical lattice calculations
of $\alpha_{s}(Q^{2})$ in the IR.  We will then discuss the results
obtained by several groups 
 which use the numerical lattice approach  to compute the QCD coupling $\alpha_s$ in the nonperturbative infrared domain.

One method for obtaining $\alpha_{s}(Q^{2})$ in the IR is to compute
the static potential, as discussed in Sections \ref{sub: large Q Lattice-QCD}
and \ref{sub:Potential-approach}.   Another approach is to compute
the coupling from vertices.  In this case, just as for the SDE formalism,
it amounts to determining correlation functions. Those were
first calculated on the lattice by Mandula and Ogilvie \cite{the:Bernard et al early mg lattice estimates}.
In the following,
we overview  representative lattice methods to obtain
$\alpha_{s}$ (valid in the IR as well as the UV domains). The examples
are restricted to the \emph{pure gauge} sector.

\subsubsection{An example of  the lattice determination of $\alpha^{gh}_{s}$}

As discussed in Section \ref{sub:Schwinger--Dyson-formalism}, $\alpha_{s}(Q^{2})$
can be defined from the ghost--gluon vertex, see Eq. (\ref{eq:alpha_s SDE ghost--gluon}).
To compute the ghost and gluon propagator dressing functions, 
discretized ghost and gluon fields need to be defined. The discretized
gluon field $\mathcal{A}_{\overrightarrow{\mu}}^{c}(x)$, where $c$
is the color index and $\overrightarrow{\mu}$ a link direction, may
be defined as the difference between a link and its adjoint: 
\begin{equation}
 \mathcal{A}_{\overrightarrow{\mu}}^{c}(x) \equiv  \frac{1}{2i} \left(U_{\overrightarrow{\mu}}(x)-
U_{\overrightarrow{\mu}}^{\dagger}(x)\right) ~ \overrightarrow{_{a \to 0}} ~ 2a\sqrt{\pi\alpha_{s}^{bare}}A_{\mu}^{c}(x).
\end{equation}
The continuum limit $a\rightarrow0$ corresponds to the physical
gluon field $A_{\mu}^{c}$.  The discretization of space-time induces a finite difference 
$\mathcal{A}_{\overrightarrow{\mu}}^{c}(x)-2a\sqrt{\pi\alpha_{s}^{bare}}A_{\mu}^{c}(x)$.
This  creates unphysical \emph{tadpole}
contributions proportional to $\alpha_{s}^{bare}$ or to $a^{3}$. Those
can be eliminated by a redefinition of the link variable or of the
action, see  Refs. \cite{the:Lepage Lattice lectures} or \cite{the:Bloch lattice SU2 (2004)}.
Other definitions of $\mathcal{A}_{\overrightarrow{\mu}}^{c}(x)$
have been used,  and are equivalent  --up to a multiplicative renormalization factor
\cite{the:Giusti 1998}. The gluon propagator from Eq. (\ref{eq:n-point function}) is 
\begin{equation} 
D_{\mu\nu}^{bc}(x-y)=\int DxA_{\mu}^{b}(x)A_{\nu}^{c}(y)e^{-iS}
\equiv\left\langle A_{\mu}^{b}(x)A_{\nu}^{c}(y)\right\rangle.
\end{equation}
It gives on the lattice
\begin{equation}
\mathcal{D}_{\overrightarrow{\mu}\overrightarrow{\nu}}^{bc}\left(x-y\right)=\left\langle \mathcal{A}_{\overrightarrow{\mu}}^{b}\left(x\right)\mathcal{A}_{\overrightarrow{\nu}}^{c}\left(y\right)\right\rangle. \end{equation}
Then, $\mathcal{D}_{\overrightarrow{\mu}\overrightarrow{\nu}}^{bc}\left(x-y\right)$
can be transformed to momentum space via a Fourier transform. This yields
the gluon propagator dressing function 
$\mathcal{Z}(Q^{2})\equiv\delta^{bc}\delta^{\mu \nu}\mathcal{D}^{bc}_{\mu \nu}(Q^{2})Q^{2}$. 
The lattice ghost propagator is obtained,  by definition, from the
inverse of the Fadeev--Popov operator $(-\partial+A)A$;  after discretization
and evaluation on the lattice, it is Fourier-transformed to momentum space.
The resulting ghost propagator dressing function is 
$\mathcal{G}=\delta^{bc}Q^{2}\mathcal{D}_{bc}^{G}\left(Q^{2}\right)$. 
The coupling $\alpha_{s}^{gh}(Q^{2})$ is then extracted
from Eq. (\ref{eq:alpha_s SDE ghost--gluon}) after applying the
ghost and gluon propagator renormalization factors, $\widetilde{Z}_{3}$
and $Z_{3}$, respectively; see Section \ref{sub:Renormalization-group}.
The validity of the calculations can be verified by comparing the UV
running of $\alpha_s^{gh}(Q^2)$  with the pQCD expectation. 

The above procedure was
applied by two different groups in \cite{the:Bloch lattice SU2 (2004)}  using different lattice simulations in the \emph{pure gauge} sector.
The results yield a SU(2) running coupling which matches the pQCD running
at $Q^{2}\gtrsim4$ GeV$^{2}$ and \emph{freezes} at $\alpha_{s}^{gh,SU(2)}(0)=5\pm1$.
It is close to the SDE SU(2) result found in \cite{the:Bloch S-D alpha_s 2003}: $\alpha_{s}^{gh,SU(2)}(0)=5.24$.
This result is extended to SU(3)-color QCD by scaling it by 
2/3: it  becomes $\alpha_{s}^{gh}(0)=3.3\pm0.7$.
Since $\alpha_{s}^{gh}(\mu^{2})$ has been determined in \cite{the:Bloch lattice SU2 (2004)} 
by matching the lattice result to the 2-\emph{loop} pQCD
expectation, the prediction is scheme-independent, but it  is dependent
on the order of truncation order of the perturbative $\beta$  series; it 
yields $\Lambda_{2-loop}=1.1$ GeV in the UV.

\subsubsection{An example of the determination of $\alpha^{3g}_{s}$}

Another approach, first proposed in  Ref.~\cite{Parrinello 1994},  is to measure
the coupling by calculating
the 3- and 2-point gluon correlation functions, $$  G_{3,\mu_{1}\mu_{2}\mu_{3}}(p_{1},p_{2},p_{3})\equiv \langle A_{\mu_{1}}(p_{1})A_{\mu_{2}}(p_{2})A_{\mu_{3}}(p_{3})\rangle$$
and $D(p)=Z(p_{2})/p_{2},$ respectively. The calculations are typically carried out
in the Landau gauge and MOM RS.  The 3-gluon vertex $G_{3}$
can be expressed in term of three gluon propagators $D$ and a 3-point
vertex function $\Gamma_{3}$ which is naturally identified with the
coupling:
\begin{equation}
G_{3,\mu_{1}\mu_{2}\mu_{3}}(p_{1},p_{2},p_{3})=\Gamma_{3,\nu_{1}\nu_{2}\nu_{3}}(p_{1},p_{2},p_{3})D_{\mu_{1}\nu_{1}}(p_{1})D_{\mu_{2}\nu_{2}}(p_{2})D_{\mu_{3}\nu_{3}}(p_{3}).\label{eq:3-gluon vertex}
\end{equation}
The vertex function is then extracted as:
\begin{equation}
\Gamma_{3,\nu_{1}\nu_{2}\nu_{3}}(p_{1},p_{2},p_{3})=G_{3,\mu_{1}\mu_{2}\mu_{3}}(p_{1},p_{2},p_{3})\prod_{n=1}^{3}D^{-1 \mu_{n}}_{\nu_{n}}(p_{n}).\label{eq:Gamma_3 truncation}
\end{equation}
The coupling is normally defined at the symmetric Euclidean point, $\alpha_{s}(Q^{2}) \equiv \Gamma_{3}(p_{1},p_{2},p_{3})$
for $p_{1}^{2}=p_{2}^{2}=p_{3}^{2}=Q^{2}$ and $p_{1}+p_{2}+p_{3}=0$.
Such conditions correspond to the MOM scheme.  In contrast, the minimal MOM scheme
$\widetilde{\mbox{MOM}}$ is preferred on the lattice, corresponding to
the condition $p_{1}=p_{2}+p_{3}=0$.

\subsection{Lattice results for $\alpha_{s}$ in the IR}

Lattice calculations for $\alpha_{V}$ defined by the static $\mbox{Q--}\overline{\mbox{Q}}$ potential leads to 
an IR-\emph{freezing} coupling \cite{the:Bali 1995 et al.}. However, for $\alpha_{s}$ defined using  \emph{Green's function}s, both
IR-vanishing and  \emph{freezing}  solutions exist,  as for the SDE 
calculations.  Calculations by Bloch and collaborators \cite{the:Bloch lattice scaling sol. 2003}, and by Furui and 
Nakajima \cite{the:Furui Nakajima} result in the \emph{scaling solution}.
On the other hand, a majority of calculations
\cite{the:Boucaud SDE massive gluon prop., the:Skullerud 2002, 
the:Oliveira-Silva 2007, the:Bornyakov 2009,
the:Ilgenfritz 2011}, indicate that in the Landau gauge and MOM 
RS, $\alpha^{gh}_{s}$  vanishes when 
$Q^{2}\rightarrow0$. 
Before concluding that these results correspond to the \emph{decoupling solution}
discussed in Section \ref{sub:Classes-of-solutions in IF domain}, one needs
to explore more mundane possibilities.  One  is that $G_{3}$ has
been  
obtained from Eq. (\ref{eq:Gamma_3 truncation}) using the
perturbative expression of the gluon propagator $D(Q^{2})$, which
diverges as $Q^{2}\rightarrow0$.  In this case, the IR-vanishing originates from an artificial 
divergence originating from pQCD which is irrelevant in the IR regime. Another possible reason is unphysical
lattice finite-size effects~\cite{the:Fischer lattice artifact study with SDE}.
Finally, the IR-behavior obviously depends on the choice of definition
of the coupling, which can lead to diverging, \emph{freezing}  or vanishing
$\alpha_{s}$. Nevertheless, it is now clear that
the \emph{decoupling solution} seen in the SDE framework also genuinely exists on the
lattice \cite{the: Sternbeck 2013}, where it is dominant and where it was in fact first discovered
\cite{the:Boucaud decoupling sol. discovery}.

Lattice data for the gluon and ghost propagators in SU(2) and SU(3)
from Ref. \cite{the:gluon ghost props lattice-1} have been used by Aguilar,
Binosi and Papavassiliou \cite{the:Aguilar Lattice 2010} to form 
$\alpha_{s}^{gh}$ from Eq. (\ref{eq:alpha_s SDE ghost--gluon}),
and $\alpha_{gse}$ using the  \emph{pinch} technique.
As already discussed for other articles from these authors,
a massive gluon formula  is assumed for the gluon propagator at
tree-level, $\Delta(Q^{2})\propto1/[Q^{2}+m_{g}^{2}(Q^{2})]$, rather than
a massless propagator. This accounts for
the \emph{freezing}  of their couplings in the IR rather than vanishing couplings.
Assuming $m_{g}(0)=0.5$ GeV, the
Landau gauge coupling \emph{freezes} at $\alpha_{s}^{gh}(0)=4.45$; 
for $m_{g}(0)=0.6$ GeV, the freezing value is  $\alpha_{s}^{gh}(0)=6.40$. 
The gauge-invariant coupling defined in the  \emph{pinch} technique \emph{freezes} at 
$\alpha_{gse}(0) \simeq 0.5$ for $m_{g}(0)=0.5$ GeV and 
$\alpha_{gse}(0) \simeq 0.4$ for $m_{g}(0)=0.6$ GeV.

Furui and Nakajima have used lattice gauge theory to calculate $\alpha_{s}^{gh}(Q^{2})$ 
in the Landau gauge, $\widetilde{\mbox{MOM}}$ RS and the \emph{quenched} approximation.
Quarks are defined as domain wall fermions. The  resulting coupling
\emph{freezes} at $\alpha_{s}^{gh}(0)=2.5\pm0.5$ \cite{the:Furui Nakajima}.
In determining this value, the apparent IR-vanishing trend of $\alpha_{s}^{gh}$
was ignored since it was identified to be due to finite size effects
\cite{the:Furui on finite size effects}.

Ayala and collaborators studied the influence of dynamical quarks
on $\alpha_{s}^{gh}$ by performing \emph{quenched}, and $n_{f}=2$ and 2+1+1
\emph{unquenched} lattice simulations \cite{the:Ayala 2012}.  
 These authors assume masses of order 
20-50 MeV for the light quarks,  95 MeV for the $s$
quark mass, and 1.51 GeV for the $c$ quark mass. The Landau gauge
and MOM RS are used. There is no direct influence of quarks on 
$\alpha_{s}^{gh}(Q^{2})$ for a fixed dynamical gluon mass scale. However, as quarks
change the value of $m_g$, they have an indirect influence.  The authors 
obtain a \emph{freezing}  value  of $\alpha_{s}^{gh}(0)\simeq3.2\pm0.3$
for $m_g=0.5$ GeV, and $\alpha_{s}^{gh}(0)\simeq4.7\pm0.4$
for $m_g=0.6$ GeV, both for $n_{f}=$ 2+1+1.

Recently Maas and collaborators have studied systematic lattice effects on the gluon
and ghost propagators
on $\alpha_{s}^{gh}(Q^{2})$ in the Landau 
gauge \cite{the:Maas (2015)}.
The study was carried in 2, 3 and 4 dimensions and yields a solution for $\alpha_{s}^{gh}$ which vanishes in the IR. 

The gluon and ghost propagators and other correlation functions have been 
evaluated on the lattice in many analyses, but without providing
the IR-behavior of $\alpha_{s}$. An example
of such results is shown in Fig. \ref{Flo:gluon propagator}. In  Ref.~\cite{the:Maas (2007)} 
the propagators have been obtained in the interpolating
gauge in order to test the gauge dependence of the correlation functions.  The authors obtained
propagators compatible with the \emph{Kugo--Ojima} and the Gribov--Zwanziger scenarios, regardless
of the choice of gauge. 
In \cite{the:Cucchieri-Mendes 2008} it is shown that the gluon propagator
in 3D and 4D is non-zero and finite in the IR, leading to an  IR-finite
coupling. The calculations are done in SU(2)-color but the conclusion 
extends to SU(3)$_C$. Indeed, a \emph{quenched} calculation done
in the Laplacian gauge for different  numbers of colors $ N_{C}$
(SU(2), SU(3) and SU(4)) shows little dependence of the gluon
propagator on $N_C$ \cite{the:Alexandrou (2002)}. This is compatible with
the expected $1/N_C$ dependence for the coupling because $\alpha_{pQCD}$
factors the expressions of $\alpha_s$ defined from propagators (Eqs. 
(\ref{eq:alpha_s SDE ghost--gluon}), (\ref{eq:alpha_s 3g vertex}) and
 (\ref{eq:alpha_s 4g vertex})), and that in the \emph{quenched} approximation, 
$\alpha_{pQCD} \propto 1/\beta_0 \propto N_C$.

\begin{figure}
\centering
\includegraphics[scale=0.45]{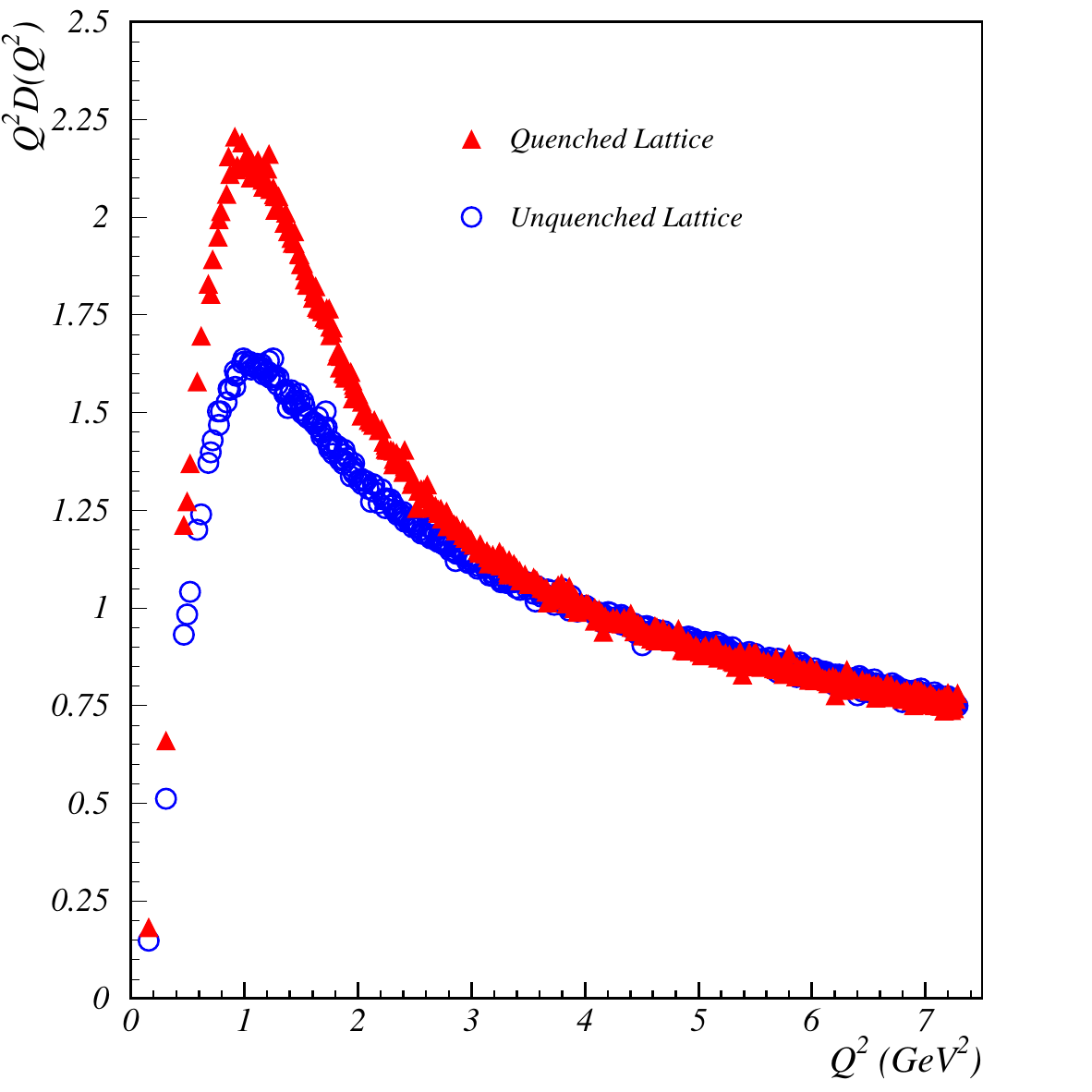}}
{\caption{ \small \label{Flo:gluon propagator}The gluon dressing function
$Z(Q^2)=Q^{2}D(Q^{2})$, from \emph{quenched} and full lattice calculations 
 from Bowman \emph{et al.}, Ref.~\cite{the:Bowman}.}
\end{figure}

Most of the propagator studies cited above are done in \emph{quenched} QCD and Landau gauge.
Calculations including dynamical quarks have also been carried out,
showing clear effects on the propagators from quark loops, but preserving the qualitative
features obtained in \emph{quenched} QCD \cite{the:Bowman}.
Calculations in other gauges also exist and display similar features
as the Landau gauge calculations, see {\it e.g.}, Ref. \cite{the:Alexandrou (2002)} for calculations done in
the Laplacian gauge. Discretization,
finite lattice size and \emph{tadpole} effects were studied in  Ref.~\cite{the:Bonnet }.

The coupling $\alpha_{s}(Q^{2})$  can be also obtained from
the static potential. In  recent results by Horsley and collaborators~\cite{the:Horsley 2014},
 $\alpha_{V}(Q^{2})$ and its \emph{$\beta$-function} are computed on a lattice
using  numerical stochastic perturbation theory; see the reviews in Refs.
~\cite{the:Vandersickel-Zwanziger Gribov Pb. review, the:Stoch. Quant. lattice} 
and Section \ref{sub:Stochastic-quantization}.
The perturbative part of the static potential, $4\alpha_{V}/3r$,
is calculated from  \emph{Wilson loops}. In the \emph{quenched} results,
$\alpha_{V}(Q^{2})$ diverges in the IR at $Q\simeq0.7\Lambda_{\overline{MS}}$.
When quarks are included, $\alpha_{V}(Q^{2})$ \emph{freezes} around $Q \simeq0.8 \, \Lambda_{\overline{MS}}$
with $\alpha_{V}(0) \simeq 0.52$. The calculations are performed for SU(3)$_C$
and include massless u and d quarks. 

Finally, another method is to use the Schr\''odinger functional; see Section
\ref{sub: large Q Lattice-QCD}. It leads to a coupling
in good agreement with pQCD down to $Q^{2}\simeq0.4$
GeV$^{2}$, but it diverges in the IR as $e^{m/Q}$ \cite{the:Luscher }.

\section{Functional  renormalization group equations \label{sub:Functional renormalization group equations}}

The  functional renormalization group (FRG) framework is similar to the
SDE approach. The central objects are  \emph{Green's function}s (although, only
dressed ones, while SDE must include both bare and dressed functions)
and the method provides an infinite system of {}``flow equations'',
or {}``exact renormalization group equations'', which fully describe
the underlying quantum field theory in continuous Euclidean space-time. Thus,
the FRG equations and SDE must in principle share the same solutions.
In practice, the different approximations needed to solve the FRG
equations and the SDE can introduce differences. Those can be used 
to assess the uncertainties on the solutions. Thereby,
just like SDE and Lattice benefited from their connection and complementarity,
the FRG approach provides yet another complementary framework. 

The starting point of the FRG method is to define an IR cut-off $k$ above which the theory is
well understood and perturbatively calculable. Then the value
of $k$ is lowered to access the nonperturbative domain. The cut-off takes the
form of a  regulator  added to the classical action. In effect,  it
adds a large, momentum-dependent, mass to the field,  suppressing the 
low modes of momenta below $k$ \cite{the:Pawlowski IR exp from FRG}.
A flow equation for the  \emph{Green's function}s expresses how those depend
on $k$ \cite{the:Wetterich-Morris FRG}. This flow equation is obtained
by first partially deriving the generating functional of the quantum
field theory which describes its dependence with respect to $k$. 
  This yields the flow equation for the generating
functional $\Gamma_k[\phi]$ 
\begin{equation} \label{FRGE}
\frac{\partial}{\partial k} \Gamma_k[\phi] =   \tfrac{1}{2} {\rm Tr} \,  \frac{\frac{\partial }{\partial k}R_k}{\Gamma_k^{(2)}[\phi] + R_k},
\end{equation}
where $R_k$ is the regulator which decouples the infrared modes with momentum below the sliding scale $k$ and $\Gamma_k^{(2)}$, the second functional derivative, is the full inverse propagator modified by the presence of the regulator  $R_k$. The trace stands for the sum over momenta, fields and internal indices. The Wetterich equation (\ref{FRGE}) is an exact flow equation  for $\Gamma_k[\phi]$ which has a one-loop structure, in contrast with the multi-loop diagram structure characteristic of perturbation theory.

The system of flow equations for the  \emph{Green's function}s are 
obtained by successively deriving the flow equation of the generating
functional with respect to the source or field terms in the action.
Once the $k\rightarrow0$ limit is reached, the cut-off disappears
and all the slow modes of the quantum field theory are integrated
out in the solution. One thus has an analytic,  nonperturbative method
for investigating strongly interacting quantum field theories. The FRG method
is particularly relevant to the study of running couplings since they can be
defined from the  \emph{Green's function}s as in other approaches. 

The FRG equations thus provide the correct RG-scaling expected from the theory.
As verification, the pure Yang--Mills expectation for the UV running of
$\alpha_{s}(Q^{2})$ has been recovered within the FRG framework
\cite{the:Gies 2002 FRG}. In practice, the infinite system of equations
must be truncated.  As  it is the case of the SDE formalism,   the truncation introduces uncertainties which 
can be difficult to fully assess.   Furthermore, multiple choices of
regulator are possible, which introduces some arbitrariness. For a 
review of the  FRG method and a discussion of its predictions for the 
running QCD coupling in the IR, see {\it e.g.,} Ref. \cite{the:FRG reviews/lectures}.

Initial FRG studies which were performed down to the scale $\sim\Lambda$ 
found an IR divergent solution of the gluon propagator,   in agreement with
the scenario of IR slavery;   see Refs. \cite{the:Ellwanger linear pot in FRG,
the:Early FRG studies}. However, since the $k\rightarrow0$  limit
was not reached, the  nonperturbative effects were not fully accounted
for. More recent FRG investigations have now invalidated the IR slavery
conclusion.
The ghost and gluon propagators have been studied within the FRG formalism in the \emph{pure field} sector,
using  Landau gauge and  4D Euclidean space~\cite{the:Fischer  IR exp from FRG,the:Pawlowski IR exp from FRG}, 
assuming the \emph{scaling solution}
as observed in the lattice and SDE frameworks (Section
\ref{sub:Classes-of-solutions in IF domain}). They yield a similar value
of the \emph{IR exponent}: in \cite{the:Fischer  IR exp from FRG},
it is found that $\kappa'\simeq0.52$, which is compatible with the
SDE value of 0.595, once the dependence of $\kappa'$ with the choice
of cut-off implementation is accounted for.   
The range $0.539\leq\kappa'\leq0.595$, with a strong preference
for the maximum value $\kappa'=0.595$ was obtained in Ref.~\cite{the:Pawlowski IR exp from FRG}. 
The resulting 
\emph{freezing}  value of the ghost--gluon coupling is $\alpha_{s}^{gh}(0)=2.972$.
All of  these results are consistent with a \emph{freezing}  of the coupling due
to IR ghost dominance and the \emph{Kugo--Ojima criterion}.
The ghost and gluon propagators and the coupling $\alpha_{s}^{gh}(Q^{2})$ have also been 
studied using the FGR approach in \cite{the:Leder 2011 FRG}.
These calculations have been also performed within the \emph{pure field} sector using 
Coulomb gauge and a focus on the \emph{scaling solution}. The resulting solution
leads to a unique \emph{freezing}  coupling, although
its \emph{freezing}  value is not specified. The solution yields $\kappa'=0.35$, in
agreement with  Ref. \cite{the:Epple 2007} using a variational approach
to the Yang--Mills Schr\''odinger equation, and with the first solution
obtained using the SDE in \cite{the:Schleifenbaum 2006}. 

The behavior of the running coupling is specifically addressed using FRG in Ref.~\cite{the:Gies 2002 FRG} within
the \emph{pure gauge} field sector for both SU(2)$_C$ and SU(3)$_C$.  The predicted coupling
\emph{freezes} in the IR at the value of $7.7\pm2$ for SU(3)$_C$ and 11.3 for
SU(2)$_C$.  The calculations were performed in the Landau  gauge.
The inclusion of quarks is expected to lower the freezing values~\cite{the:Gies 2002 FRG},  in
agreement with the discussion in Section~\ref{sub:SDE: Including quarks}.

\section{The Gribov--Zwanziger approach\label{sub:Gribov--Zwanziger-approach}}

Gribov was the first to point out the effects of gauge-fixing in a Yang\textendash{}Mills
theory \cite{the:Gribov 1978}. Those are directly
relevant to the coupling when it is defined using  \emph{Green's functions}. The coupling
is invariant under the local gauge symmetry given by $\delta_{\omega}A_{\mu}^{a}=-D_{\mu}^{ab}\omega^{b}$
(with $a,b$ color indices).  However, the standard gauge-fixing relations,
such as the Coulomb, Feynman or Landau gauges, are not enough to uniquely
specify the gauge field.  The vanishing eigenvalues of the Faddeev-Popov operator 
generate copies of the fields; {\it i.e.}, physically different fields which are  related by
gauge transformations still exist, even after fixing the gauge. To avoid
these copies in the path-integral formalism, Gribov proposed to restrict the Faddeev-Popov
operator to an integration region where the operator is strictly 
positive (the so-called ``first Gribov'' region). To implement
such a restriction, Gribov introduced a non-local operator. Even with this prescription, copies still
exist within the first Gribov region~\cite{the:Friedberg 1996}. This has prompted proposals 
to restrict  the integration to an even smaller domain, the `` fundamental modular region'',  which is 
truly free of Gribov copies. It was eventually found however, that the remaining copies appear to have no influence 
on the solutions of the theory~ \cite{the:Zwanziger Stoch. quant. 2002, the:Binosi (2015),  the:Zwanziger 1994,
the:Zwanziger Stoch. Quant 2003 },
which is consistent with the finding from lattice gauge theory,
 namely that the effect of
Gribov copies (``Gribov noise'') is negligible \cite{the:Giusti 1998, the:Cucchieri small gribov noise}.

The non-locality of Gribov's operator limits the formalism to semi-classical calculations.
A local renormalizable Lagrangian restricting the integration region to
the first Gribov region was later proposed by Zwanziger \cite{the:Zwanziger 1981-1993},
enabling  loop calculations. For a review of the Gribov--Zwanziger
approach, see  Ref. \cite{the:Vandersickel-Zwanziger Gribov Pb. review}.

Since the ghost propagator is the inverse of the Faddeev-Popov operator, 
the ghost propagator has no pole in the Gribov approach, except
on the horizon of the Gribov region, which corresponds to the $Q^{2} \to 0$
limit.  In this case,  the ghost propagator diverges, following a $1/Q^4$ dipole form.
Gribov found that when the integration of the generating functional,
or any correlation function, is restricted to the first Gribov
region, the Gribov mass generated by the copies suppresses the tree-level
gluon propagator in the Landau gauge, which should then take the IR
form:
\begin{equation}
D_{\mu\nu}^{bc}(q)=\delta^{bc}\left(\eta_{\mu\nu}-\frac{q_{\mu}
q_{\nu}}{q^{2}}\right)\frac{q^{2}}{q^{4}+m_{gr}^{4}},
\label{eq:Gribov gluon propagator 3}
\end{equation}
where $m_{gr}$ is the Gribov mass. With this form, the gluon propagator vanishes in the IR, whereas as 
noted above, the ghost propagator diverges as $1/Q^{4}$. This behavior is distinct from its $1/Q^{2}$ UV-behavior. This scenario,
the Gribov--Zwanziger approach, results in an IR \emph{freezing}  of $\alpha_{s}$
analogous to the \emph{scaling solution} seen in the SDE and lattice
studies; see Section \ref{sub:Classes-of-solutions in IF domain} and Ref.
\cite{the:Dudal 2005-2008}.  However, the predicted \emph{IR exponent} is
$\kappa'=1$ in this approach, differing from that of SDE ($\kappa'\simeq 0.595$).
The IR-\emph{freezing}  value in this scenario was calculated for the ghost--gluon
coupling by Gracey to be $\alpha_{s}^{gh}(0)=16/3\pi\simeq 1.70$ (Landau
gauge, $\overline{MS}$, one \emph{loop} calculation)~\cite{the:Gracey 2006}.

The analytical properties  of the calculations using  the Gribov--Zwanziger approach allows one to investigate the underlying 
reason for the  \emph{freezing} of the QCD coupling, at least at one loop: As one expects on general dimensional considerations,
the \emph{freezing}  occurs because all dimensionful quantities cancel each other
in the IR, so the coupling becomes scale invariant.  The mechanism for such a cancellation has its origin in the
interplay of the anti-commutating Faddeev-Popov ghosts and the commutating
{}``Zwanziger ghosts''. Those arise from enforcing the Gribov
region with the local Zwanziger Lagrangian.  More recent 2-\emph{loop} calculations 
of the gluon and ghost propagators are available~\cite{the:Gracey stat pot 2009-2010, the:Gracey arXiv:1409.0455}.
Although the original Gribov--Zwanziger approach leads to the \emph{scaling solution}, 
it should be noted that the decoupling scenario  also appears to be
possible in an extended Gribov--Zwanziger approach, see  Refs. \cite{the:Dudal 2010,
the:Dudal 2008-2011}.

The Gribov--Zwanziger scenario may appear gauge-dependent since 
gauge-fixing procedures without Gribov copies exist, such as the
axial or Laplacian gauges; see the related discussion in  Ref. \cite{the:Williams 2003}.
The light-cone gauge $A^+=0$  is also free of ghosts. In addition, there are no ghosts using the stochastic gauge-fixing quantization method
(see below) which, in fact, was 
introduced in the QCD propagators/coupling context to circumvent the
Gribov copy problem. However, in spite of the apparent gauge dependence of  the Gribov--Zwanziger scenario,
equivalent mechanisms are found in the ghost-less frameworks, as we will see  in the next section.

\section{Stochastic  quantization \label{sub:Stochastic-quantization}}

Stochastic  quantization provides a method to obtain the  \emph{Green's function}s
of an Euclidean field theory \cite{the:Nelson Stochastic quantization}.
It was first used by Zwanziger \cite{the:Zwanziger Stoch. quant. 2002,
the:Zwanziger 2003} as a means to obtain and solve the SDE
without the Faddeev\textendash{}Popov procedure, thus circumventing 
the Gribov copy problem just discussed. 

The formalism of   stochastic quantization is based
on a diffusion equation analogous to the Fokker--Planck equation of statistical
mechanics. The diffusion equation acts  within the space of gauge field configurations; it 
controls the probability distribution $P(A_\mu)$ of the gluon field
$A_{\mu}$, just as the Fokker--Plank equation controls the spatial distribution
of diffusing bodies. The distribution $P(A_\mu)$ is the weight of the QCD's Euclidean generating
functional, $P(A_\mu)=e^{-S}$;  and the nomenclature {}``stochastic''
comes from the weighted random walk of $P(A_\mu)$ characterized by the
diffusion equation. The drift force term in the diffusion equation
is given by the  functional
derivative of the Yang--Mills action, $K_{YM\mu}^{~~~~a}=-\delta S_{YM}/\delta A_{\mu}^a$.
The field solution of the equation is also the solution of the Yang--Mills
theory. There is no specific choice of gauge in this approach. Instead
of a single gauge-fixing which introduces the Gribov copies, a supplemental
drift force term in the form of infinitesimal gauge transformation,
$K_{gt ~\mu}^{~~~a}=a^{-1}(\partial_\mu \delta^{ac}+f^{abc}A_{\mu}^{b})\partial \cdot A^c$,
is added to the diffusion equation.  The coefficient $a$ is a free-parameter
akin to a gauge-fixing term which controls the balance between the Yang--Mills
drift force $K_{YM}$ and a gauge restoring force $K_{gt}$.  This
approach reproduces perturbative QCD results in the UV \cite{the:Munoz-Sudupe}.
Furthermore, in the case of Landau gauge ($a\rightarrow0$),
 stochastic quantization yields the same \emph{scaling solution}  in the IR as the SDE using the Faddeev-Popov
method (Section \ref{sub:Schwinger--Dyson-formalism}), lattice
(Section \ref{sub:Low Q Lattice-QCD-calculations}) and FRG frameworks
(Section \ref{sub:Functional renormalization group equations}),
with a compatible \emph{IR exponent} $\kappa'$ for the gluon propagator
\cite{the:Llanes-Estrada 2012, the:Zwanziger 2003}.  

It has been found recently that the decoupling scenario is also a solution
of the  stochastic quantization equations in the Landau gauge \cite{the:Llanes-Estrada 2012}.
Zwanziger has shown that  the stochastic quantization formalism, constrained to the
Landau gauge limit, generates the Fadeev--Popov theory together with the Gribov
gauge-fixing prescription \cite{the:Zwanziger Stoch. quant. 2002}.
Furthermore, the diffusion equation, which can be solved exactly
in the Landau gauge limit, has the same solution
as the Fadeev--Popov theory restricted to the first Gribov region
\cite{the:Zwanziger Stoch. Quant 2003 }. Thus, $K_{gt}$ provides
an additional term which removes the Gribov copy problem \cite{the:Zwanziger Stoch. quant. 2002,
the:Vandersickel-Zwanziger Gribov Pb. review}. 

Zwanziger has also derived the  stochastic quantization SDE using the diffusion
equation.  Using truncation, the diffusion equation was solved approximately, 
but  nonperturbatively, by assuming that the gluon propagator follows a power law
in the IR. It was found by Zwanziger ~\cite{the:Zwanziger 2003}, and
confirmed by Llanes-Estrada and Williams~\cite{the:Llanes-Estrada 2012},
that the gluon propagator contains a longitudinal component which becomes large and dominant in the IR. The
transverse gluon propagator is IR-suppressed as $(Q^{2})^{-1+2k'}\simeq(Q^{2})^{0.04}$
due to longitudinal gluon loops, whereas the longitudinal gluon propagator
is IR-enhanced as $(Q^{2})^{-1-k'}\simeq(Q^{2})^{-1.52}$.  This shows
that in the  stochastic quantization approach, longitudinal gluons
take the place of ghosts. Although the longitudinal gluon propagator
vanishes as $a$ in the Landau gauge limit, it
still acts as the equivalent of a ghost since vertices behave
as $1/a$, yielding an overall finite result. This leads to the same
picture of confinement as already discussed: confinement would result from
the enhanced long-distance propagation of longitudinal gluons ( {\it  i.~e.},
ghosts in the other frameworks), while the transverse gluons ({\it i.~e.},
the gluons in the other frameworks) decouple at long distances because
of longitudinal gluons (ghosts) loops, and become irrelevant. This
implies an IR-\emph{freezing}  behavior for $\alpha_{s}$. The inclusion of quark loops  is expected to preserve this feature
and to play no role in the IR as long as they are not massless \cite{the:Zwanziger Stoch. Quant 2003 }.

Pawlowski and collaborators ~\cite{the:Pawlowski 2010} have studied the ghost and gluon propagators
in SU(2)$_C$, in 2, 3 and 4 dimensions and in the Landau gauge using the  stochastic quantization formalism 
on the lattice,  and, as for most of the other lattice results, have obtained the \emph{decoupling solution} with
an IR-vanishing coupling \cite{the:Pawlowski 2010}.

\section{ Analytic and dispersive approaches \label{sub:Analytic and Dispersive approaches}}

We have discussed gauge-dependent and RS-dependent
couplings, starting with the SDE approach.  As already noted, these arbitrary dependences are not a major obstacle, since in general
$\alpha_s$ is not an observable.
Nevertheless, one can try to restore the definition of $\alpha_s$ to a status close to an observable,
as in the case of \emph{effective charges} such as $\alpha_V$ or $\alpha_{g1}$,  by
demanding that  it satisfies causality, {\it i.e.}, that it is  analytic in the complex
$Q^{2}$-plane, except for branch points or cuts on the real time-like axis.  Since the Landau
pole on the space-like-axis violates causality, demanding analyticity and the 
causality of $\alpha_{s}$ regularizes the Landau singularity. Such
{}`` analytic'' and {}``dispersive'' approaches were first proposed
in the context of QED by Redmond \cite{the:Redmond 1958} and then
applied to QCD \cite{the:Sanda 1979, the:non-pert in analytic QCD, the:non-pert in analytic QCD loop-4}. 
Early work by Sanda~\cite{the:Sanda 1979}, in which an analytic behavior 
is forced on $\beta(\alpha_s)$ using a Borel summation technique, 
yielded a coupling \emph{freezing}  at $ \alpha_s(0) \simeq 4$.
For reviews of the  analytic and dispersive approaches, see \cite{the:Prosperi 
alpha_s review, the:Stefanis alpha_s lowq review, the:Alpha_an reviews}. 
These approaches are also discussed in the context of the SDE in 
\cite{the:Cornwall alpha_s PT}.

\subsection{Analytic  approach \label{sub:Analytic approach}}

The Shirkov \emph{et al.} {}``analytic'' approach  results
in the folding of QCD  nonperturbative effects into the coupling~\cite{
the:non-pert in analytic QCD, the:non-pert in analytic QCD loop-4, the:MPT}. The 
starting point is to define the  analytic coupling $\alpha_{an}(Q^{2})$ 
from a {}``spectral density'' $\rho(\nu)$,
following the K\''{a}ll\'{e}n--Lehman relation:
\begin{equation}
\alpha_{an}(Q^{2})=\frac{1}{\pi}\int_{0}^{\infty}d\nu\frac{\rho\left(\nu\right)}{\nu+Q^{2}},\label{eq:Kallen-Lehman}\end{equation}
where the expression of $\rho(\nu)$,
\begin{equation}
\rho\left(\nu\right)=\Im m\left(\alpha_{pQCD}^{(l)}(-\nu-i\varepsilon)\right)\alpha_{pQCD}^{(l)}(-\nu-i\varepsilon),\label{eq:alpha_s spectral function}
\end{equation}
depends on  $\alpha_{pQCD}^{(l)}$
at \emph{loop} order $\beta_{l}$.   The incorporation of the spectral function of Eq. (\ref{eq:alpha_s spectral function})
into  Eq. (\ref{eq:Kallen-Lehman}) leads to a Kramer--Kr\''{o}nig type
of relation for the coupling, which demonstrates its analyticity/causality
property. The expression of $\alpha_{an}$ at order $\beta_{0}$ is: 
\begin{equation}
\alpha_{an}^{(0)}\left(Q^{2}\right)=\frac{4\pi}{\beta_{0}}\left(\frac{1}{\mbox{ln}(Q^{2}/\Lambda^{2})}+\frac{\Lambda^{2}}{\Lambda^{2}-Q^{2}}\right).\label{eq:alpha_s from analytic QCD}
\end{equation}
The result of regulating the \emph{Landau pole} in the  IR domain leads to
an effective   $\alpha_{an}(M_Z^{2})$ compatible with the world
data for   $\alpha_{s}(M_Z^{2})$~\cite{the:Kotikov (2012)}. 
The dependence on $\Lambda$ cancels at $Q^{2}=0$ and we have 
$\alpha_{an}^{(0)}\left(0\right)=4\pi/\beta_{0}.$
This value holds at all orders \cite{the:non-pert in analytic QCD loop-4, the:Shirkov pQCD 1999, 
the:Milton alpha_an higher order}.
Near the \emph{Landau pole}, $\alpha_{an}^{(0)}\left(\Lambda^{2}(1+\varepsilon)\right)=2\pi/\beta_{0}+
\mathcal{O} \left(\epsilon^{2}\right)$.
As can be seen in Fig. \ref{Flo:alpha_s Shirkov}, $\alpha_{an}(Q^{2})$
does not \emph{freeze} in the IR but does stay finite. 
The term $\Lambda^{2}/(\Lambda^{2}-Q^{2})$  in Eq. (\ref{eq:alpha_s from analytic QCD}) is a  nonperturbative
 \emph{power law} contribution since it is independent of  higher order perturbative terms
in $\beta_{l}$. 

\begin{figure}[ht]
\centering
\includegraphics[width=10.0cm]{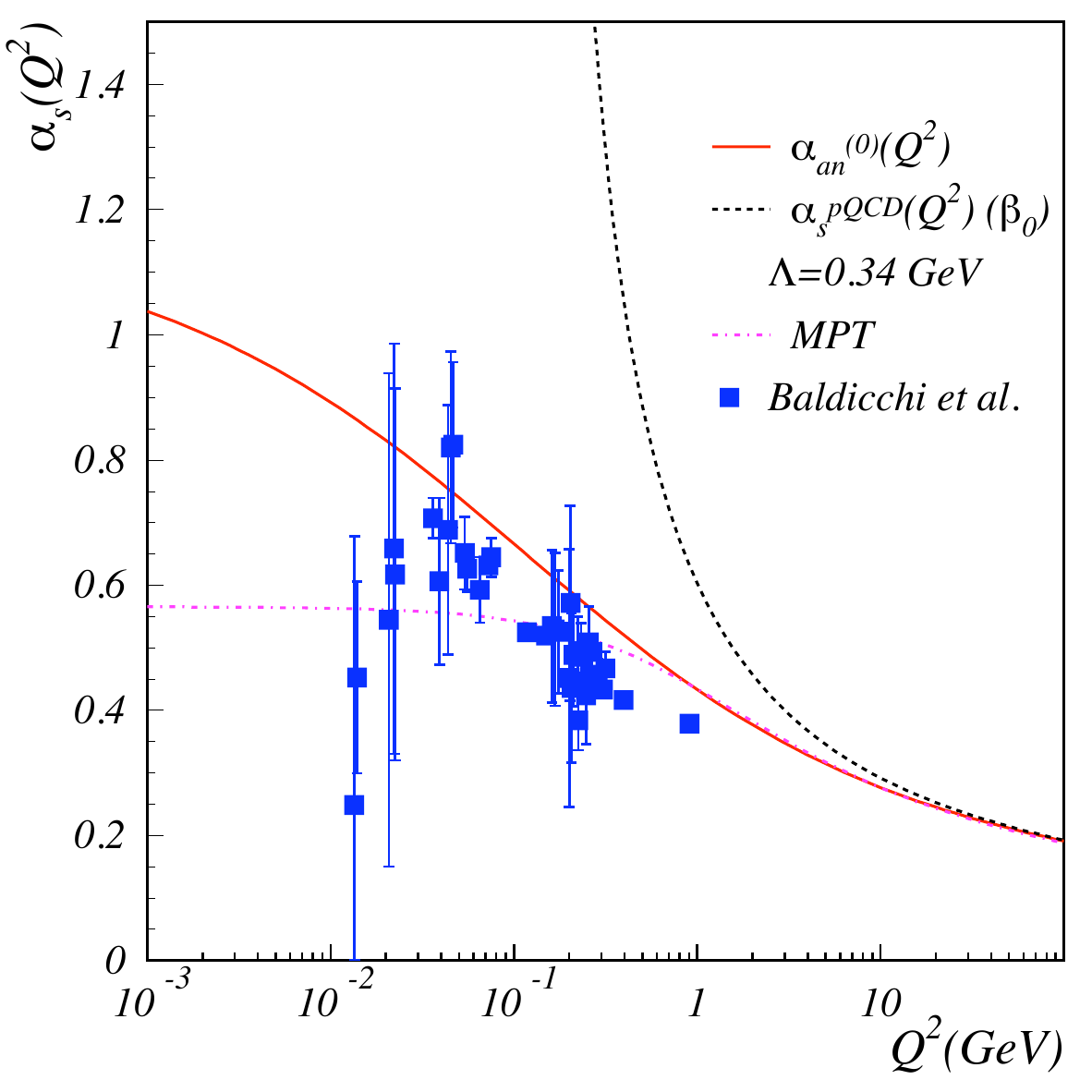}
\caption{ \small \label{Flo:alpha_s Shirkov} The strong coupling $\alpha_{an}^{(0)}(Q^{2})$
in the analytic approach (continuous line), Eq. (\ref{eq:alpha_s from analytic QCD}).
It can be compared to the pQCD expectation at leading order (dashed
line) and to the experimental extraction using the analytic definition
of the strong coupling \cite{the:Baldicchi alpha_an from quarkonium}.
The value of $\Lambda$ was taken in the $\overline{MS}$ scheme.
The dotted line is the  massive analytic perturbation theory prediction for $\Lambda=0.315$ GeV and
a glueball mass parameter $m_{gl}=0.995$ GeV.}
\end{figure}

The procedure described above is extendable to higher orders, 
\cite{the:Prosperi alpha_s review, the:Shirkov pQCD 1999}, with the same 
caveat as for the perturbative coupling: the solutions for $\beta_{2}$ and higher
orders  are either numerical or  approximately
analytic solutions. The solution at order $\beta_{1}$ involves the
Lambert function; see Section \ref{sub:pQCD evolution-equation}.
The resulting higher-order solutions $\alpha_{an}^{(l)}(Q^{2})$ are numerically
close to $\alpha_{an}^{(0)}(Q^{2})$. This is not surprising since
all $\alpha_{an}^{(l)}(Q^{2})$ have the same limit at $Q^{2}=0$; in addition, they have a similar structure at
$Q^{2}\gg\Lambda^{2}$ in the sense that
they differ only via perturbative terms $\beta_{l}$.   If nevertheless,
one needs a coupling beyond order $\beta_{0}$, convenient approximations
are given in Refs.~\cite{the:Bakulev et al. 2004, the:Kurashev et al 2003}.

The analytic approach of Shirkov  yields good   pQCD results for  large $Q^2$~\cite{the:Shirkov pQCD 1999} and  the additional  $\Lambda^{2}/\left(\Lambda^{2}-Q^{2}\right)$ term generates, for large $Q^2$, a series $(\Lambda/Q)^{2n}$ as expected from the OPE~\cite{the:Alpha_an reviews, the:Nesterenko npQCD 2000, the:Stefanis 2000, the:Shirkov 2001}; However, the results do not  encompass all nonperturbative effects. For example, a meaningful comparison with the strong coupling $\alpha_{g1}(Q^2)$, obtained from the Bjorken sum, is not possible,  see Section \ref{sub:Handling of npert. terms}. 
In addition, the decay branching ratio of the $\tau$
lepton into non-strange hadrons evaluated with $\alpha_{an}$ is significantly
underestimated \cite{the:Milton alpha_an higher order, Milton-Geshkenbein, the:Cvetic 2010-1}.

The analytic approach was used by Baldicchi\emph{ et al.} to
experimentally extract $\alpha_{s}(Q^{2})$ for $0.013<Q^{2}<0.904$
GeV$^{2}$ from quarkonium spectroscopy~\cite{the:Baldicchi alpha_an from quarkonium}, see Fig.
\ref{Flo:alpha_s Shirkov}. The result is obtained in a model-dependent
way of using the Bethe-Salpeter formalism.  The ansatz based on the  \emph{Wilson loop} correlator
used is reminiscent  of the potential approach; see Section \ref{sub:Potential-approach}.
It incorporates a Coulomb-like one-gluon exchange term proportional
to $\alpha_{s}(Q^{2})$ and a confining term proportional to $\sigma$,
the string tension. The meson mass spectrum is calculated with $\alpha_{s}(Q^{2})$
using parameters such that experimental mass values
are reproduced. The authors concluded that the meson spectrum is better reproduced with an analytic-type
coupling without \emph{freezing}  rather than with a \emph{freezing}  coupling constant
\cite{the:Baldicchi meson spectrum}. This conclusion is model-dependent 
since it depends on the methodology chosen
to  compute
the meson spectrum.  Indeed, other approaches in which $\alpha_{s}$
\emph{freezes} in the IR reproduce the meson spectrum as well,  as discussed in
Sections \ref{sub:Holographic-QCD large Q} and \ref{Sub: hadron spectrum}. The low-$Q^{2}$
results of Baldicchi \emph{et al.} are compatible with a finite $\alpha_{s}(0)$
value lower than expected from $\alpha_{an}$  (Eq. \ref{eq:alpha_s from analytic QCD}), or even a vanishing
$\alpha_{s}(0)$. The large experimental and theoretical
uncertainties prevent a firm conclusion. The coupling
is normalized in the UV  so that the 3-\emph{loop} value  $\alpha_{an}^{(2)}(M_Z^{2})$
corresponds to the average world data for $\alpha_{\overline{MS}}$.
We also note that  the experimentally
extracted $\alpha_s$ should match the $V$-scheme, so the choice to match
to $\alpha_{\overline{MS}}$ may not be optimal. G. Ganbold reproduced
the experimental  coupling $\alpha_{s}(Q^{2})$ of Baldicchi\emph{ et al.}, and
the meson mass spectrum with its Regge behavior, starting from IR-finite
quark and gluon propagators within a Bethe-Salpeter framework \cite{the:Ganbold}.

One can also improve the pQCD series by suppressing the \emph{Landau pole} and using the
analytic approach.  This method is known as 
 analytical perturbation theory (APT). For example, $\alpha_{an}$
has been applied to the Bjorken sum rule perturbative expression, Eq. (\ref{eq:Bj SR, order alpha^4}),
allowing one to push down the pQCD limit close to  the domain of validity
of  chiral perturbation theory~\cite{the:Bjorken S.R. in analytical pert theo}.

Efforts have also been made to extend the validity
of  chiral perturbation theory predictions, see {\it e.g.}, \cite{the:Burkert delta in BJ sum.},
and to systematically improve the semi-classical approximation provided
by  light-front holographic QCD \cite{the:AdS/QCD review}. Bridging the high and low $Q^2$
domains could  provide   a convenient analytical description of QCD using 
hadronic degrees of freedom at low $Q^{2}$ and partonic ones
at large $Q^{2}$.  Progress towards such bridging has been achieved recently~\cite{the:Deur Lambda AdS/QCD}.

Improved perturbation theory would also allow for more reliable extractions
of  \emph{higher-twist} terms; 
However, their meaning needs to be reinterpreted since some of their effects are
folded in $\alpha_{an}$. The
APT formalism has been extended by Bakulev and collaborators in \cite{the:Bakulev FAPT}.

In addition to the approach just described,
analytic forms for  $\alpha_{s}$ can be  obtained using effective charges and 
other methods. For example, Nesterenko \emph{et al.}
constrained the full \emph{$\beta$-function} to be analytical \cite{the:Nesterenko npQCD 2000, the:Nesterenko 2001}, yielding an {}``analytic invariant
charge''  similar to $\alpha_{an}(Q^{2})$, Eq. (\ref{eq:alpha_s from analytic QCD}).
At $\beta_{0}$ it reads: 
\begin{equation}
\alpha_{aic}(Q^{2})=\frac{4\pi}{\beta_{0}}\left(\frac{1}{\mbox{ln}(Q^{2}/\Lambda^{2})}-\frac{\Lambda^{2}}{Q^{2}\mbox{ln}(Q^{2}/\Lambda^{2})}\right)
\end{equation}
In contrast to $\alpha_{an}(Q^{2})$, $\alpha_{aic}(Q^{2})$ diverges
as $Q^{2}\rightarrow0$. This small-$Q^{2}$ behavior could account
for the linear confinement term of the static potential;  see  the discussion
in Section \ref{sub:Potential-approach}. Alternatively,  the  divergent behavior  can be
suppressed by including 
effects due to the pion mass \cite{the:Nesterenko (2005)},
yielding a value of $\alpha_{aic}(0)=0.47$ for $\Lambda=0.623$ GeV
and $n_{f}=2$. This approach has been linked to the phenomenology
of instantons \cite{the:alpha_aic and instantons}.  The compatibility
of Nesterenko's coupling to the Schwinger--Dyson formalism and  its consequences
for  chiral symmetry breaking are discussed in  Ref.~\cite{the:SDE and analytical approach},
 where IR divergent constituent quark masses are taken
as a manifestation of quark confinement, and the  value of $\Lambda=0.880$ GeV 
is constrained in order to recover the pion decay constant. 

Alekseev \emph{et al.} has introduced a {}``synthetic coupling'' \cite{the:Alekseev composite coupling}
which builds on the concept of $\alpha_{an}$, Eq. (\ref{eq:alpha_s from analytic QCD}),
by adding  nonperturbative pole terms to the coupling, or equivalently
$\delta$-functions in the spectral density $\rho(\nu)$. These terms simulate
the static  linear confinement potential
and the dynamically generated
gluon effective mass discussed in Sections \ref{sub:Potential-approach}
and \ref{sub:SDE massive gluon}. The extended definition  at order $\beta_{0}$ is:
\begin{equation}
\alpha_{syn}\left(Q^{2}\right)=\frac{4\pi}{\beta_{0}}\left(\frac{1}{\mbox{ln}(Q^{2}/\Lambda^{2})}+\frac{\Lambda^{2}}{\Lambda^{2}-Q^{2}}+\frac{c\Lambda^{2}}{Q^{2}}+\frac{\left(1-c\right)\Lambda^{2}}{Q^{2}+m_{g}^{2}}\right),\label{eq:synthetic alpha_s}
\end{equation}
where $c>1$ is a dimensionless parameter and the gluon effective
mass is $m_{g}=\Lambda/\sqrt{c-1}$. 
The term $c\Lambda^{2}/Q^{2}$ provides the linear binding
term  in the static potential with string tension $\sigma=8\pi c\Lambda^{2}/3/\beta_{0}$.
Expressions exist up to order $\beta_{3}$. 

The massive analytic perturbation theory (MPT) is another analytic
approach by Shirkov \cite{the:Bj SR in MPT} in which the suppression of the \emph{Landau pole} is
achieved by the introduction of a glueball mass $m_{gl}\simeq1$
GeV,  rather than by the $\Lambda^{2}/(\Lambda^{2}-Q^{2})$ term 
in Eq. (\ref{eq:alpha_s from analytic QCD}). The  nonperturbative coupling is essentially 
obtained from the pQCD expression of $\alpha_{s}(Q^{2})$
by substituting $Q^{2}$ by $Q^{2}+m_{gl}^{2}$. It results
in an IR-\emph{freezing}  behavior;  see Fig. \ref{Flo:alpha_s Shirkov}. MPT
was in particular used to improve the APT description of the IR experimental
data for the Bjorken sum, Eq. (\ref{eq:Bj SR, order alpha^4}), obtained
at Jefferson Lab \cite{Deur alpha_s from eg_1dvcs, the:JLab  Bj SR}.
The elimination of the \emph{Landau pole} can also be linked to a glueball mass in 
 background perturbation theory; see Section \ref{sub:Background pert theo, Simonov}.

All of these related approaches are physically-motivated to suppress the  \emph{Landau pole} which is the artifact of 
the standard perturbative approach. Several versions of analytic
couplings have been coded in Fortran by Ayala and Cvetic~\cite{the:alpha_an in fortran}.

\subsection{Dispersive  approach \label{sub:Dispersive-approach: Dok. Mar. Web.}}

The \textquoteleft{}\textquoteleft{}dispersive'' approach was proposed
by Y. L. Dokshitzer and collaborators~\cite{the:Dokshitzer dispersive approach, the:Dokshitzer/webber: HT in event shapes.}
to manage the presence of  nonperturbative $\mu_{n}(Q^{2})/Q^{n}$ \emph{power corrections}  expected from the OPE.
The parameter $\mu_{n}(Q^{2})$ depends logarithmically on $Q^{2}$;  see
{\it e.g.}, Eq. (\ref{eq:LO matching}). The goal of the resulting coupling   $\alpha_{dmw}(Q^2)$ is to 
represent the strength of the strong interaction at large distances; it is 
thus parameterized by a set of phenomenological parameters
which can be extracted from inclusive observables. Hence, the resulting coupling
incorporates \emph{power corrections},
as is the case for the \emph{effective charges} discussed in Section \ref{sub:low Q Effective-charges},
but it is designed to provide a universal coupling, without specific process
dependence. The  coupling   $\alpha_{dmw}(Q^2)$ can therefore be related to $\alpha_{V}$ and to $\alpha_{gse}$ defined
from the gluon self-energy \cite{the:Cornwall alpha_s}, see   also Ref.~\cite{the:Aguilar (2009)}.  As  it is the case for
$\alpha_{gse}$,  the coupling $\alpha_{dmw}(0)$ has an IR finite value of about $0.6$ ~\cite{the:Dokshitzer dispersive approach, the:Aguilar SDE 2001}. 
Similar approaches are
described in  the references given in~\cite{the:OPT} (Sections \ref{sub:Optimized-Perturbation-Theory}
and \ref{sub:Optimized-pertubation-theory Low-Q}) and \cite{the:Grunberg Sudakov charge}
(Section \ref{sub:Sudakov-Effective-Charges}).  It is also closely related to
Grunberg's concept of  an \emph{effective charge} (Section  \ref{sub:low Q Effective-charges}).

Dokshitzer, Marchesini and Webber have proposed
an effective coupling  which satisfies a dispersion relation:
\begin{equation}
\rho(\nu^{2})=\frac{d~\alpha_{dmw}(\nu^{2})}{d~\mbox{ln}\left(\nu^{2}\right)},
\end{equation}
where $\nu$ is a dispersive variable. Although the formalism uses
standard massless gluon fields, $\nu$ plays a role equivalent to
a small gluon mass ({}``dispersive mass'') in the calculations.
The Dokshitzer--Marchesini--Webber coupling is related to a perturbative
coupling, with a definition close to the $V$-scheme and, like the \emph{effective charges} of Sections
\ref{sub:Effective charges and CSR} and \ref{sub:low Q Effective-charges} 
or $\alpha_{gse}$ of Section  \ref{sub:SDE massive gluon}, 
it can be  viewed  as a generalization of the Gell-Mann--Low QED coupling, since it is 
the coupling stemming from the propagator of a dressed 
gluon \cite{the:Dokshitzer dispersive approach, the:Prosperi alpha_s review}.
The coupling $\alpha_{dmw}(\nu^{2})$ is also related to the  analytic coupling $\alpha_{an}$ which 
obeys the K\''{a}ll\'{e}n--Lehman dispersion
relation;  see Eq. (\ref{eq:Kallen-Lehman}):
\begin{equation}
\alpha_{an}\left(Q^{2}\right) = Q^2 \int^{\infty}_0 \alpha_{dmw}(\nu^2) \frac{d~\nu^2}{(\nu^2+Q^2)^2}  , 
\end{equation}
or
\begin{equation}
\label{eq:alpha_dmw}
\alpha_{dmw}\left(\nu^{2}\right)=sinc\left(\pi\frac{d}{d~\mbox{ln}\left(\nu^{2}\right)}\right)\alpha_{an}\left(\nu^{2}\right).\end{equation}
Since $\alpha_{an}$ is obtained from a dispersion relation, it has no \emph{Landau pole}. 
Expanding the cardinal sine to first order yields $\alpha_{dmw}$ for small $\alpha_{an}(Q^{2})$ 
regime, {\it i.e.} in the UV:
\begin{equation}
\alpha_{dmw}\left(\nu^{2}\right)\simeq\alpha_{an}\left(\nu^{2}\right)-\frac{\pi^{2}}{6}\frac{d^{2}\alpha_{an}(\nu^{2})}{d~\mbox{ln}^{2}\left(\nu^{2}\right)}+...,
\end{equation}
This effective coupling is designed to provide the \emph{power corrections}
for inclusive observables as a function of a finite set of parameters
to be determined phenomenologically. For example, the DIS structure
function is obtained by calculating the relevant Feynman diagram amplitudes with
gluons of effective mass $\nu$. At first order: 
\begin{equation}
F\left(Q^{2},x_{Bj} \right)=\int_{0}^{\infty}\frac{d\nu^{2}}{\nu^{2}}\alpha_{dmw}(\nu^{2})\dot{\mathcal{F}}\left(Q^{2},x_{Bj} \right),
\end{equation}
with $\dot{\mathcal{F}}\equiv\partial\mathcal{F}(Q^{2},x_{Bj} ,\nu^{2})/\partial \mbox{ln}(\nu^{2})$,
the {}``characteristic function'' of $F(Q^{2},x_{Bj} )$. $\dot{\mathcal{F}}$
can be calculated perturbatively; it depends only
on $x_{Bj}$ at first order,  together with the \emph{power corrections} $\mu^{2}(Q^{2})/Q^{2}$. As suggested by  the
\emph{OPE}, $F$ can also be written as:
\begin{equation}
F\left(Q^{2},x_{Bj} \right)=F^{PT}\left(Q^{2},x_{Bj} \right)+\sum\frac{\mu_{2p}\left(Q^{2},x_{Bj} \right)}{Q^{2p}},
\end{equation}
where $F^{PT}$ is the perturbative expression of $F$. The \emph{OPE} imposes the condition
that $p$ is an integer, whereas the present approach allows it to also
be a half-integer. The possible values of $p$ are predicted by the
formalism. The \emph{power correction} coefficients take the form:
\begin{equation}
\mu_{2p}=C_{1}A_{2p}+C_{2}A'_{2p}+C_{3}A{'}{'}_{2p}.
\end{equation}
The coefficients $C_{i}$ are obtained for a specific process by calculating $\mathcal{F}$
in the pQCD $\mu^{2}/Q^{2}\rightarrow0$ limit.  The  $Q^2$-dependent functions $A_{2p}$,
$A'_{2p}$ and $A{'}{ '}_{2p}$ must be obtained from measurements; 
however, they are universal. 
The  \emph{pinch} technique allows to form a 
gauge-independent formulation of $\alpha_{dmw}$ and to relate it to $\alpha_{gse}$.
This program  has been applied to a large number
of observables in addition to the applications provided in \cite{the:Dokshitzer dispersive approach}.

\section{Background  perturbation theory  \label{sub:Background pert theo, Simonov}}

Given the importance of  nonperturbative effects at low-$Q^{2}$ and the successes of the static potential
approach of  Eq. (\ref{eq:Q-Q stat pot.}), in which short- and long-distance forces are  conveniently separated,
several authors have proposed to formulate QCD with the gluon
field separated as a perturbative part and an effective  nonperturbative part \cite{the:background separation theory}.
Within this framework,  namely background perturbation theory (BPT),  Simonov computed 
the 1-\emph{loop} coupling \cite{the:Simonov 1993} in the \emph{pure field} case.  
Three possible definitions of $\alpha_s$ have been considered, including one  
based on the static potential \cite{the:Simonov 1993}.

The gluon propagator in given by its perturbative  \emph{Green's function} in
a  nonperturbative background characterized by the QCD string tension
$\sigma$. The  nonperturbative background strongly
influence the long-distance behavior of $\alpha_{s}$; it becomes
finite and \emph{freezes} at small $Q^{2}$-values.  In effect, the background
field introduces a mass term such that the argument of $\alpha_s(Q^2)$ becomes $Q^{2}+m_{2g}^{2}$
in the IR; whereas the pQCD form is retained in the UV where the influence of the  nonperturbative
background field is negligible. This replacement is valid at all \emph{loop} orders. 
For  example, the 1-\emph{loop} coupling can
be cast into the same form as first put forth by Cornwall, Eq. (\ref{eq:alpha_s cornwall}):
\begin{equation}
\alpha_{BPT}\left(Q^{2}\right)=\frac{4\pi}{\beta_{0} \, \mbox{ln}\left(\frac{m_{2g}^{2}+Q^{2}}{\Lambda^{2}}\right)}.
\end{equation}
Here, rather than being a $Q^{2}$-dependent effective gluon mass,
$m_{2g}$ is a constant.  However, it depends on the process considered 
~\cite{Badalian 1997, the:Simonov (2011)}: It can be related to  the tension of 
the fundamental QCD string $\sigma_f$: $\sqrt{2\pi\sigma_{f}}\sim1$
GeV~\cite{the:Simonov 1993}, or to  the mass of a two-gluon  bound-state
glueball $M_{2g}(0^{++})$: the tension of the adjoint string
connecting the two gluons, $m_{2g}\simeq\sqrt{2\pi\sigma_{a}}\simeq2$
GeV ~\cite{the:Simonov (2011)}. 

In this formalism the \emph{power corrections} (IR renormalon)
are again folded into the coupling. An interesting insight is that the \emph{freezing} 
value (or equivalently the value of $m_{2g}$) is not universal,  but depends on the embedding
process, as is  it is the case for the \emph{effective charges}; see Sections
\ref{sub:Effective charges and CSR} and
\ref{sub:low Q Effective-charges}. The regulator mass,
$m_{2g}\simeq1-2$ GeV --depending on the process considered--
yields a range of \emph{freezing}  values between 0.37 and 0.60 for $\Lambda_{\overline{MS}}=0.34$
GeV.
The \emph{freezing}  of $\alpha_{BPT}$ is interpreted as due
to the paramagnetic interaction of the gluon spin with the  nonperturbative
background field \cite{the:Simonov (2011)}. 

Badalian and collaborators~\cite{the:Badalian (2000)} have also used the BPT
to compute $\alpha_{BPT}$ in coordinate space. Given this coupling, the perturbative
$1/r$ static $\mbox{Q--}\overline{\mbox{Q}}$ potential can be produced \cite{the:Badalian (2000),
the:Badalian-Kuzmenko 2001}. The calculations are done up
to 3 \emph{loops}, where the values of $\Lambda_{BPT}=0.385$ GeV and
$m_{2g}=1.06$ GeV are obtained from the analysis of the bottomium fine
structure. The corresponding \emph{freezing}  values, given in Table \ref{tab:alpha_s in BPT}, 
depend on the \emph{loop} order and the number of quark flavors.

\begin{table}
\centering
\begin{tabular}{|c|c|c|c|}
\hline 
loop & 1 & 2 & 3\tabularnewline
\hline
\hline 
$n_{f}=1$ & 0.598 & 0.428 & 0.805\tabularnewline
\hline 
$n_{f}=3$ & 0.731 & 0.536 & 0.972\tabularnewline
\hline
\end{tabular}\caption{ \small Values for $\alpha_{BPT}(0)$ in   background perturbation theory
for $\Lambda_{BPT}=0.385$ GeV (taken  independently of $n_{f}$), $m_{2g}\simeq1$
GeV, and various \emph{loop} orders and $n_{f}$ values.\label{tab:alpha_s in BPT}}
\end{table}

\section{Optimized  perturbation theory \label{sub:Optimized-pertubation-theory Low-Q}}

Mattingly and Stevenson have applied OPT (see Section \ref{sub:Optimized-Perturbation-Theory})
at 3$^{rd}$ order to the $e^{+}e^{-}$ total cross-section $R_{e^{+}e^{-}}$ \cite{the:Mattingly-Stevenson}. 
This produces 
an  IR-finite optimized coupling $\alpha_{OPT}$. 
Solving the OPT equations and demanding RS-independence
yield a coupling which  \emph{freezes} below $Q^{2}\simeq0.1$ GeV$^{2}$ at
the value \cite{the:Mattingly-Stevenson, the:Prosperi alpha_s review}:
\begin{equation}
\alpha_{OPT}=\pi\frac{-\beta_{1}+\sqrt{\beta_{1}^{2}-336\beta_{0}^{2}c}}{24\beta_{0}c}=0.826
\end{equation}
with $c=10.911$. This result, which is process dependent, is
obtained at $3^{rd}$ order in the OPT and for $n_{f}=2$. The 4$^{th}$ order result yields $\alpha_{OPT}(0)=0.568$.
(The coupling does not exhibit a IR-fixed point at 2nd order.)

\section{quark--hadron duality \label{sub:Quark-hadron-duality}}

This approach by Courtoy and Liuti~\cite{Courtoy:2013qca},  is based on global quark--hadron 
duality, a phenomenon discovered in 1970 by
Bloom and Gilman~\cite{the:Bloom Gilman duality, the: H-P duality review}.
Bloom and Gilman
observed a remarkable similarity between the unpolarized proton
structure function $F_{2}^{p}$ when it is measured in DIS, and its average when it is measured
in the resonance region: in effect,  the DIS measurement appears as the 
average of the resonance contributions. This matching is interpreted as
limiting the size of the  \emph{higher-twist} contributions to $F_{2}$.
Courtoy and Liuti have noted 
that the large-$x_{Bj} $ resummation contribution
to $F_{2}^{p}$, corrected for  nonperturbative effects, requires the
knowledge of $\alpha_{s}$ in the IR \cite{Courtoy:2013qca}.
Thus  the evolution of $\alpha_{pQCD}(Q^{2})$
near $Q^{2}\simeq1$ GeV$^{2}$ must be regulated in order to explain parton--hadron duality.
The \emph{freezing} behavior of the  coupling is assumed~\cite{Courtoy:2013qca} and its  \emph{freezing}  value is determined 
to be $\alpha_{s}(Q\lesssim1\mbox{ GeV})=0.50\pm0.08$ ($\overline{\mbox{MS}}$
RS).

\section{The IR mapping of $\lambda\phi^{4}$ to Yang--Mills theories\label{sub:IR-mapping to scalar theory}}

The \emph{pure gauge} sector of QCD has been studied by Frasca by mapping it to
the scalar self-interacting $\lambda\phi^{4}$ theory~\cite{the:Frasca}.  Frasca showed that
in the IR limit or the classical limit, the mapping is complete. The
scalar boson propagator is computed, and the running coupling
can be deduced from it. The non-vanishing propagator leads to $\alpha(Q^{2})\rightarrow0$
as $Q^{2}\rightarrow0$. 

\section{The Bogoliubov compensation principle}

In  Ref.~\cite{the:Arbuzov (2013)}, Arbuzov formed a  nonperturbative coupling
by applying the Bogoliubov compensation principle \cite{the:Bogolyubov compensation principle}
to QCD.  One uses SDE constraints to modify the one-\emph{loop} pQCD expression
of $\alpha_{s}$. The first-order approximate solution to this approach
yields a gauge-invariant, RS-independent (1-\emph{loop} expression) coupling
which presents a finite maxima at the \emph{Landau pole} position, and then
vanishes in the deep IR region. The regulation of the \emph{Landau pole}
is due to a three-gluon interaction. 

\section{Curci--Ferrari  model}

Tissier and collaborators have computed the gluon and ghost propagators,
and the quark--gluon vertex with a 1-\emph{loop} model including a gluon mass
term~\cite{the:Tissier}.  They used the Curci--Ferrari model~\cite{the:Curci-Ferrari}
as an effective description. The resulting coupling is finite, vanishes
in the IR, and remains small enough to justify the use of perturbation
theory in the IR domain. When converted to the $\overline{MS}$ RS,
one finds $\alpha_{\overline{MS}}\lesssim 0.5$, except
in the Landau gauge for which the coupling is significantly larger.
The calculations were initially carried in the pure-gauge
sector, with dynamical quarks subsequently introduced.

\chapter{Comparison and discussion \label{sub:alpha_s IR: Comparison-and-discussion}}

\section{Validity of the comparison}

 In this review we have discussed
four distinct types of IR-behavior for the QCD running coupling, $\alpha_s(Q^2)$,  
 using different theoretical frameworks:  
a divergent behavior, typically behaving as $1/Q^2$; a 
 freezing to an IR fixed point; a vanishing coupling $\alpha_s(0) =0$; or IR-finite behavior
with nonzero slope.  In fact, studies which agree qualitatively with one of these
behaviors can still differ quantitatively. For example,  the models 
which predict a \emph{freezing}  of  $\alpha_s$ can disagree on its value by an
order of magnitude. These qualitative and quantitative differences can have
a number of causes. 

We itemize  below the causes that we have identified 
and illustrate them with examples.

\paragraph{Difference in the definitions}

In the perturbative domain, the coupling regulated 
 in any  RS with a massless gluon can be 
equivalently defined from the renormalization of any choice of the gluonic vertex: $\alpha_{s}^{gh} = \alpha_{s}^{3g} = \alpha_{s}^{4g} = \alpha_{s}^{qg}$. 
For example, in Section 
\ref{sub:Renormalization-group},  the coupling was defined from the quark--gluon vertex $\alpha_{s}^{qg}$; see 
Eq. (\ref{eq:z_alpha}). However, in a RS with a massive gluon, the choice of definition ceases to be equivalent (Section \ref{sub:coupling from SDE}).
In addition, these couplings can display significant gauge-dependence at small $Q^2$.

Other definitions of the coupling employ an analytical expression based on pQCD, typically Eq.  (\ref{Eq.one loop}) 
and, in one way or another, supplement it with  nonperturbative
terms while retaining gauge invariance.  The definitions can differ
from each other since different  nonperturbative contributions can
be chosen.  For example, a coupling defined as an effective  charge comprises,  by definition, all the nonperturbative
contributions (including the observable-specific ones) thus making it observable-dependent. In contrast, 
nonperturbative contributions are only partially included  in the
analytic approach, but the coupling remains observable-independent.

We will discuss   in more detail examples of adding  nonperturbative terms in Section \ref{sub:Handling of npert. terms}. 

\paragraph{Differences due to the choice of the renormalization scheme}

The dependence of the QCD coupling on the choice of the RS  can be studied 
in the UV by methods such as Commensurate Scale Relations.
However, the dependence on scheme or effective charge can remain significant in
the IR. This can explain the spread of \emph{freezing}  values seen in the 
literature~\cite{the:alpha_g_1 from AdS, deur RS-dep alpha_s}
as will be discussed in more detail in Section \ref{sub:Influence-of-the RS}.
The RS dependence  has also been investigated
in  Refs.~\cite{the:Grunberg Sudakov charge,the:Shrock scheme-dep studies,the:Grunberg 1992, 
the:Gardi-Karliner 1998}. In Ref. \cite{the:Ryttov gauge-dep}, couplings in the $\overline{MS}$,
$\widetilde{MOM}$ and the modified  regularization invariant schemes were
compared, with the conclusion that for these choices, the scheme-dependence
is moderate.  This is seemingly at odds with the results of
Refs.~\cite{the:alpha_g_1 from AdS, deur RS-dep alpha_s} for effective charges.
The importance of the RS choice --or other arbitrary choices-- on the IR-behavior of $\alpha_{s}$ clearly depends on
the IR-definition of the coupling. For example,  if the coupling is defined
from  an observable, the choice of gauge is irrelevant, whereas it is
important for couplings defined from vertices.

\paragraph{Differences in the predicted value for $\Lambda$ within a given RS}

One  expects that in a given scheme with the same value of $n_{f}$, the value
of $\Lambda$ must be universal;   In practice however, one finds  a spread
of values encountered in publications computing
$\alpha_{s}$ in the IR.

\paragraph{Difference of relativistic forms}

The confining LF harmonic oscillator potential for light quarks in the front form -- 
(Section \ref{sub:Holographic-QCD Lowq}) is equivalent to the nonrelativistic confining
linear potential in the instant form (Section \ref{sub:Potential-approach})
\cite{the:Trawinski 2014}. Thus, even if one includes the same long-distance forces
in the IR definition of $\alpha_{s}$, different 
 relativistic forms and kinematic domains can lead to  different analytic behavior of the running coupling;  {\it e.g.},
the Gaussian shape in the front form for light quarks and the $1/Q^{2}$ behavior
in the instant form for heavy static quarks.  In fact, the exponential form for the coupling obtained from holographic QCD is only valid for light quarks, while the $1/Q^{2}$ behavior
is specific to heavy static quarks.

\paragraph{Difference in gauge choices}

The effect of  the gauge choice for gauge-dependent calculations of the
coupling has been investigated by several authors~\cite{the:Llanes-Estrada 2012,
the:Maas 2009}.  However, there is no consensus on the results.
For example,  studies using linear covariant gauges indicate that the gauge-dependence
is weak for gauges chosen close to the Landau gauge 
\cite{the:Fischer Review SDE in IR, the:Fischer-Zwanziger SDE gauge-dep study, the:Alkofer-Maas-Zwanziger 2010}.
In contrast,  Aguilar and collaborators~\cite{the:Aguilar (2009),the:Aguilar Lattice 2010} 
have argued that the differences between
gauge-dependent and phenomenological (gauge-independent)  calculations of $\alpha_{s}$
can be attributed to the choice of gauge.   For example, in calculations
yielding $\alpha_{s}(0)\simeq 3$ (see Section \ref{list of results}),
the gluon propagator is typically computed in Landau gauge. In 
the phenomenological analyses, which tend
to yield $\alpha_{s}(0)\simeq0.6$, the \emph{Pinch} Technique propagator
(which is in effect similar to the Feynman gauge propagator) is relevant.   The
authors carry out Lattice QCD calculations in the two cases and 
recover the aforementioned differences.  Likewise, the coupling
defined using the Curci--Ferrari model \cite{the:Curci-Ferrari} shows
significant differences when computed in the Landau versus other gauges.
The gauge dependence
of $\alpha_{s}$ has also been discussed in Section \ref{sub:SDE Gauge-dependence}.

\paragraph{Difference due to the choice of solution}

Two solutions can be found for couplings defined from vertices, 
(see Section \ref{sub:Classes-of-solutions in IF domain}),  one leading
to a finite non-zero freezing IR value of $\alpha_s$ (scaling solution), the other
to an IR vanishing of $\alpha_s$ (decoupling solution).
Some authors have focused on one or the other solution.
For example, the authors in Ref.  \cite{the:Llanes-Estrada 2012} have shown that the decoupling
solution has  a smaller value for the action than that of the \emph{scaling solution}. This fact suggests why it is the decoupling 
solution that is more often found in lattice studies. 

\paragraph{Difference in approximations}

Approximations are often necessary to make 
calculations of  $\alpha_{s}$ tractable. Uncontrolled approximations 
will clearly produce different results. For example, in the case of the 
SDE (Section \ref{sub:Schwinger--Dyson-formalism}), integrating out  the
angles, as was done in early calculations, results in larger
\emph{freezing}  values \cite{the:Fisher S-D alpha_s}. For example,  in the pioneering
analysis in Ref. \cite{the:von-Smekal SDE alpha_gh}, one finds that $\alpha_{s}^{gh}$
\emph{freezes} at a value about 3 times higher than indicated by other determinations.

The coupling $\alpha_{s}^{gh}$ is particularly sensitive to the choice of approximations because
it is defined as the product of $Z(Q^{2},\mu)$ and $D(Q^{2},\mu)$,
the gluon and ghost propagators, respectively (see Eq. (\ref{eq:alpha_s SDE ghost--gluon})).
In the \emph{scaling solution}, $Z(Q^{2},\mu)\to 0$ and $D(Q^{2},\mu) \to \infty$
in the IR.  Thus, small differences in the approximations and the truncations employed,  the choice of gauge, and even the level of  numerical precision may lead to large differences
in the \emph{freezing}  value. 

The reliability of SDE truncation schemes was  studied
and then improved in Refs.~\cite{the:Huber SDE truncation} and 
\cite{the:Aguilar 2008}, reaching the conclusion that the standard scheme is reliable for qualitative
estimates of the propagators and the IR coupling. 

An important and
still widely used approximation is to work in the \emph{pure gauge} sector.
To appreciate its effect, one can compare for example  Eqs. (\ref{eq:gluon mass Cornwall})
and (\ref{eq:alpha_s cornwall 2}) or look at Table \ref{tab:alpha_s in BPT} page \pageref{tab:alpha_s in BPT}.
This approximation has been discussed within the SDE framework in Section \ref{sub:SDE: Including quarks}.
Unquenching the lattice calculations yields larger \emph{freezing}  values
of $\alpha_{s}$ by a factor 2, from 0.2 to 0.4 in {\it e.g.}, Ref. \cite{the:Glassner 1996},
and in the static potential approach by 40\%, from $\alpha_{V}\simeq 0.3$
(quenched) to $\alpha_{V}\simeq0.4$ ($n_{f}=2+1$) \cite{the:Bali 1995 et al.}.
In the approach of Badalian\emph{ et al.}  (BPT) in Refs.  \cite{the:Badalian (2000), the:Badalian-Kuzmenko 2001},
the $n_{f}=3$ IR \emph{freezing}  value increases by $\sim 20\%$ compared
to $n_{f}=1$ (here $\Lambda_{V}$ is fixed at $\sim0.4$ GeV). In \cite{the:Ayala 2012},
there is no direct influence from the quark loops, but they still alter
the IR-behavior of $\alpha_{s}$ by affecting the value of the effective
gluon mass. Other groups have argued that the influence of quarks  is
always small (\cite{the:Fischer SDE unquenching, the:Zwanziger Stoch. Quant 2003 }).

In lattice QCD, the discretization and finite volume approximations
may lead to unphysical artifacts, noticeably the IR-suppression of the value of  $\alpha_{s}$.
Furthermore, the choice of the mass of  light quarks has been argued to
be critical to the IR behavior of $\alpha_{s}$~\cite{the:Shirkov alpha_s low_Q review}.

\paragraph{Difference in pQCD order}

In several analyses, the pQCD expression for $\alpha_{s}$, Eq. (\ref{eq:alpha_s}),
has been used to determine the IR-behavior; see Section \ref{sub:Holographic-QCD large Q},
or to establish a modified expression for $\alpha_{s}$,
see for example Eqs. (\ref{eq:Richardson}), (\ref{eq:alpha_s cornwall})
or (\ref{eq:alpha_s from analytic QCD}). In the last case,  a low-order approximation 
is often used. As seen in Fig. \ref{Flo:alpha_s from pQCD}, the magnitude
of the coupling depends on the pQCD order. This dependence
is controlled by the choice of parameters entering the IR expression
for $\alpha_{s}$, {\it e.g.} the gluon effective mass,  or similar scale
parameters (see the enumeration page \pageref{IR mass scales}), the matching
point between the IR and UV domain, and/or the value used for $\Lambda$.

\paragraph{Difference of the medium}

A physical argument given by Brodsky and Shrock  for the origin of the  \emph{freezing}  
of $\alpha_s$~\cite{the:Brodsky & Shrock} suggests that such \emph{freezing} 
within the color confinement domain of hadrons depends on the size of the host hadron: 
the larger the hadron, the smaller the $Q^{2}$-scale at which the \emph{freezing} occurs 
 (since it is due to the long wavelength cut-off above the characteristic size of 
the given hadron) and thus the larger the value of the \emph{frozen} coupling,
since the pQCD domain in which $\alpha_{pQCD}$ grows with distance is larger.   
 Furthermore, in the specific case of $\alpha_V$, 
some medium dependence is expected, in analogy to the Bethe state-dependent 
logarithm  appearing in the Lamb Shift of hydrogenic atoms; see Section \ref{sub:Effective charges and CSR}. 

Medium dependence is also the expectation from the  
background perturbation theory (BPT); see Section \ref{sub:Background pert theo, Simonov}. Other
arguments have been put forth by Titard, Yndurain, and Pineda~\cite{the:Yndurain alpha_s and system size}.
Finally,  the inclusion of observable-dependent nonperturbative effects in the definition of the coupling,
as is the case for \emph{effective charges} 
(Sections \ref{sub:Effective charges and CSR} and \ref{sub:low Q Effective-charges}), 
will naturally introduce medium dependence. However, this dependence is sometimes suppressed as 
illustrated by the \emph{effective charge} $\alpha_{g_1}$ in Section \ref{sub: effective charge g1}.

\paragraph{Difference in group symmetries}

The strong coupling is sometimes computed in color $SU(2$), where it has a
larger value than in color $SU(3)$.  In the \emph{pure gauge} case, the comparison
is straightforward: the coupling scales as $1/N$. This 
comes from $\alpha_{pQCD}$ factoring the IR-definitions of $\alpha_{s}$,
see Eqs. (\ref{eq:alpha_s SDE ghost--gluon}), (\ref{eq:alpha_s 3g vertex})
and (\ref{eq:alpha_s 4g vertex}). For the \emph{pure gauge} case, $\alpha_{pQCD}\propto1/\beta_{0} = 3/(11N)$.
Thus, $3\alpha_{s}^{SU(3)}=N\alpha_{s}^{SU(N)}$.  This dependence
is verified by Lattice calculations \cite{the:Alexandrou (2002), the:Luscher },
the stochastic quantization approach \cite{the:Zwanziger Stoch. quant. 2002},
and the FRG framework \cite{the:Gies 2002 FRG}.

\paragraph{Difference in temperatures}

In this review, we have only considered calculations for zero temperature $T=0$.

\paragraph{Difference in space dimensionality}

Theoretical analyses of gauge couplings  have been performed in 4D, 3D or 2D  in Minkowski or Euclidean
spaces. We have focused on the 4D calculations relevant to QCD, 
and have  ignored a large body of  work done for other numbers
of dimensions, since the basic behavior of the coupling may be 
quite different.  It is important to recall that only in 4D the QCD coupling is dimensionless.

\paragraph{Ignoring Gribov copies}

Choosing a particular solution among the Gribov copies and ignoring
the others does not seem to affect the behavior of $\alpha_{s}$: several
groups using different approaches have concluded that Gribov copies have
little or no effect, see Section \ref{sub:Gribov--Zwanziger-approach}. 

\vspace{20pt}

The above list demonstrates that the different predictions for the QCD coupling in the IR cannot be compared straightforwardly.  Correcting for these differences
to obtain a valid comparison of published results 
goes well beyond the
scope of this review.  However, one can attempt to roughly assess the
effects of these differences to ascertain if a consensus on the IR-behavior
of $\alpha_{s}$ could be in sight.  First, we will discuss  the fact that couplings expressed in different RS must freeze
at different values, and how to account for this difference.

\section{Influence of the renormalization scheme \label{sub:Influence-of-the RS}}

The influence of RS can be quantitatively assessed using the light front holographic QCD approach 
\cite{deur RS-dep alpha_s}. In Section \ref{sub:Holographic-QCD large Q}, 
$\alpha_{pQCD}$ in a given RS is matched to $\alpha_{AdS}$ to determine  the perturbative QCD scale $\Lambda$
and the matching point $Q_{0}$,  which  in turn determines the quark--hadron transition.  Rather than determining $\Lambda$, one can use the known value of
$\Lambda$ and leave the RS-dependent \emph{freezing}  value $\alpha_{s}(0)$
as a free parameter, together with $Q_{0}$. This allows the IR-behavior
of $\alpha_{s}$ to be established in any RS, see Fig. \ref{Flo:different freezings}. The different \emph{freezing} 
values obtained are given in Table \ref{tab:Different-freezing-values AdS/QCD}.

\begin{table}
\centering
\begin{tabular}{|c|c|c|c|}
\hline 
$\alpha_{s}(0)$ & RS & $Q_{0}^{2}$ (GeV) & $\Lambda$ (GeV)\tabularnewline
\hline
\hline 
$1.22\pm0.04\pm0.11\pm0.09$ & $\overline{MS}$ & $0.75\pm0.03\pm0.05\pm0.04$ & $0.34\pm0.02$\tabularnewline
\hline 
$2.30\pm0.03\pm0.28\pm0.21$ & $V$ & $1.00\pm0.00\pm0.07\pm0.06$ & $0.37\pm0.02$\tabularnewline
\hline 
$3.79\pm0.06\pm0.65\pm0.46$ & $MOM$ (L)& $1.32\pm0.02\pm0.10\pm0.08$ & $0.52\pm0.03$\tabularnewline
\hline 
$3.51\pm0.14\pm0.49\pm0.35$ & $g_{1}$ & $1.14\pm0.04\pm0.08\pm0.06$ & $0.92\pm0.05$\tabularnewline
\hline
\end{tabular}\caption{\small \label{tab:Different-freezing-values AdS/QCD}Column 1: values of
$\alpha_{s}(0)$ calculated in different RS (column 2). The $MOM$ results are in the
Landau gauge. The transition
scale $Q^2_{0}$ is given in column 3. Column 4 gives the input values
used for $\Lambda$ for each  RS. The uncertainties on $\alpha_{s}(0)$
and $Q_{0}^{2}$ come (from left to right) from the uncertainty on $\alpha_{pQCD}$,
the uncertainty on the hadronic scale $\kappa$  in LFHQCD,  and an assigned 5\% uncertainty for $\Lambda$. 
\emph{Source:} The table is from Ref. \cite{deur RS-dep alpha_s}}
\end{table}

The $\overline{MS}$ \emph{freezing}  value   obtained from the matching procedure is $\alpha_{\overline{MS}}(0)=1.22\pm0.15$,
twice  as large as than the result obtained from phenomenology  (i.e. mainly from the 
spectroscopic approach, see tables \ref{tab:Summary table 1} and \ref{tab:Summary table 2}) typically
$0.5\pm0.2$.   It is closer to Cornwall's $\overline{MS}$ results
including quarks ($\alpha_{\overline{MS}}(0)=0.81$, Section \ref{sub:SDE: Including quarks}), and
comparable to the results using Shirkov's analytical coupling ($\alpha_{\overline{MS}}(0)=1.25$,
Section \ref{sub:Analytic and Dispersive approaches}), and those of
Gracey's $\overline{MS}$ calculation using the Gribov--Zwanziger approach
($\alpha_{\overline{MS}}(0)=1.70$, Section \ref{sub:Gribov--Zwanziger-approach}).
The value of $Q_{0}$ is consistent with that found by Gomez and Natale~\cite{Gomez 15}.
However, a value of $Q_{0}$ that is smaller  than 1 GeV$ $ is inconsistent with
the analysis of Badalian and collaborators of the charmonium and bottomonium
fine-structure data~\cite{the:Badalian bb and cc analyses}.  Their
coupling at $Q^{2}=1$ GeV$^{2}$, $\alpha_{\overline{MS}}(Q^{2}=1)\simeq0.4$,
is already significantly influenced by  IR effects: it
is about 50\% lower than the pQCD prediction.  However, the analysis
of Badalian and collaborators is done at 2-\emph{loops},  whereas the results reported 
in Table \ref{tab:Different-freezing-values AdS/QCD} are obtained at 4-\emph{loops}.
The $2.84\pm 0.02$ \emph{freezing}  value in the MOM scheme and Landau gauge is in close agreement
with numerous $\alpha_{s}^{gh}\simeq 3$ results obtained in the
MOM scheme and Landau gauge using SDE and lattice gauge theory.

This approximate agreement indicates that different choices of
scheme may explain a significant part of the spread of \emph{freezing}  values
seen in the literature. However, as discussed at the beginning of
this section, other factors must be  considered before a satisfactory comparison is made, 
the main factor being the particular IR-definition
of $\alpha_{s}$.  For example, correcting for the RS  does
not explain the disagreement between the \emph{effective charges} extracted
from the Bjorken sum and from $\tau$-decay: The coupling $\alpha_{\tau}(s)$
extracted from experimental data \cite{the:OPAL R_tau} in  Ref.~\cite{the: Brodsky alpha_tau}
\emph{freezes} around $\alpha_{\tau}(0)\simeq 7$. Correcting to the $g_{1}$
scheme, the \emph{freezing}  value becomes $\alpha_{(\tau)g_{1}}(0)\simeq 6$, roughly
twice the expected value of $\pi$.  This discrepancy may be
due to the unsubtracted pion pole that is part of $\alpha_{\tau}(s)$, which 
by analogy, is analogous to the elastic contribution that must be removed from
$\alpha_{g_1}$ (see Section \ref{sub:low Q Effective-charges}).

\begin{figure}
\centering
\includegraphics[width=9.0cm]{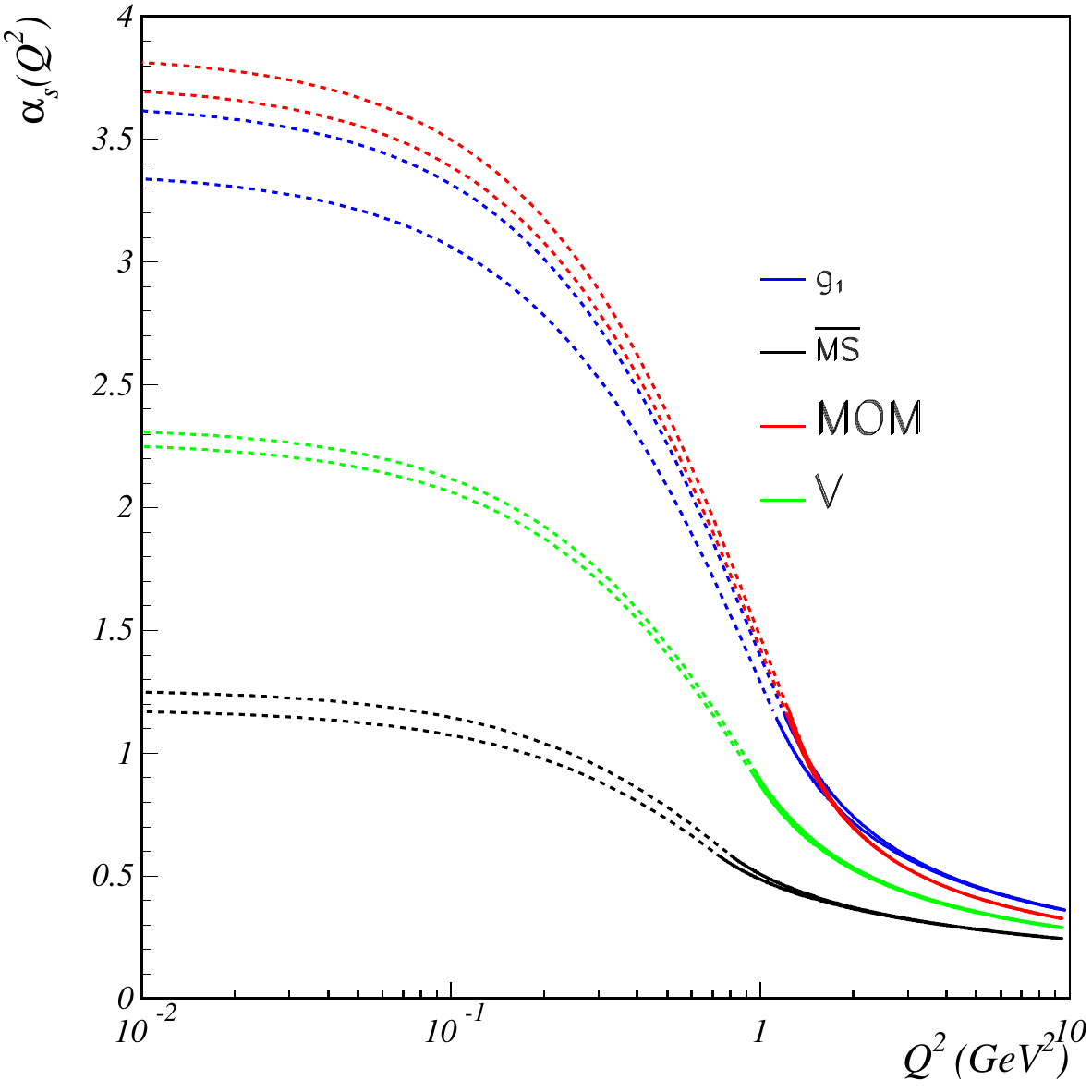}
\caption{\small \label{Flo:different freezings}How different renormalization schemes lead to different
\emph{freezing}  values for $\alpha_{s}$. The black dashed line represents
the AdS/QCD continuation of $\alpha_{pQCD}$ in the $\overline{MS}$
scheme (continuous black line), the blue line is the effective charge
$\alpha_{g_{1}}$ in the $g_{1}$ scheme (without enforcing the $\alpha_{g_{1}}(0)=\pi$
constraint), the green line is the \emph{effective charge} $\alpha_{V}$
in the potential scheme and the red line is $\alpha_{s}$ in the MOM
scheme and Landau gauge. The widths of the curves represent the uncertainty stemming
from the truncation of the pQCD $\beta$-series.}
\end{figure}

\section{The contributions of  nonperturbative terms to $\alpha_s(Q^2)$\label{sub:Handling of npert. terms}}

The QCD running coupling $\alpha_s(Q^2)$ invariably includes dynamics associated with nonperturbative physics.
This is particularly clear when one defines the coupling in terms of an \emph{effective charge} defined from a physical observable.  
In addition, the coupling at large momentum transfer will also depend on the choice of the renormalization scheme.
The reasons behind the RS-dependence of $\alpha_s(Q^2)$ are not apparent in our previous discussion. 
They can be better apprehended if the explanation is recast in  terms of non-perturbative effects. In this 
section we shall show, using two examples, how  \emph{higher-twist} terms remove the artificial pQCD 
divergence (\emph{Landau pole}) and how they shape the behavior of $\alpha_s(Q^2)$ at low $Q^2$. 
We will see in particular how the transition between the IR behaviors of $\alpha_{g_1}$ 
($g_1$ scheme) and $\alpha_{\overline{MS}}$ ($\overline{MS}$ scheme) occurs.

The role of  nonperturbative contributions and  the renormalization scheme dependence can be illustrated
using the \emph{effective charge} $\alpha_{g_{1}}$ \cite{the: Deur early alpha_g_1 proceedings}.
The integral entering the Bjorken sum rule  $\int_{0}^{1}dx_{Bj} [g_{1}^{p}(x_{Bj} ,Q^{2})-g_{1}^{n}(x_{Bj} ,Q^{2})]$
can be expanded in powers of $1/Q^2$ using guidance from the \emph{OPE}: 
\begin{equation}
\int_{0}^{1}dx_{Bj} \biggl(g_{1}^{p}\bigl(x_{Bj} ,Q^{2}\bigr)-g_{1}^{n}\bigl(x_{Bj} ,Q^{2}\bigr)\biggr)=\sum_{i=1}^{\infty}\frac{\mu_{2i}}{Q^{2i-2}}.\label{eq:gamma1 OPE}
\end{equation}
The integration includes the $x_{Bj} =1$ elastic contribution which, by definition, is \emph{higher-twist}.
The leading-twist term is given by Eq. (\ref{eq:Bj SR, order alpha^4}).
We recall it here and adopt the estimate for the $\alpha_{\overline{MS}}^{5}$
coefficient  given by Kataev~\cite{the:Kataev Bj to alpha_5}:
\begin{multline}
\mu_{2}(Q^{2})=\frac{g_{A}}{6}\bigg[1-\frac{\alpha_{\overline{MS}}}{\pi}-3.58\bigg(\frac{\alpha_{\overline{MS}}}{\pi}\bigg)^{2}-20.21\bigg(\frac{\alpha_{\overline{MS}}}{\pi}\bigg)^{3}\label{eq:Bj SR order alpha^5} \\ 
-175.7\bigg(\frac{\alpha_{\overline{MS}}}{\pi}\bigg)^{4}-893.38\bigg(\frac{\alpha_{\overline{MS}}}{\pi}\bigg)^{5}+\mathcal{O}\bigg(\alpha_{\overline{MS}}^{6}\bigg)\bigg] .
\end{multline}
From  Eqs. (\ref{eq:msbar to g_1}), (\ref{eq:gamma1 OPE}) and (\ref{eq:Bj SR order alpha^5}),  one has:
\begin{multline} 
\label{eq:msbar to g_1 with HT cor.}
\frac{\alpha_{g_{1}}(Q^{2})}{\pi}=\frac{\alpha_{\overline{MS}}}{\pi}+3.58\left(\frac{\alpha_{\overline{MS}}}{\pi}\right)^{2}+20.21\left(\frac{\alpha_{\overline{MS}}}{\pi}\right)^{3} \\
+175.7\left(\frac{\alpha_{\overline{MS}}}{\pi}\right)^{4}+893.38\left(\frac{\alpha_{\overline{MS}}}{\pi}\right)^{5}+\mathcal{O}\left(\alpha_{\overline{MS}}^{6}\right)-\frac{6}{g_{A}}\sum_{i=2}\frac{\mu_{2i}}{Q^{2i-2}}. 
\end{multline}
The  nonperturbative terms are represented by the $1/Q^{2i-2}$, $i>2$, \emph{power corrections}; 
{\it i.e.},   \emph{higher-twist} contributions of order $2i$. 
Each \emph{higher-twist} coefficient $\mu_{i}$
can be expressed as a sum of kinematical \emph{twist} terms with a  power smaller than $2i$
as well as a dynamical term of order $2i$. For example, the \emph{twist}-4 coefficient
is \cite{the: Shuryak}:
\begin{equation}\label{eq:mu_4}
\mu_{4}=\frac{M^{2}}{9}\left(a_{2}^{p-n}+4d_{2}^{p-n}+4f_{2}^{p-n}\right),
\end{equation}
where $a_{2}^{p-n}$ is the \emph{twist}-2 target-mass correction given by
the $x_{Bj} ^{2}$-weighted moment of the leading-twist $g_{1}^{p-n,LT}$
structure function: $a_{2}^{p-n}=\int_{0}^{1}{dx_{Bj} \,(x_{Bj} }^{2}g_{1}^{p-n}$).
The \emph{twist}-3 matrix element $d_{2}^{p-n}$ is given by:
\begin{equation}
d_{2}^{LT}=\int_{0}^{1}dx_{Bj} ~x_{Bj} ^{2}\left(2g_{1}^{p-n,LT}+3g_{2}^{p-n,LT}\right),
\end{equation}
where the spin structure function $g_2$ provides the twist-3 contribution. The function
$f_{2}^{p-n}$ is the pure \emph{twist}-4 contribution. The dynamical
 \emph{higher-twist} terms correspond physically to the interactions between the struck quark and
the  nucleon's quark spectators, which by definition, are nonperturbative contributions  associated with the bound-state dynamics.
The  coefficients are modified by powers of  $\log(Q^{2})$
due to DGLAP evolution.

The result of unfolding the \emph{twist}-4
and the higher order pQCD corrections to $\mu_{2}$ in  $\alpha_{g_1}$ is illustrated in
Fig. \ref{Flo:ht effects}. The blue squares give the effective charge
$\alpha_{g_1}(Q^{2})$, Eq. (\ref{eq:alpha_g_1 def.}), 
as shown in Fig. \ref{Flo:lowq alpha. exp. }. The magenta
open circles show  $\alpha_{g_1}^{\text{\scriptsize{$\neg$ RC}}}$, that is the coupling 
obtained after unfolding from $\alpha_{g_1}$ the $\mu_2$ pQCD radiative corrections (RC) up to N$^{4}$LO. 
The black stars are for $\alpha_{g_1}^{\text{\scriptsize{$\neg$ RC, HT}}}$ when the \emph{twist}-4
term is excluded  from $\alpha_{g_1}$ as well. 
In practice the three sets of data shown  in Fig. \ref{Flo:ht effects} are obtained from solving three
 different equations that contains or not the pQCD soft radiation effects and  \emph{higher-twists}: 
 The blue squares are obtained from solving Eq. (\ref{eq:alpha_g_1 def.});
 The magenta circles are obtained from solving Eq. (\ref{eq:Bj SR, order alpha^4}), or 
 equivalently from solving Eq. (\ref{eq:gamma1 OPE}) for $i_{max}=1$;  
 The black stars are obtained from solving Eq. (\ref{eq:gamma1 OPE}) for $i_{max}=2$}.

\begin{figure}
\centering
\includegraphics[width=10.0cm]{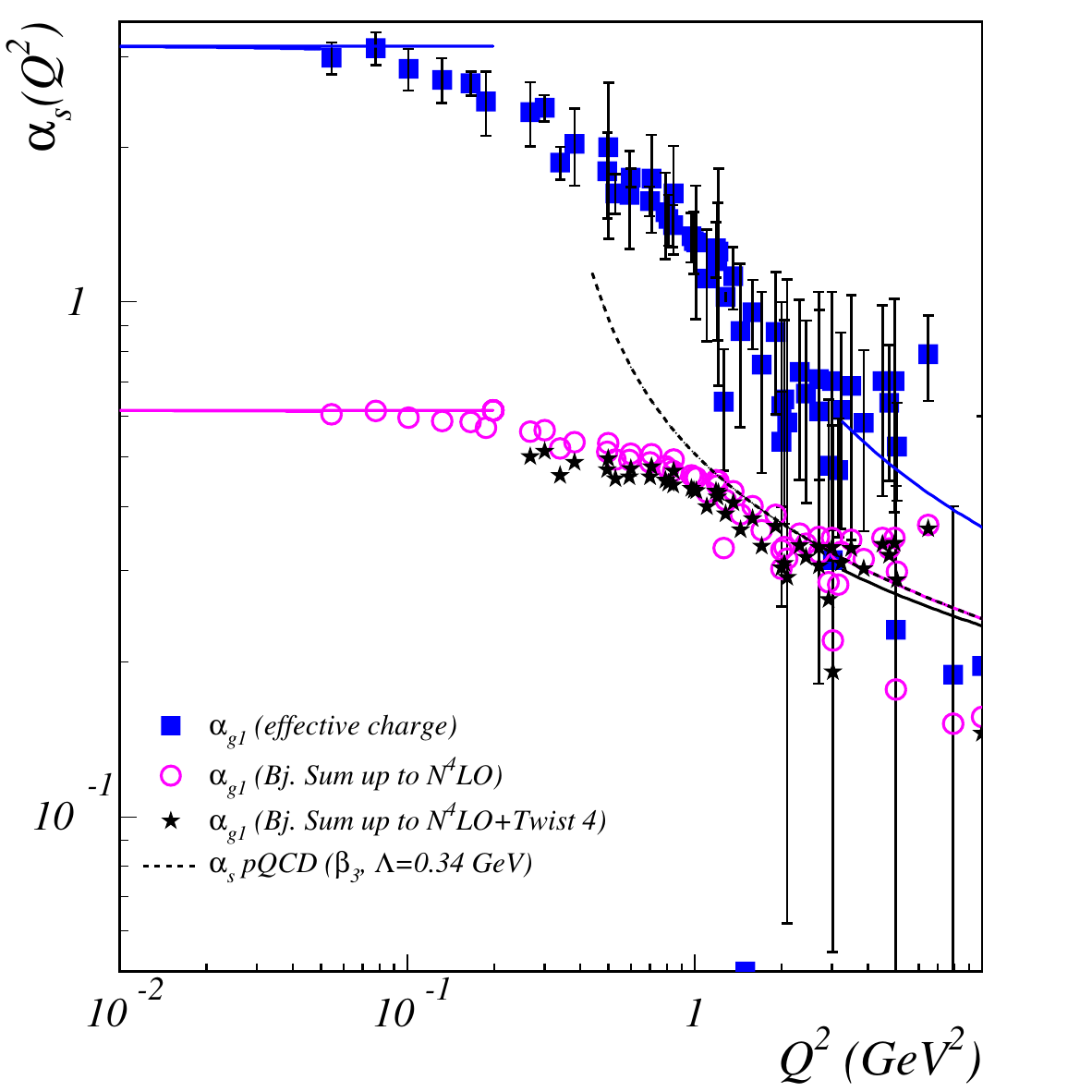}
\caption{\small \label{Flo:ht effects} Experimental data and sum rule constraints
for the \emph{effective charges} $\alpha_{g_{1}}(Q^{2})$ and $\alpha_{F_{3}}(Q^{2})$
(blue squares). The magenta circles show the effect of unfolding from
$\alpha_{g_1}$ the N$^{4}$LO pQCD corrections to $\mu_2$. The black stars
correspond to the case when the \emph{higher-twist}  coefficient $\mu_{4}$  is unfolded as well. The continuous
lines of matching colors are the corresponding predictions from the
GDH (low-$Q^{2}$ domain) and Bjorken (high-$Q^{2}$ domain) sum rules.
The dashed black line is the pQCD expectation calculated up to $\beta_{3}$.
For clarity, we show the uncertainties only for the blue square data.}
\end{figure}

To exclude the twist-4 term, we use the following procedure: A fit to the polarized parton distributions
\cite{BB pdf} can be used to determine the \emph{twist}-2 $a_{2}^{p-n}$ term in Eq. \ref{eq:mu_4}:
At $Q^{2}$= 1 GeV$^{2}$, $a_{2}^{p-n}=0.031$. The $d_{2}^{p-n}$
matrix element is obtained from  Refs.~\cite{the:E155/E155x} and \cite{d2n}
and evolved from $Q^{2}$= 5 GeV$^{2}$ to 1 GeV$^{2}$. This yields
$d_{2}^{p-n}=-0.008$.  The pure \emph{twist}-4 term $\alpha_{\overline{MS}}(Q^{2})$
is  known experimentally~\cite{Deur alpha_s from eg_1dvcs, the:JLab  Bj SR}
based on the use of the $\alpha_{\overline{MS}}(Q^{2})$ scheme.
However, for a proper
investigation of the role of  \emph{higher-twist} contributions in $\alpha_{g_{1}}(Q^{2})$, the value of
$f_{2}^{p-n}$ must be obtained independently of $\alpha_{\overline{MS}}(Q^{2})$.
(If one uses the experimentally extracted  value of $f_2^{p-n}$,  one would have 
by construction $\alpha_{g_{1}}^{\text{\scriptsize{$\neg$ RC, HT}}}(Q^{2})=\alpha_{\overline{MS}}(Q^{2})$.)
Theoretical calculations of $f_{2}$ do exist~\cite{The HT calculations} giving
$f_{2}^{p-n}=-0.03$ after subtracting an elastic contribution 
to $f_2$ of -0.02.   The \emph{power corrections} reintroduce
divergent $1/Q^{2n}$ terms.  It is normally acceptable to use the \emph{OPE} formalism down
to $Q^{2}\sim1$ GeV$^{2}$. 
In fact, the \emph{higher-twist} series may still converge at lower $Q^{2}$: 
For example, at $Q^2 =0.5$ GeV$^{2}$ the \emph{higher-twist} coefficient $\mu_{4}^{2}/Q^{2}=0.3$ is well below unity.

The continuous lines in Fig. \ref {Flo:ht effects} are the predictions from
the GDH sum rule at low-$Q^{2}$,  Eqs. (\ref{eq:GDH slope for alpha_g_1})
and (\ref{eq:alpha_g_1 IR value}), and from the Bjorken sum rule at high-$Q^{2}$,
Eq. (\ref{eq:msbar to g_1}). Once the pQCD and \emph{higher-twist} corrections
are unfolded from $\alpha_{g_1}$, the coupling  $\alpha_{g_{1}}^{\text{\scriptsize{$\neg$ RC, HT}}}$ agrees well at large
$Q^{2}$ with the pQCD expectation given by the dashed line.
At low $Q^{2}$, the \emph{freezing}  value of  $\alpha_{g_{1}}^{\text{\scriptsize{$\neg$ RC, HT}}}$
 is now about $0.6$, closer to the value found in many 
$\overline{MS}$ determinations of $\alpha_{s}$; 
see the list in the next Section and see  also Fig. \ref{Flo:plot all IR alphas}.

We note that accounting for the  \emph{twist}-4 contributions lower the \emph{freezing}  value from
0.62 to about 0.5. This is because $\mu_{4}<0$.   Experimental studies
indicate that $\mu_{6}\sim0$ and $\mu_{8}\sim-\mu_{4}>0$ \cite{Deur alpha_s from eg_1dvcs, the:JLab  Bj SR}.  
Consequently, the
addition of more \emph{higher-twist} contributions would increase the \emph{freezing} 
value, possibly close to the 1.2 value seen in Fig. \ref{Flo:different freezings}.
In addition to revealing the mechanisms behind the RS-dependence of 
$\alpha_s(0)$, namely here how one makes the transition between $\alpha_{g_1}$ and $\alpha_{\overline{MS}}$, this analysis also
shows that the failure  to account for nonperturbative
effects, behaving as $1/Q^{n}$ at intermediate $Q^2$, can lead to a \emph{Landau pole}.

As an alternative to \emph{effective charges}, we can use 
Shirkov's  analytic coupling  to study the effect of {\emph{higher-twists}. We recall
here its LO form:
\begin{equation}
\label{OPE exp of A}
\alpha_{an}^{(0)}\left(Q^{2}\right)=\frac{4\pi}{\beta_{0}}\left(\frac{1}{\mbox{ln}\left(Q^{2}/\Lambda^{2}\right)}+\frac{\Lambda^{2}}{\Lambda^{2}-Q^{2}}\right).
\end{equation}
For $ Q^{2} \gg \Lambda^{2}$, the domain where \emph{OPE} is valid, the contributions 
beyond the leading-power pQCD contributions can be expanded using $\Lambda^{2}/(\Lambda^{2}-Q^{2})=-\sum_{n=1}(\Lambda^2/Q^2)^{n}$,
in apparent agreement with the form of the \emph{power corrections}  expected from the \emph{OPE}.  We
note the negative sign which compensates for the positively diverging Landau
pole. Taking the \emph{OPE} expansion of some observable:
\begin{equation}
A\left(Q^{2}\right)=\sum_{n=0}a_{n}\alpha_{pQCD}^{n}(Q^{2})+\sum_{i=1}\frac{b_{i}(Q^{2})}{Q^{2i}}\label{eq:pQCD OPE series},
\end{equation}
and expressing it using Eq. (\ref{OPE exp of A}) {\it i.e.}, with \emph{power corrections} folded in the coupling:
\begin{equation}
A\left(Q^{2}\right)=\sum_{n=0}a'_{n}\alpha_{an}^{n}(Q^{2}),
\end{equation}
one has $a_n=a'_n$, $b_{1}=-\Lambda^{2}(a_{1}+2a_{2}\alpha_{pQCD}+ \cdots )4 \pi/\beta_0$,
$b_{2}=-\Lambda^{4}(a_{1}+a_{2}(2\alpha_{pQCD}-4 \pi/\beta_0+ \cdots)4 \pi/\beta_0, \cdots$ .
The results obtained  for the $b_{i}$
are observable-dependent (through the coefficients $a_ n$) and
have a $Q^{2}$ logarithmic dependence in the UV (through $\alpha_{pQCD}$),
as expected. However, the  $Q^2$-dependent functions $b_i$ do not include all  nonperturbative
effects. For example, $a_{n}$ and $\alpha_{pQCD}$ 
include only short-distance phenomena (gluon bremsstrahlung, vacuum polarization) whereas
some of the \emph{higher-twist} contributions represent the transverse, long-distance force responsible
for confinement \cite{the:Burkardt twists}.  Nevertheless, using
$\alpha_{an}$, rather than $\alpha_{pQCD}$, identifies     in
(\ref{eq:pQCD OPE series}) a well-defined  nonperturbative contribution
which should increase the validity range of \emph{OPE} down to
values of $Q^{2}$ close to $\Lambda^{2}$. Such an approach is discussed in more
depth in Ref.~\cite{the:Cvetic 2008-1}.

\section{Listing of the  multiple IR-behavior found in the literature \label{list of results}}

In order to facilitate the comparison between the  multiple determinations  which describe the IR-behavior
of $\alpha_{s}$ cited  in this review, we summarize the  main results in Tables \ref{tab:Summary table 1}
and \ref{tab:Summary table 2}. The corresponding predictions for  $\alpha_s(0)$ are also plotted in 
Fig. \ref{Flo:plot all IR alphas},  where vertical arrows indicate divergent couplings.
 (We do not include in  Fig. \ref{Flo:plot all IR alphas} the results from couplings defined from the 3-gluon and
4-gluon vertices.)  This figure does not correct for differences in RS, gauges or
definitions. Consequently, we are not necessarily comparing the
same quantities.  Furthermore,  we have plotted some pioneering results, such as the first
SDE determination of $\alpha_{s}^{gh}$ by von Smekal {\it et.~al.}~\cite{the:von-Smekal SDE alpha_gh}, now 
superseded by more accurate calculations. 

\begin{figure}
\centering
\includegraphics[width=10cm]{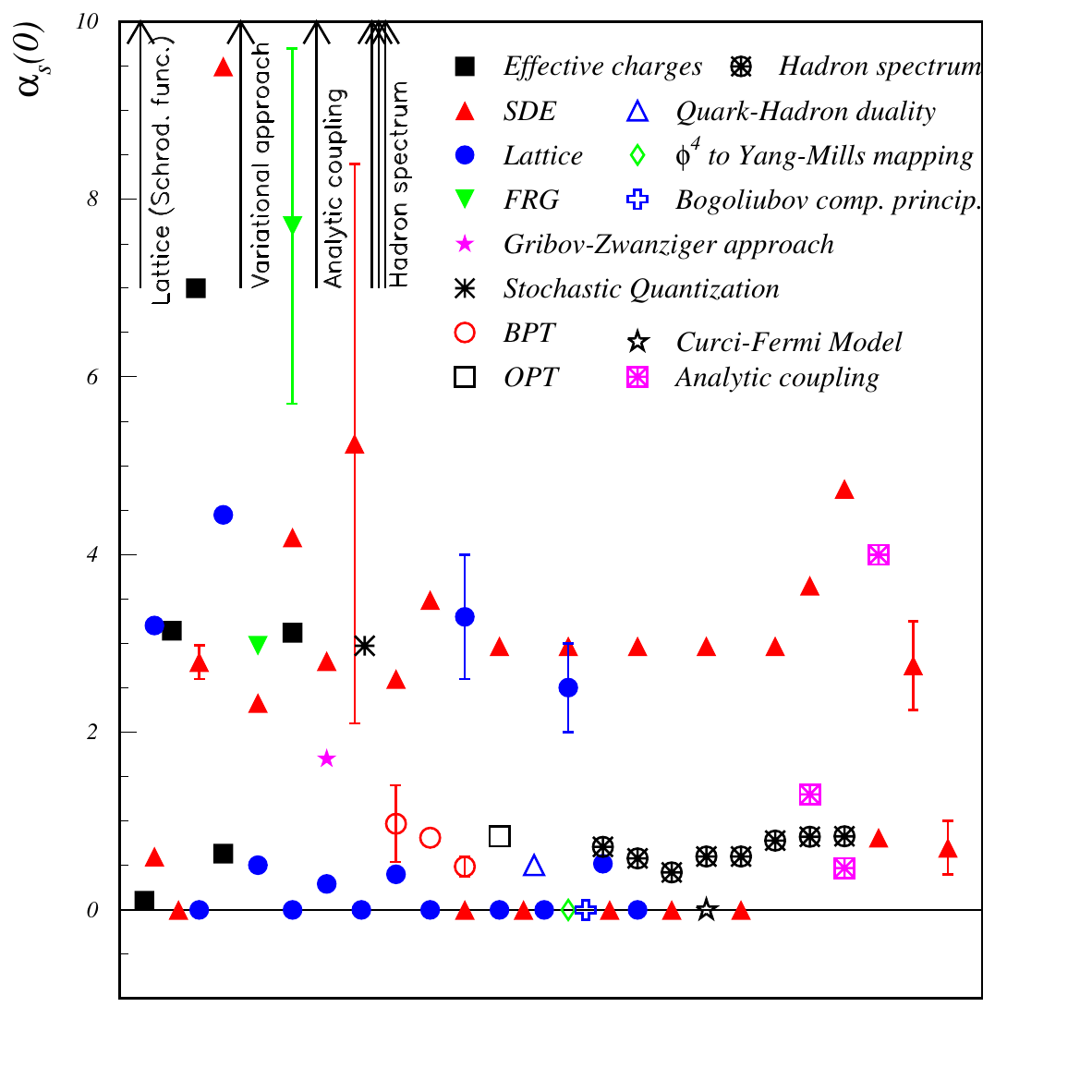}
\caption{\small \label{Flo:plot all IR alphas}  Values of $\alpha_{s}(0$) from the
literature and reported in this review. The differences in the infrared 
fixed-point value of the coupling can be due to choices of RS, gauge,
relativistic form, truncations, approximations, model dependence 
 and other points, as indicated and discussed  in the main text.
The vertical arrows indicate IR-divergent couplings.}
\end{figure}

Four main groups of results can
be identified:
\begin{itemize}
\item $\alpha_{s}(Q^{2})~\protect\overrightarrow{_{Q^{2}\to0}}~\infty$: static $\mbox{Q--}\overline{\mbox{Q}}$ potential with the string tension
folded in $\alpha_{s}$, and various  nonperturbative models.
\item $\alpha_{s}(Q^{2})\sim3$: $\alpha_{s}^{gh}(0)$ for the \emph{scaling solution} (SDE,   lattice, FRG,
  stochastic quantization), and the \emph{effective charges} $\alpha_{g_{1}}(0)$
and $\alpha_{F3}(0)$. 
\item $\alpha_{s}(Q^{2})\sim0.7$: quark models and the static $\mbox{Q--}\overline{\mbox{Q}}$ potential
without the string tension folded in $\alpha_{s}$.
\item $\alpha_{s}(Q^{2})=0$: \emph{decoupling solution} (SDE,   lattice, FRG) and various  nonperturbative
models.
\end{itemize}

\begin{table}
\vskip-1.5cm \hskip-1.5cm\begin{tabular}{|c|c|c|c|c|c|c|c|}
\hline 
{\footnotesize $\alpha_{s}(0)$} & {\footnotesize IR behavior} & {\footnotesize Definition} & {\footnotesize RS} & {\footnotesize Gauge} & {\footnotesize Quarks} & {\footnotesize Framework} & {\footnotesize References}\tabularnewline
\hline
\hline 
{\footnotesize 0.10} & {\footnotesize mono. decr.} & {\footnotesize Sudakov eff. charge} & {\footnotesize $V$} & {\footnotesize Indep.} & {\footnotesize yes} & {\footnotesize Sudakov resum.} & 
{\footnotesize \cite{the:Grunberg Sudakov charge, the:Grunberg Sudakov charge 2}}\tabularnewline
\hline 
{\footnotesize $\pi$} & {\footnotesize freezes} & {\footnotesize eff. charge} & {\footnotesize $g_{1}$} & {\footnotesize Indep.} & {\footnotesize yes} & {\footnotesize Measurements} & 
{\footnotesize \cite{the:Deur lowq alpha_s}}\tabularnewline
\hline 
{\footnotesize 3.12} & {\footnotesize freezes} & {\footnotesize eff. charge} & {\footnotesize $g_{1}$} & {\footnotesize Indep.} & {\footnotesize yes} & {\footnotesize AdS/QCD} & 
{\footnotesize \cite{the:alpha_g_1 from AdS, the:Deur Lambda AdS/QCD}}\tabularnewline
\hline 
{\footnotesize $\sim7$} & {\footnotesize freezes} & {\footnotesize eff. charge} & {\footnotesize $\tau$} & {\footnotesize Indep.} & {\footnotesize yes} & {\footnotesize Measurements} & 
{\footnotesize \cite{the: Brodsky alpha_tau}}\tabularnewline
\hline 
{\footnotesize 0.63} & {\footnotesize freezes} & {\footnotesize dressed gl. exch.} & {\footnotesize{} $\sim V$} & {\footnotesize Indep.} & {\footnotesize yes} & {\footnotesize Disp. approach} & 
{\footnotesize \cite{the:Dokshitzer dispersive approach, the:Dokshitzer/webber: HT in event shapes.} }\tabularnewline
\hline 
{\footnotesize 0.60} & {\footnotesize freezes} & {\footnotesize gl. self-energy } & 
{\footnotesize ($\sim\overline{MS}$)} & {\footnotesize Indep.} & {\footnotesize no} & {\footnotesize SDE}{\footnotesize \par}
{\footnotesize $\Lambda$=0.26 } & 
{\footnotesize \cite{the:Aguilar SDE 2001}}\tabularnewline
\hline 
{\footnotesize 0} & {\footnotesize vanishes} & {\footnotesize gl. self-energy } & {\footnotesize ($\overline{MS}$, MOM)} & {\footnotesize Landau} & {\footnotesize no} & {\footnotesize SDE}{\footnotesize \par}
{\footnotesize 0.3$\leq$$\Lambda$$\leq$0.8} & 
{\footnotesize \cite{the:Aguilar-Natale SDE 2005}}\tabularnewline
\hline 
{\footnotesize $0.7\pm0.3$} & {\footnotesize freezes} & {\footnotesize gl. self-e., gh.-gl. vert. } & {\footnotesize MOM} & {\footnotesize Landau} & {\footnotesize no} & {\footnotesize SDE  $\Lambda$=0.3 } & 
{\footnotesize \cite{the:Aguilar (2009)}}\tabularnewline
\hline 
{\footnotesize -} & {\footnotesize freezes} & {\footnotesize gl. self-energy } & 
{\footnotesize ($\sim\overline{MS}$)} & {\footnotesize Feynman} & {\footnotesize no} & {\footnotesize SDE }{\footnotesize \par}
{\footnotesize $\Lambda$=0.3 } & 
{\footnotesize \cite{the:Aguilar Papavassiliou SDE (2006)}}\tabularnewline
\hline 
{\footnotesize 2.6-2.97} & {\footnotesize freezes} & {\footnotesize gh.-gl. vertex } & {\footnotesize MOM} & {\footnotesize Landau} & {\footnotesize no} & {\footnotesize SDE} &
{\footnotesize \cite{the:Alkofer 2003} }\tabularnewline
\hline 
{\footnotesize -} & {\footnotesize freezes} & {\footnotesize 3- and 4-gl. vertex}{\footnotesize \par} & {\footnotesize MOM} & {\footnotesize Landau} & {\footnotesize no} & {\footnotesize SDE} & 
{\footnotesize \cite{the:Alkofer Definition of couplings in SDE}}\tabularnewline
\hline 
{\footnotesize Diverges} & {\footnotesize $1/Q^{2}$} & {\footnotesize quark--gl. vertex} & {\footnotesize MOM} & {\footnotesize Landau} & {\footnotesize no} & {\footnotesize SDE} & 
{\footnotesize \cite{the:Alkofer SDE & chiral sym}}\tabularnewline
\hline 
{\footnotesize $\sim2.5$} & {\footnotesize freezes} & {\footnotesize quark--gl. vertex} & {\footnotesize MOM} & {\footnotesize Landau} & {\footnotesize yes} & {\footnotesize SDE} & 
{\footnotesize \cite{the:Alkofer SDE & chiral sym}}\tabularnewline
\hline 
{\footnotesize 2.33} & {\footnotesize freezes} & {\footnotesize gh.-gl. vertex } & {\footnotesize -} & {\footnotesize Coulomb} & {\footnotesize no} & {\footnotesize SDE} & 
{\footnotesize \cite{the:Alkofer-Maas-Zwanziger 2010}}\tabularnewline
\hline 
{\footnotesize 4.19} & {\footnotesize freezes} & {\footnotesize gh.-gl. vertex } & {\footnotesize ($\sim\overline{MS}$)} & {\footnotesize Landau} & {\footnotesize no} & {\footnotesize SDE}{\footnotesize \par}
{\footnotesize $\Lambda$=0.153 } & 
\footnotesize{\cite{the:Bloch S-D alpha_s 1998}}\tabularnewline
\hline 
{\footnotesize $\sim2.8$} & {\footnotesize freezes} & {\footnotesize Int. strength} & {\footnotesize $\overline{MS}$} & {\footnotesize Indep.} & {\footnotesize no} & {\footnotesize SDE+Lattice} &
\footnotesize {\cite{the:Binosi (2015)}}\tabularnewline
\hline 
{\footnotesize 2.1 to 8.4} & {\footnotesize freezes} & {\footnotesize gh.-gl. vertex } & {\footnotesize $\overline{MS}$?} & {\footnotesize Landau} & {\footnotesize no} & {\footnotesize SDE} & 
{\footnotesize \cite{the:Bloch (2001)}}\tabularnewline
\hline 
{\footnotesize 2.6} & {\footnotesize freezes} & {\footnotesize gh.-gl. vertex } & {\footnotesize $\overline{MS}$} & {\footnotesize Landau} & {\footnotesize yes} & {\footnotesize SDE} & {\cite{the:Bloch S-D alpha_s 2002}}\tabularnewline
\hline 
{\footnotesize 3.49} & {\footnotesize freezes} & {\footnotesize gh.-gl. vertex } & {\footnotesize ($g_{1}$)}{\footnotesize \par}
 & {\footnotesize Landau} & {\footnotesize no} & {\footnotesize SDE}{\footnotesize \par}
{\footnotesize $\Lambda$=0.856 } & {\footnotesize \cite{the:Bloch S-D alpha_s 2003}}\tabularnewline
\hline 
{\footnotesize 0} & 
{\footnotesize vanishes} & {\footnotesize gh.-gl., 3-gl. vert.} & {\footnotesize MOM} & {\footnotesize Landau} & {\footnotesize no} & {\footnotesize SDE} & 
{\footnotesize \cite{the:Blum SDE 3g vertex}}\tabularnewline
\hline 
{\footnotesize $0.81$} & {\footnotesize freezes} & {\footnotesize gl. self-energy } &
{\footnotesize ($\sim\overline{MS}$)} & {\footnotesize Light cone} & {\footnotesize yes} & {\footnotesize SDE $\Lambda$=0.26} & {\footnotesize  \cite{the:Cornwall alpha_s, theCornwall alpha_s 2}}\tabularnewline
\hline
{\footnotesize 2.97}{\footnotesize \par} & {\footnotesize freezes } & {\footnotesize gh.-gl.,}{\footnotesize \par} & {\footnotesize MOM} & {\footnotesize Landau} & {\footnotesize no} & {\footnotesize SDE} & 
{\footnotesize \cite{the:Cyrol 4g alpha (2014)}}\tabularnewline
\hline
{\footnotesize $3.2\times10^{-3}$}{\footnotesize \par} & {\footnotesize freezes } & {\footnotesize 3-gl. vertex}{\footnotesize \par}
 & {\footnotesize MOM} & {\footnotesize Landau} & {\footnotesize no} & {\footnotesize SDE} & 
 {\footnotesize \cite{the:Cyrol 4g alpha (2014)}}\tabularnewline
\hline
{\footnotesize $4.2\times10^{-2}$} & {\footnotesize freezes } &
{\footnotesize 4-gl. vertex} & {\footnotesize MOM} & {\footnotesize Landau} & {\footnotesize no} & {\footnotesize SDE} &
{\footnotesize \cite{the:Cyrol 4g alpha (2014)}}\tabularnewline
\hline 
{\footnotesize 0} & {\footnotesize vanishes} & {\footnotesize gh.-gl., 3-gl., 4-gl.} & {\footnotesize MOM} & {\footnotesize Landau} & {\footnotesize no} & {\footnotesize SDE} & 
{\footnotesize \cite{the:Cyrol 4g alpha (2014)}}\tabularnewline
\hline 
{\footnotesize 2.97}{\footnotesize \par} & {\footnotesize freezes} & {\footnotesize gh.-gl. vertex } & {\footnotesize MOM} & {\footnotesize Landau} & {\footnotesize no} & {\footnotesize SDE} & 
{\footnotesize \cite{the:Eichmann SDE 3g vertex}}\tabularnewline
\hline 
{\footnotesize 0} & 
{\footnotesize vanishes} & {\footnotesize gh.-gl. vertex } & {\footnotesize MOM} & {\footnotesize Landau} & {\footnotesize no} & {\footnotesize SDE} &
{\footnotesize \cite{the:Eichmann SDE 3g vertex}}\tabularnewline
\hline 
{\footnotesize $1.6\times10^{-3}$}{\footnotesize \par} & {\footnotesize freezes} & {\footnotesize 3-gl. vertex} & {\footnotesize MOM} & {\footnotesize Landau} & {\footnotesize no} & {\footnotesize SDE} & 
{\footnotesize \cite{the:Eichmann SDE 3g vertex}}\tabularnewline
\hline 
{\footnotesize 0} &{\footnotesize vanishes} & {\footnotesize 3-gl. vertex} & {\footnotesize MOM} & {\footnotesize Landau} & {\footnotesize no} & {\footnotesize SDE} & 
{\footnotesize \cite{the:Eichmann SDE 3g vertex}}\tabularnewline
\hline 
{\footnotesize 2.97} & {\footnotesize freezes} & {\footnotesize gh.-gl. vertex } & {\footnotesize MOM} & {\footnotesize Landau} & {\footnotesize no} & {\footnotesize SDE} & 
{\footnotesize \cite{the:Fisher S-D alpha_s}}\tabularnewline
\hline 
{\footnotesize 0} & {\footnotesize vanishes} & {\footnotesize gh.-gl., 3-gl.} & {\footnotesize MOM} & {\footnotesize Lin. covar.} & {\footnotesize no} & {\footnotesize SDE} & {\footnotesize  \cite{the:Huber SDE 2015}}\tabularnewline
\hline 
{\footnotesize 2.97} & {\footnotesize freezes } & {\footnotesize gh.-gl.} & {\footnotesize MOM} & {\footnotesize Landau} & {\footnotesize no} & {\footnotesize SDE} &
{\footnotesize \cite{the:Kellermann-Fisher alpha_4g (2008)}}\tabularnewline
\hline 
{\footnotesize $2.77\times10^{-3}$} & {\footnotesize freezes } & {\footnotesize 4-gl. vertex} & {\footnotesize MOM} & {\footnotesize Landau} & {\footnotesize no} & {\footnotesize SDE} & 
{\footnotesize \cite{the:Kellermann-Fisher alpha_4g (2008)}}\tabularnewline
\hline 
{\footnotesize 3.65} & {\footnotesize freezes} & {\footnotesize gh.-gl. vertex } & {\footnotesize -} & {\footnotesize Landau} & {\footnotesize no} & {\footnotesize SDE} &
{\footnotesize \cite{the:Schleifenbaum 2006}}\tabularnewline
\hline 
{\footnotesize 4.74} & {\footnotesize freezes} & {\footnotesize gh.-gl. vertex } & {\footnotesize -} & {\footnotesize Coulomb} & {\footnotesize no} & {\footnotesize SDE} & 
{\footnotesize \cite{the:Schleifenbaum 2006}}\tabularnewline
\hline 
{\footnotesize 9.5} & {\footnotesize freezes} & {\footnotesize gh.-gl. vertex } & {\footnotesize MOM} & {\footnotesize Landau} & {\footnotesize no} & {\footnotesize SDE} & 
{\footnotesize \cite{the:von-Smekal SDE alpha_gh}}\tabularnewline
\hline 
\end{tabular}
\caption{\small Summary of the predictions for the IR behavior of $\alpha_{s}(Q^{2})$  
discussed in this review. As emphasized in the text, the 
results depend on  the different definitions of the running coupling.
The label {}``indep.'' denotes predictions which are  gauge- or RS-independent.
The expressions for  $\alpha_{s}$ are formally scheme-independent when expressed
at LO or NLO, but they remain numerically dependent on the scheme choice through
the value used for $\Lambda$. To convey this dependence, we have identified within 
parenthesis the choice of scheme corresponding to the $\Lambda$
value given in the 7th column. The sign {}``--'' indicates that information
is either unclear or unknown. 
All results are for $SU(3)_C$ in 3+1--dimensional space-time. \label{tab:Summary table 1}}
\end{table}
 \begin{table}
\hskip-2.cm\begin{tabular}{|c|c|c|c|c|c|c|c|}
\hline
{\footnotesize $\alpha_{s}(0)$} & {\footnotesize IR behavior} & {\footnotesize Definition} & {\footnotesize RS} & {\footnotesize Gauge} & {\footnotesize Quarks} & {\footnotesize Framework} & {\footnotesize References}\tabularnewline
\hline
\hline 
{\footnotesize 3.2} & {\footnotesize freezes} & {\footnotesize gh.-gl. vertex } & {\footnotesize MOM} & {\footnotesize Landau} & {\footnotesize yes} & {\footnotesize Lattice} & 
{\footnotesize \cite{the:Ayala 2012}}\tabularnewline
\hline 
{\footnotesize 4.45 } & {\footnotesize freezes} & {\footnotesize gh.-gl. vertex } & {\footnotesize MOM} & {\footnotesize Landau} & {\footnotesize no} & {\footnotesize Lattice} & 
{\footnotesize \cite{the:Aguilar Lattice 2010}}\tabularnewline
\hline 
{\footnotesize 0.5} & {\footnotesize freezes} & {\footnotesize pQCD form} & {\footnotesize MOM} & {\footnotesize Indep.} & {\footnotesize no} & {\footnotesize Lattice} & 
{\footnotesize \cite{the:Aguilar Lattice 2010}}\tabularnewline
\hline 
{\footnotesize $0.29\pm0.03$} & {\footnotesize freezes} & {\footnotesize Q--Q pot.} & {\footnotesize $V$} & {\footnotesize Indep.} & {\footnotesize no} & {\footnotesize Lattice} &
{\footnotesize \cite{the:Bali 1995 et al.}}\tabularnewline
\hline 
{\footnotesize 0.40} & {\footnotesize freezes} & {\footnotesize Q--Q pot.} & {\footnotesize $V$} & {\footnotesize Indep.} & {\footnotesize yes} & {\footnotesize Lattice} & 
{\footnotesize \par}{\footnotesize \cite{the:Bali 1995 et al.}}\tabularnewline
\hline 
{\footnotesize $3.3\pm0.7$} & {\footnotesize freezes} & {\footnotesize gh.-gl. vertex } & {\footnotesize ($\tau$)} & {\footnotesize Landau} & {\footnotesize no} & {\footnotesize Lattice}{\footnotesize \par}
{\footnotesize $\Lambda$=1.1} &{\footnotesize  \cite{the:Bloch lattice SU2 (2004)}}\tabularnewline
\hline 
{\footnotesize 0} & {\footnotesize vanishes} & {\footnotesize gh.-gl. vertex } & {\footnotesize MOM} & {\footnotesize Landau} & {\footnotesize no} & {\footnotesize Lattice} & {\footnotesize see caption}\tabularnewline
\hline 
{\footnotesize 0} & {\footnotesize vanishes} & {\footnotesize gh.-gl. vertex } & {\footnotesize MOM} & {\footnotesize Landau} & {\footnotesize yes} & {\footnotesize Lattice} & 
{\footnotesize \cite{the:Boucaud decoupling sol. discovery, the:Boucaud SDE massive gluon prop.}}\tabularnewline
\hline 
{\footnotesize $2.5\pm0.5$} & {\footnotesize freezes} & {\footnotesize gh.-gl.,  q.-gl. v.} & {\footnotesize MOM} & {\footnotesize Land. Coul.} & {\footnotesize yes} & {\footnotesize Lattice} & 
{\footnotesize \cite{the:Furui Nakajima}}\tabularnewline
\hline 
{\footnotesize 0.52} & {\footnotesize freezes} & {\footnotesize Q--Q pot.} & {\footnotesize $V$} & {\footnotesize Indep.} & {\footnotesize yes} & {\footnotesize Lat. pert. stoch. th.} &
{\footnotesize \cite{the:Horsley 2014}}\tabularnewline
\hline 
{\footnotesize diverges} & {\footnotesize $e^{m/Q}$} & {\footnotesize Schrod. func.} & {\footnotesize $\overline{MS}$} & {\footnotesize -} & {\footnotesize yes} & {\footnotesize Lattice} & 
{\footnotesize \cite{the:Luscher }}\tabularnewline
\hline
{\footnotesize 2.97} & {\footnotesize freezes} & {\footnotesize gh.-gl. vertex } & {\footnotesize -} & {\footnotesize indep.} & {\footnotesize yes} & {\footnotesize Stoch. Quant.} & 
{\footnotesize \cite{the:Zwanziger Stoch. quant. 2002, the:Zwanziger Stoch. Quant 2003 }}\tabularnewline
\hline 
{\footnotesize $\infty$} & {\footnotesize $1/Q^{2}$} & {\footnotesize pQCD form} & {\footnotesize $V$} & {\footnotesize Coulomb} & {\footnotesize yes} & {\footnotesize Variational ap.} & 
{\footnotesize \cite{the:Szczepaniak 2001}}\tabularnewline
\hline 
{\footnotesize 0.71} & {\footnotesize freezes} & {\footnotesize pQCD form} & {\footnotesize $\overline{MS}$} & {\footnotesize Indep.} & {\footnotesize yes} & {\footnotesize Pheno.} & {\footnotesize  \cite{the:Andreev 2011-2013} }\tabularnewline
\hline 
{\footnotesize $\infty$} & {\footnotesize $1/Q^{2}$} & {\footnotesize pQCD form} & {\footnotesize $V$} & {\footnotesize Indep.} & {\footnotesize yes} & {\footnotesize Pheno.} & 
{\footnotesize \cite{the:Ayala 2015}}\tabularnewline
\hline 
{\footnotesize 0.58} & {\footnotesize freezes} & {\footnotesize pQCD form} & {\footnotesize $V$} & {\footnotesize Indep.} & {\footnotesize yes} & {\footnotesize Pheno.} & 
{\footnotesize \cite{the:Badalian Q-Q static force (2005)},}\tabularnewline
\hline 
{\footnotesize $\infty$} & {\footnotesize $1/Q^{2}$} & {\footnotesize pQCD form} & {\footnotesize $\overline{MS}$} & {\footnotesize Indep.} & {\footnotesize yes} & {\footnotesize Pheno.} & 
{\footnotesize \cite{the:Belyakova 2011}}\tabularnewline
\hline 
{\footnotesize $0.42\pm0.03$} & {\footnotesize (Av. IR val.)} & {\footnotesize pQCD form} & - & {\footnotesize Indep.} & {\footnotesize yes} & {\footnotesize Pheno.} & 
{\footnotesize \cite{the:Eichten. Cornell pot.}}\tabularnewline
\hline 
{\footnotesize $\simeq0.6$} & {\footnotesize freezes} & {\footnotesize pQCD form} & - & {\footnotesize Indep.} & {\footnotesize yes} & {\footnotesize Pheno.} & 
{\footnotesize \cite{the:Eichten. Cornell pot.}}\tabularnewline
\hline 
{\footnotesize 0.78} & {\footnotesize mono. incr.} & {\footnotesize pQCD form} & {\footnotesize $\overline{MS}$} & {\footnotesize Indep.} & {\footnotesize yes} & {\footnotesize Pheno.} & {\footnotesize  \cite{the:Ganbold}}\tabularnewline
\hline 
{\footnotesize 0.60} & {\footnotesize freezes} & {\footnotesize pQCD form} & {\footnotesize ($\overline{MS}$)} & {\footnotesize Indep.} & {\footnotesize yes} & {\footnotesize Pheno. $\Lambda$=0.2} & 
{\footnotesize \cite{the:Godfrey-Isgur}}\tabularnewline
\hline 
{\footnotesize 0.82} & {\footnotesize freezes} & {\footnotesize pQCD form} & {\footnotesize  ($\overline{MS}$)} & {\footnotesize Indep.} & {\footnotesize yes} & {\footnotesize Pheno.} & 
{\footnotesize \cite{the:Bernard et al early mg lattice estimates}}\tabularnewline
\hline 
{\footnotesize diverges} & {\footnotesize $1/Q^{2}$} & {\footnotesize pQCD form} & {\footnotesize $V$} & {\footnotesize Indep.} & {\footnotesize yes} & {\footnotesize Pheno.} & {\footnotesize  \cite{the:Richardson 1979}}\tabularnewline
\hline 
{\footnotesize 0.83} & {\footnotesize freezes} & {\footnotesize pQCD form} & {\footnotesize -} & {\footnotesize Indep.} & {\footnotesize yes} & {\footnotesize Pheno.} & 
{\footnotesize  \cite{the:Zhang Koniuk quark model}}\tabularnewline
\hline 
{\footnotesize 0.83} & {\footnotesize freezes} & {\footnotesize $\alpha_{R}$} & {\footnotesize Indep. } & {\footnotesize Indep. } & {\footnotesize yes} & {\footnotesize Opt. Pert. Theo.} & 
{\footnotesize \cite{the:Mattingly-Stevenson}}\tabularnewline
\hline 
{\footnotesize 1.77} & {\footnotesize freezes} & {\footnotesize gh.-gl. vertex } & {\footnotesize $\overline{MS}$, MOM} & {\footnotesize Landau} & {\footnotesize yes} & {\footnotesize Gribov--Zwanziger} & {\footnotesize  \cite{the:Gracey 2006, the:Gracey stat pot 2009-2010}}\tabularnewline
\hline 
{\footnotesize $7.7\pm2$} & {\footnotesize freezes} & {\footnotesize -} & {\footnotesize -} & {\footnotesize Landau} & {\footnotesize no} & {\footnotesize FRG} & {\footnotesize  \cite{the:Gies 2002 FRG}}\tabularnewline
\hline 
{\footnotesize 2.97} & {\footnotesize freezes} & {\footnotesize gh.-gl. vertex} & {\footnotesize -} & {\footnotesize Landau} & {\footnotesize no} & {\footnotesize FRG} & 
{\footnotesize{} \cite{the:Pawlowski IR exp from FRG}}\tabularnewline
\hline 
{\footnotesize $\sim4$} & {\footnotesize freezes} & {\footnotesize -} & {\footnotesize Indep.} & {\footnotesize Indep.} & {\footnotesize yes} & {\footnotesize Analytic Ap.}{\footnotesize \par}
{\footnotesize $\Lambda$=0.56} & {\footnotesize \cite{the:Sanda 1979}}\tabularnewline
\hline 
{\footnotesize 1.25} & {\footnotesize mono. decr.} & {\footnotesize -} & {\footnotesize ($\overline{MS}$)} & {\footnotesize Indep.} & {\footnotesize yes} & {\footnotesize Analytic Ap.}{\footnotesize \par}
{\footnotesize $\Lambda$=0.32} & 
{\footnotesize \cite{the:MPT, the:non-pert in analytic QCD}}\tabularnewline
\hline 
{\footnotesize diverges} & {\footnotesize $1/Q^{2}$} & {\footnotesize -} & {\footnotesize Indep.} & {\footnotesize Indep.} & {\footnotesize yes} & {\footnotesize Analytic Approach} & 
{\footnotesize \cite{the:Nesterenko npQCD 2000, the:Nesterenko 2001}}\tabularnewline
\hline 
{\footnotesize 0.47} & {\footnotesize freezes} & {\footnotesize -} & {\footnotesize Indep.} & {\footnotesize Indep.} & {\footnotesize yes} & {\footnotesize Analytic Approach} & 
{\footnotesize \cite{the:Nesterenko (2005)}}\tabularnewline
\hline 
{\footnotesize 0.37-0.60} & {\footnotesize freezes} & {\footnotesize pQCD form} & {\footnotesize $\overline{MS}$} & {\footnotesize Indep. } & {\footnotesize no} & {\footnotesize BPT} & {\footnotesize  \cite{the:Simonov 1993, the:Simonov (2011)}}\tabularnewline
\hline 
{\footnotesize 0.81} & {\footnotesize freezes} & {\footnotesize pQCD form} & {\footnotesize V} & {\footnotesize Indep. } & {\footnotesize yes} & {\footnotesize BPT} & 
{\footnotesize \cite{the:Badalian (2000)} }\tabularnewline
\hline 
{\footnotesize $0.97\pm0.44$} & {\footnotesize freezes} & {\footnotesize pQCD form} & {\footnotesize V} & {\footnotesize Indep. } & {\footnotesize yes} & {\footnotesize BPT} & 
{\footnotesize \cite{the:Badalian (2000), the:Badalian-Kuzmenko 2001}}\tabularnewline
\hline 
{\footnotesize $0.50\pm0.08$} & {\footnotesize freezes} & {\footnotesize pQCD form} & {\footnotesize $\overline{MS}$} & {\footnotesize indep.} & {\footnotesize yes} & {\footnotesize duality.} & 
{\footnotesize  \cite{Courtoy:2013qca}}\tabularnewline
\hline 
{\footnotesize 0} & {\footnotesize vanishes} & {\footnotesize gh.-gl. vertex} & - & - & {\footnotesize no} & {\footnotesize $\phi^{4}$-YM mapping } & {\footnotesize  \cite{the:Frasca}}\tabularnewline
\hline 
{\footnotesize 0} & {\footnotesize vanishes} & {\footnotesize pQCD form} & {\footnotesize $\overline{MS}$} & {\footnotesize indep.} & {\footnotesize no} & {\footnotesize Bogoliubov comp.} & 
{\footnotesize \cite{the:Arbuzov (2013)}}\tabularnewline
\hline 
{\footnotesize 0} & {\footnotesize vanishes} & {\footnotesize pQCD form} & {\footnotesize $\overline{MS}$} & {\footnotesize Landau} & {\footnotesize yes} & {\footnotesize Curci-Ferrari Model} & 
{\footnotesize \cite{the:Tissier}}\tabularnewline
\hline
\end{tabular}
\caption{\small Continuation of Table \ref{tab:Summary table 1}. The  references for the seventh row are \cite{the:Bornyakov 2009, the:Ilgenfritz 2011,the:Maas (2015), the:Oliveira-Silva 2007, the:Pawlowski 2010, the:Skullerud 2002}.
\label{tab:Summary table 2}}
\end{table}
We do not  attempt to reconcile all the results,  but the preceding
discussion suggests that most can be made consistent, at least qualitatively.

\paragraph{Reconciling the $\alpha_{s}(0)\sim 0.7$ group with the $\alpha_{s}(0)\sim 3$
group}

In   Ref.~\cite{the:Aguilar Lattice 2010}, the difference
between the group $\alpha_{s}(0)\sim 0.7$ and the gauge-dependent
results pertaining to the $\alpha_{s}(0)\sim 3$ group is  attributed to different 
gauge choices: it is argued that the Feynman gauge produces results close
to the phenomenological  value $\alpha_{s}(0)\sim 0.7$, whereas the Landau
gauge yields $\alpha_{s}(0)\sim 3$. Furthermore, it was shown in  Ref.~\cite{the:alpha_g_1 from AdS},
and discussed in the beginning of this Section, that the other gauge-independent
results belonging to this group 
computed in the $g_{1}$
or $\overline{MS}$ RS  agree with the results computed in the $MOM$
RS once the RS-dependence is corrected for, see Fig. \ref{Flo:different freezings}.  The AdS/QCD continuation
of the pQCD calculation yields a higher $\alpha_{g_1}(0)$ than expected, 
but is still compatible within uncertainties with the $\pi$ constraint. 
Similarly, the \emph{freezing}  value for $\alpha_{\overline{MS}}$ obtained from the AdS/QCD--pQCD matching has a larger central value than the typical 0.7 \emph{freezing}  value. 

\paragraph{Reconciling the divergent $\alpha_{s}(Q^{2})~\protect\overrightarrow{_{Q^{2}\to0}}~\infty$
group with the $\alpha_{s}(0)\simeq 0.7$ and $\alpha_{s}(0)\simeq 3$ groups}

As we have emphasized, the QCD running coupling
does not need to be divergent in the IR in order for a theory to be confining. 
For example, the coupling for the analytically solvable \cite{Brodsky:1997de} confining theory 
QCD(1+1) is finite.

Confinement in QCD can be due to multiple gluon exchange diagrams, where
the exchanged gluons are connected by the gluon and four-gluon couplings as in
the ``H'' diagrams;  see Fig. \ref{QCD_vertex}(d). The network of gluons corresponds 
to a ``string'' or ``flux  tube'' connecting the Q and $\overline{\mbox{Q}}$.
The fact that the~H diagram becomes IR divergent as the Q  and $\overline{\mbox{Q}}$ separate
in impact space is indicative that QCD must confine color charges in order to render the theory self-consistent. 
Thus the infinite sum of such diagrams could be the source of the confining potential. For example, 
the confining harmonic oscillator  potential  between light quarks  and antiquarks in the LF Hamiltonian
derived using light-front holography and AdS/QCD may well come from the IR behavior of the sum of H graphs, 
and has been shown to be equivalent to the diverging linear potential in the instant form \cite{the:AdS/QCD review}.
Thus the apparent divergence of the running coupling  derived using the potential ($V$-scheme) 
may stem from the impossibility to consistently define 
an \emph{effective charge} from static heavy-quark interactions because
the perturbative QCD contributions of ``H diagrams''  are IR divergent, 
as first noted by Appelquist, Dine and Muzinich \cite{the:Appelquist}.
This suggests that the $\alpha_{s}(Q^{2})~\protect\overrightarrow{_{Q^{2}\to0}}~\infty$
group of schemes could be united with the $\alpha_{s}(0)\sim 0.7$ and $\alpha_{s}(0)\sim 3$
groups if compared within a consistent framework.

\paragraph{The $\alpha_{s}(0)=0$ group}

It is unclear how to reconcile the three previous groups of solutions for the running coupling with a decoupling  model in which the coupling vanishes in the IR:
$\alpha_{s}(0)=0$.  Only one solution  underlies 
hadron physics, and thus either $\alpha_{s}(0)=0$ or the three other reconcilable
groups are irrelevant.   We note that the IR-\emph{freezing}  solution
is supported by measurements of  \emph{effective charges} such as $\alpha_{g_{1}}$, 
$\alpha_{F3}$ and $\alpha_{\tau}$, and it is a necessary ingredient of the AdS/QCD
approach which provides a successful description of hadron properties. 
Furthermore,
the \emph{decoupling solution} is not compatible with the \emph{Kugo--Ojima
confinement criterion}.  Other arguments disfavoring the decoupling
solution (violation of the BRST symmetry and unconfined gluons) are
given in \cite{the:Fischer 2009 against decoupling}.
On the other hand, an argument for the \emph{decoupling solution} is that
it appears to have a lower action than the \emph{freezing}  solution \cite{the:Llanes-Estrada 2012}.
It is possible that both solutions are realized in Nature, that the \emph{decoupling solution} does not
lead to confinement or lead to a hadron of larger size than that with a coupling following the \emph{scaling solution}.
In that case, the consequence of the \emph{decoupling solution} would not be observed in spite of its
possibly lower action.

\subsubsection{What is the best definition for $\alpha_{s}$?}

An important principle of the renormalization group is that  predictions for observables cannot
depend on theoretical conventions such as the RS, the
initial scale, or the choice of effective charge. Predictions for observables
must thus also be independent of the choice of the definition for $\alpha_s$.  This  implies that 
different choices for the running coupling must be related to each other; for example 
``\emph{Commensurate  scale relations}'' interconnect different couplings in the high $Q^2$ pQCD regime.

The optimal choice for the definition of $\alpha_{s}(Q^{2})$ at all scales
is an unsettled question: there are many ways to define a coupling which
satisfies the RGE.
This is also the case in QED;  nevertheless, the Gell Mann--Low definition --the effective charge
defined from elastic scattering of heavy charges-- is universally used.
However, it would in principle be possible to  choose an alternative formal choice of coupling in QED based on dimensional regularization such as  $\alpha_{\overline{MS}}$.

To help in assessing the optimal choice of IR-definition for $\alpha_{s}(Q^{2})$, we summarize
in Table \ref{tab:summary of definitions of alpha_s} various definitions
for $\alpha_{s}$, together with their properties and relative advantages.
In addition to this table, we can list other desirable, but not necessarily fundamental, properties:
\begin{itemize}
\item Interpretability: the coupling should be easily interpretable, as is the case for QED.
\item Simplicity: it  should be relatively easy to produce the coupling.
\item Finiteness: the coupling should be finite at all scales. 
Since predictions for observables must be finite, this condition implies
that if $\alpha_{s}(Q^{2})$ diverges, it must be compensated by other  divergent  factors or terms not
included in its primary definition.   Although the cancellation of infinities is theoretically possible, 
the identification of the individual diverging effects cannot be done solely from physical criteria. 
In addition, from the practical point of view, the cancellation of large quantities greatly increases 
systematic  uncertainties.
\end{itemize}

This discussion  does not imply that only one definition or approach should
be used to study $\alpha_{s}$ in the IR  domain.  A
variety of approaches is useful since different techniques
often give different perspectives.  However,  it would be desirable
to present  results in a commonly agreed definition of the QCD coupling (and scheme and
$N_{C}$ and $n_{f}$ values) in order to assure the clarity and consistency
of this field of study. Table \ref{tab:summary of definitions of alpha_s} may help 
to choose a convenient definition for such reference. 

{\footnotesize
\begin{table}
\hskip-1.cm\begin{tabular}{|c|c|c|c|c|c|c|}
\hline 
{\footnotesize Definition}	 & {\footnotesize Analytic} & {\footnotesize RS-} & {\footnotesize Gauge-} & {\footnotesize Universal} & {\footnotesize Based on} & {\footnotesize IR-}\tabularnewline
 	 &   & {\footnotesize indep.} & {\footnotesize indep.} & {\footnotesize} & {\footnotesize first principles} & {\footnotesize finite}\tabularnewline
\hline
\hline 
{\footnotesize pQCD, $\alpha_{pQCD}$} & yes & no & depends$^{1}$ & yes & yes & {\footnotesize no (\emph{Landau pole})}\tabularnewline
\hline 
{\footnotesize Eff. charge, $\alpha_{g_{1}}$, $\alpha_{\tau}$, $\alpha_{AdS}$} & yes & yes & yes & yes$^{2}$ & yes & {\footnotesize yes (freezes)}\tabularnewline
\hline 
{\footnotesize Static quark potential} & yes & yes & yes & yes & no &{\footnotesize no}\tabularnewline
\hline 
{\footnotesize Vertices, $\alpha_{s}^{gh}$}, & yes & no & no & in & yes & {\footnotesize yes (freezes or}\tabularnewline
 {\footnotesize $\alpha_{s}^{Q-g}$, $\alpha_{s}^{3g}$, $\alpha_{s}^{4g}$} &  &  &  & principle$^{3}$ &  & {\footnotesize vanishes)}\tabularnewline
\hline 
{\footnotesize Quark model/Spectroscopy} & yes & no & yes & yes & no & {\footnotesize yes (freezes)}\tabularnewline
\hline 
{\footnotesize Analytical coupling} & yes & no & yes & yes & no$^{4}$ & {\footnotesize yes (freezes)}\tabularnewline
\hline 
{\footnotesize Dispersive coupling} & yes & no & yes & yes & yes & {\footnotesize yes (freezes)}\tabularnewline
\hline 
{\footnotesize OPT} & yes & yes & yes & unclear & no & {\footnotesize yes (freezes)}\tabularnewline
\hline 
{\footnotesize Duality} & no & no & yes & unclear & no & {\footnotesize yes (freezes)}\tabularnewline
\hline 
{\footnotesize $\lambda\phi^{4}\rightarrow$ Yang Mills mapping} & yes & no & unclear & yes & no & {\footnotesize yes (vanishes)}\tabularnewline
\hline 
{\footnotesize Curci-Ferrari Model} & yes & no & no & no & yes & {\footnotesize yes (vanishes)}\tabularnewline
\hline

\end{tabular}

\caption{\small Summary of various definitions of $\alpha_s$.
``Universal'' refers to a unique coupling able to describe any
observable or vertex. \protect\\
Notes:\protect\\
$^1$ Depends on the RS. 
 {\it E.g.} the coupling in $\overline{MS}$ is gauge-independent but not necessarily the one in a MOM gauge.\protect\\
$^2$ \emph{Effective charges} can be related in the perturbative domain using the \emph{CSR}, see 
Section \ref{sub:Effective charges and CSR}. Those relations are continued 
in the IR using the method described in Section \ref{sub:Influence-of-the RS}.\protect\\
$^3$ In principle, $\alpha_{s}^{gh}$, $\alpha_{s}^{Q-g}$, $\alpha_{s}^{3g}$,
and $\alpha_{s}^{4g}$ can be related. Also, two solutions (scaling
and decoupling, see Section \ref{sub:Classes-of-solutions in IF domain}) exist, without consensus on which
one is realized in Nature.\protect\\
$^4$ Demanding causality as a criterion to define the coupling is not a first
principle since the coupling is not necessarily an observable. 
\label{tab:summary of definitions of alpha_s}}
\end{table}
}

\chapter{Conclusions}

The QCD coupling $\alpha_s(Q^2) $ plays a fundamental role in hadron, nuclear, and particle physics, setting the strength of  quark and gluon interactions over the entire range of momentum transfer $Q. $ The analytic dependence of the coupling in $Q^2$ is determined by its logarithmic derivative, the QCD \emph{$\beta$-function}, which
not only incorporates the physics of asymptotic freedom at large momentum transfer, but also the nonperturbative dynamics underlying color confinement. 
It is possible that the QCD and electroweak running couplings could merge at very high momentum transfers, reflecting the unifying physics of a grand unified theory~\cite{the:Binger-Brodsky}.
 
In this article, we have reviewed the  present theoretical and phenomenological understanding of $\alpha_{s}$, both at   high- and  low-momentum transfer. We attempted to be both pedagogical and comprehensive, although we had to leave out interesting topics such as the coupling in the $Q^2<0$ time-like domain, in space-times of dimension other than 4, or studies of the coupling at non-zero temperature.  

Remarkable progress has been made in the last decade  determining the QCD coupling's strength at high momentum transfer using phenomenological input from collider experiments, together with extensive high-order theoretical perturbative computations.  
Perturbative QCD calculations typically use dimensional regularization to control ultraviolet divergences of loop integrals. 
It has thus become conventional to adopt the  $\overline {MS } $ definition of the coupling.   The value of $\alpha_{\overline {MS}}(Q^2 = M^2_Z)$   has been determined from various experiments   to remarkable precision
and using different processes, providing  an important test of the validity of QCD.  High precision theoretical determinations of $\alpha_{s}$
from lattice gauge theory have also been obtained.
Although some tension still exists between various determinations, it probably reflects  too
optimistic estimates of systematic uncertainties, rather than new physics or a problem associated with QCD itself.

As we have emphasized, the choice of the $\overline {MS }$ scheme is only a convention.   Other renormalization schemes are possible, such as the MOM scheme, although its use is problematic because of its dependence on the choice  of gauge.    

It has also become conventional to guess the \emph{renormalization scale} $\mu_R$ 
for the $\overline {MS }$ coupling scheme. The most common choice is to set $Q^{2}=\mu_R^{2}$
because it yields the simplest perturbative expansions of observables. The setbacks 
are convergence issues with the expansion series and an unintuitive interpretation 
of the coupling. Although the choice of the simplicity criterium is arbitrary, one expects 
that the dependence of the pQCD predictions on this guess become diminished at high order.  
However, this expectation conflicts with the $\alpha^n_s \beta_0^n n! $ 
divergent \emph{renormalon} growth of the pQCD series.   The fixed order pQCD 
predictions are scheme-dependent if one uses an arbitrary scale for $\mu_R$.   
This  is at odds with an important principle -- ``renormalization group invariance'':  theoretical predictions cannot depend on theoretical conventions, such as the choice of scheme or  the choice of the  \emph{renormalization scale}.     

As we have discussed, none of these problems appear when one uses the ``Principle of 
Maximum Conformality''  (\emph{PMC}) to set the \emph{renormalization scale} 
order-by-order in perturbation theory. As in QED,  all terms in the pQCD series  involving the 
$\beta$-function can be summed into the running coupling by shifting the 
\emph{renormalization scale} at each order;  this sets the argument of the coupling at each 
order, and the resulting coefficients of the series then match the coefficients of the 
corresponding \emph{conformal} theory with $\beta=0$. The $\beta_i$ terms can be 
identified unambiguously at each order by introducing an extra parameter $\delta$ 
in the subtraction that defines the $\overline {MS }$ scheme.   

The use of the \emph{PMC} thus eliminates renormalization-scale ambiguities, eliminates 
\emph{renormalon} divergences, and --most important-- gives predictions which are 
independent of the choice of renormalization scheme.  Also, as in QED, the \emph{PMC} 
\emph{renormalization scales}  reflect the virtuality of the gluon propagators and correctly set 
the number $n_f$ of contributing quark flavors  at each order.   
For example, the reduced virtuality of the two $s$-channel
gluons in the $q \bar q \to g g \to t \bar t$ amplitude can account for  the
large  $t \bar t$  asymmetry in the $\bar p p \to t \bar t X$ observed  at
colliders \cite{Brodsky:2012ik, Wang:2015lna}.
Nonzero quark masses can be retained so that the $\beta$-function remains analytic as it passes through each flavor threshold~\cite{the:Binger-Brodsky}.

Applying the \emph{PMC} also provides ``\emph{Commensurate Scale Relations},''  between QCD observables which have no \emph{renormalization scale} or scheme ambiguities. A classic example is the ``Generalized Crewther Relation''  \cite{the: Crewther}
 which connects the QCD corrections to the Bjorken sum rule, which is measured in polarized deep inelastic lepton--nucleon scattering, to the QCD corrections to the annihilation cross section ratio $R_{e^+ e^-} (s)$.   Again there are no \emph{renormalization scale} ambiguities.

In the second part of the review, we have 
discussed a much more complex problem, the behavior of the QCD coupling $\alpha_{s}(Q^2)$ at low momentum transfers, 
where even its definition is not universally agreed upon.      We have  described  a number of nonperturbative approaches
for defining and computing  the coupling at small $Q^2$, such as approximately solving the Schwinger--Dyson equations and the use of different assumptions for its analytic form.
The various analyses   reported in the literature yield predictions 
for $\alpha_{s}(Q^2=0)$ which range  from zero to infinity.

The different theoretical and analytical approaches for analyzing the QCD coupling can be classified within 
three groups:  models where $\alpha_{s}(Q^2=0)$  vanishes, models where 
$\alpha_{s}(Q^2) $ diverges as $Q^{2}\rightarrow0$, and models  where 
$\alpha_{s}(Q^2) $ remains  roughly constant and of moderate value  in the IR.
These models reflect the range of possible definitions of 
$\alpha_{s}$ itself, the choice of renormalization scheme, and whether or not one 
introduces the concept of an effective nonzero gluon mass  or equivalent momentum scales.

An important tool for resolving the ambiguity in defining $\alpha_{s}(Q^2)$ is the use of ``\emph{effective charges}'':   a coupling defined directly from a perturbatively calculable observable. By definition an \emph{effective charge} is finite at all scales.
A natural choice would be to define  the QCD coupling from the heavy-quark potential as a generalization of the traditional Gell-Mann--Low coupling used in QED.    However, this coupling, called  $\alpha _V(Q^2)$, requires an infrared cutoff because of ``H'' graph contributions arising from non-Abelian corrections to multi-gluon exchange diagrams.

A satisfactory choice for defining  the QCD coupling is the \emph{effective charge} $\alpha_{g_1} (Q^2)$  which is obtained from the sum of QCD radiative corrections to   the Bjorken sum rule. It is infrared finite and well measured, both at low and high $Q^2$. The behavior of  $\alpha_{g_1} (Q^2)$ at high $Q^2$ is known to four \emph{loops} in pQCD. 

Remarkably, all of the measurements of  $\alpha_{g_1} (Q^2)$ for $Q^2 < 1~GeV^2$ are consistent with the nonperturbative prediction from 
AdS/QCD and light-front holography:
$${\alpha_{g_1} (Q^2)\over \pi}  =   \exp {\left(-Q^2\over 4 \kappa^2\right)},$$
where $\kappa$ can be determined by the proton mass, $\kappa = M_p/2 $, or other
hadron masses. 
The same theory successfully predicts the Regge spectroscopy of the light mesons and baryons, as well as  their dynamical structure, such as the light-front wavefunctions underlying the hadron from factors, structure functions, and distribution amplitudes. The value of $\kappa$  also determines the confining potential for light quarks in the frame-independent light-front Hamiltonian.

It is also remarkable that by matching  at a scale $Q_0$ the value of $\alpha_{g_1} (Q^2)$ and its derivative obtained from light-front holographic QCD to the predictions of pQCD, one can determine 
$\Lambda_{\overline {MS}} $ from the value of $\kappa$. The result of  this matching is in agreement with the value of $\Lambda_{\overline {MS}} $ determined from high energy physics phenomenology.  The value of the matching scale: $Q^2_0  \simeq 1.25~GeV^2$ in the $g_1$ scheme can be interpreted as the transition scale between the perturbative and nonperturbative domains of QCD.  It is thus  natural to adopt $Q^2_0$  as the starting scale for the DGLAP evolution of structure functions and the ERBL evolution of hadronic distribution amplitudes. The light front holographic QCD analysis can also be used to complement and constrain other nonperturbative approaches to QCD.

Thus QCD and the study of its running coupling have  now entered a new domain, where predictions for hadronic phenomena can be made over all scales.  We hope that this review has illuminated the physics underlying the remarkable progress in this field.

\paragraph{Acknowledgments}

We thank Hans Guenter Dosch, David d'Enterria, John A. Gracey, Andrei L. Kataev, Cedric Lorc\'{e}, Matin Mojaza, Christian Weiss, 
Xing-Gang Wu and Yang Ma for instructive discussions on $\alpha_{s}$
and related topics. We are grateful to A. Faessler for his invitation
to write this review. This material is based upon work supported by
the U.S. Department of Energy, Office of Science, Office of Nuclear
Physics under contract DE--AC05--06OR23177 and DE--AC02--76SF00515.  
SLAC-PUB-16448.

\appendix

\chapter{Lexicon}

To make this review accessible to non-specialists,
we list here some of the specific vocabulary  associated with  studies
of the QCD coupling, with brief explanations and references to where
it is first discussed in the review. 
\begin{itemize}
\item Asymptotic series; Poincar\'{e} series. See also ``renormalons''.
A series that converges up to an order 
$k$ and then diverges. The series reaches its best convergence at
order $N_{b}$ and then diverges for orders $N\gtrsim N_{b}+\sqrt{N_{b}}$. 
Quantum Field Theory series typically are asymptotic and converge 
up to an order $N_b \simeq 1/a$, with $a$ the expansion coefficient. 
IR \emph{renormalons}  generate an $n!\beta^{n}$ factorial growth
of the $n$th coefficients in \emph{nonconformal} ($\beta \neq 0$) theories. Perturbative calculation
to high order ($\alpha_{s}^{20}$) has been performed on the Lattice 
\cite{the:Renormalon growth} to check 
the asymptotic behavior of QCD series. Factorial growth is seen up
to the 20th order of the calculated series. 
\item $\beta$-function. The logarithmic derivative of $\alpha_{s}$: $\beta\left(\mu^{2}\right)=\frac{d\alpha_{s}\left(\mu\right)}{d\mbox{\footnotesize{ln}}(\mu)}$
where $\mu$ is the \emph{subtraction point}. In the perturbative
domain, $\beta$ can be expressed as a perturbative series \textbf{$\beta=-\frac{1}{4\pi}\sum_{n=0}\left(\frac{\alpha_{s}}{4\pi}\right)^{n}\beta_{n}$}. (See Section \ref{sub:pQCD evolution-equation}.)
\item Commensurate Scale Relations (CSR). Relations linking two effective
charges obtained from different observables or two \emph{effective couplings}
expressed in different renormalization schemes. 
Since BLM/\emph{PMC} scale setting is used,  all terms involving the   coefficients $\beta_{n}$ do not appear
in the \emph{CSR} series coefficients, eliminating the \emph{renormalon}  divergence 
(see ``Asymptotic series'' and  ``renormalons'').  The convergence property is thus
superior compared to the more straightforward relations obtained by
equating the perturbative expressions of the two couplings.
(See Section \ref{sub:Effective charges and CSR}.)
\item Condensate (or Vacuum Expectation Value, VEV). The vacuum expectation value of a given local operator.
  Condensates allow one to parameterize the  nonperturbative \emph{OPE}'s power
corrections.    Condensates and vacuum loop diagrams do not appear in the frame-independent 
light-front Hamiltonian since all lines have $k^+ = k^0 + k^3 \ge 0$ 
and the sum of $+$ momenta is conserved at every vertex.
In the light-front formalism condensates are associated with physics of the hadron 
wave function and are called ``in-hadron''  condensates, which  refers to physics 
possibly contained in the higher LF  Fock states of the hadrons \cite{Casher:1974xd}. In the case of the Higgs theory,
the usual Higgs VEV of the instant form Hamiltonian is replaced by a ``zero mode'', a background field with $k^+=0$ \cite{Brodsky:2012ku}.
\item Conformal behavior/theory. The behavior of a quantity or a theory
that is scale invariant. In a conformal theory the \emph{$\beta$-function} vanishes.  More rigorously, a conformal  theory  is invariant under  both  dilatation and the special conformal transformations which involve inversion. (See Sections \ref{sub:Holographic-QCD large Q} and \ref{sub:Holographic-QCD Lowq}.)
\item Couplant: the normalized coupling, defined as $\alpha_{s}/\pi$.
\item Decoupling solution. See also \emph{scaling solution}. One of the two classes of solutions for couplings defined using vertices. The decoupling
solution leads to an IR-vanishing $\alpha_{s}$. (See Section \ref{sub:Classes-of-solutions in IF domain}.)
\item Dimensional transmutation: The emergence of a mass or momentum scale in the quantum theory with 
a classical Lagrangian devoid of  explicit mass or energy parameters \cite{the: dim. trans.}. (See Section \ref{Purpose of the running coupling}.)
\item Effective coupling. The renormalized (running) coupling, in contrast
with the constant bare coupling. (See Section \ref{sec:Phenomenological-introduction:}.)
\item Effective charge. An effective coupling defined from a perturbatively calculable observable. It  includes
all perturbative and relevant nonperturbative effects.
(See Sections \ref{sub:Effective charges and CSR} and ~\ref{sub:low Q Effective-charges}.)
\item Freezing. The loss of scale dependence of finite $\alpha_{s}$ in the infrared.  See
also conformal behavior. 
\item Green's function, $n$-point Green's function. The function which describes the propagation of a field between $n$ space-time points.
\item Higher-twist  (see also ``Twist''), $1/Q^{n}$ \emph{power corrections}, typically derived from the \emph{OPE} 
analysis of  the nonperturbative effects of multiparton interactions.
Higher-twist is sometimes interpretable as kinematical phenomena, {\it e.g.} the mass $M$ of a 
nucleon introduces a \emph{power correction} beyond the pQCD scaling violations, or as dynamical 
phenomena, {\it e.g.} the intermediate distance transverse forces that confine 
quarks \cite{the:Burkardt twists}. (See Section \ref{sub:low Q Effective-charges}.)
%
\item Infrared fixed point. The value of the momentum transfer scale $\mu$ where $\beta(\mu)=0$.
This implies either that $\alpha_{s}$ either freezes in the IR ($\beta(\mu)$
remains 0 beyond the IR fixed point), or that it vanishes ($\beta(\mu)$
becomes positive passed the IR fixed point).
\item Infrared exponent. The power-law exponent which describes how the ghost and gluon propagators
scale with $Q^{2}$. (See Section \ref{sub:Classes-of-solutions in IF domain}.)
\item Kugo--Ojima confinement criterion \cite{the:Kugo--Ojima}. A condition
that ensures the conservation of the global color charge. It is a
necessary but not a sufficient criterion for confinement. It implies
that the Fadeev--Popov ghost propagator behavior in the low energy
regime is more singular than $1/Q^{2}$,  and that the gluon propagator  is less
singular;  see {\it e.g.},   Refs.~\cite{the: Greensite conf., the:Alkofer von smekal SDE review}. 
In such a case, the physical space of states can contain only color singlets, {\it i.e.}, the theory is confining.
\item Landau pole; Landau singularity; Landau ghost. The point where a perturbative coupling
diverges. At first order (1-loop) in pQCD, this occurs at the \emph{scale parameter}
$\Lambda$. The value can depend on the choice of renormalization scheme, the order $\beta_{i}$
at which the coupling series is estimated, the number of flavors $n_{f}$
and the approximation chosen to solve Eq. (\ref{eq:alpha_s beta series})
for orders higher than $\beta_{1}$. The Landau pole is unphysical. (See Section \ref{sub:Landau-Pole}.)
\item ($n$-)Loops.  The order of perturbation theory appearing in the 
coupling calculation; {\it i.e.},  the order of   the $\beta$-series. 
In the case of QED, it is the order of vacuum polarization loops contributing to the renormalization of the coupling.
The $n$-\emph{loop} approximation corresponds
to a $\beta$-series truncated to  order $\beta_{n-1}$.  (See Section \ref{sub:beta-coefficients calc.}.)
\item Operator Product Expansion (OPE). See also higher-twist.  The \emph{OPE} uses the \emph{twist} of effective operators to predict the power-law fall-off of an amplitude.
It thus can be used to distinguish 
logarithmic leading \emph{twist} perturbative corrections from the $1/Q^{n}$ \emph{power corrections}. The \emph{OPE}
typically does not provide values for the nonperturbative \emph{power correction} coefficients.
 (See Section \ref{sub:low Q Effective-charges}.)
\item Pinch technique. A method which adds external on-shell legs to ensure the gauge invariance of an amplitude.
For example, the calculation of the  three-gluon vertex becomes gauge  invariant if the coupling to three on-shell external quark lines are included. See Refs. \cite{the:Binger-Brodsky, the:Binosi Pinch technic review}
\item Principle of Maximal Conformality (PMC).  A method used to set the \emph{renormalization scale}, order-by-order in perturbation theory, by shifting all $\beta$ terms  in the pQCD series  into the \emph{renormalization scale} of the running QCD coupling at each order.  The resulting coefficients of the series then match the coefficients of the corresponding \emph{conformal} theory with $\beta=0$.   The PMC generalizes the Brodsky Lepage Mackenzie BLM method to all orders.  In the Abelian $N_C\to 0$ limit, the PMC reduces to the standard Gell-Mann--Low method used for scale setting in QED \cite{Brodsky:1997jk}.
\item Power corrections. See ``Higher-twist'' and ``Renormalons''.
\item Pure gauge sector; pure Yang Mill; pure field. Non Abelian field theory
without fermions. See also \emph{quenched} approximation.
\item Quenched approximation. Calculations where the fermion loops are neglected.
It differs from the \emph{pure gauge}, pure Yang Mills case in that heavy (static)
quarks are present. 
\item Renormalization scale.  The argument of the running coupling. See also ``Subtraction point''.
\item Renormalon. The residual between the physical value of an observable
and the \emph{Asymptotic series} of the observable at its best convergence
order $n\simeq1/\alpha_{s}$.  The terms of a pQCD calculation which involve the \emph{$\beta$-function} typically diverge as $n!$;  {\it i.e.}, as a renormalon.
Borel summation techniques indicate that IR  renormalons  can often be interpreted as \emph{power
corrections}. Thus, IR  renormalons  should be related to the  \emph{higher
twist} corrections of the \emph{OPE} formalism \cite{the:Renormalons}.
The existence of IR  renormalons  in \emph{pure gauge} QCD is supported by Lattice 
QCD \cite{the:Renormalon growth}.  See also ``Asymptotic series''.
\item Scale-fixing. The \emph{renormalization scale} at each order of perturbation theory is set by the BLM/\emph{PMC} method by absorbing the contributing  $\beta$ terms. In the case of 
QED, scale-fixing resums the vacuum polarization contributions to the photon propagator.
\item   QCD Scale parameter $\Lambda$. The UV scale ruling the energy-dependence
of $\alpha_{s}$. It also provides the
scale at which $\alpha_{s}$ is expected to be large and  nonperturbative
treatment of QCD is required. (See Section \ref{sub:pQCD evolution-equation}.)
\item Scaling solution. See also  \emph{decoupling solution}. One of the two classes of solutions for couplings defined using vertices. 
The scaling solution results in an IR-\emph{freezing}  of $\alpha_{s}$. (See Section \ref{sub:Classes-of-solutions in IF domain}.)
\item Slavnov--Taylor identities \cite{the:Slavnov--Taylor id.}: The non-Abelian
generalization of the Ward--Takahashi identities. (See Section \ref{sub:beta-coefficients calc.}.)
\item  Subtraction point $\mu$. The  scale at which the renormalization
procedure subtracts the UV divergences.   
(See Section \ref{sub:pQCD evolution-equation}.)
\item Tadpole corrections. In the context of lattice QCD, tadpole terms
are unphysical contributions to the lattice   action which arise from
the discretization of space-time. They contribute at NLO of the bare
coupling $g^{bare}=\sqrt{4\pi\alpha_{s}^{bare}}$ to the expression
of the gauge link variable $U_{\overrightarrow{\mu}}$. (The LO corresponds
to the continuum limit.) To suppress these contributions, one can redefine
the lattice action by adding larger  \emph{Wilson loops} or by rescaling the
link variable. 
\item Twist.  The twist of an elementary operator is given by its dimension minus its spin. 
For example, the  quark operator $\psi$  has dimension $3$, spin $1/2$ and thus twist $=1$.   It is also the number of constituents of a hadron.
See ``Higher-twists'' .
\item Unquenched QCD. See  \emph{pure gauge} sector and \emph{quenched} approximation.
\item Wilson Loops. Closed paths linking various sites in a lattice \cite{Wilson lattice}. They are used to define the lattice   action. 
(See Section \ref{sub: large Q Lattice-QCD}.)
\end{itemize}

\end{document}